\documentclass[oneside,12pt,tighten]{book} 
\usepackage{latexsym,cite,subeqn,oldgerm,amsfonts}
\input amssym.def 
\input amssym.tex
\input epsf
\makeatletter  

\@addtoreset{equation}{chapter}
\makeatother

\font\msym=msbm10
\def\Com{{\mathop{\hbox{\msym \char  '103}}}}

\def\P{\mathbb{P}} 
 
\def\T{{\mathop{\hbox{\msym \char  '124}}}}
\def\S{\mathbb{S}}
\def\T{\mathbb{T}}
\def\D{\mathbb{D}}
\def\M{\mathbb{M}}
\def\G{\mathbb{G}}

\makeatletter
\renewcommand{\@makecaption}[2]{\small
  \vspace{\abovecaptionskip}
  \sbox{\@tempboxa}{#1: #2}
  \ifdim \wd\@tempboxa > \linewidth
    #1: #2\par
  \else
    \centering\usebox{\@tempboxa}\par
  \fi
  \vspace{\belowcaptionskip}}
\makeatother

\newcommand{\tselea}[1]{\label{#1}}
\newcommand{\tseleq}[1]{\label{#1}}

\newcommand{\tseref}[1]{\ref{#1}} 

\newcommand{\vect}[1]{\bf #1}

\newcommand{\cali}[1]{\cal #1}

\font\sans=cmss12
\font\upright=cmu10 scaled\magstep1
\newcommand{\ssf}{\sans}

\newcommand{\R}{\hbox{\upright\rlap{I}\kern 1.7pt R}}
\newcommand{\Z}{\hbox{\upright\rlap{\ssf Z}\kern 2.7pt {\ssf Z}}}

\newcommand{\1}{\hbox{\upright\rlap{I}\kern -2.5pt 1}}

\begin{document}
\pagestyle{empty}
\vspace{3cm}
\begin{center}
\large{\textbf{Equilibrium and Non-equilibrium Quantum Field Theory}}\\
~\\
\large{\textbf{Petr Jizba}}\\
~{}\\
\large{\textbf{Summary}}
\end{center}

This dissertation is concerned with various aspects of equilibrium and
non--equilibrium quantum field theory.
\vspace{2mm}

We first  focus   in  Chapter \ref{dcoset}  on infrared    effects  in
finite--temperature quantum   field theory.    We  propose  a   simple
mathematical method  (based  on  the  largest--time equation  and  the
Dyson--Schwinger equations) which allows systematic calculations of the
change  of  density  in  energy/particles    in heat--bath   during  a
scattering/decay of external   particles   within the  heat bath.    A
careful   analysis reveals that  the  resulting changes  in the energy
density  are finite even in the  case of  massless heat--bath particles
(no infrared catastrophe). 
\vspace{2mm}

As a  next point we re--consider in  Chapter \ref{PE} the usual method
of pressure  calculations. We use  the  so called hydrostatic pressure
(or pressure at a  point)  which is  defined via the  energy--momentum
tensor.  We go through all delicate points which one must deal with in
the context of quantum   field theory.  Namely the  renormalisation of
composite  operators and  the  vital  role of  renormalisation  for  a
consistent quantum    field theoretical  definition  of pressure   are
discussed.  We  finally apply  the  whole procedure  to a   toy  model
system; $\lambda  \Phi^{4}$ theory  with $O(N)$ internal  symmetry. In
the case  of  the large--$N$ limit  (also Hartree--Fock approximation)
the pressure  is  an exactly   solvable  quantity.  Using the   Mellin
transform technique  we then  perform the large--temperature expansion
of the pressure to all orders. 
\vspace{2mm}

The hydrostatic pressure can be naturally extended to non--equilibrium
systems. Using the Jaynes--Gibbs  principle of maximal entropy and the
(non--equilibrium)   Dyson--Schwinger  equations we derive generalised
Kubo--Martin--Schwinger equations and  set  up a  calculational scheme
for pressure calculations away from thermal equilibrium. As an example
we explicitly evaluate in    Chapter  \ref{PN} the pressure for    our
$O(N)\;\lambda \Phi^{4}$ theory in the large--$N$ limit in the case of
two translationally invariant non--equilibrium systems. 
\vspace{2mm}

There  follow  five  appendices  which collect  together  much of  the
background material   required in the main body    of the thesis.  The
important part is the detailed  analysis in Appendix \ref{A1} of the
Dyson--Schwinger equations.  The   derivation  there  shows  how   the
Dyson--Schwinger equations may be recast into a very useful functional
form.   In  Appendices \ref{A2}  and  \ref{A3}  we  clarify some finer
mathematical manipulations   needed  in    Chapter    \ref{PE}.    The
fundamentals of  the information or Shannon  entropy  are presented in
Appendix  \ref{A4}.    Appendix  \ref{MF}  covers   the   elements  of
dimensional regularisation and  special functions  which underlie much
of the material presented in the earlier chapters.

\newpage
\setlength{\evensidemargin}{0.1cm}
\setlength{\oddsidemargin}{0.1cm}
\baselineskip24pt
\begin{titlepage}
\title{\textbf{Equilibrium and Non-equilibrium Quantum Field Theory} \\~~~\\
\textbf{}\\
~~\\
~~ }
\vspace{3.0cm}
\author{\textbf{Petr Jizba}\\
~~~\\
Fitzwilliam College\\
~~\\
~~\\
~~\\
~~\\
~~\\
~~\\
~~\\
A dissertation  submitted for the degree of Doctor of Philosophy\\
\\
University of Cambridge\\
~~\\
~~\\
March, 1999}
\date{}
\maketitle
\end{titlepage}

\newpage
\vspace{3cm}
\begin{center}
\Large{\textbf{Declaration of Originality}}
\end{center}
This dissertation contains the results of research  carried out in the
Department of Applied  Mathematics and Theoretical Physics, University
of Cambridge, between October, 1995 and March, 1999.\newline 
\newline
Excluding introductory section the research described in this
dissertation  is original unless where  explicit reference  is made to
the  work    of others.   Some  of   this  work was   carried   out in
collaboration.  I further  state that no  part of this dissertation or
anything   substantially   the  same  has    been   submitted for  any
qualification other than  the  degree of  Doctor of  Philosophy at the
University of Cambridge.  Some   of the research in  this dissertation
has been, or is to be, published.  The results in Chapter \ref{dcoset}
appeared in [J1].  The material presented in Chapter \ref{PE} has been
accepted to Phys.~Rev.~{\bf{D}} [J2].  Chapter \ref{PN} results from a
collaboration with Dr.~E.S.~Tututi [JT1], [JT2], [JT3].\newline 
\newline 

\begin{tabular}{ll}
$\mbox{[J1]}$ & P.~Jizba, Phys.~Rev.~{\textbf{D57}}: 3634, 1998.\\
~&\\
$\mbox{[J2]}$ & P.~Jizba, {\texttt{hep-th/9801197}},
Phys.~Rev.~{\bf{D}} (in press).\\
~&\\
$\mbox{[JT1]}$&  P.~Jizba and E.S.~Tututi, {\texttt{hep-th/9809110}}, in
{\textit{Proceedings of the 5th Int.}}\\
~&\\
$~$ & \hspace{0.2cm}{\textit{Workshop on Thermal Field  and Their Applications, Regensburg,
1998.}}\\
~&\\
$\mbox{[JT2]}$&  P.~Jizba and E.S.~Tututi, Phys.~Rev.~{\bf{D60}}:
105013, 1999.\\
~&\\
$\mbox{[JT3]}$& E.S.~Tututi and P.~Jizba, {\texttt{DAMTP-1998-163}}, in \textit{Proceedings of
the VIII}\\
~&\\
$~$ &  \hspace{0.2cm}{\textit{Mexican School of the Particles and Fields, Mexico city, 1998.}}
\end{tabular}

\newpage
\vspace{3cm}
\begin{center}
\Large{\textbf{Acknowledgement}}
\end{center}
I  would  like  to  express  my  respectful   thanks  to  my  research
supervisor, Prof.~P.V.~Landshoff,   for  his   careful   guidance
throughout my whole Ph.D., from the primitive
inception    of  the    research   to   its   writing--up  in     this
dissertation. I wish  to extend  my  gratitude to  Dr.~E.S.~Tututi  and
Dr.~M.~Blasone with  whom I have  enjoyed a relaxed and stimulating
collaboration  and  who, among others, shared the many up--and--downs
of my professional and personal life.\newline
\newline
My fundamental gratitude to my mother goes beyond the limit words can
capture. I would like to dedicate this dissertation to her.\newline
\newline 
Finally, it     is      a  pleasure  to  thank to    all     my
friends. Particularly the   encouragement   and support that    I 
gained  from  Marketa   Mazakova  was   invaluable.   I   also
acknowledge the financial assistance   that I have received  from  the
Cambridge  Overseas Trust,    Fitzwilliam College  and  the Board   of
Graduate Studies, University of Cambridge. 

\newpage
~~\\
~~\\
~~\\
~~\\
~~\\
\begin{center}
\large{\textit{\textbf{To my mother}}}
\end{center}
~~\\
~~\\
~~\\
\begin{center}
\large{\textit{The scientist does not  study nature because it
is useful; he studies it because he delights in it, and he delights in
it because it is beautiful. If nature were not beautiful, it would not
be worth knowing, and if nature were not worth knowing, life would not
be worth living.}} 

\vspace{6mm}

{\textit{Henri Poincar\'e}}

\end{center}

\newpage
\baselineskip26pt
\pagestyle{plain}
\setcounter{page}{1}
\pagenumbering{roman}
\tableofcontents

\newpage
\baselineskip26pt
\pagestyle{plain}
\begin{center}
{\bf{\Huge List of figures}}
\end{center}
\vspace{1cm}
\begin{tabular}{llr}
FIG.2.1 & \parbox[t]{10cm}{A one loop triangle diagram.} &
\parbox[b]{2cm}{~~~~~p.15} \\
~ & ~&~\\
FIG.2.2 & \parbox[t]{10cm}
{An example of a cut
diagram in the $\varphi ^{3}$ theory which does not contribute to the RHS's
of (\tseref{CE4})--(\tseref{CE5}).  Arrows indicate the flow of
energy.} 
& \parbox[b]{2cm}{~~~~~p.19} \\
~ & ~&~\\
FIG.2.3 & \parbox[t]{10cm}{Generic form of the cut diagram at the $T=0$. Shadow is on the
2nd type vertex area.} &\parbox[b]{2cm}{~~~~~p.19} \\
~ & ~&~\\
FIG.2.4 &\parbox[t]{10cm}{An example of non--vanishing cut diagrams at the $T\not =0$.
The heat--bath consists of two different particles. External particles are
not thermalized.} & \parbox[b]{2cm}{~~~~~p.25} \\
~ & ~&~\\
FIG.2.5 & \parbox[t]{10cm}{The cut diagram from
FIG.\ref{fig4} c) demonstrates that the cut can be defined in many ways but
the number of crossed lines is still the same.} &\parbox[b]{2cm}{~~~~~p.25}\\
~ & ~&~\\
FIG.2.6 & \parbox[t]{10cm}{The numerator of
(\tseref{dN}) and (\tseref{N}) can be calculated using the modified cut 
diagrams for $\langle {\T}^{\dag} {\P} \,\T \rangle_{p_{1} p_{2}}$.
As an example we depict all the possible contributions to the numerator 
derived from the cut diagram on Fig.\ref{fig.3} c). The wavy lines and
thin lines describe the heat-bath particles. The crossed lines denote the
substituted propagators, in this case we wish to calculate the thin--line
particle number spectrum.} &\parbox[b]{2cm}{~~~~~p.35}\\
~ & ~&~\\
FIG.2.7 & \parbox[t]{10cm}{The modified cut diagrams involved in an 
order--$e^{2}$ contribution
to the photon number spectrum. Dashed lines: photons. Solid lines: $\phi$,
$\phi^{\dag}$ particles. Bold lines: $\Phi$ particles.}
&\parbox[t]{2cm}{~~~~~p.38}\\
~ & ~&~\\
FIG.2.8 & \parbox[t]{10cm}{The diagram a) with a corresponding
kinematics.} &\parbox[b]{2cm}{~~~~~p.38}\\
~ & ~&~\\
FIG.2.9 & \parbox[t]{10cm}{The lowest--order cut diagram for $\langle
\T{\P}{\T}^{\dag}\rangle_{p_{1}p_{2}}$.} &\parbox[b]{2cm}{
~~~~~p.41}\\
\end{tabular}

\newpage
\baselineskip26pt
\pagestyle{plain}
\vspace{5mm}
\begin{tabular}{llr}
FIG.2.10 & \parbox[t]{10cm}{The generating thermal
diagrams involved in an order--$e^{4}$ contribution to the electron number
spectrum. Dashed lines: photons. Thin lines: $\phi$, $\phi^{\dag}$  
particles. Bold lines: $\Phi$ particles. Half--bold lines:
electrons.} & \parbox[b]{2cm}{~~~~~p.46}\\
~ & ~&~\\
FIG.2.11 & \parbox[t]{10cm}{The non--vanishing modified cut diagrams from
FIG.\ref{PJ1}c).}&\parbox[b]{2cm}{~~~~~p.47}\\
~ & ~&~\\
FIG.3.1 & \parbox[t]{10cm}{The graphical representation of $D^{\mu \nu}(p^{n}|p)$.} &
\parbox[b]{2cm}{~~~~~p.59} \\
~ & ~&~\\
FIG.3.2 & \parbox[t]{10cm}{Counterterm renormalisation of the last two diagrams in
Eq.(2.17). (Cut legs indicate amputations.)} &\parbox[b]{2cm}{~~~~~p.67} \\
~ & ~&~\\
FIG.3.3 &\parbox[t]{10cm}{ The Keldysh--Schwinger time path.} &\parbox[b]{2cm}{ ~~~~~p.81} \\
~ & ~&~\\
FIG.3.4 &\parbox[t]{10cm}{ First few bubble diagrams in the ${\textswab{M}}$
expansion.} & \parbox[b]{2cm}{~~~~~p.83} \\
~ & ~&~\\
FIG.4.1 & \parbox[t]{10cm}{A plot  of  $\langle  \beta\rangle$ vs.  ${\cali{M}}$   at
$\sigma = 100$ \mbox{MeV}.} &\parbox[b]{2cm}{~~~~~p.119} \\
~ & ~&~\\
FIG.4.2 & \parbox[t]{10cm}{ A plot of the Eq.(\tseref{trans1}): a) the general shape, b)
a small $x$ behaviour.} &\parbox[b]{2cm}{~~~~~p.121} \\
~ & ~&~\\
FIG.4.3 & \parbox[t]{10cm}{A plot  of pressure  as a function   of $T$,  $\sigma$  for
$m_{r}=  100\mbox{MeV}$.   The   gray line  corresponds   to equilibrium
pressure, the black line corresponds to pressure (\tseref{bb6}).}
&\parbox[b]{2cm}{~~~~~p.122}
\\
~ & ~&~\\
FIG.4.4 & \parbox[t]{10cm}{A  plot   showing   the  difference of   equilibrium    and
non-equilibrium pressures for $m_{r}=100\mbox{MeV}$.}
&\parbox[b]{2cm}{~~~~~p.123}\\
~ & ~&~\\
FIG.4.5 & \parbox[t]{10cm}{Behaviour of the pressure  (\tseref{precal4}) as a  function   
of $\alpha$ and $T_{0}$ at $m_r=100 \mbox{MeV}$.} &
\parbox[b]{2cm}{~~~~~p.126}\\
~ & ~&~\\
FIG.A.1 & \parbox[t]{10cm}{ Diagrammatic equivalent of Eq.(\tseref{S-D12}). The cut   
 separates areas constructed out of $F[\psi]$ and
$G[\psi]$.} & \parbox[b]{2cm}{~~~~~p.141} 
\end{tabular}

\newpage
\baselineskip10pt
\pagestyle{plain}
\pagenumbering{roman}
\chapter*{}
\begin{center}
{\bf{ \Large{Notation}}}
\end{center}
 
\vspace{1cm}
\parbox[t]{15.5cm}{Natural units $\hbar = c = k_{B} = 1$ will be used
throughout this dissertation. The following sub and superscripts will
be used to label various quantities in the following text.}
\vspace{1cm}

\begin{tabular}{ll}
Subscripts & ~ \\
${~}_{s}$ & \hspace{1.8cm} Schr\"odinger picture\\
${~}_{H}$ & \hspace{1.8cm} Heisenberg picture \\
${~}_{C}$ & \hspace{1.8cm} Closed--time path (also the
Keldysch--Schwinger path)\\
${~}_{fix}$ & \hspace{1.8cm} Gauge--fixing term \\
${~}_{in}$ & \hspace{1.8cm} Interaction component \\
${~}_{hb}$ & \hspace{1.8cm} Heat--bath component \\
${~}_{r}$ & \hspace{1.8cm} Renormalised object (operators, vertex
functions, etc.)\\ 
${~}_{V}$ & \hspace{1.8cm} Volume \\
~&~ \\
Superscripts & ~ \\
~ & ~ \\
${~}^{\dag}$ & \hspace{1.8cm} Hermitian conjugation \\
${~}^{*}$ & \hspace{1.8cm} Complex conjugation (except for
the ${\cali{T}}^{*}$--ordering)\\
${~}^{T}$ & \hspace{1.8cm} 4--dimensional transverse components \\
${~}^{L}$ & \hspace{1.8cm} 4--dimensional longitudinal components \\ 
${~}^{D}$ & \hspace{1.8cm} Symbol for a dimension \\
~&~ \\
Either & ~\\ 
~& ~ \\
$\mu, \nu, \lambda$ & \hspace{1.8cm} 4--dimensional spacetime index \\
$i, j$ & \hspace{1.8cm} 3--dimensional space index \\
$\alpha, \beta$ & \hspace{1.8cm} \parbox[t]{9.9cm}{Closed--time path indices :$\alpha,
\beta = \{+, -\} $ for non-equilibrium, and $\alpha, \beta =\{1, 2\}$
for equilibrium}\\
$a, b, c, d$ & \hspace{1.8cm} Internal symmetry indices \\
\end{tabular}
\vspace{4cm}

\parbox[t]{15.5cm}{Some of the important fields and functionals which will be
used are.}   

\vspace{6mm}
\begin{tabular}{ll}
$H = \int d^{3}x \, {\cali{H}}(x)$ & \hspace{1.8cm} Hamiltonian \\
${\cali{L}}(x)$ & \hspace{1.8cm} Lagrange density \\
$\P$ & \hspace{1.8cm} Projection operator onto final particle states \\
$\S$ & \hspace{1.8cm} $\S$--matrix ($\S = \1 + i\T$)\\
$\T$ & \hspace{1.8cm} $\T$--matrix \\
$U(t;t')$ & \hspace{1.8cm} Ket evolution operator \\
$\rho$ & \hspace{1.8cm} Density matrix \\
$\Theta^{\mu \nu}$ & \hspace{1.8cm} General energy--momentum tensor\\
$\Theta^{\mu \nu}_{c}$ & \hspace{1.8cm} Canonical energy--momentum
tensor\\
$\Phi , \, \phi, \, \psi$ & \hspace{1.8cm} General fields\\
\end{tabular}
\vspace{1cm}

\parbox[t]{15.5cm}{Various other notations that will be used are.}
\vspace{6mm}

\begin{tabular}{ll}
$\D$ & \hspace{1.8cm} \parbox[t]{9.9cm}{Full thermal propagator for
scalar fields in the closed-time formalism (i.e. $2 \times 2$ matrix)}\\
$\D^{c}$ & \hspace{1.8cm} Connected thermal 2--point Green's function,
i.e. connected part of $\D_{F}$\\ 
$\D_{F}$ & \hspace{1.8cm} \parbox[t]{9.9cm}{Free thermal propagator
for scalar fields in the closed-time formalism (i.e. $2 \times 2$
matrix)}\\
$\D^{(n)}$ & \hspace{1.8cm} Full thermal $n$--point Green's function\\ 
$G^{(n)}$ & \hspace{1.8cm} Non--equilibrium $n$--point connected Green's function \\
$\G$ & \hspace{1.8cm} \parbox[t]{9.9cm}{Non--equilibrium propagator in the
closed--time
path formalism (i.e. $2 \times 2$ matrix)}\\
$f_{B}$ & \hspace{1.8cm} Bose--Einstein distribution \\
$f_{F}$ & \hspace{1.8cm} Fermi--Dirac distribution \\
$\textswab{H}$ & \hspace{1.8cm} Information content \\
$I(\ldots)$ & \hspace{1.8cm} Shannon (information) entropy\\
$\Im(\ldots)$ & \hspace{1.8cm} Amount of information conveyed by a
single message \\
$\M$ & \hspace{1.8cm} Bogoliubov matrix \\
$\textswab{M}$   & \hspace{1.8cm} Massieu function $= -\beta \times$
(Helmholtz) free energy\\
${\cali{O}}(z)$ & \hspace{1.8cm} Order $z$\\
$p(T)$ & \hspace{1.8cm} Thermodynamic pressure at the temperature $T$\\ 
$p(x,T)$ & \hspace{1.8cm} Hydrostatic pressure  at the temperature $T$\\
${\cali{P}}(T)$ & \hspace{1.8cm} Hydrostatic pressure for
translationally invariant media \\
$S_{F}$ & \hspace{1.8cm} Feynman (causal) propagator for
spin--$\frac{1}{2}$ particles; ($T=0$)\\
$S^{+}$ & \hspace{1.8cm} Positive energy part of $S_{F}$\\
$S^{-}$ & \hspace{1.8cm} Negative energy part of $S_{F}$\\
$S_{G}$  & \hspace{1.8cm} von Neumann--Gibbs entropy
\end{tabular}

\begin{tabular}{ll}
${\cali{T}}$ & \hspace{1.8cm} Time ordering symbol\\
${\cali{T}}^{*}$ & \hspace{1.8cm} The ${\cali{T}}^{*}$ (or covariant)
ordering\\
$W$ & \hspace{1.8cm} Generating functional for $G^{(n)}$\\
$Z$ & \hspace{1.8cm} (Grand) partition function\\
${\cali{Z}}$ & \hspace{1.8cm} The Jaynes--Gibbs partition function \\
$Z_{\Phi}$ & \hspace{1.8cm} Wave function renormalisation\\
$\beta$ & \hspace{1.8cm} Inverse temperature $1/T$\\
$\Gamma$ & \hspace{1.8cm} \parbox[t]{9.9cm}{Generating functional for
$\Gamma^{(n)}$ (i.e.
the effective action)}\\
$\Gamma^{(n)}$ & \hspace{1.8cm} n--point vertex function (i.e. 1PI
n--point Green's function)\\
$\frac{\stackrel{\rightarrow}{\delta}}{\delta \psi(z)}$ &
\hspace{1.8cm} Left--handed variation \\
$\frac{\stackrel{\leftarrow}{\delta}}{\delta \psi(z)}$ &
\hspace{1.8cm} Right--handed variation \\
$\delta^{\pm}(\ldots)$ & \hspace{1.8cm} \parbox[t]{9.9cm}{Positive ($+$),
negative ($-$)
frequency parts of Dirac's $\delta$--function}\\
$\Delta_{F}$ & \hspace{1.8cm} Feynman (causal) propagator for scalar
fields; ($T=0$) \\
$\Delta^{+}$ & \hspace{1.8cm} Positive energy part of $\Delta_{F}$\\
$\Delta^{-}$ & \hspace{1.8cm} Negative energy part of $\Delta_{F}$\\
$\mu$ & \hspace{1.8cm} Chemical potential\\
$\phi_{\alpha}^{a}$ & \hspace{1.8cm} \parbox[t]{9.9cm}{Expectation value
of the
Heisenberg field operator $\Phi_{\alpha}^{a}$ in the presence of a
c--number source $J$ (i.e. $\phi_{\alpha}^{a} = \langle \Phi_{\alpha}^{a}
\rangle$)}\\
$\Sigma$ & \hspace{1.8cm} Proper self--energy for the scalar theory; ($T=0$)\\   
${\vect{\Sigma}}$ & \hspace{1.8cm} \parbox[t]{9.9cm}{Proper self--energy for the
scalar theory at finite
temperature  in the closed-time path formalism (i.e. $2 \times 2$
matrix)}\\
${\tilde{\vect{\Sigma}}}$ & \hspace{1.8cm} Self--energy for the scalar
theory at finite temperature \\
$\Omega$ & \hspace{1.8cm} \parbox[t]{9.9cm}{Grand (canonical)
potential. In the case
when the canonical ensemble is in question, $\Omega$ is (Helmholtz)
free energy}\\
$\langle \ldots \rangle$ & \hspace{1.8cm} Expectation
value \\
$\langle \ldots \rangle_{\{p_{k} \}}$ & \hspace{1.8cm}
\parbox[t]{9.9cm}{Thermal expectation value. Particles with momenta
$\{p_{k} \}$ are unheated } 
\end{tabular}
\vspace{7mm}
\begin{itemize}

\item The Minkowski metric used throughout is $g^{\mu \nu} = g_{\mu
\nu} = \mbox{diag}(1,-1,-1,-1)$ 
\end{itemize}
\vspace{-2mm}
\begin{displaymath}
g_{\mu \nu}g^{\mu \nu} = \delta^{\mu}_{\,\, \mu} = \mbox{Tr}(g)= 4, \,
\,\, g_{\mu \mu} = g^{\mu \mu} = -2
\end{displaymath}
\begin{itemize}
\vspace{2mm}
\item Derivatives with respect to $x^{\mu}$ or
$x_{\mu}$ are abbreviated as 
\end{itemize}
\vspace{-2mm}
\begin{displaymath}
\partial_{\mu} \equiv \partial / \partial x^{\mu}, \; \; \; \partial^{\mu} \equiv
\partial / \partial x_{\mu}
\end{displaymath}

\newpage
\baselineskip24pt
\setcounter{page}{1}
\pagenumbering{arabic}
\pagestyle{headings}
\chapter{Introduction and overview}
\vspace{1cm}
 
The     development of the    theory of   quantised  fields at  finite
temperature and density (also  thermal quantum field theory)  over the
past   forty years  or  so has   led  to  fundamental changes in   the
understanding of a wide number of  physical phenomena. Among those one
may  mention symmetry   restoration during   high--temperature   phase
transitions\cite{AL,Wein,DJ} which   has found significant application
in early  universe cosmology, and  we may also mention applications in
neutron--star\cite{CP}  and supernova\cite{PH} astrophysics.  However,
perhaps  one  of the most  important   applications of thermal quantum
field theory nowadays is to quantum chromodynamics (QCD). 

The reason for this interest may be traced to the mid--70's when the notion
of asymptotic freedom of QCD started to emerge.   At zero temperature and
chemical potential the low--energy and/or momentum transfer behaviour of
QCD is  characterised  by confinement (i.e.   strong interaction among
QCD   constituents). The  internal  scale that   determines  the boundary
between small and large  energy in QCD  is $\sim \, \Lambda_{QCD} \sim
0.2\mbox{GeV}$.  As the energy and/or momentum transfer increases, QCD
begins to be  characterised by asymptotic  freedom, i.e., the  coupling
evolves as

\begin{displaymath}
\lambda_{QCD}(Q^{2},     T=0=\mu)      \stackrel{Q^{2}     \rightarrow
\infty}{\rightarrow} 0, 
\end{displaymath}

\noindent ($Q^{\mu}$ is the four--momentum transfer, $Q^{2}>0$) and so
quarks and    gluons   behave  like  weakly    interacting,   massless
particles\footnote{Because quarks    become  massless, the  deconfined
phase   leads     to  chiral--symmetry  restoration     (under  normal
circumstances  the chiral flavour  group $SU_{L}(N)\times  SU_{R}(N)$,
with $N$ the  number of quark flavours,  is  broken to a  vector--like
subgroup $SU_{V}(N)$).  The chiral  phase transition is expected to be
particularly       interesting   at     high    temperatures    and/or
densities\cite{JBe,MAHa}.}  in    high--energy and/or  large   $Q^{2}$
processes.  This behaviour  is usually tested experimentally  by means
of deep inelastic electron--nucleon scattering. 

If one  starts to study QCD at  finite temperature and/or finite baryon
density one automatically introduces  new (intensive) variables,
namely the
temperature $T$ and the quark chemical potential $\mu$. These bring an
additional  mass scale  with which the  coupling  $\lambda_{QCD}$ can
run. It was Collins and Perry who first showed in\cite{CP} that strong
interactions  become weak not  only at high energy--momentum transfer,
as  in the  deep inelastic  scattering,  but  also  at  very high  baryon
density.    This    reasoning   was     quickly  extended  to   finite
temperature\cite{DJ,JIK} where it was shown that 

\begin{displaymath} 
\lambda_{QCD}(Q^{2}=0,  T, \mu ) \rightarrow 0,
\end{displaymath}

\noindent  provided   that $T   \gg  \Lambda_{QCD}$  and/or   $\mu \gg
\Lambda_{QCD}$. Thus  at a sufficiently  high temperature  and/or
baryon number density QCD systems consist of free quarks and gluons,
regardless of  the energy--momentum transfer.   This  ``deconfined''
phase  of QCD   is called the  quark--gluon  plasma.   As  was just
mentioned, the temperature  and/or chemical potential must be  greater
than the QCD fundamental mass  scale $\Lambda_{QCD}$. In practice this 
means  that the temperature for creation  of  the quark--gluon plasma
must    be at   least    of  the   order  $\sim   0.2\mbox{GeV}   \sim
10^{12}\mbox{K}$  and/or  the baryon number  density  must be of order
$\sim    \Lambda_{QCD}^{3}           \sim       0.8\mbox{(GeV)}^{3} \sim
10^{42}\,\mbox{cm$^{-3}$}$.   Such temperatures prevailed  in the very
early stage  of the universe ($\sim 10^{-6}\mbox{s}-10^{-5}\mbox{s}$
after the  Big--Bang) and such  densities are  expected to be present,
for instance, in the core of neutron stars\cite{CP}. 

Apart from the   cosmological   and astrophysical implications,    the
quark--gluon  plasma naturally offers  an important laboratory for the
study of asymptotic freedom. This fact has led  to the construction of
a  Relativistic  Heavy  Ion Collider   (RHIC)  at Brookhaven  National
Laboratory, where it is expected  that two on--head colliding beams of
$^{197}$Au will generate  a sufficient centre--of--mass energy density
which,  when  properly  thermalized, will  allow  the  formation  of a
quark--gluon  plasma. Similar  experiments  are planned  in the  Large
Hadron Collider (LHC) at CERN with colliding Pb beams.

So far we  have only outlined possible  applications of QFT to systems
which are in  thermodynamical  equilibrium.  However, during the  last
several years there has  been  a steadily  growing awareness that  the
usual  equilibrium (or thermodynamical)  approaches  in quantum  field
theory fail to   describe   correctly such fundamental    phenomena as
realistic  phase transitions (both  in condensed matter physics and in
high--energy physics), quantum electrodynamical (QED) plasma evolution
(and  the related problem  of hot thermonuclear fusion), heat transfer
in   stars,  etc.   Presently, the   most   important  applications of
non--equilibrium  QFT,   however, are  in the    physics of the  early
universe. The reason why it is so  important to go beyond conventional
equilibrium   approaches  in describing the   universe  is the extreme
conditions  present in the  early  period of its evolution\cite{EWKo}.
It is  now well recognised  that at the  very early times the universe
was very hot with energy densities approaching the Planck mass density
$  \sim  5.2 \times  10^{93}\,\mbox{g~cm$^{-3}$}$,  and it then cooled
rapidly due to the  expansion.  Because of  this fact, it  is expected
that the  matter fields went  through a hierarchy of phase transitions
before  reaching the present  status  quo\cite{TWBKi2}.  It is obvious
that if the matter fields involved in the phase transitions interacted
at a rate much  smaller than the  typical cooling (or relaxation) rate
of  the      universe,  the   evolution    would  proceed     in    an
out--of--equilibrium  manner.  Such  non--equilibrium dynamics may, in
turn, lead to many crucial physical consequences.  One such example is
the baryon asymmetry\cite{RuSh,EWKo,EWKo1}.   If one  assumes that the
relative abundance of baryonic matter  over antibaryonic matter is not
created {\em{a priori}}   via the initial--time conditions,  then  one
must find a physical mechanism which could generate such an asymmetry.
It was A.D.~Sakharov\cite{AS} who  proposed more than thirty years ago
that,  in  order  to  explain  a biased   production   of baryons over
anti--baryons, one   needs to take  into account  the non--equilibrium
evolution of the universe since baryons and anti--baryons are produced
in  equal  number in  any  equilibrium process.   This   is one of the
celebrated Sakharov criteria\cite{AS,RuSh,ArMt} for baryogenesis. 

Another   extremely      interesting  example   of      application of
non--equilibrium   QFT in cosmology   is the production of topological
defects during  phase transitions.  Interest  in this area  stems from
the belief that  topological defects,  such  as cosmic strings,  might
provide an explanation of structure formation and the cosmic microwave
background   radiation anisotropies in   the universe\cite{TWBKi}.  The
domain structure of  a ferromagnet is  known to become very  different
when the sample is cooled adiabatically  through the Curie temperature
compared to when it is cooled rapidly. In  analogy one may expect that
the production of topological  defects and their evolution will depend
strongly on  the  cooling  rate of the   universe.  The   formation of
defects  and their  dynamics can  be  beautifully mimicked  in quantum
liquids\cite{TK,HL}  such as helium   $^{3}$He or $^{4}$He, in  liquid
nematic  crystals\cite{NT}  or in  superconductors\cite{WZ}.    Recent
experiments\cite{TK} with $^{3}$He have  confirmed that the density of
vortex lines  nucleated  during  the  phase transition  from a  normal
$^{3}$He liquid to its superfluid B--phase depends considerably on the
cooling rate. 

In the  framework of non--equilibrium  applications of  QFT one should
not forget  the  currently very important  application to relativistic
heavy--ion collisions, which,  as we have  mentioned, are expected  to
produce    a   sufficient  thermal environment for      creation  of a
quark--gluon plasma.  The   major indication that a   non--equilibrium
description seems to be necessary comes from  the expectation that the
time scale at which the quark--gluon plasma should exhibit itself will
be too short for its macroscopic equilibration.  The point is that the
energy   density  in the  reaction  zone,  once  the  plasma begins to
thermalise,  will create  a pressure  which will  consequently lead to
explosive  expansion of the  plasma.  The associated expansion time is
estimated  to be only 10 or  100 times longer  than  the time scale of
microscopic equilibration  processes\cite{UH}. This introduces a basic
uncertainty as to whether an equilibrium treatment may be used safely.
On the other hand,  if a non--equilibrium description proves  crucial,
then one must face the question to  what extent a small, short--lived,
fast  expanding system  of   quarks  and  gluons   can  be called    a
quark--gluon  plasma. One  should  honestly admit that  there does not
presently   exist   any   generally    accepted  model  which    would
satisfactorily describe the quark--gluon plasma dynamics. 

All the mentioned non--equilibrium processes  are characterised by the
rate at  which a  system changes  (e.g.  expansion   rate, dissipative
rate, the  rate at which    particles are exchanged with  an  external
environment) which is comparable to or  greater than the rate at which
microscopic  interactions (i.e.    the  equilibrating mechanism)   are
happening. In practice this means that the relaxation time scale (i.e.
the mean  time  in which  the system relaxes   to equilibrium) is much
longer than the  time scale at which  observations  are performed, and
consequently a system does not  evolve adiabatically (i.e. it does not
go  through   a  sequence of  states   each of  which  is   at thermal
equilibrium). It is  needless to say that many  of phenomena in Nature
are precisely of this character. 

Whereas for the description of matter in  equilibrium one has at one's
disposal a systematic and unified approach based upon the formalism of
the Gibbs (micro-- or  grand--)  canonical density matrices, no   such
simple way seems to exist in  a non--equilibrium theory because of the
variety  of   phenomena and   of the    complexity  of the   evolution
processes. This is to be expected, as there is basically no limitation
on the  amount  of questions that  may be  asked about time--dependent
phenomena  unless one clearly  specifies which degrees  of freedom (or
what  degree of reduction)  one will adopt for  the description of the
evolution  of a system.  Within  such  a reduced description, however,
the dynamics has   been shown to be  capable  of prediction.  In  this
framework  there have steadily crystallised   two major (and  mutually
distinct) approaches\cite{JDo,WTGr,Bal}. 

The first one,  which we  mostly refer to  in  our thesis, is the,  so
called, `Maxent school'    (or  maximal entropy school)  founded    by
E.T.Jaynes   {\em   {et al.}}\cite{Jayn,Jayn2,Jayn3,BM1}.  The   basic
philosophy  behind this  approach  is rooted   in the  fact that   the
prediction of  the future macroscopic evolution  of a system cannot be
done  with  certainty on  the basis   of the initial  macroscopic data
because of the existing correlations  with the parameters or data that
are discarded  in  the given  reduced description.   The corresponding
statistical  inference about  the   system  is  not deduced from   the
underlying microscopic dynamics    but instead  is rather   based   on
information  theory.  There  the  algorithm of  entropy maximalisation
leads  to  the density  matrix  (or probability distribution) with the
least informative content subject to the prior knowledge which one has
about  the system.  Generally   one  may,  at different times,   adopt
different  macroscopic  parameters describing   the  system  (i.e.  at
different times one may  choose a different level  or reduction in the
description).  If  this  is the case,  the  maximalisation  of entropy
naturally leads to  a more complex form of  the density matrix.   This
branch of  the Maxent school  (and its various modifications)  is also
known as the {\em {projection operator technique}}\cite{RBa}. 

The second school, the so called `Brussels school', is associated with
I.Prigogine, R.Bales- cu  {\em   {et al.}}\cite{IPr,RB,Bal}.   The basic
emphasis  is  here put on the   microscopic  dynamics (i.e.  Hamilton,
Schr\"odinger or  Liouville equations), and  all other  non--dynamical
approaches,  such    as  those coming  from   information  theory, are
discarded: everything should be derived  from the dynamics alone.  The
ultimate aim  is a  dynamical   separation into the  parameters  to be
retained  on   the desired  level  of description,   and  those to  be
discarded.  This separation, if  it  exists,  should emerge  the  from
dynamics  as   an asymptotic   property   valid for  the   large (i.e.
observational)  time scales.  It  is   well known  since the time   of
Boltzmann that  the latter may  be  achieved by introducing  a certain
hypothesis  about the  microscopic  behaviour   of  the system;   e.g.
Boltzmann's           ``Stosszahlansatz''       (random      collision
hypothesis)\cite{Tol},        Ehrenfest's      coarse         graining
hypothesis\cite{PEh}, or  Bogoliubov's no  initial--time  correlations
hypothesis\cite{NNBo}.  This  line of reasoning is basically inherited
and progressed  by the Brussels school.  It  should  be noted that all
the  statistical inferences    or hypotheses  here  have a    strictly
dynamical    nature (they are  directly  motivated   by the underlying
dynamics), but on the other hand they are in a certain respect {\em{ad
hoc}}, because only certain probabilistic  features of the microscopic
dynamics are emphasised.

\vspace{5mm}
\rule[0.05in]{12cm}{0.1mm}~~~~{\bf{\large{Synopsis}}}
\vspace{2mm}

We   first  focus in Chapter   \ref{dcoset}   on infrared  effects  in
finite--temperature  QFT.   We propose   a simple mathematical  method
(based   on   the  largest--time  equation    and the Dyson--Schwinger
equations) which allows systematic calculations of the change of energy
density (or particle  density)   in a heat--bath  during a   decay (or
scattering) of  the external  particle(s) within the  heat  bath.  The
applied method naturally leads  to an interpretation  of the change of
density  of    energy  (particles)    in  terms  of   three   additive
contributions: stimulated emission, absorption and fluctuations of the
heat--bath particles.  This result is completely non--perturbative.  A
careful analysis  reveals that  the  resulting  change in  the  energy
density    is   finite  even in    the   case of   massless heat--bath
particles. This means that there is no infrared catastrophe. 
\vspace{2mm}

As the next point  we re--consider in Chapter  \ref{PE} the problem of
calculating pressure.  We  use the so--called hydrostatic pressure (or
pressure at a point) which is defined via the energy--momentum tensor.
The obvious advantage is a  possible extension into a non--equilibrium
medium.  We go through all the delicate points that  must be dealt with
in the context of quantum  field theory.  Renormalisation of composite
operators  and  in general the   vital  role of renormalisation for  a
consistent quantum   field--theoretical  definition  of  pressure   is
discussed. We   finally  apply the whole  procedure   to  a toy--model
system: $\lambda \Phi^{4}$ theory with  $O(N)$ internal symmetry.   In
the   case   of the      large--$N$ limit  (also    the  Hartree--Fock
approximation)  the pressure is   exactly solvable.  Using  the Mellin
transform technique we perform the large--temperature expansion of the
pressure to all orders in $T$. 
\vspace{2mm}

The hydrostatic pressure can be naturally extended to non--equilibrium
systems. Using the Jaynes--Gibbs principle of  maximal entropy and the
(non--equilibrium) Dyson--Schwinger equations  we derive in
Chapter \ref{PN} the generalised Kubo--Martin--Schwinger equations and
set   up a calculational   scheme for pressure  calculations away from
thermal equilibrium. As an example we  explicitly evaluate pressure for
the $O(N)\;\lambda  \Phi^{4}$ theory in  the  large--$N$ limit  in two
cases of translationally invariant non--equilibrium systems. 
\vspace{2mm}

There  follow five  appendices which comprise  much  of the background
material required in the main body of the  thesis.  The important part
is the detailed analysis  in Appendix \ref{A1} of the Dyson--Schwinger
equations.  The   derivation  there shows  how  the   Dyson--Schwinger
equations may be formulated in a very useful  functional form.  We also
outline  the  connection with   the more conventional   approaches. In
Appendices   \ref{A2} and \ref{A3}  we  clarify some finer mathematical
manipulations    needed in  Chapter   \ref{PE}.   The  fundamentals of
information or Shannon   entropy are presented  in Appendix  \ref{A4}.
Appendix \ref{MF} covers  the elements  of dimensional  regularisation
and special functions which underlie much of the material presented in
the earlier chapters.

\vspace{5mm}
\rule[0 in]{12cm}{0.1mm}~~~~{\bf{{\large{Epilogue}}}}
\vspace{2mm}

A reader of this dissertation  might be disappointed  by the fact that
he  or she  will  not  find  a usual pedagogical  introduction  to the
subject.  The omission of such  an introduction was dictated mainly by
two  considerations.   Firstly,  we have  not   felt very competent  to
provide a good account  of  the fundamentals  of both  equilibrium and
non--equilibrium quantum field theory.    The subjects themselves  are
currently immensely  vast and the number  of problems involved quickly
approaches  the ``thermodynamic  limit''.  We  therefore  take a  more
pragmatic point   of  view and  proceed  by   sampling a few  definite
problems, which we develop and analyse  in great detail.  Secondly, and
most importantly, we  personally hold the  opinion  that a dissertation
should reflect student's ability to cope with a subject and creatively
apply it   to practical problems    rather  than write   an essay   on
mathematical or physical foundations of the  subject in question.  Any
such attempts would lead (at least in our case) to pure epigonism since
we do not feel that we could add anything  substantially new to existing
(and pedagogically excellent) textbooks and  review articles; see  for
example Refs.\cite{LB,AD,LW,Cub,TA,Gibbs,KB,FW,PR,Tol} for equilibrium
quantum   field    theory    and     statistical     physics       and
Refs.\cite{Zub,RB,KCC,Bal} for  non--equilibrium  quantum field theory
and statistical physics.

To end, we wish to make one more  remark.  We are perfectly aware that
the presented work is  incomplete in many  respects.  Our ignorance  or
lack of understanding  of many important topics  is of course  in part
responsible   for  this weakness.   In particular,   we refer here to
discussions and   applications    of   such  crucial areas     as  the
imaginary--time (Euclidean or Matsubara) formalism, (non--equilibrium)
thermofield   dynamics,   hard    thermal  loops,    the  theory    of
temperature--induced   phase    transitions,  linear response  theory,
stochastic  approaches to non--equilibrium dynamics, quantum transport
equations, theory  of transport coefficients and  many more, which are
definitely missing in   this dissertation.  Although we  had originally
planned to  include some of  the  aforementioned issues  (namely those
which directly concern or  resonate with our present research), neither
space nor time has allowed us to fulfil this wish.







\chapter{Heat--bath particle number spectrum\label{dcoset}}

In recent years much theoretical effort has been invested in the understanding 
relativistic heavy ion collisions as these can create critical energy
densities which are large enough to produce the quark--gluon plasma (the
deconfined phase of quarks and gluons) \cite{LB, LW}.

A natural tool for testing the quark--gluon plasma properties
would be to look for the particle number spectrum formed when a particle
decays within the plasma itself.  As the plasma created during heavy ion
collisions is, to a very good approximation, in thermodynamical
equilibrium \cite{LB} (somewhat like a ``microwave oven'' or a heat
bath), one can use the whole machinery of statistical physics and QFT in
order to predict the final plasma number spectrum. Such calculations,
derived from first principles, were carried out by Landshoff and Taylor
\cite{PVLJT}.

Our aim  is to  find a  sufficiently easy    mathematical formalism
allowing us to perform mentioned   calculations to any order.  Because
unstable particles treated in \cite{PVLJT} can not naturally appear in
asymptotic  states, we  demonstrate our  approach  on a mathematically
more correct (but from practical point of view less relevant) process;
namely on the scattering  of two particles inside of  a heat bath. The
method presented here  however, might  be  applied as well to  a decay
itself (provided that the corresponding  decay rate is much less  than
any of the characteristic energies of the process). In this chapter we
formulate the basic diagrammatic rules for the methodical perturbative
calculus  of plasma   particle  number  spectrum  $d\langle  N(\omega)
\rangle/d\omega $ and  discuss it  in the simple  case of  a heat bath
comprised  of photons and electrons,  which are for simplicity treated
as scalar particles.

In Section \ref{bt} we review the basic concepts and techniques needed
from the theory  of the largest-time  equation (both for $T=0$  and $T
\not=0$)    and the Dyson--Schwinger   equations.  Rules for  the cut
diagrams at finite--temperature  are derived and subsequently extended
to the  case when unheated fields are  present. It was already pointed
out in \cite{PL2}   that the thermal  cut diagrams  are virtually  the
Kobes--Semenoff  diagrams    \cite{LB}   in  the  Keldysh    formalism
\cite{K1}. This observation will allow us  to identify type 1 vertices
in the  real  time finite---temperature  diagrams  with the  uncircled
vertices used in the (thermal)    cut diagrams, and similarly type   2
vertices will be identified with the circled, cut diagram vertices. As
we  want to  restrict  our  attention to only   some particular  final
particle  states, further restrictions  on  the possible cut  diagrams
must  be included. We shall study  these restrictions in the last part
of Section \ref{bt}.

As we  shall show in Section \ref{HBP},  the heat--bath particle number
spectrum  can be conveniently  expressed  as a  fraction. Whilst it is
possible to   compute the denominator  by  means  of  the  thermal cut
diagrams developed   in  Section  \ref{bt},  the  calculation  of  the
numerator requires more  care. Using the Dyson--Schwinger equations, we
shall see  in Section  \ref{MCD} that  it  can be  calculated  through
modified thermal  cut   diagrams. The   modification consists of   the
substitution   in turn of  each heat   bath particle  propagator by an
altered one. We also show that there must be only one modification per
diagram. From  this we conclude that from  each individual cut diagram
we get  $n$ modified  ones   ($n$  stands for    the total number   of
heat--bath  particle propagators in  the diagram). Furthermore, in the
case when more types of the heat bath particles are present, one might
be  only  interested in  the  number spectrum of    some of these. The
construction of the  modified cut  diagrams in  such cases follow  the
same procedure as  in the  previous  situation. We find  that only the
propagators affiliated to the desired fields must be altered.

In Section \ref{MP}  the presented approach is  applied to a toy model
in  which a  gluon   plasma is  simulated by  scalar  photons, and  we
calculate the resulting changes in the number spectrum of the `plasma'
particles.  Section \ref{MP} ends with a qualitative discussion of the
quark--gluon plasma simulated by scalar photons and electrons.

For reader's convenience,  the  chapter is accompanied with   Appendix
\ref{A1}  where we derive, directly  from the  thermal Wick's theorem,
the   (thermal) Dyson--Schwinger  equations as   well   as other useful
functional identities valid at finite temperature. 

\vspace{5mm}

\section{Basic tools \label{bt}}
\subsection{Mean statistical value \label{bt1}}

The central idea of thermal QFT is based on the fact that one can not take the
expectation value of an observable $A$ with respect to some pure state as
generally all states have non-zero probability to be populated and
consequently one must consider instead a mixture of states generally
described by the density matrix $\rho$. The mean statistical value
of $A$ is then

\begin{equation}
\langle A \rangle = Tr(\rho A),
\tseleq{mean value}
\end{equation}

\vspace{2mm}

\noindent where the trace has to be taken over a complete set of {\em
physical} states. For a statistical system in thermodynamical equilibrium
$\rho$ is given by the Gibbs (grand--) canonical distribution

\begin{equation} \rho = \frac{e^{-\beta (H - \mu N)}}{Tr(e^{-\beta (H - \mu
N)})} = \frac{e^{-\beta K }}{Z},
\end{equation}

\vspace{1mm}

\noindent here $Z$ is the partition function, $H$ is the Hamiltonian, $N$
is the conserved charge (e.g. baryon or lepton number), $\mu$ is the
chemical potential, $K = H - \mu N$, and $\beta$ is the inverse
temperature: $\beta = 1/T$.

\vspace{3mm}

\subsection{Largest--time equation at T=0 \label{bt2}}

An important property inherited from zero--temperature QFT is {\em the
largest--time equation} (LTE) \cite{PN, V, TV}. Although the following
sections will mainly hinge on the {\em thermal} LTE, it is instructive to
start first with the zero--temperature one. The LTE at $T=0$ is a
generally valid identity which holds for any individual diagram
constructed with propagators satisfying certain simple properties. For
instance, for the scalar theory with a coupling constant $g$ one can
define the following rules:

\begin{figure}[h]
\epsfxsize=6cm 
\centerline{\epsffile{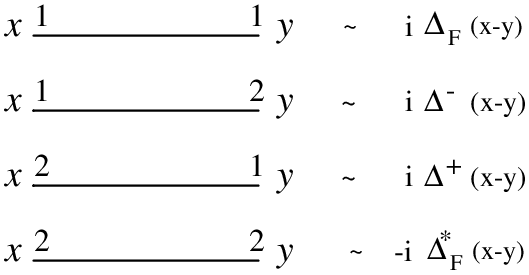}}
\vspace{4mm}
\end{figure}

\begin{figure}[h]
\vspace{3mm}
\epsfxsize=4cm
\centerline{\epsffile{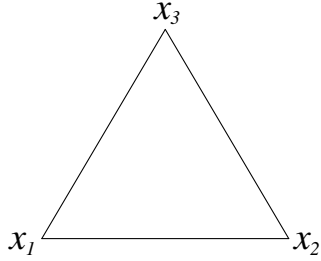}}
\caption{{ A one loop triangle diagram.}}
\label{fig20}
\vspace{3mm}
\end{figure}

\noindent Here $i\Delta_{F}$ is the Feynman propagator, $i\Delta^{+}$
($i\Delta^{-}$) is corresponding positive (negative) energy part of
$i\Delta_{F}$, the `$*$' means complex conjugation and index $1$ ($2$)
denotes {\em type--1} ({\em type--2}) vertex; type--1 vertex has attached a 
factor $-ig$ whilst type--2 bears a factor $ig$. Using this prescription,
we can construct diagrams in configuration space.  With each diagram then
can be associated a function $F(x_{1},\ldots x_{n})$ having all the 2nd
type vertices underlined.  For example, for the triangle diagram in
FIG.\ref{fig20} we have

\begin{eqnarray} 
F(x_{1},x_{2},x_{3}) &=&
(-ig)^{3} \, i\Delta_{F}(x_{1}-x_{2}) \, i\Delta_{F}(x_{1}-x_{3})
\, i\Delta_{F} (x_{2}-x_{3})\nonumber\\
F({\underline{x_{1}}},x_{2},x_{3}) &=&
(-ig)^{2}(ig) \, i\Delta^{+}(x_{1}-x_{2})
\, i\Delta^{+}(x_{1}-x_{3})  \, i\Delta_{F} (x_{2}-x_{3})\nonumber\\ 
F({\underline{x_{1}}},{\underline{x_{2}}},x_{3}) &=&
(ig)^{2}(-ig)
\, (-i)\Delta_{F}^{*}(x_{1}-x_{2}) \, i\Delta^{+}(x_{1}-x_{3}) \, i\Delta^{+}
(x_{2}-x_{3})\nonumber\\
F({\underline{x_{1}}},{\underline{x_{2}}},{\underline{x_{3}}}) &=&
(ig)^{3}
\,
(-i)\Delta_{F}^{*}(x_{1}-x_{2})
\, (-i)\Delta_{F}^{*}(x_{1}-x_{3}) \, (-i)\Delta_{F}^{*}
(x_{2}-x_{3}).\nonumber\\
\mbox{etc.}\mbox{\hspace{1.3cm}}&&\nonumber
\end{eqnarray}

\noindent The LTE then states that for a function $F(x_{1}, \ldots, x_{n})$
corresponding to some diagram with $n$ vertices

\vspace{-1mm}
\begin{equation} F( \ldots, {\underline{x_{i}}}, \ldots ) + F( \ldots,
x_{i}, \ldots )= 0, \tseleq{LTE20} \end{equation}
\vspace{1mm}

\noindent provided that $x_{i0}$ is the largest time and all other
underlinings in $F$ are the same. The proof of Eq.(\tseref{LTE20}) is
based on an observation that the propagator $i\Delta_{F}(x)$ can be
decomposed into positive and negative energy parts, i.e.

\begin{eqnarray}
i\Delta_{F}(x) &=& \theta(x_{0})i\Delta^{+}(x) +
\theta(-x_{0})i\Delta^{-}(x),\\
i\Delta^{\pm}(x) &=& \int \frac{d^{4}k}{(2\pi)^{3}}e^{-ikx}\theta(\pm
k_{0})\delta(k^{2}-m^{2}). \tseleq{decomp} \end{eqnarray}
  
\noindent Incidentally, for $x_{i0}$ being the largest time this
directly implies

\vspace{-3mm}
\begin{eqnarray}
i\Delta_{F}(x_{j}-x_{i}) &=&
i\Delta^{-}(x_{j}-x_{i}),\nonumber\\
-i\Delta_{F}^{*}(x_{i}-x_{j}) &=&
i\Delta^{-}(x_{i}-x_{j}),\nonumber\\
i\Delta_{F}(0) &=& -i\Delta_{F}^{*}(0).
\tselea{decomp1}
\end{eqnarray}

\noindent As $F(\ldots , {\underline{x_{i}}}, \ldots)$ differs from
$F(\ldots , x_{i}, \ldots)$ only in the propagators
directly connected to $x_{i}$ - which are equal (see
Eq.(\tseref{decomp1})) - and in the sign of the $x_{i}$ vertex, they must
mutually cancel.

Summing up Eq.(\tseref{LTE20}) for all possible underlinings
(excluding $x_{i}$), we get the LTE where the special r{\^{o}}le of the
largest time is not manifest any more, namely
  
\begin{equation}
\sum_{index}F(x_{1}, x_{2}, \ldots , x_{n}) = 0.\tseleq{LTE19}\end{equation}

\vspace{2mm} 

\noindent The sum $\sum_{index}$ means summing over all possible
distributions of indices 1 and 2 (or equivalently over all possible
underlinings). The zero--temperature LTE can be easily reformulated for the
$\T$--matrices. Let us remind that the Feynman diagrams for the $\S$--matrix
($\S=\1+i\T$) can be obtained by multiplying the corresponding $F(x_{1},
\ldots, x_{n})$ with the plane waves for the incoming and outgoing
particles, and subsequently integrate over $x_{1}\ldots x_{n}$. Thus, in
fixed volume quantisation a typical $n$--vertex Feynman diagram is given by

\begin{equation}
 \int
\prod_{i=1}^{n}dx_{i} \prod_{j}
\frac{e^{-ip_{j}x_{m_{j}}}}{\sqrt{2{\omega}_{p_{j}}V}} \prod_{k}
\frac{e^{iq_{k}x_{m_{k}}}}{\sqrt{2{\omega_{q_{k}}}V}}F(x_{1}, \ldots,
x_{n}). \tseleq{Tmatrix1}
\end{equation}

\vspace{2mm}

\noindent Here the momenta $\{ p_{j}\}$ are attached to incoming particles
at the vertices $\{x_{m_{j}}\}$, while momenta $\{q_{k}\}$ are attached to
outgoing particles at the vertices $\{x_{m_{k}}\}$. In order to
distinguish among various functions $F(x_{1}, \ldots, x_{n})$ with the
same variables $x_{1}, \ldots, x_{n}$, we shall attach a subscript $l_{n}$
to each function $F$. For instance, the function $F_{1_{4}}(x_{1}, \ldots,
x_{4})$ corresponding to the diagram

\begin{figure}[h]
\vspace{4mm}
\epsfxsize=7cm
\centerline{\epsffile{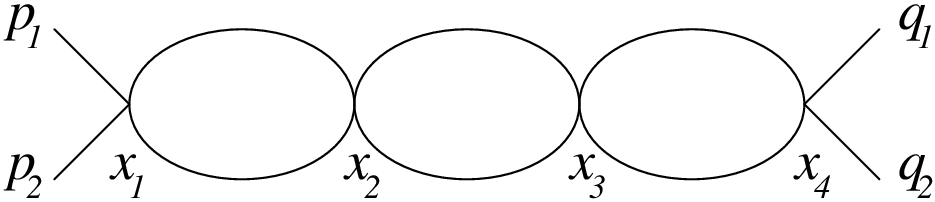}}
\vspace{4mm}
\end{figure}

\vspace{-2mm}

\noindent contributes to $\langle q_{1}q_{2}| i\T | p_{1}p_{2}\rangle$ by

\vspace{2mm}

\begin{eqnarray}
\int \prod_{i=1}^{4}dx_{i} \frac{e^{-i(p_{1}+p_{2})x_{1}}}{V\sqrt{4
\omega_{p_{1}} \omega_{p_{2}}}} \frac{e^{i(q_{1}+q_{2})x_{4}}}{V \sqrt{4
\omega_{q_{1}}
\omega_{q_{2}}}}(i\Delta_{F}(x_{1}-x_{2}))^{2}(i\Delta_{F}(x_{2}-x_{3}))^{2}
(i\Delta_{F}(x_{3}-x_{4}))^{2},\nonumber
\end{eqnarray}

\vspace{2mm}

\noindent similarly, the function $F_{2_{4}}(x_{1}, \ldots, x_{4})$
corresponding to the diagram

\begin{figure}[h]
\epsfxsize=7cm
\centerline{\epsffile{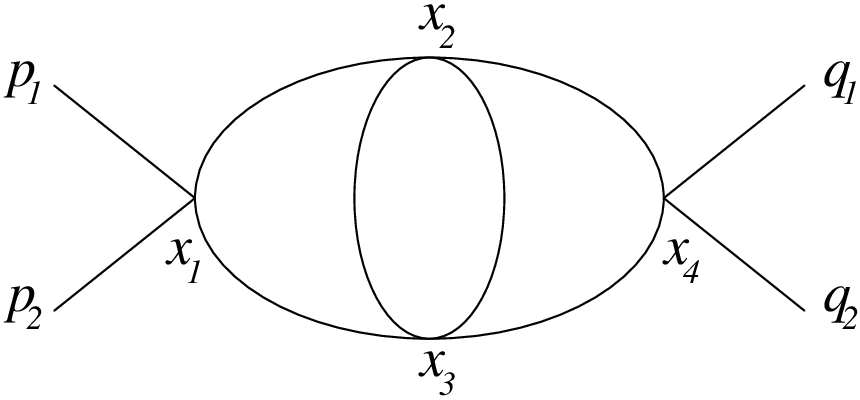}}
\end{figure}

\vspace{3mm}

\noindent contributes to $\langle q_{1}q_{2}| i\T | p_{1}p_{2}\rangle$ by

\begin{eqnarray*}
&&\int \prod_{i=1}^{4}dx_{i} \frac{e^{-i(p_{1}+p_{2})x_{1}}}{V\sqrt{4
\omega_{p_{1}} \omega_{p_{2}}}} \frac{e^{i(q_{1}+q_{2})x_{4}}}{V \sqrt{4
\omega_{q_{1}}
\omega_{q_{2}}}}i\Delta_{F}(x_{1}-x_{2})i\Delta_{F}(x_{1}-x_{3})
(i\Delta_{F}(x_{2}-x_{3}))^{2}\\
&&\hspace{5.5cm}\times ~i\Delta_{F}(x_{4}-x_{3})i\Delta_{F}
(x_{4}-x_{2}),\\
&&\mbox{\hspace{-16.5mm}etc.}
\end{eqnarray*}

\noindent This can be summarised as

\begin{equation}
\langle \{q_{k}\} | i\T | \{p_{j}\} \rangle = \sum_{n} \int \ldots \int
\prod_{i=1}^{n}dx_{i} \sum_{l_{n}}\prod_{j}
\frac{e^{-ip_{j}x_{m_{j}}}}{\sqrt{2{\omega}_{p_{j}}V}} \prod_{k}
\frac{e^{iq_{k}x_{m_{k}}}}{\sqrt{2{\omega_{q_{k}}}V}}F_{l_{n}}(x_{1},
\ldots, x_{n}). \tseleq{Tmatrix} 
\end{equation}

\vspace{2mm}

\noindent Consider now the case $|\{p_{j}\} \rangle = |\{q_{k}\} \rangle$
(let us call it $| a \rangle $). From the unitarity condition:
$\T-{\T}^{\dag} = i{\T}^{\dag}\T$, we get

\begin{equation}
\langle a|\T |a \rangle - \langle a|\T|a \rangle^{*} = i \langle a|
{\T}^{\dag}\T|a \rangle. \tselea{LTE45} \end{equation}

\vspace{2mm}

\noindent On the other hand, by construction $F({\underline{x_{1}}}, \ldots,
{\underline{x_{n}}})= F^{*}(x_{1}, \ldots, x_{n})$, and thus (see
(\tseref{LTE19}))

\begin{equation}
F(x_{1}, \ldots, x_{n}) + F^{*}(x_{1}, \ldots, x_{n}) = -
\sum_{index^{'}}F(x_{1}, \ldots, x_{n}).   
\tseleq{CE1}
\end{equation}

\noindent The prime over {\em index} in (\tseref{CE1}) indicates that we
sum neither over diagrams with all type 1 vertices nor diagrams with all
type 2 vertices. Using (\tseref{Tmatrix}), and identifying
$|\{q_{k}\}\rangle$ with $|\{p_{k}\}\rangle$ ($ = |a \rangle$) we get

\begin{equation}
\langle a|\T|a \rangle - \langle a|\T|a \rangle^{*} = - \sum_{index^{'}}
\langle a|\T|a \rangle,
\tseleq{CE4}
\end{equation}  

\vspace{-3mm}

\noindent or (see (\tseref{LTE45}))
\vspace{-2mm}

\begin{equation}
\langle a |{\T}^{\dag}\T| a \rangle = i\sum_{index^{'}}
\langle a |\T| a \rangle. \tseleq{CE5}
\end{equation}
\vspace{1mm}

\noindent Eq.(\tseref{CE4}) is the special case of the LTE for the
$\T$--matrices. The finite--temperature extension of (\tseref{CE5}) will prove
crucial in Section \ref{MCD}.

\begin{figure}[h]
\epsfxsize=9.5cm
\centerline{\epsffile{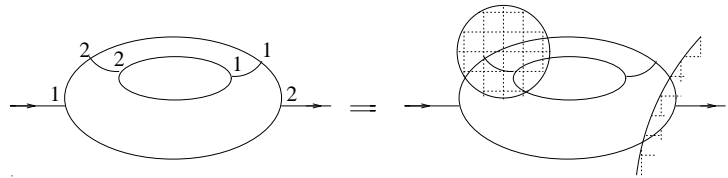}}
\caption{{ An example of a cut
diagram in the $\varphi ^{3}$ theory which does not contribute to the RHS's
of (\tseref{CE4})--(\tseref{CE5}).  Arrows indicate the flow of
energy.}}
\label{fig.2}
\end{figure}

Owing to the $\theta(\pm k_{0})$ in $\Delta^{\pm}(x)$ (see
Eq.(\tseref{decomp})), energy is forced to flow only towards type 2  
vertices. From both the energy--momentum conservation in each vertex and
from the energy flow on the external lines, a sizable class of the diagrams
on the RHS's of (\tseref{CE4})-(\tseref{CE5}) will be automatically zero.
Particularly regions of either 1st or 2nd type vertices which are not
connected to any external line violate the energy conservation and thus do
not contribute (no islands of vertices), see FIG.\ref{fig.2}.
Consequently, the only surviving diagrams are those whose any 1st type
vertex area is connected to incoming particles and any 2nd type vertex
area is connected to outgoing ones. From historical reasons the border
between two regions with different type of vertices is called {\em cut}
and corresponding diagrams are called {\em cut diagrams}.

\vspace{3mm}

\begin{figure}[h]
\vspace{3mm}
\epsfxsize=5cm
\centerline{\epsffile{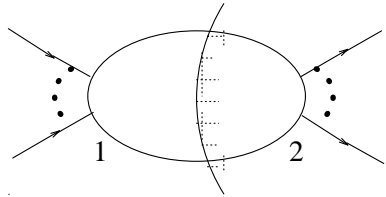}}
\caption{{ Generic form of the cut diagram at the $T=0$. Shadow is on the
2nd type vertex area. }} \label{fig4}
\vspace{3mm}
\end{figure}

\noindent We have just proved a typical feature of $T=0 $ QFT, namely
any cut diagram is divided by the cut into two areas only, see
FIG.\ref{fig4}. Eq.(\tseref{CE4}), rewritten in terms of the cuttings is so
called {\em cutting equation} (or Cutkosky's cutting rules) \cite{PN,
V, TV}.

\vspace{2mm}

\noindent One point should be added. Inserting the completeness relation
$\sum_{f} |f\rangle \langle f| =1$ into the LHS of (\tseref{CE5}), we get

\begin{equation}
\sum_{f}\langle a|{\T}^{\dag}|f \rangle \langle f|\T|a \rangle = i
\sum_{cuts} \langle a|\T|a \rangle.
\tselea{MLTE}
\end{equation}

\noindent It may be shown \cite{PL2, PN} that all the intermediate
particles in $|f \rangle$ correspond to cut lines. This has a natural
extension when $\langle a|{\T}^{\dag}\T|a \rangle \rightarrow \langle
a|{\T}^{\dag}\P \,\T|a \rangle$ with $\P$ being a projection
operator (${\P}={\P}^{\dag}={\P}^{2}$) which eliminates
some of the states $|f \rangle$. It is easy to see that in such case

\begin{equation}
\langle a|{\T}^{\dag}{\P}\,\T| a \rangle = i{\tilde{\sum_{cuts}}}
\langle a|\T|a \rangle,
\tselea{MLTE1}
\end{equation}

\vspace{2mm}

\noindent where tilde over the $\sum_{cuts}$ indicates that one sums over
the diagrams which do not have the cut lines corresponding to particles
removed by $\P$.

There is no difficulty in applying the previous results to
spin--$\frac{1}{2}$ \cite{PN, V}. The LTE follows as before: the diagram
with only $iS_{F}$ propagators (and $-ig$ per each vertex) plus the
diagram with only $\hat{(iS_{F})}$ propagator {\footnote{The function
$i{\hat{S}}_{F}(x)$, similarly as $(i\Delta_{F})^{*}(x)$, interchanges the
r{\^{o}}le $S^{+}$ and $S^{-}$. Unlike bosons, for fermions
$i{\hat{S}}_{F}(x)$ is not equal to $(iS_{F})^{*}(x)$. Despite that,
Eq.(\tseref{CE4}) still holds \cite{PN}.}} (and $ig$ per
each vertex) equals to minus the sum of all diagrams with one up to $n-1$
the type 2 vertices ($n$ being the total number of vertices). For gauge
fields more care is needed. Using the Ward identities one can show   
\cite{PN} that type 1 and type 2 vertices in
(\tseref{CE4})--(\tseref{CE5}) may be mutually connected only by {\em
physical particle} propagators (i.e. neither through the propagators
corresponding to particles with non--physical polarisations or Fadeev--Popov
ghosts and antighosts).

\vspace{3mm}

\subsection{Thermal Wick's theorem (the Dyson--Schwinger equation) \label{bt3}}

The key observation at finite temperature is that for systems of {\em
non--interacting} particles in thermodynamical equilibrium Wick's theorem is
still valid, i.e. one can decompose the $2n$--point (free) thermal Green
function into a product of two--point (free) thermal Green functions. This
may be defined recursively by

\begin{equation} \langle {\cali{T}} ( \psi (x_{1}) \ldots \psi (x_{2n}) )
{\rangle}= \sum_{\stackrel{j}{j \not= i}} \varepsilon_{P}\langle  
{\cali{T}} ( \psi (x_{i})\psi (x_{j}) ) {\rangle}~\langle 
{\cali{T}} ( \prod_{k \not= i;j}\psi(x_{k})) {\rangle},
\tseleq{wick} \end{equation}

\noindent where $\varepsilon_{P}$  is the signature of the permutation
of fermion operators ($=1$  for  boson operators) and ${\cali{T}}$  is
the  standard   time ordering  symbol     (for generalisation to   the
contour--time--path ordering see Appendix   \ref{A1}).  Note that   the
choice  of ``$i$" in  (\tseref{wick})   is completely arbitrary.   The
proof can be found for example in \cite{LB,Mills,Evans}.  Similarly as
at $T=0$, Wick's  theorem can also  be written for the (free)  thermal
Wightman functions \cite{Mills, IZ}, i.e. 

\begin{equation}
\langle \psi (x_{1}) \ldots \psi (x_{2n})
{\rangle} =
\sum_{\stackrel{j}{j \not= 1}} \varepsilon_{P}\langle \psi (x_{1})\psi
(x_{j}) {\rangle} ~  \langle
\prod_{k \not= 1;j}\psi(x_{k}) \rangle.
\tseleq{wick2}
\end{equation}

\noindent  A particularly advantageous form of   this is the so called
Dyson--Schwinger equation (see Appendix \ref{A1})   which, at the   $T
\not= 0$, reads 

\begin{equation} \langle G[\psi] \psi(x) F[\psi] \rangle = \int dz
\langle \psi(x) \psi(z) \rangle \left\langle G[\psi]
\frac{{\stackrel{\rightarrow}{\delta}}
F[\psi] }{\delta \psi(z)} \right\rangle + \int dz \langle \psi(z)
\psi(x) \rangle \left\langle \frac{
G[\psi]\stackrel{\leftarrow}{\delta}}{\delta
\psi(z)} F[\psi] \right\rangle, \tseleq{S-D2} \end{equation}

\vspace{2mm}

\noindent  where   $\psi(x)$  is  an  interaction--picture  field  and
$G[\ldots]$ and $F[\ldots]$ are   functionals of $\psi$. The   arrowed
variations   $\frac{\delta}{\delta \psi(z)}$  are   defined  as formal
operations satisfying two conditions, namely:

\begin{equation}  
\frac{\stackrel{\rightarrow}{\delta}}{\delta \psi_{n}(z)}(\psi_{m}(x)
\psi_{q}(y)) =
\frac{\delta \psi_{m}(x)}{\delta
\psi_{n}(z)}\psi_{q}(y)+(-1)^{p}\psi_{m}(x)\frac{\delta
\psi_{q}(y)}{\delta \psi_{n}(z)}, \nonumber
\tseleq{var1}
\end{equation}

\noindent or

\vspace{-2mm}

\begin{equation}
(\psi_{m}(x) \psi_{q}(y))\frac{\stackrel{\leftarrow}{\delta}}{\delta
\psi_{n}(z)} = (-1)^{p}\frac{\delta \psi_{m}(x)}{\delta
\psi_{n}(z)}\psi_{q}(y)+\psi_{m}(x)\frac{\delta
\psi_{q}(y)}{\delta
\psi_{n}(z)}, \nonumber
\tseleq{var12}
\end{equation}

\noindent with

\vspace{-2mm}

\begin{equation}  
\frac{\delta \psi_{m}(x)}{\delta \psi_{n}(y)} =
\delta(x-y)\delta_{mn}.\nonumber \tseleq{var2}
\end{equation}

\vspace{2mm}

\noindent The ``$p$'' is $0$ for bosons and $1$ for fermions; subscripts
$m,n$ suggest that several types of fields can be generally present. Note,
for bosons $\frac{{\stackrel{\rightarrow}{\delta}}F}{\delta \psi}
=\frac{F\stackrel{\leftarrow}{\delta}}{\delta \psi}$ which we shall denote
as $\frac{\delta F}{\delta \psi}$. For more details see Appendix \ref{A1}.

\vspace{4mm}

\subsection{Thermal largest--time equation \label{bt4}}

The LTE (\tseref{CE5}) can be extended to the finite--temperature case,
too.  Summing up in (\tseref{CE5}) over all the eigenstates of $K$ ($= H-\mu
N$) with the weight factor $e^{-\beta K_{i}}$ ($i$ labels the
eigenstates), we get

\begin{equation}
\langle \T{\T}^{\dag} \rangle = i \sum_{index^{'}} \langle \T
\rangle.
\tseleq{TLTE2}
\end{equation}    

\vspace{2mm}

\noindent Let us consider the RHS of (\tseref{TLTE2}) first. The
corresponding thermal LTE and diagrammatic rules (Kobes--Semenoff rules
\cite{LB}) can be derived precisely the same way as at $T=0$ using
the previous, largest--time argumentation \cite{LB, KS}. It turns out 
that these rules have basically identical form as those in the previous   
section, with an exception that now $\langle 0 | \ldots |0 \rangle
\rightarrow \langle \ldots \rangle$. Note that labelling vertices  
by 1 and 2 we have naturally got a doubling of the number of degrees of
freedom. This is a typical feature of the {\em real--time formalism} in
thermal QFT (here, in so called {\em Keldysh version} \cite{LB}).

We should also emphasise that it may happen some fields are not   
thermalized. For example, external particles entering a heat bath or
particles describing non--physical degrees of freedom \cite{LR}.
Particularly, if some particles (with momenta $\{ p_{j} \}$) enter the
heat bath, the mean statistical value of an observable $A$ is then

\begin{eqnarray}
\sum_{i} \frac{e^{-\beta K_{i}}}{Z} \langle i; \{ p_{j} \}|A|i; \{ p_{j}
\} \rangle &=& Z^{-1} Tr(\rho_{\{ p_{j} \}} \otimes e^{-\beta K}
A),\nonumber\\    
\rho_{\{ p_{j} \}} &=& |\{ p_{j} \}\rangle \langle \{ p_{j} \} |,\nonumber
\end{eqnarray}

\noindent which we shall denote as $\langle A \rangle_{\{ p_{j} \}}$.
From this easily follows the generalisation of (\tseref{TLTE2})

\begin{equation}
\langle \T{\T}^{\dag} \rangle_{\{p_{k}\}} = i \sum_{index^{'}} \langle \T
\rangle_{\{p_{k}\}}. \tseleq{TLTE3}
\vspace{1mm}
\end{equation}

\noindent Unlike  $T=0$,  we   find   that  the cut    diagrams   have
disconnected vertex areas   and no kinematic  reasonings   used in the
previous section can, in general, get rid of them. This is because the
thermal part   of  $\langle   \varphi(x)  \varphi(y)  \rangle$
describes\footnote{Note that $\langle  \varphi(x) \varphi(y) \rangle
 = \langle : \varphi(x)  \varphi(y): \rangle + \langle
0|   \varphi(x) \varphi(y)  |0  \rangle  $ and   $\langle : \varphi(x)
\varphi(y)          :        \rangle            =         \int
\frac{d^{4}k}{(2\pi)^{3}}f_{B}(k_{0})        \delta      (k^{2}-m^{2})
e^{-ik(x-y)}$,  with $f_{B}(k_{0})=(e^{\beta |k_{0}|} -1)^{-1}. $} the
absorption of on shell particle from the  heat bath or the emission of
one into it. Thus, at $T \not = 0 $, there is no definite direction of
transfer of energy from type 1 vertex to type 2 one as energy flows in
both directions. Some  cut diagrams nevertheless  vanish. It is simple
to see that only those  diagrams survive in which the non--thermalized
external  particles ``enter'' a diagram via  the 1st type vertices and
``leave''  it via the 2nd  type  ones. We might   deduce this from the
definition of $\langle \T \rangle_{\{ p_{j} \}}$, indeed 

\begin{equation}
\sum_{index^{'}} \langle \T \rangle_{\{ p_{j} \}}= \sum_{index^{'}}
\sum_{i} \frac{e^{-\beta K_{i}}}{Z}
\langle i;\{ p_{j} \}|\T|i; \{ p_{j} \} \rangle.\nonumber
\end{equation}
\vspace{0.5mm}

\noindent Note, we get the same set of thermal cut diagrams interchanging
the summation $\sum_{index^{'}}$ with $\sum_{i}$. It is useful to start
then with $\sum_{index^{'}} \langle i;\{ p_{j} \}|\T|i; \{ p_{j} \}   
\rangle$.  This is, as usual, described by the ($T=0$) cutting rules. In
the last section we learned that the general structure of the corresponding
cut diagrams is depicted in FIG.\ref{fig4}, particularly the external
particles enter the cut diagram via type 1 vertices and leave it via type
2 ones. Multiplying each diagram (with the external particles in the state
$|i; \{p_{j}\} \rangle$) with the pre--factor $\frac{e^{-\beta K_{i}}}{Z}$ 
and summing subsequently over $i$, we again retrieve the thermal cut
diagrams, though now it becomes evident that the particles $\{ p_{j} \}$
enter such diagram only via type 1 vertices and move off only through type
2 ones, since the summation of the ($T=0$) cut diagrams from which it was 
derived does not touch lines corresponding to unheated particles. Note,
the latter analysis naturally explains why the unheated particles obey the
($T=0$) LTE diagrammatic rules even in the thermal diagrams

\begin{figure}[h]
\epsfxsize=15.3cm
\centerline{\epsffile{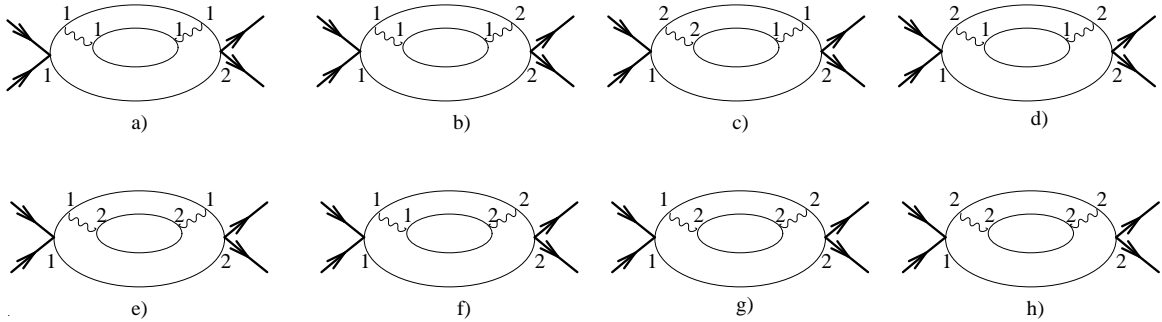}}
\caption{{ An example of non--vanishing cut diagrams at the $T\not =0$.  
The heat--bath consists of two different particles. External particles are
not thermalized.}}
\label{fig.3}
\end{figure}

\begin{figure}[h]  
\epsfxsize=15.3cm
\centerline{\epsffile{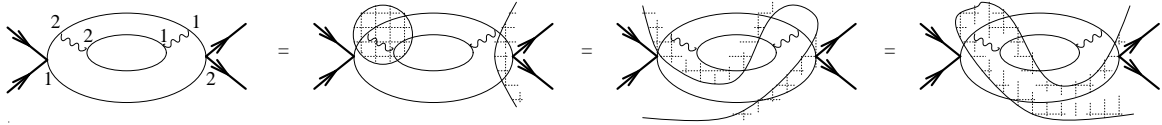}} 
\caption{{ The cut diagram from   
Fig.\ref{fig4} c) demonstrates that the cut can be defined in many ways but
the number of crossed lines is still the same.}} 
\label{61} 
\vspace{3mm}
\end{figure}

Another vanishing  comes  from kinematic reasons.   Namely  three--leg
vertices with all the on shell particles (1--2 lines) can not conserve
energy--momentum and consequently  the whole cut diagram  is zero.  As
an illustration let  us consider all the non--vanishing, topologically
equivalent cut  diagrams of   given  type involved  in  a  three--loop
contribution to $i\sum_{index^{'}} \langle \T \rangle_{pq}$ (see
FIG.\ref{fig.3})\footnote{Let us emphasise that  originally we had the
64 possible cut diagrams.}. Let us  stress one more point. In contrast
with $T=0$, at finite temperature the cut itself neither is unique nor
defines topologically   equivalent areas,  see FIG.\ref{61},  only the
number of crossed legs  is, by definition, invariant.  This  ambiguity
shows  that  the  concept of  the  cut  is not very  useful  at finite
temperature and in the following we  

\vspace{3mm} 
\vspace{2mm}

\noindent shall refrain from using it.  

In Section \ref{MCD} it
will prove useful to have  an analogy of (\tseref{TLTE3}) for $\langle
{\T}^{\dag}{\P}\,\T  \rangle$. Here $\P$
\vspace{5mm}

\vspace{2mm}

\noindent is the projection operator defined as 

\begin{equation}
{\P} = \sum_{j} |a;j\rangle  \langle a;j|\, ,
\tseleq{proj}
\vspace{1mm}
\end{equation}

\noindent where ``$j$'' denotes the physical states for the heat--bath
particles and ``$a$'' labels the physical states for the outgoing,
non--thermalized particles. Let us deal with $\langle {\T}^{\dag}{\P}\,\T
\rangle$. Using (\tseref{MLTE1}), we acquire

\vspace{-1mm}
\begin{equation} \langle {\T}^{\dag}{\P}\, \T\rangle = i\sum_{l}   
\frac{e^{-\beta K_{l}}}{Z} {\tilde{\sum_{index^{'}}}} \langle l|\T|l
\rangle \, . 
\end{equation}

\noindent Interchanging the summations, we finally arrive at

\begin{equation}
\langle {\T}^{\dag} {\P} \,\T \rangle = i{\tilde{\sum_{index^{'}}}}
\langle \T \rangle,
\tseleq{proj2}
\vspace{1mm}
\end{equation}

\noindent where tilde over the $\sum_{index^{'}}$ means that we are
restricted to consider the cut diagrams, with only (1--2)--particle lines
corresponding to the ``$a$'' and ``$j$''particles (i.e. $ \langle 0|
\varphi(x) \varphi(y) |0 \rangle$ and $\langle \psi(x) \psi(y)
\rangle$, respectively). The extension of Eq.(\tseref{proj2}) to
the case where some external, non--thermalized particles $\{p_{k}\}$ are
present is obvious, and reads

\begin{equation}
\langle {\T}^{\dag} {\P} \,\T \rangle_{\{p_{k}\}} =
i{\tilde{\sum_{index^{'}}}} \langle \T \rangle_{\{p_{k}\}}.
\tseleq{proj4}
\end{equation}

\noindent Finally, let us note that using the LTE, one may extend the
previous treatment to various Green functions. The LTE
for Green's functions is then a useful starting point for dispersion
relations, see e.g.  \cite{LB, KS}.

\vspace{5mm}

\section{Heat--Bath particle number spectrum:\\
 general framework \label{HBP}}

The cutting equation (\tseref{proj4}) can be fruitfully used for both the
partition function $Z$ and the heat--bath particle number spectrum
$d\langle N(\omega) \rangle /d \omega $ calculations. To see that,
let us for simplicity assume that two particles (say $\Phi_{1}, \Phi_{2}$)
scatter inside a heat bath. We are interested in the heat--bath number
spectrum after two different particles (say $\phi_{1}, \phi_{2})$ appear in
the final state. Except for the condition that the external particles are
different from the heat bath ones, no additional assumption about their 
nature is needed at this stage.

The initial density matrix $\rho_{i}$ (i.e. the density matrix
describing the physical situation before we introduce the particles
$\Phi_{1}(p_{1}), \Phi_{2}(p_{2})$ into the oven) can be written as

\begin{equation} \vspace{2mm} \rho_{i} = Z_{i}^{-1} \sum_{j} e^{-\beta
K_{j}}|j;p_{1}, p_{2} \rangle \langle j; p_{1}, p_{2}|, \tseleq{rhoi}
\end{equation}

\noindent where ``$j$'' denotes the set  of occupation numbers for the
heat--bath particles.  A  long time after   the  scattering the  final
density matrix $\rho_{f}$ reads 

\begin{equation}   
\rho_{f} = Z_{f}^{-1} \sum_{j} e^{-\beta K_{j}} {\P}\, \S|j; p_{1}, p_{2}
\rangle \langle j; p_{1}, p_{2}|{\S}^{\dag} {\P}^{\dag},
\tseleq{rhof}
\vspace{1mm}
\end{equation}

\noindent here ${\P}$ is the projection operator projecting out all  
the non--heat--bath final states except of $\phi_{1}(q_{1}),
\phi_{2}(q_{2})$ ones. The $\S$--matrix in (\tseref{rhof}) is defined in a
standard way: $\S=\1+i\T$. The $Z_{f}$ in (\tseref{rhof}) must be different
from $Z_{i}$ as otherwise $\rho_{f}$ would not be normalised to unity. In
order that $\rho_{f}$ satisfy the normalisation condition $Tr(\rho_{f}) =
1$, one finds

\begin{equation}
\vspace{1mm}
Z_{f}= \sum_{j} e^{-\beta K_{j}} \langle j; p_{1}, p_{2}|{\S}^{\dag} {\P}
\,\S|j; p_{1},
p_{2} \rangle = {}\langle {\S}^{\dag}{\P}\,\S \rangle_{p_{1}
p_{2}} ~ Z_{i}  = \langle {\T}^{\dag}{\P} \,\T \rangle_{p_{1}
p_{2}} ~ Z_{i}.
\tseleq{Zf}
\end{equation}

\noindent   The key point  is that  we have  used in (\tseref{Zf}) the
$\T$--matrix because   the initial state   $|\Phi_{1} (p_{1}), \Phi_{2}
(p_{2})  \rangle$  is, by  definition,  different from  the  final one
$|\phi    (q_{1}),\phi_{2}    (q_{2})    \rangle$  and    consequently
${\P}\,\S$  can be replaced  by  $i{\P}\,\T$. This allows us  to
calculate $Z_{f}$ using directly  the diagrammatic technique  outlined
in the preceding section.

From (\tseref{mean value}) and (\tseref{rhof}) one can directly
read off that the number spectrum of the heat--bath particles is:

\begin{eqnarray} \frac{d \langle N_{l}( \omega )\rangle_{f}}{ d\omega} & = &
\int \frac{ d^{3}{\vect{k}}}{(2 \pi)^{3}} {\delta}^{+}(
{\omega}^{2} - {\vect{k}}^{2} - m_{l}^{2})
\sum_{f} \langle f | a^{\dag}_{l}({\vect{k}}; \omega)a_{l}({\vect{k}};
\omega) \rho_{f}| f \rangle \nonumber  \\
 & = & \int \frac{d^{3}{\vect{k}}}{(2 \pi)^{3}} \delta^{+}(
\omega^{2}-{\vect{k}}^{2} -m_{l}^{2}) \frac{{}\langle
{\T}^{\dag}{\P}
a^{\dag}_{l}({\vect{k}}; \omega)a_{l}({\vect{k}}; \omega) \T  
\rangle_{p_{1} p_{2}}}{{} \langle {\T}^{\dag} {\P}\,\T \rangle_{p_{1} p_{2}}},
\tselea{dN}
\end{eqnarray}

\noindent and consequently

\begin{equation} \langle N_{l} \rangle_{f} = \int \frac{d^{4}k}{(2
\pi)^{3}} \delta^{+}(k^{2}-m_{l}^{2})  \frac{{}\langle
{\T}^{\dag}{\P}
a^{\dag}_{l}(k)a_{l}(k) \T \rangle_{p_{1} p_{2}}}{{} \langle {\T}^{\dag} {\P}\,\T \rangle_{p_{1} p_{2}}},
\vspace{1mm}
\tseleq{N}
\end{equation}

\noindent where we have used the completeness relation for the final
states $|f \rangle$ and $[{\P};a^{\dag}a]=0$. The subscript ``$l$''
denotes which type of heat--bath particles we are interested in. In the
following the index will be mostly suppressed.

\vspace{5mm}

\section{Modified cut diagrams \label{MCD}}

To proceed further with (\tseref{dN}) and (\tseref{N}), we expand the
$\T$--matrix in terms of time--ordered  interaction--picture fields, i.e.

\begin{equation} \T[\psi] = \sum_{n} \int dx_{1} \ldots \int dx_{n}
\alpha_{n}(x_{1}\ldots x_{n}) {\cali{T}}(\psi(x_{1})\ldots \psi(x_{n})).
\tseleq{Hin} \vspace{1mm} \end{equation}

\noindent Here $\psi$ represents a heat--bath field in the interaction
picture. Other fields (i.e. ${\overline{\phi}}$,$\phi$ and $\Phi$) are
included\footnote{ When Fermi  fields are involved,  we  have, for the
sake  of  compactness,   included  in  the   argument of  $\psi$   the
space--time coordinate,  the Dirac index, and  a  discrete index which
distinguishes  $\psi_{\alpha}$ from ${\overline{\psi}}_{\alpha}$.}  in
the  $\alpha_{n}$. An extension of   (\tseref{Hin}) to the case  where
different   heat--bath  fields are    present  is natural.   Employing
(\tseref{Hin}) in $\langle {\T}^{\dag}{\P} \, \T\rangle_{p_{1}p_{2}}$,
one can readily  see  that this factorises  out  in each term   of the
expansion a {\em pure}  thermal  mean value $\langle \ldots  \rangle$.
The general structure of each   such thermal mean value is:  ${\langle
G_{m}[\psi] F_{n}[\psi]  \rangle}$, where $F_{n}[\ldots]$  and
$G_{m}[\ldots]$ are the operators with ``$n$" chronological and ``$m$"
anti--chronological     time      ordered       (heat--bath)   fields,
respectively. Analogous factorisation is true in the expansion of $~{}
\langle   {\T}^{\dag}{\P}^{\dag}a^{\dag}     a          {\P}      \,\T
\rangle_{p_{1}p_{2}}$. The only   difference is that the  pure thermal
mean value has  the  form $\langle G_{m}[\psi]  a^{\dag}a  F_{n}[\psi]
\rangle$  instead\footnote{Remember   that  ${\P}=  {\P}^{'}  \otimes
{\P}^{''} =   |q_{1},   q_{2}\rangle \langle  q_{1},    q_{2}| \otimes
\sum_{j}|j\rangle \langle  j|$.  Here ${\P}^{''}=  \sum_{j} |j \rangle
\langle    j|$ behaves as  an identity   in  the subspace of heat--bath
states.}. In   case when various heat-bath fields   are present,  $m =
m_{1}+m_{2}+ \ldots  +m_{n}$, with ``$m_{l}$''  denoting the number of
the heat--bath fields of $l$--th type.

Applying the Dyson--Schwinger equation to $\langle
G_{m}[\psi]a^{\dag}a F_{n}[\psi] \rangle$ twice and summing over 
``$n$" and ``$m$", we get cheaply the following expression (c.f. also
(\tseref{S-D8}))

\vspace{-2mm}
\begin{eqnarray}
\lefteqn{\langle {\T}^{\dag}{\P} a_{l}^{\dag}a_{l} \T
\rangle_{p_{1}p_{2}} =}\nonumber \\ &= & \int
dxdy \{ \langle \psi_{l}(x)a_{l}^{\dag} \rangle \langle
a_{l} \psi_{l}(y) \rangle + (-1)^{p}\langle \psi_{l}(x) a_{l} 
\rangle \langle a_{l}^{\dag} \psi_{l}(y) \rangle\}\,
\left\langle
\frac{{\T}^{\dag} {\stackrel{\leftarrow}{\delta}}}{\delta \psi_{l}(x)}{\P}
\frac{{\stackrel{\rightarrow}{\delta}} \T}{\delta \psi_{l}(y)}
\right\rangle_{p_{1} p_{2}} 
\nonumber \\
&+& \int
\frac{dxdy}{2}\{\langle \psi_{l}(x)a_{l} \rangle\langle
\psi_{l}(y)a_{l}^{\dag} \rangle +(-1)^{p} \langle \psi_{l}(x) a_{l}^{\dag}
\rangle\langle \psi_{l}(y)a_{l}
\rangle\} \left\langle
\frac{{\T}^{\dag}{\stackrel{\leftarrow}{\delta^{2}}}}{\delta \psi_{l}(y) \delta
\psi_{l}(x)}{\P}\,\T
\right\rangle_{p_{1} p_{2}} \nonumber\\ &+& \int
\frac{dxdy}{2}\{\langle a_{l}\psi_{l}(x) \rangle \langle
a_{l}^{\dag} \psi_{l}(y) \rangle + (-1)^{p} \langle a_{l}^{\dag}
\psi_{l}(x) \rangle \langle a_{l}\psi_{l}(y)
\rangle \} \left\langle
{\T}^{\dag} {\P}\frac{{\stackrel{\rightarrow}{\delta^{2}}}\T}{\delta
\psi_{l}(y) \delta \psi_{l}(x)} \right\rangle_{p_{1} p_{2}}
\nonumber\\  
&+& \langle a_{l}^{\dag}a_{l} \rangle \langle {\T}^{\dag}{\P}\,\T
\rangle_{p_{1}p_{2}}, \tselea{average1}
\end{eqnarray}

\noindent  A similar   decomposition for $\langle  {\T}^{\dag}{\P}\,\T
\rangle_{p_{1}p_{2}}$  would not be very useful (cf.(\tseref{S-D13}));
instead     we   define        $\langle      ({\T}^{\dag}{\P}\,\T)^{'}
\rangle_{p_{1}p_{2}}$  having the    same  expansion    as    $\langle
{\T}^{\dag}{\P}\,\T \rangle_{p_{1}p_{2}}$   except         for     the
$\alpha_{n}(\ldots){\P}  \alpha_{m}^{\dag}(\ldots)$  are  replaced  by
$\alpha_{n}(\ldots){\P}\alpha_{m}^{\dag}(\ldots)\frac{n_{l}+m_{l}}{2}$.
In             this            formalism           $\langle          (
{\T}^{\dag}{\P}\,\T)^{'}\rangle_{p_{1}p_{2}}$ reads 

\vspace{-2mm}
\begin{eqnarray}
\lefteqn{\langle ({\T}^{\dag}{\P} \,\T)^{'} \rangle_{p_{1}p_{2}}}
\nonumber\\ &=& \int dxdy {\langle
\psi_{l}(x)\psi_{l}(y) \rangle} \left\langle   
\frac{{\T}^{\dag} {\stackrel{\leftarrow}{\delta}}}{\delta \psi_{l}(x)}{\P}
\frac{{\stackrel{\rightarrow}{\delta}} \T}{\delta \psi_{l}(y)}
\right\rangle_{p_{1} p_{2}} \nonumber\\ &+& \int
\frac{dxdy}{2}{\langle {\overline{\cali{T}}} ( \psi_{l}(x)\psi_{l}(y) )
\rangle} \left\langle
\frac{{\T}^{\dag}{\stackrel{\leftarrow}{\delta^{2}}}}{\delta \psi_{l}(y) \delta
\psi_{l}(x))}{\P}\,\T \right\rangle_{p_{1}p_{2}}
\nonumber\\ &+& \int
\frac{dxdy}{2}\langle {\cali{T}} ( \psi_{l}(x)\psi_{l}(y) )
\rangle \left\langle
{\T}^{\dag} {\P}\frac{{\stackrel{\rightarrow}{\delta^{2}}}\T}{\delta
\psi_{l}(y) \delta \psi_{l}(x)} \right\rangle_{p_{1}p_{2}},     
\tselea{average2}
\end{eqnarray}

\vspace{1mm}

\noindent    with     the     ${\overline{\cali{T}}}$     being    the
anti--chronological   ordering  symbol.  Comparing (\tseref{average2})
with (\tseref{S-D12}), we can interpret the RHS of (\tseref{average2})
as a   sum   over {\em  all}  possible   distributions  of  one   line
(corresponding to $\psi_{l}$) inside of  each given ($T\not= 0$ !) cut
diagram  constructed     out    of   $\langle      {\T}^{\dag}{\P}\,\T
\rangle_{p_{1}p_{2}}$. As (\tseref{average2})  has precisely the  same
diagrammatical structure as 

\begin{displaymath}
\langle          {\T}^{\dag}{\P}a^{\dag}a  \T
\rangle_{p_{1}p_{2}}        -\langle         a^{\dag}a
\rangle~\langle         {\T}^{\dag}{\P}\,\T
\rangle_{p_{1}p_{2}}
\end{displaymath}

\noindent (cf.(\tseref{average1})), it shows that
in  order   to  compute{\footnote{Here   $\frac{d\langle   N(\omega)
\rangle_{i}}{d\omega}     =     \int\frac{d^{3}{\vect{k}}}{(2\pi)^{3}}
\delta^{+}(\omega^{2}-{\vect{k}}^{2} - m^{2}) \langle a^{\dag}(\omega,
{\vect{k}})a(\omega,         {\vect{k}})             \rangle$,
(cf.    (\tseref{dN})).}} the  numerator    of $\frac{d\Delta  \langle
N(\omega)     \rangle}{d\omega}  =      \frac{d\langle       N(\omega)
\rangle_{f}}{d\omega}-\frac{d\langle  N(\omega) \rangle_{i}}{d\omega}$
one      can     simply        modify     the     usual       $\langle
{\T}^{\dag}{\P}\,\T\rangle_{p_{1}p_{2}}$ cut  diagrams  by the
following one--line replacements (cf.(\tseref{dN})). 

\vspace{4mm}

\noindent {\bf (i) For neutral scalar bosons:}

\begin{eqnarray}
\langle  \varphi(x)\varphi(y) \rangle &\rightarrow &
\int \frac{d^{3}{\vect{k}}}{(2\pi)^{3}}\; \delta^{+}(\omega^{2}
-{\vect{k}}^{2}-m^{2}) \{
\langle \varphi(x)a^{\dag}({\vect{k}};\omega) \rangle \langle
a({\vect{k}};\omega ) \varphi(y) \rangle\nonumber\\
 && +\; \langle \varphi(x)
a({\vect{k}};\omega) \rangle\langle
a^{\dag}({\vect{k}};\omega) \varphi(y) \rangle \}\nonumber \\
&&\mbox{\vspace{5mm}}\nonumber\\
&=& \int
\frac{d^{4}k}{(2\pi)^{3}}\; \delta(k^{2}-m^{2}) \left\{
f_{B}(\omega)(f_{B}(\omega)
+ 1) \right. \nonumber \\
&& \times
~(\delta^{-}(k_{0}+\omega)+\delta^{+}(k_{0}-\omega))\nonumber\\
&& +\left. \delta^{+}(k_{0}-\omega)(1 +
f_{B}(\omega)) - \delta^{-}(k_{0}+\omega)f_{B}(\omega) \right\}
e^{-ik(x-y)},
\tselea{new1}
\end{eqnarray}
\vspace{-1mm}

\noindent where $f_{B}(\omega)$ is the Bose--Einstein distribution:
$f_{B}(\omega)= \frac{1}{e^{\beta|\omega|}-1}$. Term
$\theta(-k_{0})f_{B}(\omega)$ describes the absorption of a heat--bath
particle, so reduces the number spectrum, that is why the negative sign
appears in front of it. Analogously,

\vspace{-1mm}

\begin{eqnarray}
\langle {\cali{T}} ( \varphi(x)\varphi(y) ) \rangle
&\rightarrow &
\int \frac{d^{3}{\vect{k}}}{(2\pi)^{3}}\;\delta^{+}(\omega^{2}
-{\vect{k}}^{2}-m^{2}) \{\langle a^{\dag}({\vect{k}};\omega) \varphi(x)
\rangle \langle a({\vect{k}};\omega)\varphi(y)
\rangle\nonumber\\
&&+ \;\langle a({\vect{k}};\omega)\varphi(x) \rangle \langle
a^{\dag}({\vect{k}};\omega) \varphi(y) \rangle\}\nonumber\\
&&\mbox{\vspace{5mm}}\nonumber\\
&=&
\int \frac{d^{4}k}{(2\pi)^{3}}
\delta(k^{2}-m^{2})(1 +
f_{B}(\omega))f_{B}(\omega)e^{-ik(x-y)}\nonumber\\
&& \times ~(\delta^{+}(k_{0}-\omega)+\delta^{-}(k_{0}+\omega)).
\tselea{new4} 
\end{eqnarray}

\vspace{2mm}

\noindent Similarly, for $\Delta \langle N \rangle$ one needs the
following replacements (cf.(\tseref{N}))

\begin{eqnarray}
\langle  \varphi(x)\varphi(y) \rangle &\rightarrow & \int
\frac{d^{4}k}{(2\pi)^{3}}\;\delta(k^{2}-m^{2})\left\{
f_{B}(\omega_{k})(f_{B}(\omega_{k}) +
1) \right.
\nonumber \\ &&\left. +~ \theta(k_{0})(1 + f_{B}(\omega_{k})) -
\theta(-k_{0})f_{B}(\omega_{k})\right\}e^{-ik(x-y)}\nonumber,\\
&&\mbox{\vspace{1cm}}\nonumber\\
\langle {\cali{T}} ( \varphi(x)\varphi(y) ) \rangle
&\rightarrow &
\int \frac{d^{4}k}{(2\pi)^{3}}\;
\delta(k^{2} - m^{2})(1 +
f_{B}(\omega_{k}))f_{B}(\omega_{k})e^{-ik(x-y)},\nonumber\\
\tselea{new2}
\end{eqnarray}

\vspace{2mm}

\noindent       with         the        dispersion            relation
$\omega_{k}=\sqrt{{\vect{k}}^{2}-m^{2}}$. 

\vspace{1cm}

\noindent {\bf (ii) For Dirac fermions:}

\vspace{3mm}

\noindent The Dirac field is comprised of two different types of
excitations (mutually connected via charge conjugation),
so the corresponding number operator $N(\omega) = N_{b}(\omega) +
N_{d}(\omega)$ with

\begin{eqnarray}
N_{b}(\omega) &=& \sum_{\alpha =1,2} \int
\frac{d^{3}{\vect{k}}}{(2\pi)^{3}}\;\delta^{+}(\omega^{2}
-{\vect{k}}^{2}-m^{2})
b_{\alpha}^{\dag}({\vect{k}};\omega)b_{\alpha}({\vect{k}};\omega)\nonumber\\
N_{d}(\omega) &=& \sum_{\alpha =1,2} \int
\frac{d^{3}{\vect{k}}}{(2\pi)^{3}}\;\delta^{+}(\omega^{2}
-{\vect{k}}^{2}-m^{2})
d_{\alpha}^{\dag}({\vect{k}};\omega)d_{\alpha}({\vect{k}};\omega).\nonumber
\end{eqnarray}

\noindent Thus, the one--line replacements needed for $d\Delta \langle
N_{b}(\omega) \rangle /d\omega$ are

\begin{eqnarray}
\langle  \psi_{\rho}(x){\overline{\psi}}_{\sigma}(y)
\rangle &\rightarrow & \sum_{\alpha =1,2} \int
\frac{d^{3}{\vect{k}}}{(2\pi)^{3}}\;\delta^{+}(\omega^{2}
-{\vect{k}}^{2}-m^{2}) \{
\langle \psi_{\rho}(x)b_{\alpha}^{ \dag}({\vect{k}};\omega) \rangle
\langle b_{\alpha}({\vect{k}};\omega ) {\overline{\psi}}_{\sigma}(y)
\rangle\nonumber\\
 && - \; \langle \psi_{\rho}(x)
b_{\alpha}({\vect{k}};\omega) \rangle \langle
b_{\alpha}^{\dag}({\vect{k}};\omega) {\overline{\psi}}_{\sigma}(y)
\rangle \}\nonumber \\ &&\mbox{\vspace{5mm}}\nonumber\\
&=& \; \int \frac{d^{4}k}{(2\pi)^{3}}\; \delta^{+}(k^{2}-m^{2})\; \delta
(k_{0}-\omega)\;(\not k + m)_{\rho \sigma} \nonumber\\
&& \; \times \;\{(1-f_{F}(\omega)) -
f_{F}(\omega)(1 - f_{F}(\omega))\}e^{-ik(x-y)},\nonumber\\
\tselea{4.85}
\end{eqnarray}

\noindent where $f_{F}(\omega)$ is the Fermi--Dirac distribution:
$f_{F}(\omega)=\frac{1}{e^{\beta (|\omega|-\mu)}+1}$, and

\begin{eqnarray}
\langle {\cali{T}} ( \psi_{\rho}(x){\overline{\psi}}_{\sigma}(y))
\rangle &\rightarrow & \sum_{\alpha =1,2} \int
\frac{d^{3}{\vect{k}}}{(2\pi)^{3}}\;\delta^{+}(\omega^{2}
-{\vect{k}}^{2}-m^{2}) \;\{ \langle
b_{\alpha}({\vect{k}};\omega)\psi_{\rho}(x) \rangle
\langle b_{\alpha}^{\dag}({\vect{k}};\omega){\overline{\psi}}_{\sigma}(y)
\rangle \nonumber\\
&& - \; \langle b_{\alpha}^{\dag}({\vect{k}};\omega)
\psi_{\rho}(x) \rangle \langle
b_{\alpha}({\vect{k}};\omega){\overline{\psi}}_{\sigma}(y)
\rangle\}\nonumber \\  
&&\mbox{\vspace{5mm}}\nonumber\\
&=& - \; \int \frac{d^{4}k}{(2\pi)^{3}}\; \delta^{+}(k^{2}-m^{2})\; \delta
(k_{0}-\omega)\;(\not k + m)_{\rho \sigma}\nonumber\\
&& \; \times \; f_{F}(\omega)(1 - f_{F}(\omega))e^{-ik(x-y)}.\nonumber\\
\tselea{4.9}
\end{eqnarray}

\noindent Correspondingly, for $\Delta \langle N_{b} \rangle$ we need

\begin{eqnarray}
\langle  \psi_{\rho}(x){\overline{\psi}}_{\sigma}(y)
\rangle &\rightarrow &  \int \frac{d^{4}k}{(2\pi)^{3}}\;
\delta^{+}(k^{2}-m^{2})\;(\not k + m)_{\rho \sigma} \nonumber\\
&&\times \;\{(1-f_{F}(\omega)) -
f_{F}(\omega)(1 - f_{F}(\omega)\}e^{-ik(x-y)} \nonumber\\
&&\mbox{\vspace{1cm}}\nonumber\\
\langle {\cali{T}} ( \psi_{\rho}(x){\overline{\psi}}_{\sigma}(y))
\rangle &\rightarrow& - \; \int \frac{d^{4}k}{(2\pi)^{3}}\;
\delta^{+}(k^{2}-m^{2})\;(\not k + m)_{\rho \sigma}
f_{F}(\omega)(1 - f_{F}(\omega))e^{-ik(x-y)}.\nonumber\\   
\tselea{4.10}
\end{eqnarray}

\noindent For the $d$--type excitations the prescription is very similar,
actually, in order to get $\frac{d\Delta \langle N_{d}(\omega)
\rangle}{d\omega}$, the following substitutions must be performed in
(\tseref{4.85})--(\tseref{4.10}): $\theta(k_{0}) \rightarrow
\theta(-k_{0})$, $f_{F}\rightarrow (1-f_{F})$ and $\mu \rightarrow -\mu$.

\vspace{1cm}  

\noindent {\bf (iii) For gauge fields in the axial temporal gauge
($A^{0}=0$):}   

\vspace{4mm}

\noindent The temporal gauge is generally incorporated in the gauge fixing
sector of the Lagrangian and particularly

\begin{equation}
{\cali{L}}_{fix}=-\frac{1}{2\alpha}(A_{0})^{2}; \alpha
\rightarrow 0. \tseleq{gauge fixing}
\end{equation}

\noindent The principal advantage of the axial gauges arises from the
decoupling the F--P ghosts in the theory. This statement is of course
trivial in QED as any linear gauge (both for covariant and non--covariant
case) brings this decoupling automatically \cite{LB}. Particular
advantage of the temporal gauge comes from an elimination of non--physical
scalar photons from the very beginning.

Let us decompose a gauge field $A_{i}, i=1,2,3$ into the
transverse and longitudinal part, i.e. $A_{i}=A_{i}^{T}+A_{i}^{L}$ with

\begin{equation}
A_{i}^{T}=\left( \delta_{ij} - \frac{\partial_{i}
\partial_{j}}{{\vec{\partial}}^{2}} \right) A_{j} \;\; \mbox{and}\;\;
A_{i}^{L}=\frac{\partial_{i}
\partial_{j}}{{\vec{\partial}}^{2}} A_{j}, \nonumber
\end{equation}

\vspace{3mm}
\noindent and use the sum over gauge--particle polarisations

\begin{equation}
\sum_{\lambda = 1}^{2} {\varepsilon}^{(\lambda)}_{i}(k)
{\varepsilon}^{(\lambda)}_{j}(k) = {\delta}_{ij} -
\frac{k_{i}k_{j}}{{\vect{k}}^{2}},\nonumber
\end{equation}

\noindent with ${\varepsilon}^{(\lambda)}(k)$ being polarisation vectors,
then for $d\Delta \langle N^{T}(\omega) \rangle / d\omega$ we get the
following one--line replacements

\begin{eqnarray}
\langle A^{T}_{i}(x)A^{T}_{j}(y) {\rangle} &\rightarrow&
\sum_{\lambda = 1}^{2}
\int
\frac{d^{3}{\vect{k}}}{(2\pi)^{3}}\;\delta^{+}(\omega^{2}
-{\vect{k}}^{2}-m^{2}) \langle
A^{T}_{i}(x)a^{\dag}_{\lambda}({\vect{k}};\omega)
{\rangle} \langle a_{\lambda}({\vect{k}};\omega)A^{T}_{j}(y)
{\rangle}\nonumber \\
&&+ \;\langle A^{T}_{i}(x)a_{\lambda}({\vect{k}};\omega)
{\rangle} \langle
a^{\dag}_{\lambda}({\vect{k}};\omega)A^{T}_{j}(y) {\rangle}_{\beta}
\}\nonumber\\ &&\mbox{\vspace{5mm}}\nonumber\\
&=&\left( \delta_{ij} - \frac{\partial_{i}
\partial_{j}}{{\vec{\partial}}^{2}} \right)
(\mbox{Eq.(\tseref{new1})})
\nonumber\\ &&\mbox{\vspace{1.5cm}}\nonumber\\
\langle {\cali{T}} (A^{T}_{i}(x)A^{T}_{j}(y) ) {\rangle}
&\rightarrow &
\sum_{\lambda = 1}^{2} \int
\frac{d^{3}{\vect{k}}}{(2\pi)^{3}}\;\delta^{+}(\omega^{2}
-{\vect{k}}^{2}-m^{2}) \{ \langle
A^{T}_{i}(x)a^{\dag}_{\lambda}({\vect{k}};\omega){\rangle}
\langle A^{T}_{i}(y)a_{\lambda}({\vect{k}};\omega)
{\rangle}\nonumber \\ &&+\; \langle
A^{T}_{i}(x)a_{\lambda}({\vect{k}};\omega){\rangle}\langle
A^{T}_{i}(x)a^{\dag}_{\lambda}({\vect{k}};\omega) {\rangle}\nonumber\\
&&\mbox{\vspace{5mm}}\nonumber\\
&=&\left( \delta_{ij} - \frac{\partial_{i}
\partial_{j}}{{\vec{\partial}}^{2}} \right) (\mbox{Eq.(\tseref{new4})}).
\tselea{4.11}
\end{eqnarray}

\noindent The replacements needed for $\Delta \langle N^{T} \rangle$ can
be concisely expressed as

\begin{eqnarray}
\langle \ldots \rangle \rightarrow \left( \delta_{ij} -
\frac{\partial_{i}
\partial_{j}}{{\vec{\partial}}^{2}}\right) (\mbox{Eq.(\tseref{new2})})
\tselea{4.12}
\end{eqnarray}

\noindent As for the longitudinal (non--physical) degrees of freedom, it is
obvious that

\begin{equation}
\langle A^{L}_{i}(x)A^{L}_{j}(y){\rangle};
 \; \;\langle
{\cali{T}} (A^{L}_{i}(x)
A^{L}_{j}(y) )  {\rangle} \rightarrow 0. \end{equation}   

\vspace{4mm}

\noindent Eqs.(\tseref{new1})--(\tseref{4.12}) can be most easily derived in
the finite--volume limit, e.g. for a scalar field we reformulate
$\varphi(x)$ as

\begin{displaymath}
\varphi(x) = \sum_{r} \frac{A_{r}}{\sqrt{2E_{r}V}}e^{-iE_{r}t +
i{\vect{k}}_{r}{\vect{x}}} +
\frac{A^{\dag}_{r}}{\sqrt{2E_{r}V}}e^{iE_{r}t-i{\vect{k}}_{r}{\vect{x}}},
\end{displaymath}

\noindent rescaling the annihilation and creation operators by defining
$a(k) = \sqrt{2E_{k}V} A_{k}$ in such a way that $[A_{k};A^{\dag}_{k^{'}}] =
\delta_{kk^{'}}$ (so that $ {\langle A^{\dag}_{k}A_{k^{'}}
\rangle} = \delta_{kk^{'}}f_{B}(k_{0})$), while $\int
\frac{d^{3}{\vect{k}}}{(2\pi)^{3}} \rightarrow \frac{1}{V}\sum_{\vect{k}}$.

\begin{figure}[h] 
\vspace{3mm} 
\epsfxsize=11.5cm
\centerline{\epsffile{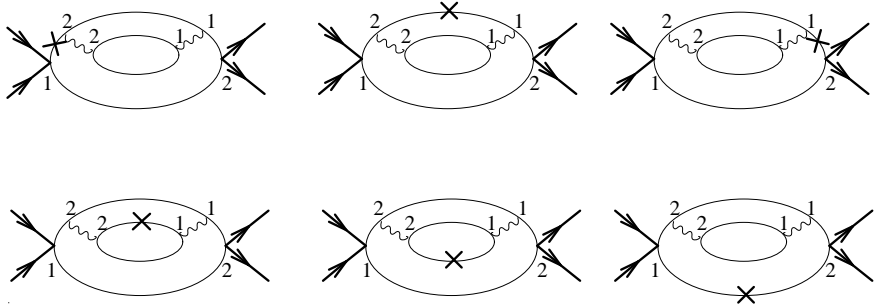}} 
\caption{{ The numerator of
(\tseref{dN}) and (\tseref{N}) can be calculated using the modified cut
diagrams for $\langle {\T}^{\dag} {\P} \,\T \rangle_{p_{1} p_{2}}$.
As an example we depict all the possible contributions to the numerator  
derived from the cut diagram on Fig.\ref{fig.3} c). The wavy lines and
thin lines describe the heat--bath particles. The crossed lines denote the
substituted propagators, in this case we wish to calculate the thin--line
particle number spectrum.}} 
\label{fig7} 
\vspace{1mm} 
\end{figure}

The replacements (\tseref{new1})--(\tseref{4.12}) are meant in
the following sense: firstly one constructs all the $T \not= 0$ diagrams
for ${}\langle {\T}^{\dag}{\P}\,\T \rangle_{p_{1}p_{2}} $, using
the LTE (\tseref{proj4}) and the rules mentioned therein. In order to
calculate the numerator of (\tseref{dN}) or (\tseref{N}) we simply replace
(using corresponding prescriptions) $one$ heat--bath particle line in each
cut diagram and this replacement must sum for all the possible heat--bath
particle lines in the diagram. If more types of heat--bath particles are
present, we replace only those lines which correspond to particles whose   
number spectrum we want to compute (see FIG.\ref{fig7}).

\vspace{2mm}

\noindent The terms in the replacements (\tseref{new1})--(\tseref{4.12})
have a direct physical interpretation. The $f(\omega_{k})$ and
$(1+(-1)^{p}f(\omega_{k}))$ can be viewed as the absorption and emission
of the heat--bath particles respectively \cite{PVLJT}. The term
$f(\omega_{k})(1+(-1)^{p}f(\omega_{k}))$ describes the fluctuations of the
heat bath particles. This is because for the non--interacting heat--bath 
particles $\langle (n_{k}-\langle n_{k}\rangle)^{2}
\rangle = f(\omega_{k})(1+(-1)^{p}f(\omega_{k}))$. The substituted
propagators can be therefore schematically depicted as

\vspace{1mm}
\begin{figure}[h]
\hspace{-2cm}
\epsfxsize=4cm
\centerline{\epsffile{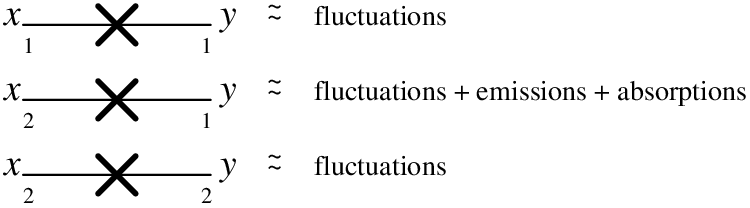}}
\label{fig16}
\end{figure}
\vspace{2mm}

\noindent Collecting all the contributions from emissions, absorptions and
fluctuations separately, one can schematically write

\begin{equation}
\frac{d\langle N(\omega)\rangle_{f}}{d\omega}=\frac{d\langle
N(\omega)\rangle_{i}}{d\omega} + F^{emission}(\omega) +
F^{absorption}(\omega)+F^{fluc}(\omega),
\tseleq{L-T}
\end{equation}

\vspace{1mm}
\noindent where, for instance for neutral scalar bosons

\vspace{-3mm}

\begin{eqnarray*}
F^{emission}(\omega)= Z^{-1}_{f}\int \frac{d^{4}k}{(2\pi)^{3}}
\delta^{+}(k^{2}-m^{2})\delta(k_{0}-\omega)(1+f_{B}(\omega))\left\langle
\frac{{\T}^{\dag} {\stackrel{\leftarrow}{\delta}}}{\delta
\psi(x)}{\P}
\frac{{\stackrel{\rightarrow}{\delta}} \T}{\delta \psi(y)}
\right\rangle_{p_{1} p_{2}}.
\end{eqnarray*}

\noindent Using (\tseref{new4}), it is easy to write down the analogous
expressions for the $F^{absorption}$ and $F^{fluc}$. To the lowest
perturbative order, the form (\tseref{L-T}) was obtained by Landshoff and
Taylor \cite{PVLJT}.

\vspace{1.5cm}

\section{Model process \label{MP}}
\vspace{5mm}
\subsection{Basic assumptions \label{MP1}}

To illustrate the modified cut diagram technique, we shall restrict
ourselves to a toy model, namely to a scattering of two neutral scalar
particles $\Phi$ (pions) within a photon heat bath, with a pair of scalar
charged particles {$\phi, {\overline\phi}$} (`muon' and `antimuon') left
as a final product. Both initial and final particles are supposed to be
unheated. We further assume that the heat--bath photons $A$ are scalars,
i.e. the heat--bath Hamiltonian has form

\vspace{-1mm}
\begin{displaymath} H^{hb} = \frac{1}{2} (\partial_{\nu}A)^{2}
- \frac{m_{\gamma}^{2}}{2}A^{2}. \vspace{2mm} \tseleq{Hamilt} \end{displaymath}

\noindent In order to mimic scalar electrodynamic, we have chosen the
interacting Hamiltonian entering in the $\T$--matrix as

\begin{displaymath}
H_{in} = \frac{\lambda}{2}\Phi^{2}\phi {\phi}^{\dag} + (eA +
\frac{e^{2}}{2}A^{2})\phi {\phi}^{\dag}.
\end{displaymath}

\vspace{4mm}

\subsection{Calculations \label{MP2}}

\begin{figure}[h]
\epsfxsize=11cm
\centerline{\epsffile{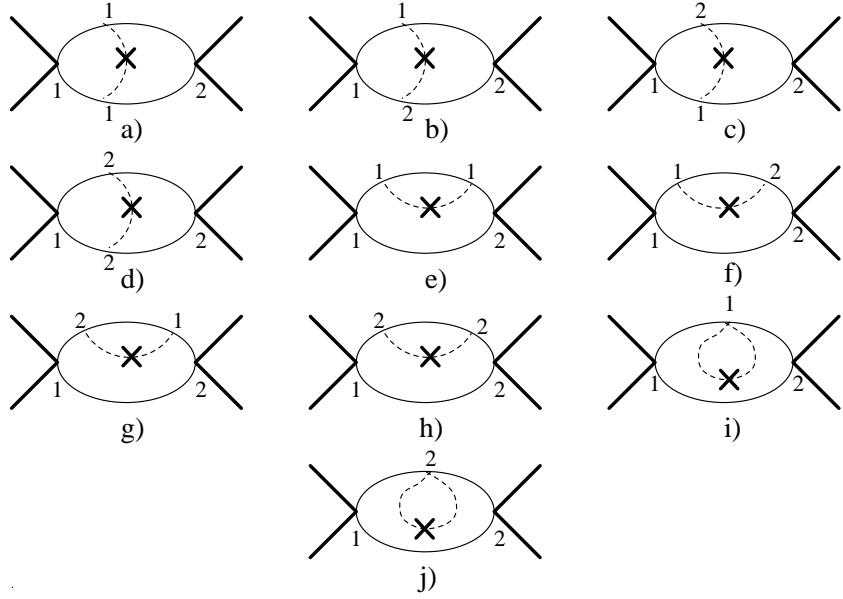}}
\caption{ The modified cut diagrams involved in an order--$e^{2}$ contribution
to the photon number spectrum. Dashed lines: photons. Solid lines: $\phi$,
$\phi^{\dag}$ particles. Bold lines: $\Phi$ particles.}
\label{fig35}
\end{figure}

We    can  now   compute   an   order--$e^{2}$   contribution to   the
$\frac{d\Delta \langle  N_{\gamma}(\omega)     \rangle}{d\omega}$. The
evaluation    of    the   $\frac{d\Delta  \langle   N_{\gamma}(\omega)
\rangle}{d\omega}$ is straightforward.  In FIG.\ref{fig35} we list all
the modified cut diagrams  contributing to an order--$e^{2}$. Note that
diagrams  b) and c)  are topologically identical.  Similarly, diagrams
e), f),  h),  i) and j)  should  be taken with combinatorial  factor 2
(corresponding diagrams  with a heat-bath  particle line on the bottom
solid  line are  not   shown). Of   course,  diagram  g) vanishes  for
kinematic reasons.  

\vspace{1.5cm}

\vspace{5mm}

\begin{figure}[h]
\epsfxsize=7cm
\centerline{\epsffile{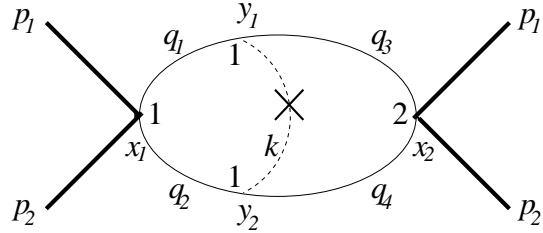}}
\caption{ The diagram a) with a corresponding kinematics.}
\label{fig100} 
\end{figure}

For instance, in order to calculate the
contribution from diagram a) (see also FIG.\ref{fig100}) 
\vspace{3mm}

\vspace{2mm}

\noindent we go back   to Eq.(\tseref{Tmatrix}) and   to prescriptions
(\tseref{new1})--(\tseref{new4}), so we get 

\begin{eqnarray}
\mbox{a)}=&& \frac{- \lambda^{2}e^{2}}{V^{2}4\omega_{p_{1}}\omega_{p_{2}}}
\int d^{4}x_{1}d^{4}x_{2}d^{4}y_{1}d^{4}y_{2}\;
e^{-i(p_{1}+p_{2})x_{1}}\;e^{i(p_{1}+p_{2})x_{2}}\;
i\Delta_{F}(y_{1}-x_{1})\;
i\Delta_{F}(y_{2}-x_{1})\nonumber\\
&\times & \;i\Delta^{-}(y_{1}-x_{2})\;i\Delta^{-}(y_{2}-x_{2})\;\int
\frac{d^{4}k}{(2\pi)^{3}} \delta(k^{2}-m^{2})(1 +
f_{B}(\omega))f_{B}(\omega)\nonumber \\
&\times & \;
(\delta^{+}(k_{0}-\omega)+\delta^{-}(k_{0}+\omega))\;
e^{-ik(y_{1}-y_{2})}\nonumber\\
&&\mbox{\hspace{3mm}}\nonumber \\
=&& \frac{\lambda^{2} e^{2} t}{V 4 \omega_{p_{1}} \omega_{p_{2}} (2 \pi)^{5}}
f_{B}(\omega)(1+f_{B}(\omega))\; \int d^{4}k \;d^{4}q_{3} \;d^{4}q_{4}\;
\delta^{+}(q_{3}^{2}-m_{\mu}^{2})\;
\delta^{+}(q_{4}^{2}-m_{\mu}^{2})\nonumber \\
&\times & \delta(k_{0}-\omega)\;\left\{\frac{1}{-2q_{3}k + m_{\gamma}^{2}}
\; \frac{1}{2q_{4}k+m_{\gamma}^{2}} + \frac{1}{2q_{3}k + m_{\gamma}^{2}}\;
\frac{1}{-2q_{4}k+m_{\gamma}^{2}}\right\} \nonumber \\
&\times &  \delta(k^{2}-
m_{\gamma}^{2})\;\delta^{4}(-p_{1}-p_{2}+q_{1}+q_{2}). \tselea{DI1}
\end{eqnarray}

\vspace{2mm}
\noindent We have dropped the $i\epsilon$ prescription in the
propagators since adding/ subtracting an on--shell momenta $q_{1;2}$
to/from an on--shell momenta $k$ we can not fulfil the condition $(k\pm
q_{1;2})^{2}=m^{2}_{\mu}$. As it is usual, we have assumed that our
interaction is enclosed in a `time' and volume box ($t$ and $V$
respectively). Analogously one can calculate contributions from other
diagrams in FIG.\ref{fig35}. Let us emphasise that it is necessary to give
sense to graphs e), h), i) and j) as these suffer with the pinch
singularity; the muon--particle propagator $(p_{1;2}^{2}-m^{2})^{-1}$ has
to be evaluated at its pole because of the presence of an on--shell line
(1--2 line) with the same momenta. Some regularisation is obviously
necessary. Using the formal identity \cite{LB}

\begin{equation}
\frac{1}{x \pm i\epsilon}\delta(x)=
-\frac{1}{2}\delta^{'}(x) \mp i\pi(\delta(x))^{2},
\end{equation}

\vspace{3mm}

\noindent we discover that the unwanted $\delta^{2}$ mutually cancel
between e) and h) diagrams (similarly for i) and j) diagrams). An  
alternative (but lengthier) way of dealing with the latter pinch
singularity; i.e. switching off the interaction with a heat bath in the
remote past and future, is discussed in \cite{PVLJ}. Evaluating all the
diagrams (note, we should attach to each digram the factor $\frac{1}{2!}$
coming from a Taylor expansion of the $\T$--matrix), we are left with (c.f.
Eq.(\tseref{L-T})):
\vspace{1cm}

\begin{eqnarray}
F^{emission}(\omega)\;\;\;\; + \mbox{\hspace{-3mm}}&&F^{absorption}(\omega)\nonumber\\
&=&\frac{t\lambda^{2}e^{2}}{\langle
\T\,{\P}\,{\T}^{\dag}\rangle_{p_{1}p_{2}}\,V 8
\omega_{p_{1}}\omega_{p_{2}}(2\pi)^{5}}\int d^{4}k \;
\delta(k^{2}-m_{\gamma}^{2}) \;
\delta(k_{0}-\omega)\nonumber\\
&&\times \int  d^{4}q_{1}d^{4}q_{2} \; \delta^{+}(q_{1}^{2}-m_{\mu}^{2})
\; \delta^{+}(q_{2}^{2}-m_{\mu}^{2})\nonumber\\  
&&\times ~\left\{
K_{1}(1+f_{B}(\omega)) \; \delta^{4}(-Q+q_{1}+q_{2}+k)\right. \nonumber\\
&&\left.-K_{2}f_{B}(\omega)
\; \delta^{4}(-Q+q_{1}+q_{2}-k)\right\}
\tselea{30}   
\end{eqnarray}

\noindent and

\begin{eqnarray}
F^{fluct}(\omega)&=&\frac{t\lambda^{2}e^{2}f_{B}(\omega)(1+f_{B}(\omega))}{\langle
\T\,{\P}\,{\T}^{\dag}\rangle_{p_{1}p_{2}}\,V 8
\omega_{p_{1}}\omega_{p_{2}}(2\pi)^{5}}\int d^{4}k
\; \delta(k^{2}-m_{\gamma}^{2})\; \delta(k_{0}-\omega)\nonumber\\
&&\times \int  d^{4}q_{1}d^{4}q_{2} \; \delta^{+}(q_{1}^{2}-m_{\mu}^{2})\;
\delta^{+}(q_{2}^{2}-m_{\mu}^{2})\nonumber\\
&&\times \left\{
\delta^{4}(-Q+q_{1}+q_{2}+k)K_{1}+\delta^{4}(-Q+q_{1}+q_{2}-k)K_{2}\right.\nonumber\\
&&\left.-~2\delta^{4}(-Q+q_{1}+q_{2})K_{3} \right\}\nonumber\\
&&\mbox{\vspace{5mm}}\nonumber\\
&+&~\frac{t\lambda^{2}e^{2}f_{B}(\omega)(1+f_{B}(\omega))}{\langle
\T\,{\P}\,{\T}^{\dag}\rangle_{p_{1}p_{2}}\,V 8
\omega_{p_{1}}\omega_{p_{2}}(2\pi)^{5}}\int d^{4}k
\; \delta(k^{2}-m_{\gamma}^{2}) \; \delta(k_{0}-\omega)\nonumber\\
&&\times \int
d^{4}q_{1}d^{4}q_{2} \; \delta^{4}(-Q+q_{1}+q_{2})\nonumber\\
&&\times \left\{
\left(1-\frac{1}{2q_{1}k   
-m^{2}_{\gamma}}+
\frac{1}{2q_{1}k +
m^{2}_{\gamma}}\right)\delta^{+}(q_{2}^{2}-m_{\mu}^{2})\frac{\partial}{\partial
m^{2}_{\mu}}\; \delta^{+}(q_{1}^{2}-m_{\mu}^{2})\right. \nonumber\\
&&\left. + (q_{1} \leftrightarrow
q_{2})\right\}
\tselea{31}
\end{eqnarray}  

\begin{figure}[h]
\epsfxsize=7cm  
\centerline{\epsffile{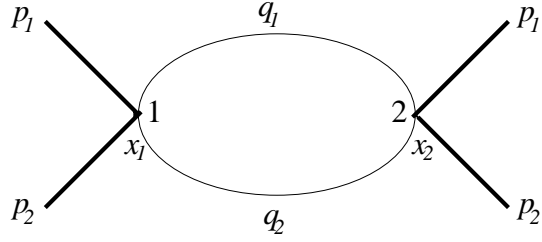}}
\caption{The lowest-order cut diagram for $\langle
\T{\P}{\T}^{\dag}\rangle_{p_{1}p_{2}}$.} \label{fig101}
\end{figure}

\vspace{4mm}

\noindent with $K_{1}=\left( \frac{1}{2q_{1}k + m^{2}_{\gamma}}+ 
\frac{1}{2q_{2}k + m^{2}_{\gamma}}\right)^{2}$, $K_{2}=\left(
\frac{1}{2q_{1}k - m^{2}_{\gamma}}+ \frac{1}{2q_{2}k -
m^{2}_{\gamma}}\right)^{2}$,
$K_{3}=\frac{2}{(2q_{1}k-m^{2}_{\gamma})(2q_{2}k+m^{2}_{\gamma})}$ and
$Q=p_{1}+ p_{2}$. The relevant (i.e. order--$e^{0}$) term (see FIG.\ref{fig101})

\vspace{1cm}

\vspace{2mm}

\noindent for $\langle   \T\,    {\P}\,{\T}^{\dag}\rangle_{p_{1}p_{2}}$ reads 

\vspace{-2mm}
\begin{eqnarray}
\langle \T\, {\P}\,{\T}^{\dag}\rangle_{p_{1}p_{2}}&=&
\frac{\lambda^{2}t}{16\;V\omega_{p_{1}}\omega_{p_{2}}(2\pi)^{2}}\int
d^{4}q_{1}d^{4}q_{2} \; \delta^{+}(q_{1}^{2}-m^{2}_{\mu})\;
\delta^{+}(q_{2}^{2}-m_{\mu}^{2})\; \delta^{4}(-Q+q_{1}+q_{2})\nonumber\\
&=& \frac{\lambda^{2}t}{64V\omega_{p_{1}}\omega_{p_{2}}|Q|
(2\pi)}\sqrt{Q^{2}-4m_{\mu}^{2}}.
\tselea{EX}
\end{eqnarray}
\vspace{0.5mm}

\noindent Eqs.(\tseref{30}) and  (\tseref{31})  are  analogous to  the
result obtained in \cite{PVLJT} for  the decay. In order to understand
their  structure, let us  deal with the number spectrum\footnote{So we
implicitly assume that  the  photon mass $m_{\gamma}$ is  sufficiently
small.} for small  $\omega$'s. To do this,  we change the  integration
variables 
\begin{eqnarray}
&&q_{1}\rightarrow q_{1}\mp \frac{1}{2}k\nonumber \\
&&q_{2}\rightarrow q_{2}\mp \frac{1}{2}k.
\tselea{TR}
\end{eqnarray}

\noindent These changes lead to

\begin{eqnarray}
&&(2q_{i}k \pm m^{2}_{\gamma})\; \delta^{+}(q_{1}^{2}-m^{2}_{\mu})\;
\delta^{+}(q_{2}-m^{2}_{\mu})\; \delta^{4}(-Q + q_{1} + q_{2} \pm
k)\nonumber\\
&&\mbox{\hspace{1cm}} \longrightarrow \; 2q_{i}k \;
\delta^{+}(q_{1}^{2}-M^{2}\mp X)\; \delta^{+}(q_{2}^{2}-M^{2}\mp Y)\;
\delta^{4}(-Q+q_{1}+q_{2}),
\tselea{EX1}
\end{eqnarray}

\vspace{3mm}

\noindent where $M^{2}=m^{2}_{\mu}-\frac{1}{4}m_{\gamma}^{2}$, $X=q_{1}k$ and
$Y = q_{2}k$. In addition, transformations (\tseref{TR}) have unite Jacobian.
If one Taylor expands (\tseref{EX1}) in terms of $X$ and $Y$ then one gets
successively higher $\omega$--contributions to (\tseref{30})--(\tseref{31}).
Expanding (\tseref{30}) to the first order in $X$ and $Y$, and keeping only
temperature--dependent pieces, we have

\begin{eqnarray}
&& \langle \T\, {\P} \,{\T}^{\dag} \rangle_{p_{1}p_{2}}\,\frac{8 V
\omega_{p_{1}}\omega_{p_{2}}}{t}\;(F^{emission}(\omega)+F^{absorption}(\omega))\nonumber\\
&& \mbox{\hspace{2.5cm}} \sim \frac{\lambda^{2} e^{2}}{(2
\pi)^{5}}f_{B}(\omega)\; \int d^{4}q_{1} d^{4}q_{2}\;
\delta(k^{2}-m_{\gamma}^{2})\; \delta(k_{0}-\omega)\; A,
\tselea{EX2}
\end{eqnarray}

\vspace{1mm} 

\noindent with

\begin{displaymath}
A = \frac{\partial}{\partial_{M^{2}_{1}}} \int d^{4}q_{1}d^{4}q_{2} \;
\delta^{+}(q_{1}^{2}-M^{2}_{1})\; \delta^{+}(q_{2}^{2}-M^{2}_{2})\; (4KX)\;
\delta^{4}(Q-q_{1}-q_{2})\left|_{{~}_{M_{1}=M_{2}=M}}\right.
.
\end{displaymath}

\vspace{3mm}

\noindent Here $K=\left(\frac{1}{2q_{1}k} + \frac{1}{2q_{2}k}\right)^{2}$  
(we have performed transformation $q_{1} \leftrightarrow q_{2}$ in order
to express (\tseref{EX2}) solely in terms of $X$). As  $A$ is a
Lorentz scalar, it must depend on $k$ only via product $(kQ)$. One can
thus evaluate $A$ in the frame where $Q=(Q_{0}, {\vect{0}})$ and then
replace $\omega Q_{0}$ by $(kQ)$ (see also \cite{PVLJT}).
Straightforward calculations show that

\begin{equation}
A =
\frac{-(2\pi)(kQ)^{3}}{|Q|\sqrt{\frac{Q^{2}}{4}-M^{2}}
\;\left(\frac{M}{|Q|}(kQ)^{2} + m^{2}_{\gamma}
\left( \frac{|Q|^{3}}{4}-MQ^{2}\right) \right)^{2}}.
\tseleq{EX3}
\end{equation}

\vspace{3mm}

\noindent Recalling (\tseref{EX}), we get

\begin{eqnarray}
&&\mbox{\hspace{-5.5mm}} F^{emission}(\omega)+F^{absorption}(\omega)\nonumber \\
&& \mbox{\vspace{-4.5mm}}\nonumber \\
&& \mbox{\hspace{-4.5mm}} \sim \frac{Q^{2}f_{B}(\omega) e^{2}}{\pi^{2}
M^{2} \sqrt{Q_{0}^{2}-Q^{2}} \sqrt{Q^{2}-4M^{2}}\sqrt{Q^{2}-4m^{2}_{\mu}}}
\nonumber \\
&& \mbox{\hspace{-4.5mm}} \times
\left\{ \mbox{ln} \left( \frac{(\omega Q_{0} + |{\vect{k}}|
|{\vect{Q}}|)^{2}  +
m^{2}_{\gamma}\frac{Q^{2}}{M^{2}}\left(\frac{Q^{2}}{4}-
M^{2}\right)}{(\omega
Q_{0} - |{\vect{k}}| |{\vect{Q}}|)^{2}  +
m^{2}_{\gamma}\frac{Q^{2}}{M^{2}}\left(\frac{Q^{2}}{4}-
M^{2}\right)}\right)
\right. + \frac{m^{2}_{\gamma} \left(
\frac{Q^{2}}{4}   - M^{2} \right)}{\frac{M^{2}}{Q^{2}}(\omega
Q_{0} - |{\vect{k}}| |{\vect{Q}}|)^{2}  + m^{2}_{\gamma} \left(
\frac{Q^{2}}{4} - M^{2}  \right) }\nonumber \\
&& \left. \mbox{\hspace{-4.5mm}}  - \frac{m^{2}_{\gamma}
\left( \frac{Q^{2}}{4}
- M^{2} \right)}{\frac{M^{2}}{Q^{2}}(\omega
Q_{0} + |{\vect{k}}| |{\vect{Q}}|)^{2}  + m^{2}_{\gamma} \left(
\frac{Q^{2}}{4} - M^{2}  \right) } \right\}
\tselea{EX4} \end{eqnarray}

\vspace{3mm}

\noindent with $|{\vect{k}}| = \sqrt{\omega^{2}-m^{2}_{\gamma}}$ and 
$|{\vect{Q}}|= \sqrt{Q_{0}^{2}-Q^{2}}$. Eq.(\tseref{EX4}) takes a
particularly simple form if $m_{\gamma}$ is negligibly small (i.e. if
$m_{\gamma} \ll \omega $), then

\begin{eqnarray}
&&F^{emission}(\omega)+F^{absorption}(\omega)\nonumber \\
&& \mbox{\hspace{1.6cm}} \sim
\frac{2f_{B}(\omega)e^{2}}{\pi^{2}}
\frac{Q^{2}}{m^{2}_{\mu}\;\sqrt{Q_{0}^{2}-Q^{2}}\; (Q^{2}-4m^{2}_{\mu})}
\mbox{ln}
\left(\frac{Q_{0}+\sqrt{Q^{2}_{0}-Q^{2}}}{Q_{0}-\sqrt{Q^{2}_{0}-Q^{2}}}\right).
\tselea{EX211}
\end{eqnarray}

\vspace{4mm}

\noindent Similarly as in the previous case we can evaluate
$F^{fluct}$. Performing transformation (\tseref{TR}), and expanding
(\tseref{31}) to the first order in $X$ and $Y$, we get

\begin{eqnarray}
&& \langle \T\, {\P} \,{\T}^{\dag} \rangle_{p_{1}p_{2}}\,\frac{8 V
\omega_{p_{1}}\omega_{p_{2}}}{t}\;F^{fluct}(\omega)\nonumber \\
&& \mbox{\hspace{1cm}} \sim \;\frac{\lambda^{2} e^{2}}{(2
\pi)^{5}}f_{B}(\omega)(1+f_{B}(\omega))\; \int d^{4}q_{1} d^{4}q_{2}\;
\delta(k^{2}-m_{\gamma}^{2})\; \delta(k_{0}-\omega)\; B,
\tselea{EX25} 
\end{eqnarray}

\noindent with

\begin{eqnarray*}
B &=&
\int
d^{4}q_{1}d^{4}q_{2}\;\delta^{4}(Q-q_{1}-q_{2})
\;\left(\frac{\partial}{\partial m^{2}_{\mu}}\right)
\delta^{+}(q_{1}^{2}-m^{2}_{\mu})\; \delta^{+}(q_{2}^{2}-m^{2}_{\mu})\\
&&\mbox{\vspace{2mm}}\\
&+&\int
d^{4}q_{1}d^{4}q_{2}\;\delta^{4}(Q-q_{1}-q_{2})
\;\left\{ \left(\frac{\partial}{\partial M}\right)^{2}\right. \\
 &-& 2 \left.
\left(\frac{Qk}{q_{1}k}\right) \frac{\partial^{2}}{\partial
M_{1}\partial M_{2}}\right\}
\delta^{+}(q_{1}^{2}-M^{2}_{1})\; \delta^{+}(q_{2}^{2}-M^{2}_{2})
\left|_{{~}_{M_{1}=M_{2}=M}}\right.
.
\end{eqnarray*}

\vspace{3mm}

\noindent Direct calculations lead to

\vspace{1mm}

\begin{eqnarray*}
B &=& \frac{2\pi\; (kQ)^{2}}{Q^{2}(Q^{2}-4M^{2})^{\frac{3}{2}}}
\left\{
\frac{ M^{2}}{\left(
\frac{ M^{2}}{Q^{2}}(kQ)^{2} + m^{2}_{\gamma}\left( \frac{Q^{2}}{4}-M^{2}
\right) \right)} -
\frac{(\frac{Q^{2}}{4}-M^{2})(2M^{2}-Q^{2})\;m^{2}_{\gamma}}{\left(
\frac{ M^{2}}{Q^{2}}(kQ)^{2} + m^{2}_{\gamma}\left( \frac{Q^{2}}{4}-M^{2}
\right) \right)^{2}}\right\} \\
 &&- \frac{\pi}{|Q| \sqrt{Q^{2}-4M^{2}}}
- \frac{2\pi}{|Q| ({Q^{2}-4M^{2}})^{\frac{3}{2}}}.
\end{eqnarray*}

\vspace{2mm}

\noindent After some analysis we finally get

\begin{eqnarray}
 F^{fluct}(\omega) &\sim &\;
\frac{f_{B}(\omega)(1+f_{B}(\omega))\; m_{\gamma} \;
e^{2} }{4\pi^{2} \;M^{2}\; \sqrt{Q_{0}^{2}-Q^{2}}\;
\sqrt{Q^{2}-4m^{2}_{\mu}}}
\left\{ \frac{|Q|}{M} \left[
\mbox{arctg} \left( \frac{ \frac{M}{|Q|} (\omega Q_{0}
+|{\vect{k}}||\vect{Q}|)}{m_{\gamma} \sqrt{ \frac{Q^{2}}{4}-M^{2}
}} \right) \right. \right. \nonumber \\
&&\mbox{\vspace{2mm}}\nonumber \\
&& \left. \left. - \; \mbox{arctg} \left( \frac{ \frac{M}{|Q|} (\omega
Q_{0} -|{\vect{k}}||\vect{Q}|)}{m_{\gamma} \sqrt{ \frac{Q^{2}}{4}-M^{2}
}} \right)\right] \right. \nonumber\\
&&\mbox{\vspace{7mm}}\nonumber\\
&&\mbox{\vspace{7mm}}\nonumber\\
&& + \; \frac{(2M^{2}-Q^{2})\; m_{\gamma}}{2\sqrt{Q^{2}-4M^{2}}}
\left[ \frac{\omega Q_{0} +
|{\vect{k}}||{\vect{Q}}|}{\frac{M^{2}}{Q^{2}}(\omega
Q_{0}+|{\vect{k}}||{\vect{Q}}|)^{2} +
m^{2}_{\gamma}(\frac{Q^{2}}{4}-M^{2})}\right. \nonumber\\
&& \left. \left.- \; \frac{\omega Q_{0} -
|{\vect{k}}||{\vect{Q}}|}{\frac{M^{2}}{Q^{2}}(\omega
Q_{0}-|{\vect{k}}||{\vect{Q}}|)^{2} +
m^{2}_{\gamma}(\frac{Q^{2}}{4}-M^{2})}\right] \right\}\nonumber\\
&&\mbox{\vspace{7mm}}\nonumber\\
&&\mbox{\vspace{7mm}}\nonumber\\
&&- \;
\frac{f_{B}(\omega)(1+f_{B}(\omega)) |{\vect{k}}|\;e^{2}}{
\pi^{2}\; (Q^{2}-4m_{\mu}^{2})}.
\tselea{EX20}
\end{eqnarray}

\vspace{3mm}

\noindent Expression (\tseref{EX20}) considerably simplifies in the limit
$m_{\gamma} \rightarrow 0$. In the latter case

\begin{equation}
F^{fluct} \sim \; -\; \frac{f_{B}(\omega)(1+f_{B}(\omega))\;
\omega \;e^{2}}{ \pi^{2}(Q^{2}-4m^{2}_{\mu})} ,
\tseleq{EX21}
\end{equation}

\noindent so the leading behaviour for $F^{fluct}$ at small $\omega$ and
$m_{\gamma} \ll \omega$ is dominated by $\omega^{-1}$. Note that separate
contributions to the $0$--th order of a Taylor expansion of $F^{fluc}$
behave as $\omega^{-2}$ but they cancel between themselves leaving behind
parts proportional at worst to $\omega^{-1}$. The minus sign in
(\tseref{EX21}) reflects the fact that the fluctuations tend to suppress an
increase in the particle number spectrum when $\omega$ is small. On the
other hand, from (\tseref{EX211}) we see that the emissions and absorptions
stimulate an increase in the particle number spectrum for small $\omega$.

A result similar to (\tseref{EX211}) and (\tseref{EX21}) has
been derived by Landshoff and Taylor\cite{PVLJT} for a decay using
proper scalar electrodynamics, though in their case a contribution from the
emission and absorption dominated over fluctuations for small $\omega$.  
Note that in our model both contributions are of comparable size at
$\omega \sim 0$. The former feature is inherently connected with the fact
that our `photons' are scalar particles. If photons were vector particles
an additional photon momentum $k_{\mu}$ would go with each three--line
photon--muon vertex and so one might expect that the contributions
(\tseref{EX211}) and (\tseref{EX21}) would be `soften' at small
$\omega$. We have checked explicitly that for zero--mass photons in the
axial temporal gauge (i.e. $A^{0}=0$) this is indeed the case, and
it was found that $F^{emission}+F^{absorption} \propto \omega^{-1}$ whilst
$F^{fluct} \propto \omega$.

Until now we have supposed that our heat bath contains only
(scalar) photons in thermal equilibrium. However, one could similarly  
treat a heat bath which is comprised of photons and charged particles, let 
say electrons, mutually coexisting in thermal equilibrium. To be more
specific, let us assume that the heat--bath photons $A$ and electrons
$\Psi$ are both scalars so the heat--bath Hamiltonian takes form
\vspace{1mm}
\begin{eqnarray} H^{hb} &=& H^{\gamma} + H^{e} + eA\Psi {\Psi}^{\dag}
+\frac{e^{2}}{2}A^{2} \Psi {\Psi}^{\dag} \nonumber \\ H^{e} &=&
\partial_{\nu} \Psi \partial^{\nu} {\Psi}^{\dag} - m_{e}^{2} \Psi
{\Psi}^{\dag}\nonumber \\ H^{\gamma} &=& \frac{1}{2} (\partial_{\nu}A)^{2}
- \frac{m_{\gamma}^{2}}{2}A^{2}, \vspace{2mm} \tselea{Hamilt2} \end{eqnarray}

\noindent and the $\T$--matrix interacting Hamiltonian $H_{in}$ reads

\begin{displaymath}
H_{in} = \frac{\lambda}{2}\Phi^{2}\phi {\phi}^{\dag} + (eA +
\frac{e^{2}}{2}A^{2})\Psi {\Psi}^{\dag} + (eA + \frac{e^{2}}{2}A^{2})\phi 
{\phi}^{\dag}.
\end{displaymath}

\noindent It is usually argued \cite{NS, EP} that the interacting
pieces in $H^{hb}$ can be dropped provided that $t_{i} \rightarrow   
-\infty$ and $t_{f} \rightarrow \infty$. Since we assume that `pions' are
prepared in the remote past and `muons' are measured in the remote future, 
we shall accept in the following this omission.

\begin{figure}[h] 
\epsfxsize=11cm
\centerline{\epsffile{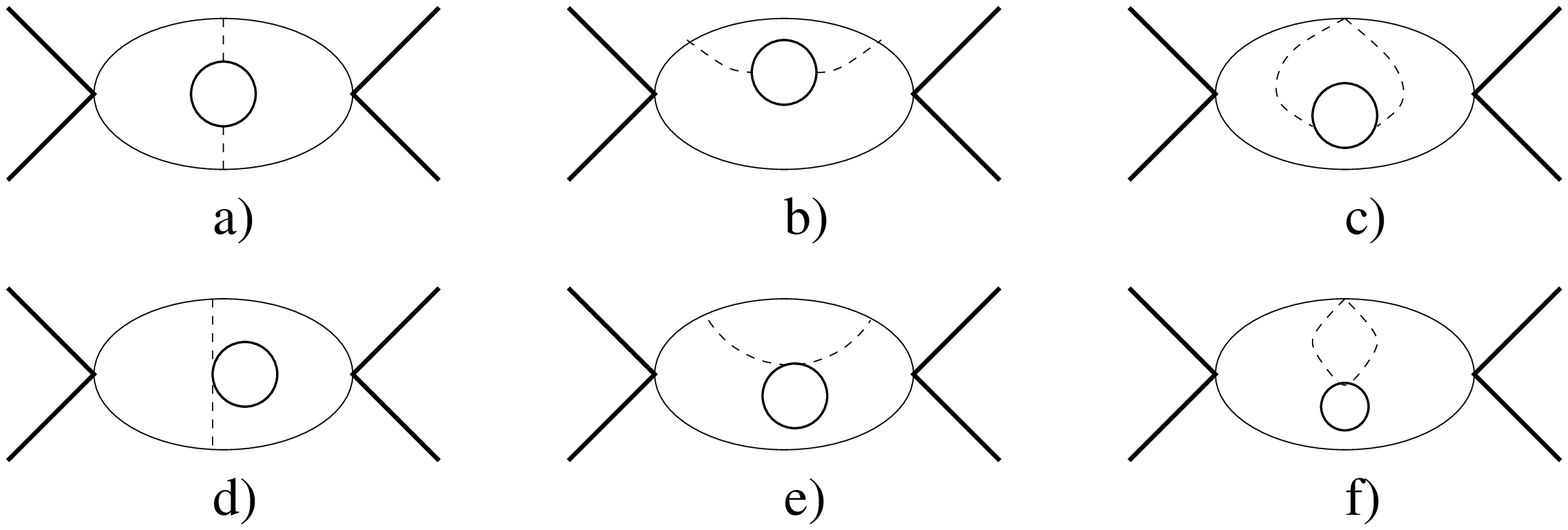}} 
\caption{The generating thermal
diagrams involved in an order--$e^{4}$ contribution to the electron number
spectrum. Dashed lines: photons. Thin lines: $\phi$, $\phi^{\dag}$
particles. Bold lines: $\Phi$ particles. Half--bold lines:  electrons.}  
\label{PJ1} 
\end{figure}

We can now approach to calculate both the photon and electron
number spectrum, i.e. $\frac{d\Delta \langle N_{\gamma}(\omega)  
\rangle}{d\omega}$ and $\frac{d\Delta \langle N_{e}(\omega)
\rangle}{d\omega}$ respectively.  As for $\frac{d\Delta \langle
N_{\gamma}(\omega) \rangle}{d\omega}$, an order--$e^{2}$ contribution is
clearly done only by diagrams in Fig.\ref{fig35} as there are no relevant
graphs with electron vertices contributing to this order, so
(\tseref{EX4}) and (\tseref{EX20}) still remain true. On the other hand,
there is no order--$e^{2}$ contribution to $\frac{d\Delta \langle
N_{e}(\omega) \rangle}{d\omega}$. The lowest order in $e$ (keeping
$\lambda^{2}$ fixed) is $e^{4}$. This brings richer diagrammatic structure
then in the photon case. In FIG.\ref{PJ1} we list all the generating
thermal diagrams contributing to an order--$e^{4}$.

\vspace{2mm} 
\vspace{2mm}

\begin{figure}[h] \epsfxsize=14.7cm \centerline{\epsffile{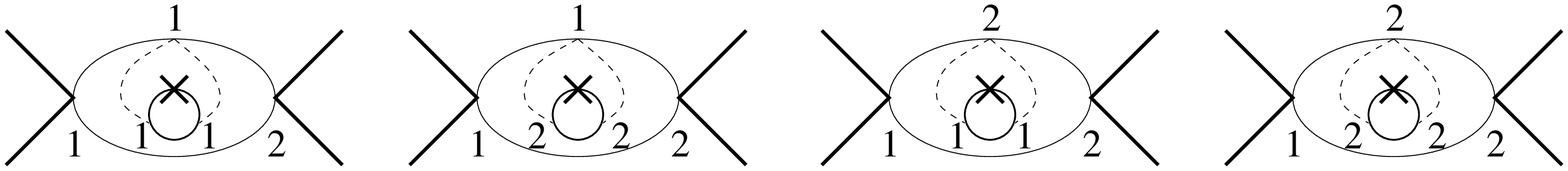}}
\caption{ The
 non--vanishing modified cut diagrams from FIG.\ref{fig55}c).} \label{fig56}
\end{figure}

\noindent It is easy to see that out of these 6 generating thermal
diagrams we get 43 non--vanishing and topologically inequivalent modified
cut diagrams; for example from FIG.\ref{PJ1}c) we have only those
diagrams which are depicted in FIG.\ref{fig56}. 

\vspace{4mm}

\noindent Note, the graphs of FIG.\ref{fig56}  must be multiplied by a
factor of four as there are two equivalent  insertions of the modified
electron line and  two  equivalent distributions of  the  photon--muon
vertex (so together with $\frac{1}{2!}$ from a Taylor expansion of the
$\T$--matrix  we get  the  symmetry factor  2).  Analogously we get 10
inequivalent modified cut diagrams from  FIG.\ref{PJ1}a); 7 from b); 8
from d); 6 from e) and 8 from f).) The actual electron number spectrum
calculations   are thus  rather   involved. Nevertheless,   one  might
evaluate                      fairly                           quickly
$F^{emission}(\omega_{e})+F^{absorption}(\omega_{e})$  as   there  are
only three diagrams which contribute, namely:

\vspace{4mm}
 \begin{figure}[h] \epsfxsize=14.7cm
\centerline{\epsffile{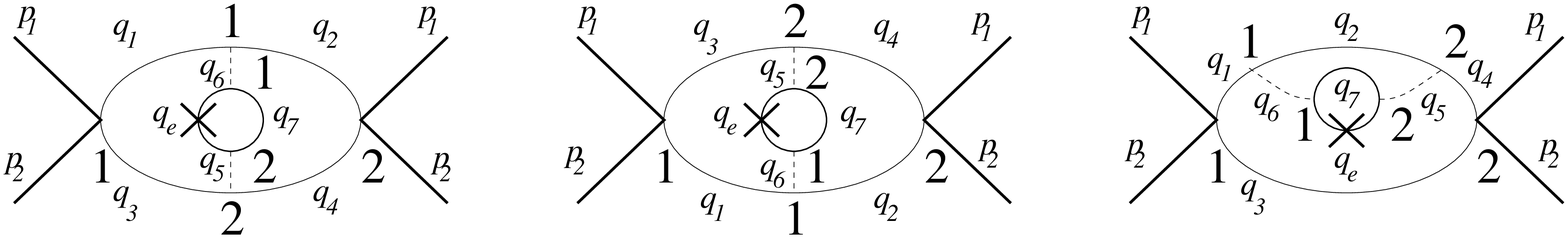}}
\end{figure} \vspace{2mm}

\noindent Let us remind that in the finale state we must have,   
apart from the heat-bath particles, only two `muons', and so the
diagram

\vspace{1.5cm}

 \begin{figure}[h] \epsfxsize=5cm
\centerline{\epsffile{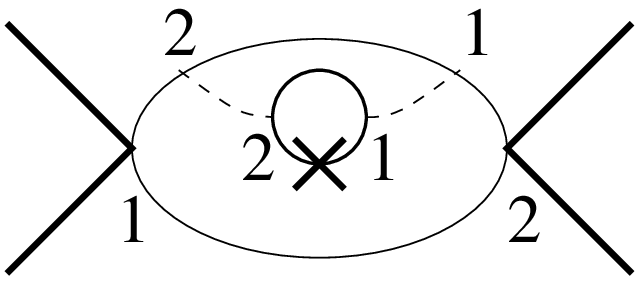}}
\end{figure} \vspace{2mm}

\noindent can not contribute to $\frac{d\Delta \langle N_{e}(\omega)
\rangle}{d\omega}$. Subtracting a temperature independent part, we are
left with

\begin{eqnarray}
&&F^{emission}(\omega_{e})+F^{absorption}(\omega_{e})\nonumber\\
&&\mbox{\hspace{7mm}} = \frac{t\lambda^{2}e^{4}\;f_{B}(\omega_{e})}{\langle
T{\cali{P}}T^{\dag}\rangle_{p_{1}p_{2}}\,V
\omega_{p_{1}}\omega_{p_{2}}(2\pi)^{8}}\int d^{4}q_{7}\; d^{4}q_{e}\;
\delta^{+}(q_{7}^{2}-m^{2}_{e}) \; \delta(q^{2}_{e}-m_{e}^{2}) \;
\delta(q_{e}^{0}-\omega_{e})\nonumber\\
&&\mbox{\hspace{1cm}} \times \int  d^{4}q_{2}\;d^{4}q_{3} \;
\delta^{+}(q_{2}^{2}-m_{\mu}^{2})
\; \delta^{+}(q_{3}^{2}-m_{\mu}^{2})~\left\{
K_{1} \; \delta^{4}(-Q+q_{2}+q_{3}+q_{7}-q_{e})\right. \nonumber\\
&&\mbox{\hspace{1cm}} \left.-K_{2}\;
\; \delta^{4}(-Q+q_{2}+q_{3}+q_{7}+q_{e})\right\},
\tselea{PEHB1}
\end{eqnarray}

\noindent with

\begin{eqnarray*}
&&K_{1}=\frac{1}{(-q_{2}Q + Q^{2}+i\epsilon)}\; \frac{1}{(-q_{3}Q + 
Q^{2}-i\epsilon)}\;
\frac{1}{(-2q_{7}q_{e}+2m^{2}_{e}-m^{2}_{\gamma})^{2}}\nonumber\\
&&K_{2}=K_{1}(q_{e} \rightarrow -q_{e}).
\end{eqnarray*}

\noindent If we are interested in the qualitative behaviour of
(\tseref{PEHB1}) at small $\omega$'s, we need to  perform an integration
over $p_{e}$ only. In order to keep our calculations as simple as
possible, let us assume that $m_{e}=m_{\gamma}=0$. Eq.(\tseref{PEHB1})
can now be handled in a similar way as in the photon heat bath case. We
first perform a transformation

\vspace{-2mm}
\begin{eqnarray*}
&&q_{7} \rightarrow q_{7} \mp q_{e},\\
&&q_{e} \rightarrow q_{e}.
\end{eqnarray*}

\noindent So (\tseref{PEHB1}) now reads

\begin{eqnarray}
\mbox{(\tseref{PEHB1})}&=&\frac{t\lambda^{2}e^{4}\;f_{B}(\omega_{e})}{\langle
\T{\P}{\T}^{\dag}\rangle_{p_{1}p_{2}}\,V
\omega_{p_{1}}\omega_{p_{2}}(2\pi)^{8}}\;\int\;d^{4}q_{2}\;d^{4}q_{3} \;
\delta^{+}(q_{2}^{2}-m_{\mu}^{2})
\; \delta^{+}(q_{3}^{2}-m_{\mu}^{2})\nonumber\\
&&\times \;\frac{1}{(-q_{2}Q + Q^{2}+i\epsilon)}\; \frac{1}{(-q_{3}Q +
Q^{2}-i\epsilon)}\;B,
\end{eqnarray}

\noindent where

\begin{displaymath}
B=\int \frac{d^{4}q_{7}\;d^{4}q_{e}}{(2q_{7}q_{e})^{2}}\left\{
\delta^{+}(q_{7}^{2}+X)-\delta^{+}(q_{7}^{2}-X)\right\}\;
\delta(q_{e}^{2})\;\delta(q_{7}^{0}-\omega_{e})\;
\delta^{4}(-Q+q_{2}+q_{3}+q_{7}),
\end{displaymath}

\noindent with $X=2q_{7}q_{e}$. As before we might expand $B$ in terms of
$X$. First surviving term reads

\begin{eqnarray}
B &\sim & \int d^{4}q_{7}\;d^{4}q_{e}\; \left( \partial_{q_{7}^{2}} \delta   
(q_{7}^{2})\right ) \delta^{4}(-Q +q_{2}+q_{3}+q_{7})\;
\frac{2X}{(2q_{7}q_{e})^{2}}\nonumber\\
&=& \left. -\omega^{0}_{e}\; \partial_{m^{2}}\int d^{4}q_{7}\;
\delta(q_{7}^{2}-m^{2})\; \delta^{4}(-Q + q_{2}+q_{3}+q_{7})\;
\frac{1}{|{\vect{q}}_{7}|} \mbox{ln} \left(
\frac{q_{7}^{0}-|{\vect{q}}_{7}|}{q_{7}^{0}+|{\vect{q}}_{7}|}\right)
\right|_{m=0},\nonumber\\ \end{eqnarray}

\noindent so $B \propto \omega_{e}^{0}$, and consequently
$F^{emission}(\omega_{e})+F^{absorption}(\omega_{e}) \propto  
\omega_{e}^{-1}$. Straightforward application of the previous mathematical
operations to $F^{fluct}(\omega_{e})$ reveals that
$F^{fluct}(\omega_{e})\propto \omega_{e}^{-1}$ as well. Let us mention
that the separate contributions present in $F^{emission}(\omega_{e})$,
$F^{absorption}(\omega_{e})$ and $F^{fluct}(\omega_{e})$ behave as
$\omega^{-2}_{e}$ but they mutually cancel leaving behind terms
proportional at worst to $\omega^{-1}_{e}$.

Surprisingly enough, we have found that, for small $\omega$, our
heat bath (\tseref{Hamilt2}) changes due to scattering $\Phi\Phi
\rightarrow \phi{\bar{\phi}}$ in such a way that the rate of change in the
electron number spectrum has qualitatively similar behaviour (i.e.
$\omega^{-1}$) as the rate of change in the photon number spectrum. This
is so provided one assumes that both electrons and photons are massless
particles.  Clearly, $\omega^{-2}$ behaviour would be disastrous as it
would suggest that the energy density $\omega dN / d\omega$ of the
heat--bath particles behaves as $\omega^{-1}$ which would, if integrated,
produce an infinite contribution to the total energy carried off by the
heat--bath particles.

\vspace{5mm}

\section{Conclusions \label{C1}}

In this chapter we have formulated a systematic method for studying the
heat--bath particle number spectrum using modified cut diagrams. In
particular, for the quark--gluon plasma in thermodynamical equilibrium our
approach should be useful as an effective alternative to the Landshoff and
Taylor \cite{PVLJT} approach. The method used in \cite{PVLJT} (i.e.
to start from first principles) suffers from the lack of a systematic
computational approach for higher orders in coupling constants. One of the
corner stones of our formalism is the largest--time equation (LTE). We have
shown how the zero--temperature LTE can be extended to finite temperature.
During the course of this analysis, we have emphasised some important
aspects of the finite--temperature extension which are worth mentioning.
Firstly, many of kinematic rules valid for zero--temperature diagrams can
not be directly used in the finite--temperature ones. This is because the
emission or absorption of heat--bath particles make it impossible to fix
some particular direction to a diagrammatic line. It turns out that one
finds more diagrams then one used to have at zero temperature. The most
important reductions of the diagrams have been proved. The rather
complicated structure of the finite--temperature diagrams brings into play
another complication: uncutable diagrams. It is well known that at zero  
temperature one can always make only one cut in each cut diagram (this can be
viewed as a consequence of the unitarity condition). This is not true
however at finite temperature. We have found it as useful to start fully
with the LTE analysis which is in terms of type 1 and type 2 vertices.
This language allows us to construct systematically all the cut diagrams.
We have refrained from an explicit use of the cuts in finite--temperature
diagrams as those are ambiguous and therefore rather obscure the
analysis.

The second, rather technical, corner stone are the (functional) thermal
Dyson--Schwinger equations. We have developed a formalism of the {\em  
arrowed} variations acting directly on field operators. This provides an  
elegant technique for dealing in a practical fashion with expectation
values (both thermal and vacuum) whenever functions or functionals of
fields admit the decomposition (\tseref{Hin2}). The merit of the
Dyson--Schwinger equations is that they allow us to rewrite an expectation   
value of some functional of field in terms of expectation values of less
complicated functionals. Some illustrations of this and further thermal 
functional identities are derived in Appendix \ref{A1}.

When we have studied the heat--bath particle number spectrum, we
applied the Dyson--Schwinger equations both to numerator and denominator of
corresponding expression. The results were almost the same. The simple
modification of one propagator rendered both equal. We could reflect this
on a diagrammatical level very easily as the denominator was fully
expressible in terms of thermal cut diagrams. Our final rule for the  
heat--bath particle spectrum is

\begin{equation}
\frac{d \Delta \langle N(\omega) \rangle}{d\omega} = \frac{\langle
{\T}^{\dag}{\P}\,\T \rangle_{p_{1}p_{2}}^{M}}{\langle
{\T}^{\dag}{\P}\,\T \rangle_{p_{1}p_{2}}},\nonumber
\end{equation}

\noindent with $\T$ being the $\T$--matrix, ${\P}$ being the projection
operator onto final states, $p_{1}, p_{2}$ being the momenta of particles
in the initial state, $\beta$ being the inverse temperature and $M$ being
abbreviation for the modified diagrams. Modification of the cut diagrams
consist of the substitution in turn of each heat--bath particle line by an 
altered one. This substitution must be done in each cut diagram.
Replacement must be only one per modified diagram. Our approach is
demonstrated on a simple model where two scalar particles (`pions')
scatter, within a photon heat bath, into a pair of charged
particles (`muon' and `antimuon') and we explicitly calculate the
resulting changes in the number spectra of the photons and. It is also
discussed how the results will change if the photon heat bath is replaced
with photon--electron one.

\chapter{Pressure at thermal equilibrium\label{large-N1}  \label{PE}}
\section{Introduction\label{large-N2} \label{PE1}}

A  significant quantity  of  physical interest  that  one  may want to
calculate in field theory at finite temperature, either at equilibrium
or out of  equilibrium, is pressure.  In thermal quantum  field theory
(both  in the real-- and  imaginary--time formalism) where one usually
deals    with  systems  in thermal  equilibrium      there is an  easy
prescription for  a pressure calculation.  The latter  is based on the
observation that for     thermally  equilibrated systems     the grand
canonical partition function $Z$ is given as 

\begin{equation}
Z = e^{-\beta \Omega} = Tr(e^{-\beta(H-\mu_{i}N_{i})}),
\tselea{1}
\end{equation}

\vspace{2mm}
\noindent where $\Omega$ is the grand canonical  potential, $H$ is the
Hamiltonian,    $N_{i}$   are   conserved  charges,   $\mu_{i}$    are
corresponding   chemical  potentials, and    $\beta$  is   the inverse
temperature:   $\beta      =   1/T$.    Using       identity    $\beta
\frac{\partial}{\partial  \beta}=  -T    \frac{\partial}{\partial  T}$
together with (\tseref{1}) one gets 

\begin{equation}
 T \left( \frac{\partial \Omega}{\partial T} \right)_{\mu_{i},V} =
\Omega -E + \mu_{i} N_{i},
\tselea{2}
\end{equation}

\vspace{2mm}
\noindent with $E$ and $V$ being the averaged energy and volume of the 
system respectively. A comparison of (\tseref{2}) with a corresponding 
thermodynamic expression for the grand canonical potential \cite{LW,GM,Cub} requires that entropy $S=- \left(\frac{\partial \Omega}{\partial T}
\right)_{\mu_{i}, V}$, so that

\begin{equation}
d\Omega = -SdT - pdV -N_{i}d\mu_{i}\; \Rightarrow \; p= -\left(
\frac{\partial \Omega}{\partial V} \; \right)_{\mu_{i}, T}.
\tselea{3}
\end{equation}

\vspace{2mm}

\noindent For large systems one can usually neglect surface effects so $E$
and $N_{i}$ become extensive quantities. Eq.(\tseref{2}) then immediately
implies that $\Omega$ is extensive quantity as well, so (\tseref{3})
simplifies to   
\begin{equation}
p = -\frac{\Omega}{V} = \frac{\mbox{ln}Z}{\beta V}.
\tselea{4}
\end{equation}

\vspace{3mm}
\noindent The pressure defined by Eq.(\tseref{4}) is the so called
thermodynamic pressure.

Since $\mbox{ln}Z$ can be systematically calculated summing up
all connected closed diagrams (i.e. bubble diagrams) \cite{LW,LL,DJ},
the pressure calculated via (\tseref{4}) enjoys a considerable popularity
\cite{ID1,LW,LB,ID}.  Unfortunately, the latter procedure can not
be extended to out of equilibrium as there is, in general, no definition
of the partition function $Z$ nor grand canonical potential $\Omega$ away
from an equilibrium.

Yet another, alternative definition of the pressure not hinging on
thermodynamics can be provided; namely the hydrostatic pressure which is
formulated through the energy--momentum tensor $\Theta^{\mu \nu}$. The
formal argument leading to the hydrostatic pressure in $D$ space--time
dimensions is based on the observation that $\langle \Theta^{0 j}(x)
\rangle$ is the mean (or macroscopic) density of momenta ${\vect{p}}^{j}$
at the point $x^{\mu}$. Let ${\vect{P}}$ be the mean total
$(D-1)$--momentum of an infinitesimal volume $V^{(D-1)}$ centred at
${\vect{x}}$, then the rate of change of $j$--component of ${\vect{P}}$
reads

\begin{equation}
\frac{d{\vect{P}}^{j}(x)}{dt} = \int_{V^{(D-1)}}d^{D-1}{\vect{x}}' \;
\frac{\partial}{\partial
x^{0}} \langle \Theta^{0 j}(x^{0},{\vect{x}}' ) \rangle =
- \sum_{i=1}^{D-1}\int_{\partial
V^{(D-1)}} d{\vect{s}}^{i} \; \langle \Theta^{i j}
\rangle.
\tselea{5}
\end{equation}

\vspace{2mm}

\noindent In the second equality we have exploited the continuity equation
for $\langle \Theta^{\mu j} \rangle$ and successively we have used Gauss's
theorem\footnote{The macroscopic conservation law for $\langle
\Theta^{\mu \nu} \rangle$ (i.e. the continuity equation) has to be
postulated. For some systems, however, the later can be directly derived
from the corresponding microscopic conservation law \cite{DG}.}. The
$\partial V^{(D-1)}$ corresponds to the surface of $V^{(D-1)}$.

Anticipating     a  system out  of     equilibrium,  we must assume  a
non--trivial    distribution  of   the  mean  particle  four--velocity
$U^{\mu}(x)$ (hydrodynamic velocity). Now, a pressure is by definition
a scalar quantity.  This particularly means  that it should not depend
on the hydrodynamic velocity. We must thus go  to the local rest frame
and evaluate pressure there. However,  in the local rest frame, unlike
the equilibrium,   the notion of   a  pressure acting equally   in all
directions  is  lost.   In  order to  retain  the scalar  character of
pressure, one customarily  defines the {\em pressure  at a  point} (in
the following denoted as   $p(x)$)  \cite{Bach}, which is simply   the
`averaged pressure' {\footnote {To  be  precise, we should  talk about
averaging the normal   components  of stress \cite{Bach}.}} over   all
directions at a given point.  In  the local rest frame Eq.(\tseref{5})
describes  $j$--component of the force  exerted  by the medium on  the
infinitesimal  volume  $V^{(D-1)}$.    (By definition,  there   is  no
contribution   to  $d{\vect{P}}^{j}(x) / dt$    caused by the particle
convection   through  $\partial  V^{(D-1)}$.)   Averaging the  LHS  of
(\tseref{5})   over      all      directions    of     the      normal
${\vect{n}}({\vect{x}})$, we   get\footnote{The  angular average    is
standardly   defined  for      scalars   (say, $A$)    as;    $\int  A
\;d\Omega({\vect{n}})/  \int   d\Omega({\vect{n}})$, and  for  vectors
(say, ${\vect{A}}^{i}$)       as;  $\sum_{j}\int {\vect{A}}^{j}     \;
{\vect{n}}^{j}\;       d\Omega               ({\vect{n}})/        \int
d\Omega({\vect{n}})$.   Similarly we  might    write down  the angular
averages for tensors of a higher rank.}

\begin{eqnarray}
\mbox{$\frac{1}{\left(S^{D-2}_{1}\right)}$}\sum_{j=1}^{D-1}\int \;
\frac{d{\vect{P}}^{j}(x)}{dt}\; {\vect{n}}^{j}
\; d
\Omega({\vect{n}}) &=&
- \mbox{$\frac{1}{\left(S^{D-2}_{1}\right)}$}\sum_{j,i=1}^{D-1}\int_{\partial
V^{(D-1)}}
ds\; \langle \Theta^{ij}(x') \rangle \; \int d\Omega({\vect{n}})\;
{\vect{n}}^{i}{\vect{n}}^{j}\nonumber\\
&=& \frac{1}{(D-1)}\sum_{i=1}^{D-1} \int_{\partial V^{(D-1)}}ds \;
\langle \Theta^{i}_{\; i}(x')
\rangle,
\tselea{EMT22}
\end{eqnarray}

\vspace{3mm}

\noindent where $d\Omega({\vect{n}})$ is an element of solid angle about
${\vect{n}}$ and $S^{D-2}_{1}$ is the surface of $(D-2)$--sphere with unit
radius ($\int d\Omega({\vect{n}}) = S^{D-2}_{1} = 2
\pi^{\frac{D-1}{2}}/\Gamma (\mbox{$\frac{D-1}{2}$})$) . On the other hand,
from the definition of the pressure at a point $x^{\mu}$ we might write

\vspace{2mm}

\begin{equation}
\left(S^{D-2}_{1}\right)^{-1}\sum_{j=1}^{D-1}\int \;
\frac{d{\vect{P}}^{j}(x)}{dt}\; {\vect{n}}^{j}
\; d
\Omega({\vect{n}}) = - p(x) \; \int_{\partial V^{(D-1)}}ds,
\tseleq{EMT23}
\end{equation}

\vspace{3mm}

\noindent here the minus sign reflects that the force responsible for a
compression (conventionally assigned as a positive pressure) has reversed
orientation than the surface normals ${\vect{n}}$ (pointing outward). In
order to keep track with the standard text--book definition of a sign of
a pressure \cite{Cub,Bach} we have used in (\tseref{EMT23}) the
normal ${\vect{n}}$ in a contravariant notation (note, ${\vect{n}}^{i} = -
{\vect{n}}_{i}$). Comparing (\tseref{EMT22}) with (\tseref{EMT23}) we can
write for a sufficiently small volume $V^{(D-1)}$

\begin{equation}
p(x)= - \frac{1}{(D-1)} \sum_{i=1}^{D-1}\langle \Theta^{i}_{\;i}(x)
\rangle. \tseleq{EMT24}
\end{equation}

\vspace{3mm}

\noindent We should point out that in equilibrium the thermodynamic
pressure is usually identified with the hydrostatic one via the virial
theorem \cite{LW,Zub}. In the remainder of this chapter we shall deal
with the hydrostatic pressure at equilibrium. We shall denote the
foregoing as ${\cali{P}}(T)$, where $T$ stands for temperature. We consider
the non--equilibrium case in the next chapter.

The plan of this chapter is as follows. In Section \ref{PE2} we review
the necessary mathematical framework needed for the renormalisation of  
the energy--momentum tensor.(For an extensive review see also
refs.\cite{LW,Collins,Brown}.)  The latter is discussed on the
$O(N)\; \Phi^{4}$ theory. As a byproduct we renormalise
$\Phi_{a}^{2}$, $\Phi_{a}\Phi_{b}$ and $\Theta^{\mu \nu}$
operators. The corresponding QFT extension of (\tseref{EMT24}) is
obtained.

Resumed form for the pressure in the large--$N$ limit,
together with the discussion of both coupling constant and mass
renormalisation is worked out in Section \ref{PE3}. The discussion is
substantially simplified by means of the Dyson--Schwinger equations.

In Section \ref{HTE} we end up with the high--temperature expansion of
the pressure. Calculations are performed for $D=4$ (both for massive and
massless fields) and the result is expressed in terms of renormalised
masses $m_{r}(0)$ and $m_{r}(T)$. The former is done by means of the
Mellin transform technique.

This chapter is furnished with two appendices. In Appendix \ref{A2} we
clarify some mathematical manipulations  needed in Section \ref{PE3}.  For the
completeness'   sake    we   compute   in    Appendix    \ref{A3}  the
high--temperature   expansion  of  the  thermal--mass  shift  $\delta
m^{2}(T)$ which will prove useful in Section \ref{HTE}. 

\vspace{5mm}

\section{Renormalisation \label{PE2}}

If we proceed with (\tseref{EMT24}) to QFT this leads to the notorious
difficulties connected with the fact that $\Theta^{\mu \nu}$ is a
(local) composite operator.  If only a free theory would be in
question then the normal ordering prescription would be sufficient to
render $\langle \Theta^{\mu \nu} \rangle$ finite. In the general case,
when the interacting theory is of interest, one must work with the
Zimmerman `normal' ordering prescription instead. Let us demonstrate
the latter on the $O(N)\; \Phi^{4}$ theory. (In this section we
keep $N$ arbitrary.) Such a theory is defined by the bare Lagrange
function

\begin{equation}
{\cali{L}}= \frac{1}{2}\sum_{a=1}^{N}\left( (\partial
\Phi_{a})^{2}-m_{0}^{2}\Phi_{a}^{2} \right) -
\frac{\lambda_{0}}{8N}\left( \sum_{a=1}^{N} (\Phi_{a})^{2} \right)^{2},
\tseleq{6}
\end{equation}

\vspace{2mm}

\noindent we assume that $m^{2}_{0}>0$. The corresponding canonical
energy--momentum tensor is given by

\begin{equation}
\Theta^{\mu \nu}_{c} =
\sum_{a}\partial^{\mu}\Phi_{a}\partial^{\nu}\Phi_{a} - g^{\mu \nu}
{\cali{L}}\, .
\tseleq{7}
\end{equation}

\vspace{2mm}

\noindent The Feynman rules for Green's functions with the energy--momentum
insertion can be easily explained in momentum space. In the reasonings to
follow we shall need the (thermal) composite Green's
function\footnote{By $\Phi$ we shall mean the field in the Heisenberg
picture. The subscript $H$ will be introduced in cases when a possible
ambiguity could occur.}

\begin{equation}
D^{\mu \nu}(x^{n}|y) = \langle {\cali{T}}^{*}\left\{ \Phi_{r}(x_{1}) \ldots
\Phi_{r}(x_{n}) \Theta^{\mu \nu}_{c}(y) \right\} \rangle\, .
\tseleq{8}
\end{equation}

\vspace{2mm}

\noindent Here  the subscript $r$ denotes   the renormalised fields in
the  Heisenberg picture  (the  internal indices   are suppressed)  and
${\cali{T}}^{*}$  is so called  ${\cali{T}}^{*}$ product (or covariant
${\cali{T}}$    product)  \cite{N,RJ,CCR,IZ}.    The  ${\cali{T}}^{*}$
product  is defined in  such a way that it  is simply the ${\cali{T}}$
product with all  differential operators ${\cali{D}}_{\mu_{i}}$ pulled
out of the ${\cali{T}}$--ordering symbol, i.e.

\begin{equation}
{\cali{T}}^{*}\{ {\cali{D}}_{\mu_{1}}^{x_{1}}\Phi(x_{1})\ldots
{\cali{D}}_{\mu_{n}}^{x_{n}}\Phi(x_{n})
\} = {\cali{D}}(i\partial_{\{\mu\}}) {\cali{T}}\{\Phi(x_{1})\ldots
\Phi(x_{n})\}\, ,
\tseleq{TP1}
\end{equation}
\vspace{1mm}

\noindent  where ${\cali{D}}(i\partial_{\{\mu\}})$  is  just a  useful
 short--hand                        notation                        for
 ${\cali{D}}_{\mu_{1}}^{x_{1}}{\cali{D}}_{\mu_{2}}^{x_{2}}\ldots
 {\cali{D}}_{\mu_{n}}^{x_{n}}$.  In   the   case  of  thermal  Green's
 functions,   the  ${\cali{T}}$ might  be as  well  a contour--ordering
 symbol.  It is the mean  value of the ${\cali{T}}^{*}$ ordered fields
 rather than the ${\cali{T}}$ ones,  which corresponds at $T=0$ and at
 equilibrium to  the  Feynman path integral representation  of Green's
 functions \cite{CCR,JS}.

A typical contribution to $\Theta^{\mu \nu}_{c}(y)$ can be   
written as

\begin{equation}
{\cali{D}}_{\mu_{1}}\Phi(y)
\; {\cali{D}}_{\mu_{2}}\Phi(y) \ldots {\cali{D}}_{\mu_{n}}\Phi(y)\, ,
\tseleq{CO1}
\end{equation}

\vspace{2mm}

\noindent so the typical term in (\tseref{8}) is

\begin{displaymath}
{\cali{D}}(i\partial_{\{\mu\}})\; \langle {\cali{T}}^{*}\left\{
\Phi_{r}(x_{1})\ldots   
\Phi_{r}(x_{n})\Phi(y_{1})\ldots \Phi(y_{k}) \right\}\rangle \left.
\right|_{y_{i}=y}.
\end{displaymath}

\vspace{3mm}

\noindent Performing the Fourier transform in (\tseref{8}) we get

\begin{equation}
D^{\mu \nu}(p^{n}|p)= \sum_{k=\{2,4\}}\; \int  \left( \prod_{i=1}^{k}  
\frac{d^{D}q_{i}}{(2\pi)^{D}}\right) (2\pi)^{D}\;
\delta^{D}(p-\sum_{j=1}^{k}q_{j})\; {\cali{D}}_{(k)}^{\mu \nu}(q_{\{ \mu
\}}) \; D(p^{n}|q^{k}),
\tseleq{11}
\end{equation}

\vspace{3mm}  

\noindent where ${\cali{D}}_{(k)}^{\mu \nu}(\ldots)$ is a Fourier
transformed differential operator corresponding to the quadratic (k=2) and
quartic (k=4) terms in $\Theta^{\mu \nu}_{c}$. Denoting the new vertex
corresponding to ${\cali{D}}_{(k)}^{\mu \nu}(\ldots)$ as $\otimes$, we can
graphically represent (\tseref{8}) through (\tseref{11}) as

\vspace{5mm}

\begin{figure}[h] 
\begin{center}   
\leavevmode
\hbox{%
\epsfxsize=7.5cm
\epsffile{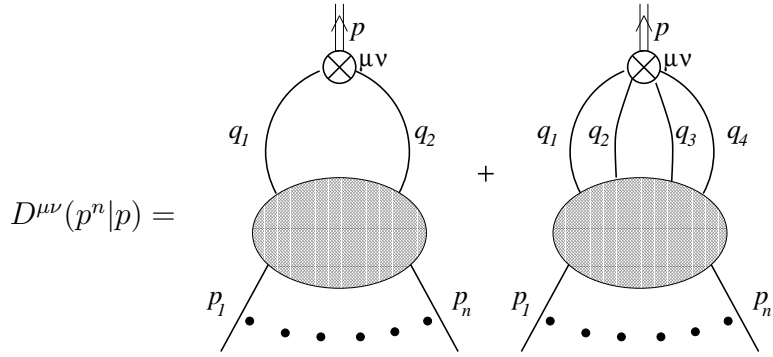}}
\caption{ The graphical representation of $D^{\mu \nu}(p^{n}|p)$.}
\label{fig1}
\end{center}
\begin{picture}(10,10)
\put(50,100){$D^{\mu \nu}(p^{n}|p)=$}
\end{picture}
\end{figure}
\noindent For the case at hand one can easily read off from (\tseref{7})
an explicit form of the bare composite vertices, the foregoing are

\vspace{3mm}

\begin{figure}[h]
\begin{flushleft}
\leavevmode
\hbox{%
\epsfxsize=2cm
\epsffile{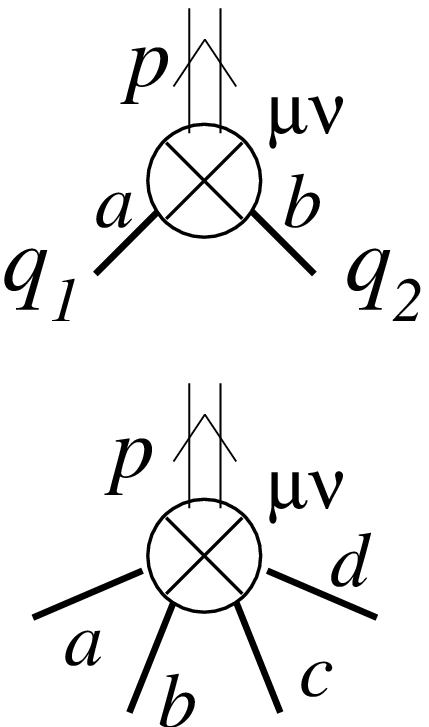}}
\end{flushleft}   
\begin{picture}(20,5)
\put(65,105){$\sim \;{\cali{D}}_{(2)}^{\mu \nu}(q_{\{ \mu \}}) =
\frac{1}{2}\;\delta_{ab}\;\{ 2(q_{1}-p)^{\mu}q_{1}^{\nu}-g^{\mu
\nu}((q_{1}-p)_{\lambda}q_{1}^{\lambda}-m_{0}^{2})\}$}
\put(65,50){$\sim \;{\cali{D}}_{(4)}^{\mu \nu}(q_{\{ \mu \}}) = \frac{g^{\mu
\nu} \lambda_{0}}{8N}\{
2(\delta_{ab}\delta_{cd}+\delta_{ac}\delta_{bd}+\delta_{ad}\delta_{bc})
- 5 \delta_{ab}\delta_{cd}\delta_{ac}\}$}
\end{picture}
\end{figure}
\noindent (For the  internal    indices we do not    adopt  Einstein's
summation convention.) We  have tacitly assumed in FIG.\ref{fig1} that
the vacuum  bubble  diagrams present in the   shaded blobs are divided
out. We have  also   implicitly assumed that summation  over  internal
indices  is  understood.  Note that  in the  case of thermal composite
Green's function, the new vertices are clearly of type--1 as the fields
from which they are deduced have all a real--time argument\footnote{For
a brief introduction to the real--time formalism in thermal QFT see for
example \cite{TA}.} (type--1 fields). 

\vspace{7mm}  
\noindent {\it Renormalisation of $\Phi_{a}(x)\Phi_{b}(x)$}
\vspace{2mm}      

\noindent Now, if there would be no $\Theta^{\mu \nu}_{c}$ insertion in
(\tseref{8}), the latter would be finite, and  so it is natural to define the
renormalised energy--momentum tensor $[\Theta^{\mu \nu}_{c}]$ (or Zimmermann
normal ordering) in such a way that

\begin{displaymath}
D_{r}^{\mu \nu}(x^{n}|y) = \langle {\cali{T}}^{*}\left\{ \Phi_{r}(x_{1}) \ldots
\Phi_{r}(x_{n})\; [\Theta^{\mu \nu}_{c}] \right\} \rangle,
\end{displaymath}

\noindent is finite for any $n > 0$. To see what is involved, we
illustrate the mechanism of the composite operator renormalisation on   
$\Phi_{a}(x)\Phi_{b}(x)$. We shall use the mass--independent
renormalisation (or minimal subtraction scheme -- (MS)) which is
particularly suitable for this purpose.  In MS we can expand the bare 
parameters into the Laurent series which has a simple form
\cite{Brown,IZ,JS}, namely

\begin{equation}
\lambda_{0} = \mu^{4-D}\; \lambda_{r} \left( 1 + \sum_{k=1}^{\infty}
\frac{a_{k}(\lambda_{r}; D)}{(D-4)^{k}}\right)
\tseleq{CO2}
\end{equation}    
\begin{equation}
m_{0}^{2} = m_{r}^{2} \left( 1 + \sum_{k=1}^{\infty}\frac{b_{k}(\lambda_{r};
D)}{(D-4)^{k}} \right)\, .
\tseleq{CO3}
\end{equation}

\vspace{2mm}

\noindent Here $a_{0}$ and $b_{0}$ are analytic in $D=4$. The parameter
$\mu$ is the scale introduced by the renormalisation in order to keep
$\lambda_{r}$ dimensionless. An important point is that both $a_{k}$'s and
$b_{k}$'s are mass, temperature and momentum independent.

It was Zimmermann who first realized that the forest formula
known from the ordinary Green's function renormalisation
\cite{Collins,IZ} can be also utilised for the composite Green's
functions rendering them finite \cite{Collins,Zimm}. That is, we start
with Feynman diagrams expressed in terms of physical (i.e. finite)
coupling constants and masses. As we calculate diagrams to a given order,
we meet UV divergences which might be cancelled by adding counterterm
diagrams. The forest formula then prescribes how to systematically cancel
all the UV loop divergences by counterterms to all orders. However, in
contrast to the coupling constant renormalisation, the composite vertex
need not to be renormalised multiplicatively. We shall illustrate this
fact in the sequel. Let us also observe that in the lowest order (no loop)
the renormalised composite vertex equals to the bare one, and so to that
order $A=[A]$, for any composite operator $A$.

Now, from (\tseref{CO2}) and (\tseref{CO3}) follows that for any
function $F= F(m_{r}, \lambda_{r})$ we have
\vspace{2mm}

\begin{displaymath}
\frac{\partial F}{\partial m^{2}_{r}} = \frac{\partial
m_{0}^{2}}{\partial m_{r}^{2}}\; \frac{\partial F}{\partial m_{0}^{2}} =
\frac{
m_{0}^{2}}{ m_{r}^{2}}\; \frac{\partial F}{\partial m_{0}^{2}}\, .
\end{displaymath}

\vspace{2mm}

\noindent So particularly for

\begin{displaymath}
F= D(x_{1}, \ldots , x_{n}) = \langle {\cali{T}}^{*} \{ \Phi_{r}(x_{1}) \ldots 
\Phi_{r}(x_{n}) \} \rangle,
\end{displaymath}

\noindent one reads
\begin{eqnarray}
&&m^{2}_{r}\; \frac{\partial}{\partial m^{2}_{r}} D(x_{1}, \ldots , x_{n})
= m^{2}_{0}\; \frac{\partial}{\partial m^{2}_{0}} D(x_{1}, \ldots ,
x_{n})\nonumber \\
&&\mbox{\hspace{1.5cm}}= \left( - \frac{i}{2} \right) {\cali{N}}\; \int   
d^{D}x\;\sum_{a=1}^{N}\;
\int {\cali{D}}\phi\; \phi_{r}(x_{1}) \ldots \phi_{r}(x_{n})\;
m_{0}^{2}\phi_{a}^{2}(x)\; \mbox{exp}(iS[\phi, T])\nonumber \\
&&\mbox{\hspace{1.5cm}}= \left( - \frac{i}{2} \right) \; \int
d^{D}x\;\sum_{a=1}^{N}\;D_{a}(x_{1}, \ldots, x_{n}|x; m_{0}^{2})\, .
\tselea{CO5}
\end{eqnarray}

\vspace{2mm}

\noindent Here ${\cali{N}}^{-1}$ is the standard denominator of the path
integral representation of Green's function. We should apply the
derivative also on ${\cali{N}}$ but this would produce disconnected graphs
with bubble diagrams. The former precisely cancel the very same
disconnected graphs in the first term, so we are finally left with no
bubble diagrams in (\tseref{CO5}). In the Fourier space (\tseref{CO5})
reads

\begin{equation}   
m^{2}_{r}\; \frac{\partial}{\partial m^{2}_{r}} D(p_{1}, \ldots ,p_{n}) =
\left( - \frac{i}{2}  \right) \sum_{a=1}^{N}\; D_{a}(p_{1}, \ldots ,
p_{n}|0;m_{0}^{2})\, .
\tseleq{CO6}
\end{equation}

\vspace{2mm}

\noindent As the LHS is finite, there cannot be any pole terms on the RHS
either, and so $\sum_{a} m^{2}_{0}\Phi_{a}^{2}$ is by itself a renormalised
composite operator. We see that $m_{0}^{2}$ precisely compensates the
singularity of $\sum_{a=1}^{N} \Phi_{a}^{2}$.

Now, it is well known that any second--rank tensor (say $M_{a
b}$) can be generally decomposed into three irreducible tensors; an
antisymmetric tensor, a symmetric traceless tensor and an invariant
tensor. Let us set $M_{ab}=\Phi_{a}\Phi_{b}$, so the symmetric traceless  
tensor $K_{a b}$ reads

\vspace{-2mm}
\begin{equation}
K_{ab}(x) = \Phi_{a}(x)\Phi_{b}(x)-\delta_{a
b}/N\;\sum_{c=1}^{n}\Phi^{2}_{c}(x)\, ,
\tseleq{CO64}      
\end{equation}

\noindent whilst the invariant tensor $I_{a b}$ is

\begin{displaymath}
I_{a b}(x) = \delta_{a b}/N \sum_{c=1}^{N}\Phi_{c}^{2}(x)\, .
\end{displaymath}

\noindent Because the renormalised composite operators have to preserve a
tensorial structure of the bare ones, we immediately have that

\begin{equation}
K_{a b} = A_{1}[K_{a b}]\;\;\; \mbox{and}\;\;\; I_{a b} = A_{2}[I_{a
b}]\, ,
\tseleq{CO66}
\end{equation}

\vspace{2mm}

\noindent where both $A_{1}$ and $A_{2}$ must have structure $(1 + \sum   
(\mbox{poles}))$. The foregoing guarantees that to the lowest order $K_{a
b} = [K_{a b}]$ and $I_{a b}= [I_{a b}]$. As we saw in (\tseref{CO6}),
$m_{0}^{2}I_{a b}$ is renormalised, and so from (\tseref{CO66}) follows
that $m_{0}^{2}I_{a b} = C\; [I_{a b}]$. Here $C$ has dimension $[m^{2}]$
and is analytic in $D=4$.  We can uniquely set $C=m^{2}_{r}$ because only
this choice fulfils the lowest order condition $I_{ab}=[I_{ab}]$ (c.f.
Eq.(\tseref{CO3})). Collecting our results together we might write

\begin{equation}
\sum_{c} \Phi_{c}^{2} = Z_{\Sigma \Phi^{2}}\; \left[ \sum_{c}
\Phi_{c}^{2} \right] = Z_{\Sigma \Phi^{2}} \sum_{c} [\Phi_{c}^{2}]\, ,
\tseleq{CO67}
\end{equation}

\noindent with $Z_{\Sigma\Phi^{2}} = A_{2}= \frac{m^{2}{r}}{m^{2}_{0}}$.
In the second equality we have used an obvious linearity \cite{Collins}
of $[\ldots]$. From (\tseref{CO64}) and (\tseref{CO67}) follows that

\begin{equation}
\Phi_{a}(x)\Phi_{b}(x) = A_{1}[\Phi_{a}(x)\Phi_{b}(x)] -
\frac{\delta_{ab}}{N}(A_{1} - Z_{\Sigma \Phi^{2}})\;
\sum_{c=1}^{N}[\phi_{c}^{2}(x)]\, . \tseleq{CO57}
\end{equation}

\noindent So particularly for $\Phi^{2}_{a}$ one reads

\begin{equation}
\Phi^{2}_{a} = \frac{1}{N}\left( (N-1)A_{1} + Z_{\Sigma \Phi^{2}} \right)
\; [\Phi^{2}_{a}] - \frac{1}{N} \left( A_{1} - Z_{\Sigma \Phi^{2}}
\right) \sum_{c \not = a } [\Phi_{c}^{2}]\, .
\tseleq{CO58}
\end{equation}

\vspace{3mm}

\noindent From the discussion above it does not seem to be possible to
obtain more information about $A_{1}$ without doing an explicit
perturbative calculations, however, it is easy to demonstrate that $A_{1} 
\not= Z_{\Sigma \Phi^{2}}$. To show this, let us consider the simplest
non--trivial case; i.e. N=2, and calculate $A_{1}$ to order $\lambda_{r}$.
For that we need to discuss the renormalisation of the $n$-point composite
Green's function with, say, $\Phi^{2}_{1}$ insertion. To do that, it
suffices to discuss the renormalisation of the corresponding 1PI $n$--point
Green's function. The perturbative expansion for the composite vertex to
order $\lambda_{r}$ can be easily generated via the Dyson--Schwinger
equation \cite{PC} and it reads

\vspace{8mm}

\begin{figure}[h]
\begin{center}
\leavevmode
\hbox{%
\epsfxsize=12.3cm
\epsffile{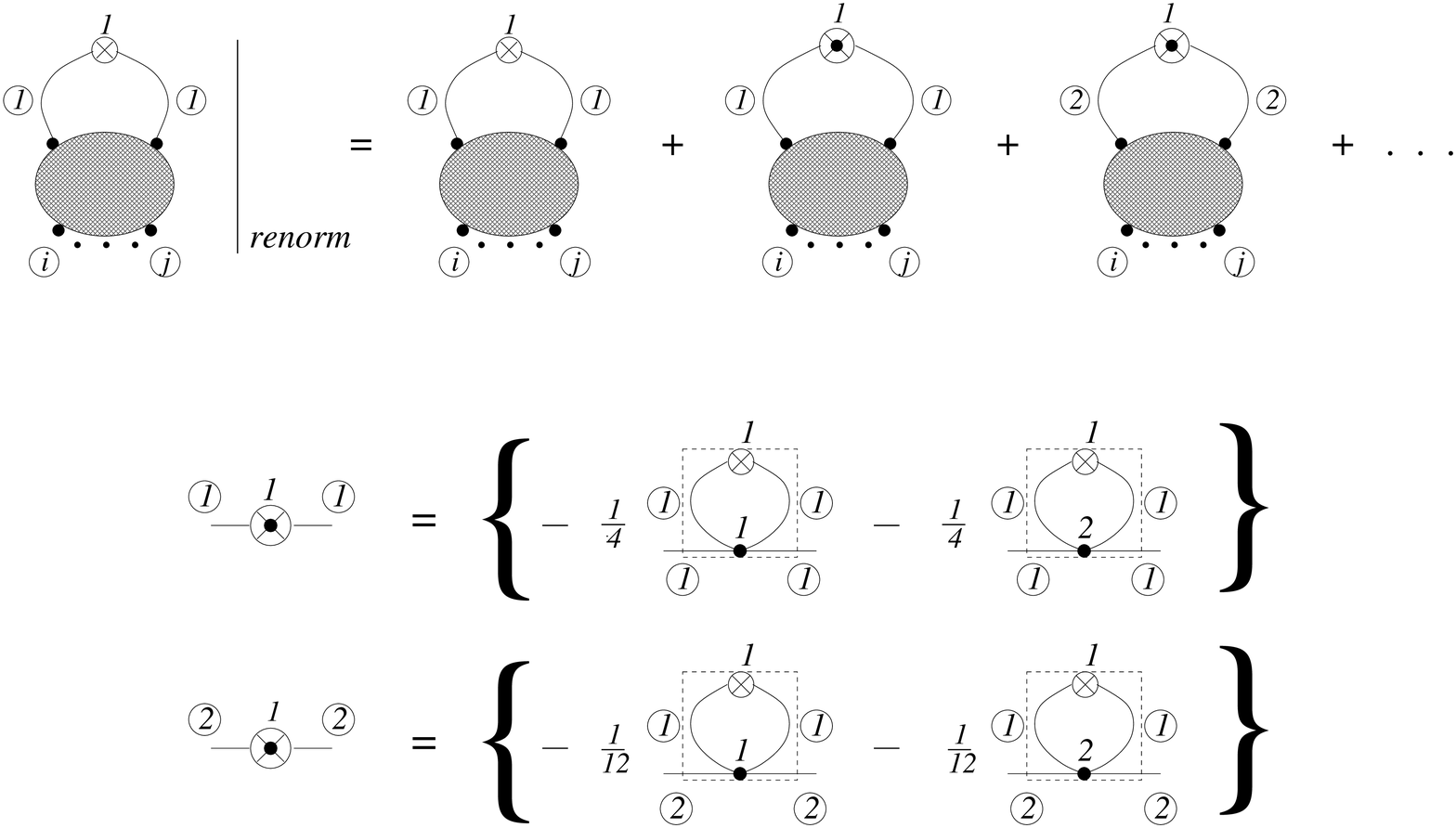}}
\end{center}  
\setlength{\unitlength}{1mm}
\begin{picture}(20,7)
\put(0,40){where}
\put(154,71){(3.24)}
\end{picture}
\end{figure}

\addtocounter{equation}{1}

\noindent Here cross--hatched blobs refer to (renormalised) 1PI
$(n+2)$--point Green's function, circled indices mark a type of the field  
propagated on the indicated line, and uncircled numbers refer to thermal
indices (we explicitly indicate only relevant thermal indices). The
counterterms, symbolised by a heavy dot, are extracted from the boxed
diagrams (elementary Zimmermann forests).  In MS scheme one gets the
following results:

\vspace{8mm}

\begin{figure}[h]
\begin{flushleft}
\leavevmode
\hbox{%
\epsfxsize=2cm
\epsffile{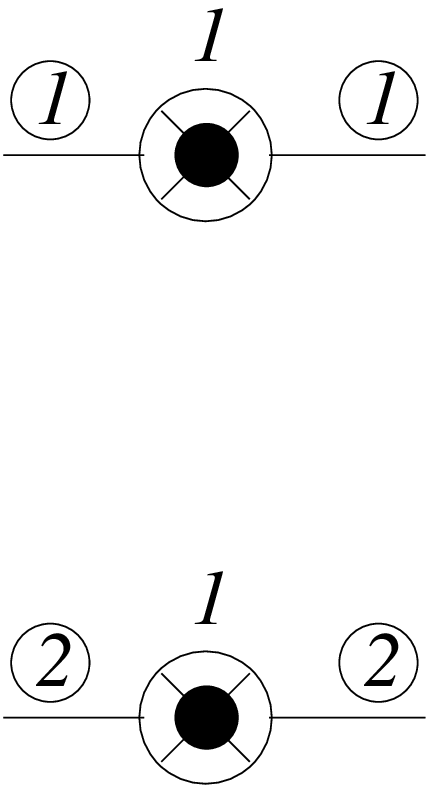}}
\end{flushleft}
\setlength{\unitlength}{1mm}
\begin{picture}(15,10)
\put(25,43){$= \frac{i\;\lambda_{r}\mu^{4-D}}{4}\int
\frac{d^{D}q}{(2pi)^{D}} \left\{ \D_{11}(q)\D_{11}(-q) -
\D_{12}(q)\D_{12}(-q) \right\}|_{\mbox{{\scriptsize MS pole term}}}$}
\put(25,32){$= - \frac{1}{4} \partial_{m_{r}^{2}}\left(
\frac{\Gamma\left(1-\frac{D}{2}\right)}{(4 \pi)^{\frac{D}{2}}} \;
\lambda_{r}\; \mu^{4-D}\; m_{r}^{D-2}\right)|_{\mbox{\scriptsize MS}} =
- \lambda_{r} \mu^{4-D} /2\; (D-4)\; (4 \pi)^{2}$}
\put(25,17){$ = - \lambda_{r} \mu^{4-D} / 6\; (D-4)\; (4 \pi)^{2}$.}
\end{picture}
\end{figure}
\noindent Here $\D_{11}$ and $\D_{12}$ are the usual thermal propagators
in the real--time formalism \cite{LW,LB,TA} (see also Section \ref{PE3}).
From (4.24) we can directly read off that
\vspace{1mm}

\begin{displaymath}
[\Phi^{2}_{1}] = \left(1 - \frac{\lambda_{r} \mu^{4-D}}{2\;(D-4)\; (4
\pi)^{2}} +
{\cali{O}}(\lambda_{r}^{2})\right)\;\Phi_{1}^{2} + \left(-
\frac{\lambda_{r} \mu^{4-D}}{6\; (D-4)\; (4 \pi)^{2}} +
{\cali{O}}(\lambda_{r}^{2})\right)\;\Phi_{2}^{2}\, .
\end{displaymath}

\vspace{4mm}

\noindent As the coefficient before $\Phi_{2}^{2}$ is not zero, we
conclude that $A_{1} \not= Z_{\Sigma \Phi^{2}}$. It is not a great   
challenge to repeat the previous calculations for the $\Phi_{1}\Phi_{2}$
insertion. The latter gives
\vspace{1mm}

\begin{displaymath}
 A_{1} = 1 - \frac{\lambda_{r} \mu^{4-D}}{3\; (D-4)\;
(4 \pi)^{2}} +
{\cali{O}}(\lambda_{r}^{2})\, .
\end{displaymath}

\vspace{3mm}

\noindent Eq.(\tseref{CO58}) exhibits the so called operator mixing
\cite{IZ}; the renormalisation of $\Phi_{a}^{2}$ cannot be considered
independently of the renormalisation of $\Phi_{c}^{2}$ ($c \not= a$). The
latter is a general feature of composite operator renormalisation.  Note,
however, that $\Phi_{a}\Phi_{b}$ ($a \not=b$) do not mix by
renormalisation, i.e. they renormalise multiplicatively. It can be shown
that composite operators mix under renormalisation only with those
composite operators which have dimension less or equal \cite{Collins,IZ,Zimm}.

Unfortunately, if we apply the previous arguments to $n=0$, the result
is  not finite; another additional  renormalisation must be performed.
The fact that  the expectation values  of $[\ldots]$ are generally  UV
divergent,   in spite  of  being  finite  for  the  composite  Green's
functions\footnote{Also called  the matrix elements  of  $[\ldots]$.},
can be nicely illustrated  with the composite operator $[\Phi^{2}]$ in
the  $N=1$ theory.  Taking the   diagrams for  $D(0|0)$  and  applying
successively the (unrenormalised) Dyson--Schwinger equation  \cite{PC}
we get

\vspace{4cm}

\begin{figure}[h]
\begin{center}
\leavevmode
\hbox{%
\epsfxsize=12cm
\epsffile{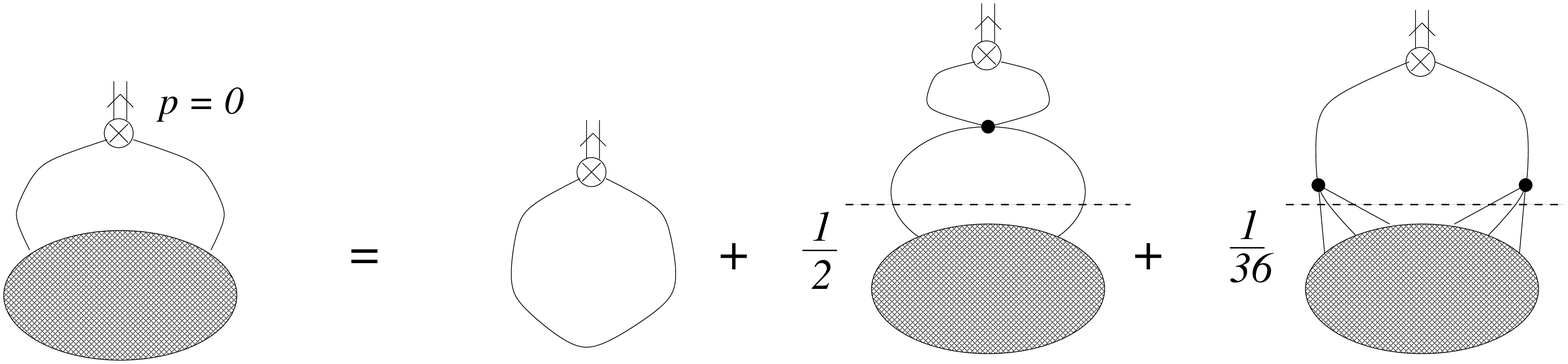}}
\end{center}
\setlength{\unitlength}{1mm}
\begin{picture}(20,7)
\put(154,20){(3.25)}
\end{picture}
\end{figure}

\addtocounter{equation}{1}

\noindent Eq.(3.25) might be rewritten as

\begin{eqnarray}
D(0|0) &=& D(0|0)|_{\lambda_{r}^{0}}\nonumber \\
&& + \frac{1}{2} \; \int
\frac{d^{D}q_{1}}{(2\pi)^{D}}\frac{d^{D}q_{2}}{(2\pi)^{D}}\;
\delta^{D}(q_{1}+ q_{2})\;
D^{amp}(q^{2}|0)|_{\lambda_{r}}\; D(q^{2})\nonumber \\
&& + \frac{1}{36} \; \int \prod_{i=1}^{6}
\frac{d^{D}q_{i}}{(2\pi)^{D}}\; \delta^{D}(\sum_{j=1}^{6}q_{j})\;
D^{amp}(q^{6}|0)|_{\lambda_{r}^{2}}\; D(q^{6})\, ,
\tselea{CO8}
\end{eqnarray}

\vspace{3mm}

\noindent where $D^{amp}(q^{m}|0)|_{\lambda_{r}^{k}}$ is the
$m$--point amputated composite Green's function to order $\lambda_{r}^{k}$,
and $D(q^{m})$ is the full $m$--point Green's function.  The crucial point is
that we can write $D(0|0)$ as a sum of terms, which, apart
from the first (free field) diagram, are factorised to the product of the
composite Green's function with $n > 0$ and the full Green's function.
(The factorisation is represented in (2.17) by the dashed lines. )

\begin{figure}[h]
\begin{center}
\leavevmode
\hbox{%
\epsfxsize=14cm
\epsffile{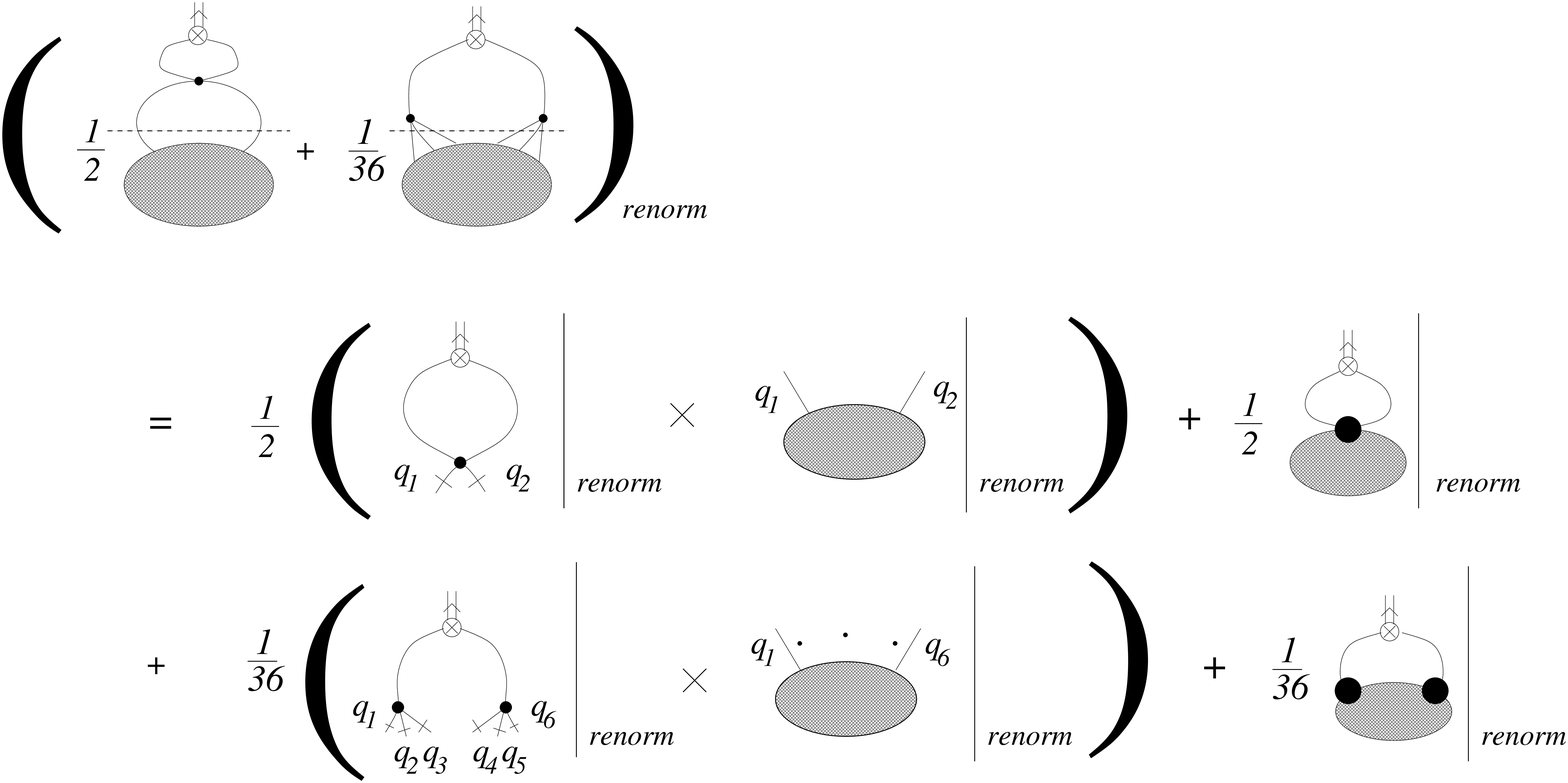}}
\caption{ Counterterm renormalisation of the last two diagrams in
Eq.(3.25). (Cut legs indicate amputations.)} \label{fig11}
\end{center}
\setlength{\unitlength}{1mm}
\begin{picture}(10,7)
\end{picture}
\end{figure}

Now, utilising the counterterm renormalisation to the last two
diagrams in (3.25) we get situation depicted in FIG.\ref{fig11}. Terms
inside of the parentheses are finite, this is because both the composite
Green's functions ($n \ge 2$ !) and the full Green's functions are finite
after renormalisation. The counterterm diagrams, which appear on the RHS
of the parentheses, precisely cancel the UV divergences coming from the
loop integrations over momenta $q_{1}\ldots q_{i}$ which must be finally
performed.  The heavy dots schematically indicates the corresponding
counterterms. In the spirit of the counterterm renormalisation we should
finally subtract the counterterm associated with the overall
superficial divergence\footnote{A simple power counting in the $\Phi^{4}$
theory reveals \cite{IZ} that for a composite operator $A$ with
dimension $\omega_{A}$ the superficial degree of divergence $\omega$
corresponding to an $n$-point diagram is $\omega = \omega_{A}-n$.} related
to the diagrams in question. But as we saw this is not necessary;
individual counterterm diagrams (Zimmermann forests) mutually cancel their
divergences leaving behind a finite result.

\vspace{11mm}

\vspace{3mm}

\noindent So the only UV divergence in Eq.(3.25) which cannot be cured by
existing counterterms is that coming from the first (i.e.  free field or  
ring) diagram. The foregoing divergence is evidently temperature
independent (to see that, simply use an explicit form of the free thermal
propagator $\D_{11}$). Hence, if we define

\begin{equation}
\langle \Phi^{2} \rangle_{\mbox{\footnotesize{renorm}}} = \langle [\Phi^{2}]
\rangle - \langle 0| [\Phi^{2}] |0 \rangle,
\tseleq{Norm}  
\end{equation}
\noindent or, alternatively
\begin{equation}
\langle \Phi^{2} \rangle_{\mbox{\footnotesize{renorm}}} = \langle [\Phi^{2}]
\rangle - \langle [\Phi^{2}] \rangle|_{\mbox{\footnotesize{free
fields}}}\, ,
\tseleq{Norm2}
\end{equation}

\vspace{2mm} 

\noindent we get  finite quantities, as desired. On the other hand, we
should emphasise that

\begin{equation}
\langle \Phi^{2} \rangle - \langle 0| \Phi^{2} |0 \rangle =
Z_{\Phi^{2}}\left\{ \langle [\Phi^{2}] \rangle - \langle
0| [\Phi^{2}] |0 \rangle \right\} \not= \mbox{finite in $D$=4}\, .
\end{equation}

\vspace{2mm}

\noindent An extension of the previous reasonings to any $N>1$ is
straightforward, only difference is that we must deal with operator
mixing which makes (\tseref{Norm}) and (\tseref{Norm2}) less trivial.

The important lesson which we have learnt here is that the
naive ``double dotted'' normal product (i.e. subtraction of the vacuum
expectation value from a given operator) does not generally give
a finite result. The former is perfectly suited for the free theory
($Z_{\Sigma \Phi^{2}} = 1$) but in the
interacting case we must resort to the prescription (\tseref{Norm}) or
(\tseref{Norm2}) instead.

\vspace{6mm}
\noindent{\it Renormalisation of the energy--momentum tensor}
\vspace{3mm}

\noindent In order to  calculate the hydrostatic  pressure, we need to
find        such             $\langle        \Theta^{\mu      \nu}_{c}
\rangle|_{\mbox{\footnotesize{renorm}}}$ which apart from being finite
is also consistent with   our derivation of the  hydrostatic  pressure
performed   in the introductory  section.  In   view  of the  previous
treatment, we cannot, however, expect  that $\Theta^{\mu \nu}_{c}$ will
be renormalised multiplicatively. Instead new  terms with a  different
structure than $\Theta^{\mu \nu}_{c}$ itself  will be generated during
renormalisation.  The latter must add  up to $\Theta^{\mu \nu}_{c}$ in
order to render $D^{\mu \nu}(x^{n}|y)$ finite\footnote{In fact it can
be shown \cite{LW,Collins} that the Noether currents corresponding to
a  given internal symmetry    are renormalised, i.e  $J^{a}= [J^{a}]$,
however, this  is not the case for  the Noether currents corresponding
to external symmetries (like $\Theta^{\mu \nu}_{c}$ is).}.

Now, the key ingredient exploited in Eq.(\tseref{5}) is the
conservation law (continuity equation). It is well known that one can 
`modify' $\Theta^{\mu \nu}_{c}$ in such a way that the new tensor
$\Theta^{\mu \nu}$ preserves the convergence properties of $\Theta^{\mu
\nu}_{c}$. Such a modification (the Pauli transformation) reads
\begin{eqnarray}
&&\Theta^{\mu \nu} = \Theta^{\mu \nu}_{C} +
\partial_{\lambda}X^{\lambda \mu \nu}\nonumber \\
&& X^{\lambda \mu \nu} = -X^{\mu \lambda \nu}\, .
\tselea{pp1}
\end{eqnarray}

\vspace{2mm}

\noindent For scalar fields, (\tseref{pp1}) is the only transformation 
which neither changes the divergence properties of $\Theta^{\mu \nu}_{c}$
nor the generators of the Poincare group constructed out of $\Theta^{\mu
\nu}_{c}$ \cite{LW,DG,RJ,IZ}. Because the renormalised (or improved)
energy momentum tensor must be conserved (otherwise theory would be
anomalous), it has to mix with $\Theta^{\mu \nu}_{c}$ under
renormalisation only via the Pauli transformation, i.e.

\begin{equation}
[\Theta^{\mu \nu}_{c}] = \Theta^{\mu \nu}_{c} +
\partial_{\lambda}X^{\lambda \mu \nu}\nonumber \, .
\tseleq{ppp1}
\end{equation}

\vspace{2mm}

\vspace{3mm}

\noindent In order to determine $X^{\lambda \mu \nu}$, we should
realize that its role is to cancel divergences present in $\Theta^{\mu
\nu}_{c}$.  Such a cancellation can be, however, performed only by
means of composite operators which are even in the number of fields
(note that $\Theta^{\mu \nu}_{c}$ is even in fields and Green's functions
with the odd number of fields vanish).  Recalling the condition that
renormalisation can mix only operators with dimension less or equal,   
we see that the dimension of $X^{\lambda \mu \nu}$ must be $D-1$, and
that $X^{\lambda \mu \nu}$ must be quadratic in fields. The only
possible form which is compatible with tensorial  structure
(\tseref{pp1}) is then

\begin{equation}
X^{\lambda \mu \nu} = \sum_{a, b = 1}^{N} c_{a b}(\lambda_{r};D)\; \left(
\partial^{\mu} g^{\lambda \nu} - \partial^{\lambda} g^{\mu \nu}
\right) \; \Phi_{a}\Phi_{b}\, .
\tseleq{ppp5}
\end{equation}

\vspace{2mm}  

\noindent From the fact that both $\Theta^{\mu \nu}_{c}$ and $[\Theta^{\mu
\nu}_{c}]$ are $O(N)$ invariant
(see Eq.(\tseref{7})), $\partial _{\lambda} X^{\lambda \mu \nu}$ must be
also $O(N)$ invariant, so $c_{a b} = \delta_{ab} c$. Thus, finally we can
write

\begin{equation}
[\Theta^{\mu\nu}_{c}] = \Theta^{\mu \nu}_{c} + c(\lambda_{r};D)\; \sum_{a
=1}^{N} \left( \partial^{\mu}\partial^{\nu} - g^{\mu
\nu} \partial^{2}\right) \; \Phi_{a}^{2}\, ,
\tseleq{ppp6}
\end{equation}

\vspace{3mm}

\noindent with $c = c_{0} + \sum (\mbox{poles})$, here $c_{0}$ is analytic
in $D$. Structure of $c(\lambda_{r};D)$ could be further determined,
similarly as in the $N=1$ theory, employing a renormalisation group
equation \cite{Brown}. We do not intend to do that as the detailed
structure of $c$ will show totally irrelevant for the following
discussion, however, it turns out to be important in non-equilibrium case.

Now, similarly as before, $[\Theta^{\mu \nu}_{c}]$ gives the
finite composite Green's functions if $n >0$ but the expectation value
$\langle [\Theta^{\mu \nu}_{c}] \rangle$ is divergent (discussion for the 
$N=1$ theory can be found in Brown \cite{Brown}). The
unrenormalised Dyson--Schwinger equation  for $D^{\mu \nu}(0|0)$
reads

\vspace{8mm}

\begin{figure}[h]
\vspace{3mm}
\begin{center}
\leavevmode
\hbox{%
\epsfxsize=14cm
\epsffile{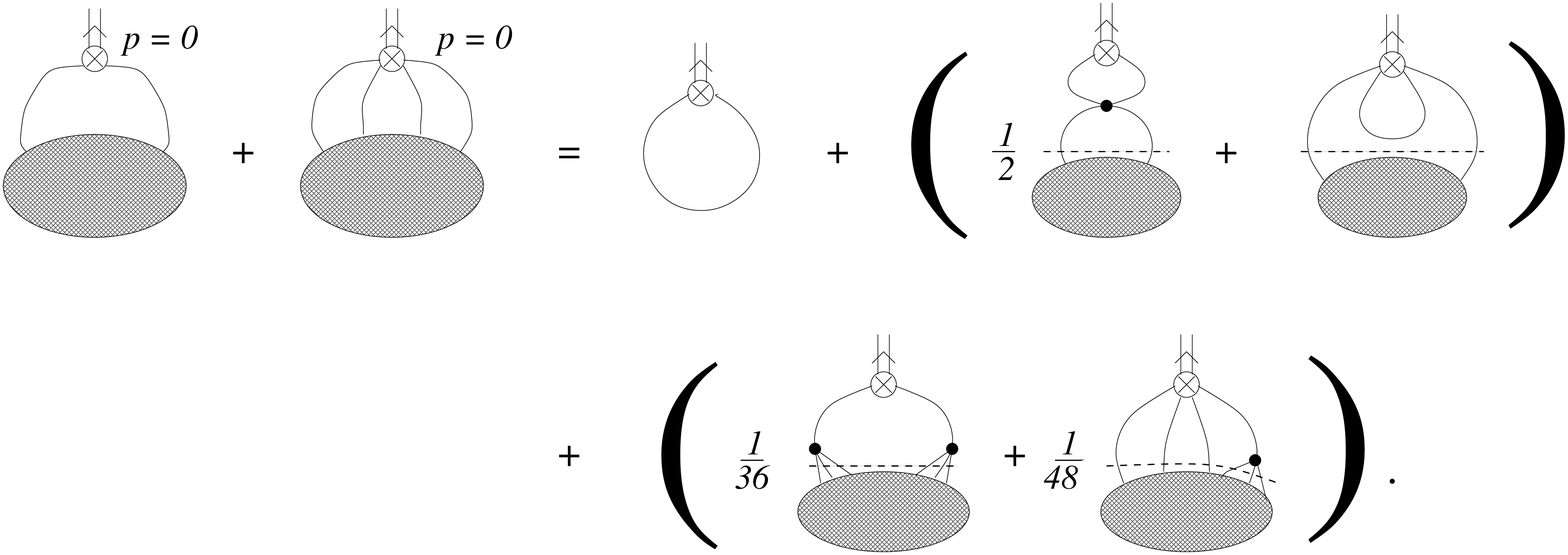}}
\end{center}
\setlength{\unitlength}{1mm}
\begin{picture}(20,7)
\put(152.5,20){(3.34)}
\end{picture}
\end{figure}

\addtocounter{equation}{1}

\noindent The structure of the composite vertices in (3.34) is that
described at the beginning of this section. Note that the amputated   
composite Green's functions in individual parentheses are of the same 
order in $\lambda_{r}$. Performing the counterterm renormalisation as in  
the case of $\langle [\Phi^{2}] \rangle$, we factorise the graphs inside
of parentheses into the product of the renormalised 2-- (and 6--) point
composite Green's function and the renormalised full 2-- (and 6--) point
Green's function. The latter are finite. The UV divergences arisen during
the integrations over momenta connecting both composite and full Green's
functions are precisely cancelled by the remaining counterterm diagrams.
The only divergence comes from the free--field contribution, more precisely
from the $T=0$ ring diagram. Defining

\begin{equation} \langle \Theta^{\mu \nu}_{c} \rangle
|_{\mbox{\footnotesize{renorm}}} = \langle [\Theta^{\mu \nu}_{c}] \rangle
- \langle 0| [\Theta^{\mu \nu}_{c}]|0 \rangle, \tseleq{ppp2}
\end{equation} \noindent or \begin{equation} \langle \Theta^{\mu \nu}_{c}
\rangle |_{\mbox{\footnotesize{renorm}}} = \langle [\Theta^{\mu \nu}_{c}]
\rangle - \langle [\Theta^{\mu \nu}_{c}] \rangle
|_{\mbox{\footnotesize{free field}}}, \tseleq{ppp3} \end{equation}

\vspace{2mm} 

\noindent we get the finite expressions. Note that the conservation law is
manifest in both cases. In equilibrium (and in $T=0$) we can, due to
space--time translational invariance of $\langle \ldots \rangle$, write

\begin{equation}
\langle [\Theta^{\mu \nu}_{c}] \rangle = \langle \Theta^{\mu
\nu}_{c}\rangle +
\partial_{\lambda} \langle X^{\lambda \mu \nu } \rangle = \langle
\Theta^{\mu \nu}_{c}\rangle\, .
\tseleq{ppp8}
\end{equation}

\vspace{2mm}

\noindent Using (\tseref{ppp2}) or (\tseref{ppp3}) we get either  the
thermal interaction pressure or the interaction pressure,
respectively. This can be explicitly written as

\begin{equation}
{\cali{P}}_{\mbox{\footnotesize{th.int.}}}(T) = {\cali{P}}(T) -
{\cali{P}}(0) =
-\frac{1}{(D-1)} \sum_{i=1}^{D-1}\left\{ \langle \Theta_{c\; i}^{i}
\rangle - \langle 0| \Theta_{c\;i}^{i} | 0 \rangle \right\},
\tseleq{ppp9}
\end{equation}
\noindent or
\begin{equation}
{\cali{P}}_{\mbox{\footnotesize{int.}}}(T) = {\cali{P}}(T) -
{\cali{P}}_{\mbox{\footnotesize{free
field}}}(T) = -\frac{1}{(D-1)} \sum_{i=1}^{D-1} \left\{ \langle \Theta_{c\;
i}^{i} \rangle -
\langle \Theta_{c\; i}^{i} \rangle|_{\mbox{\footnotesize{free field}}}
 \right\}\, .
\tseleq{ppp10}
\end{equation}

\vspace{2mm}  

\noindent In order  to   keep  connection with  calculations  done  by
Drummond {\em et al.} in  \cite{ID1} we shall in  the sequel deal with
the thermal interaction pressure only. If instead of an equilibrium, a
non--equilibrium medium would be in question, translational invariance
of  $\langle  \ldots \rangle$  might  be  lost,   in that  case either
prescription  (\tseref{ppp2})  or  (\tseref{ppp3})  is obligatory, and
consequently $c(\lambda_{r};D)$   in  (\tseref{ppp6}) must  be further
specified. 

\vspace{5mm}

\section{Hydrostatic pressure \label{PE3}}


In the previous Section we have prepared ground for a hydrostatic
pressure calculations in the $O(N)\; \Phi^{4}$ theory. In this section
we aim to apply the previous results to the massive $O(N)\;  \Phi^{4}$
theory in the large--$N$ limit. Anticipating an out of equilibrium
application, we shall use the real--time formalism even if the
imaginary--time one is more natural in the equilibrium context. As we
aim to evaluate the hydrostatic pressure in 4 dimensions, we use here,
similarly as in the previous section, the usual dimensional
regularisation to regulate the theory (i.e. here and throughout we
keep $D$ slightly away from the physical value $D=4$).

Let us start first with some essentials of our model system
at finite temperature.

\vspace{3mm}

\subsection{Mass renormalisation \label{PE31}}

In the Dyson multiplicative renormalisation the fact that the complete
propagator has a pole at the physical mass leads to the usual
mass renormalisation prescription \cite {IZ}:

\begin{equation}
m_{r}^{2}= m_{0}^{2} + \Sigma(m^{2}_{r})\, ,
\tseleq{c1}
\end{equation}

\vspace{2mm}

\noindent where $m_{r}$ is renormalised mass and $\Sigma(m_{r}^{2})$ is
the proper self--energy evaluated at the mass shell; $p^{2}= m_{r}^{2}$. In
fact, Eq.(\tseref{c1}) is nothing but the statement that 2--point vertex
function $\Gamma^{(2)}_{r}$ evaluated at the mass--shell must vanish. The
Dyson--Schwinger equations corresponding to the proper self--energies read
\cite{PC,EM,PJ,HS}:

\vspace{5mm}

\begin{figure}[h]
\begin{center}
\leavevmode
\hbox{%
\epsfxsize=9.5cm
\epsffile{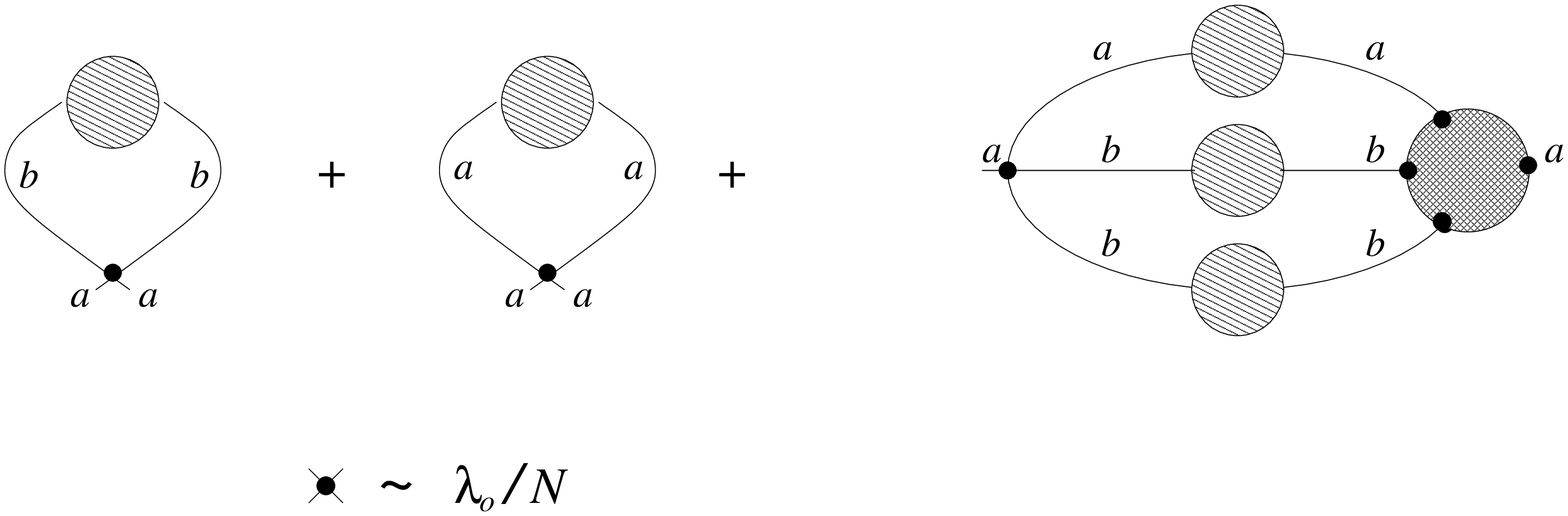}}
\label{fig5}
\end{center}
\setlength{\unitlength}{1mm}
\begin{picture}(10,7)
\put(0,31){$\Sigma^{aa} =$}
\put(16.5,31){$\frac{1}{2}\sum_{b=1}^{N}$}
\put(79,31.4){$\frac{i}{2}\sum_{b=1}^{N}$} 
\put(0,11){$\Sigma^{ac}|_{a\not= c}\; = 0\; \; \; ;$}
\put(154,31){(3.41)}
\end{picture}
\end{figure}

\addtocounter{equation}{1}

\noindent where hatched blobs represent 2--point connected Green's
functions whilst cross-hatched blobs represent proper vertices
$\Gamma^{(4)}_{r}$ (i.e. 1PI 4--point Green's functions). As $\Sigma^{aa}$
are the same for all $a$, we shall simplify notation and write $\Sigma$
instead. In the sequel the following convention is accepted:

\vspace{7mm}

\begin{figure}[h]
\begin{center}
\leavevmode
\hbox{%
\epsfxsize=2cm
\epsffile{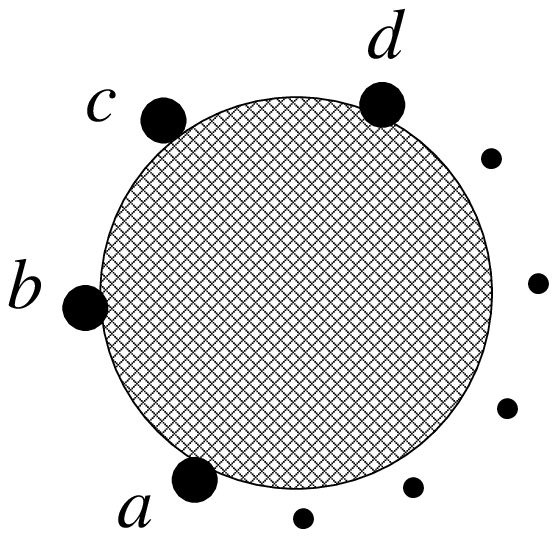}}
\end{center}
\setlength{\unitlength}{1mm}
\begin{picture}(10,7)
\put(96,20){$ =\; \Gamma^{(n)\; abcd \ldots}_{r}$}
\end{picture}
\end{figure}

\noindent Note that the second term in (3.41) does not contribute in the
large--$N$ limit. It is easy to see that the third term does not contribute
either. This is because each hatched blob behaves at most as $N^{0}$
whilst $\Gamma^{(4)}$ goes maximally\footnote{ In the
$\Phi^{4}$ theory there is a simple relation between the number of loops
($L$), vertices ($V$) and external lines ($E$); $4V = 2I +E$. Together
with the Euler relation for connected graphs; $L= I -V +1$ (here $I$ is
the number of internal lines), we have $L-V = \frac{2-E}{2}$. As each loop
carries maximally a factor of $N$ (this is saturated only for `tadpole'
loops) and each vertex carries a factor of $N^{-1}$, the overall blob
contribution behaves at most as $N^{L-V} = N^{\frac{2-E}{2}}.$} as $N^{-1}$ .
Consequently, various contributions from the first graph in (3.41)
contribute at most $N^{0}$, whereas in the second graph the contributions
contribute up to order $N^{-1}$. So the first diagram dominates, provided
we retain only such 2--point connected Green's functions which are
proportional to $N^{0}$ (as mentioned in the footnote, these are comprised
only of `tadpole' loops.).  After neglecting the `setting sun' graph,
Eg.(3.41) generates upon iterating the so called superdaisy diagrams
\cite{ID1,EM,CJT}.

Let us now define $\Sigma(m^{2}_{r}) = \lambda_{0}\;
{\cali{M}}(m_{r}^{2})$. Because the `tadpole' diagram in (3.2) can be
easily resumed we observe that

\begin{equation}
{\cali{M}}(m_{r}^{2}) = \frac{1}{2}\int \frac{d^{D}q}{(2 \pi)^{D}} \;
\frac{i}{q^{2}-m_{0}^{2}-\Sigma(m^{2}_{r})+i\epsilon} = \frac{1}{2}\int
\frac{d^{D}q}{(2 \pi)^{D}} \;
\frac{i}{q^{2}-m^{2}_{r}+i\epsilon}\, ,
\tseleq{b45}
\end{equation}

\vspace{4mm}

\noindent hence we see that $\Sigma$ is external--momentum independent. So
if we had started with the renormalisation prescription:
$i\Gamma^{(2)}_{r}(p^{2}=0)= - m^{2}_{r}$, we would arrived at
(\tseref{c1}) as well (this is not the case for $N=1$!).

At finite temperature the strategy is analogous. Due to a
doubling of degrees of freedom, the full propagator is a $2\times 2$
matrix. The latter satisfies, similarly as at $T=0$, Dyson's equation
\begin{equation}
\D = \D_{F} + \D_{F} \left(-i{\vect{\Sigma}}
\right) \D \, .
\tseleq{b5}
\end{equation}  

\vspace{2mm}

\noindent An important point is that there exists a real, non--singular
matrix $\M $ (Bogoliubov matrix) \cite{LW,LB,TA} having a property
that

\begin{equation}
\D_{F} = \M \left( \begin{array}{cc}
                                   i\Delta_{F} & 0 \\
                                        0      & -i\Delta^{*}_{F}
                                    \end{array} \right) \M \, .
\tseleq{m8}
\end{equation}

\vspace{2mm}

\noindent Here $\Delta_{F}$ is the standard Feynman propagator and `*'
denotes the complex conjugation. Consequently, the full matrix propagator
may be written as

\begin{equation}
\D = \M \left( \begin{array}{cc}
              \frac{i}{p^{2}-m_{0}^{2}-\Sigma_{T}+i\epsilon} & 0 \\
                             0 &
\frac{-i}{p^{2}-m_{0}^{2}-\Sigma^{*}_{T}-i\epsilon}
                  \end{array} \right) \M \, .
\tseleq{m9}
\end{equation}  

\vspace{4mm}
                                        
\noindent Similarly as in many body systems, the position of the
(real) pole of $\D$ in $p^{2}$ fixes the temperature--dependent effective
mass $m_{r}(T)$ \cite{LB,FW}. The latter is determined by the equation

\begin{equation}
m_{r}^{2}(T) = m_{0}^{2} +
\mbox{Re}\left(\Sigma_{T}(m^{2}_{r}(T))\right)\, .
\tseleq{m10}
\end{equation}

\vspace{2mm}

\noindent From the   explicit form  of $\M$ it   is possible  to  show
\cite{LW,LB}                                                      that
$\mbox{Re}{\vect{\Sigma}}_{11}=\mbox{Re}\Sigma_{T}$.  As  before,  the
structure of the   proper    self--energy can be  deduced   from   the
corresponding   Dyson--Schwinger    equation. Following   the    usual
real--time  formalism convention (type--1 vertex $\sim -i\lambda_{0}$,
type--2 vertex $\sim i\lambda_{0}$ ), the former reads: 

\vspace{8mm}

\begin{figure}[h]
\begin{center}
\leavevmode
\hbox{%
\epsfxsize=12.4cm
\epsffile{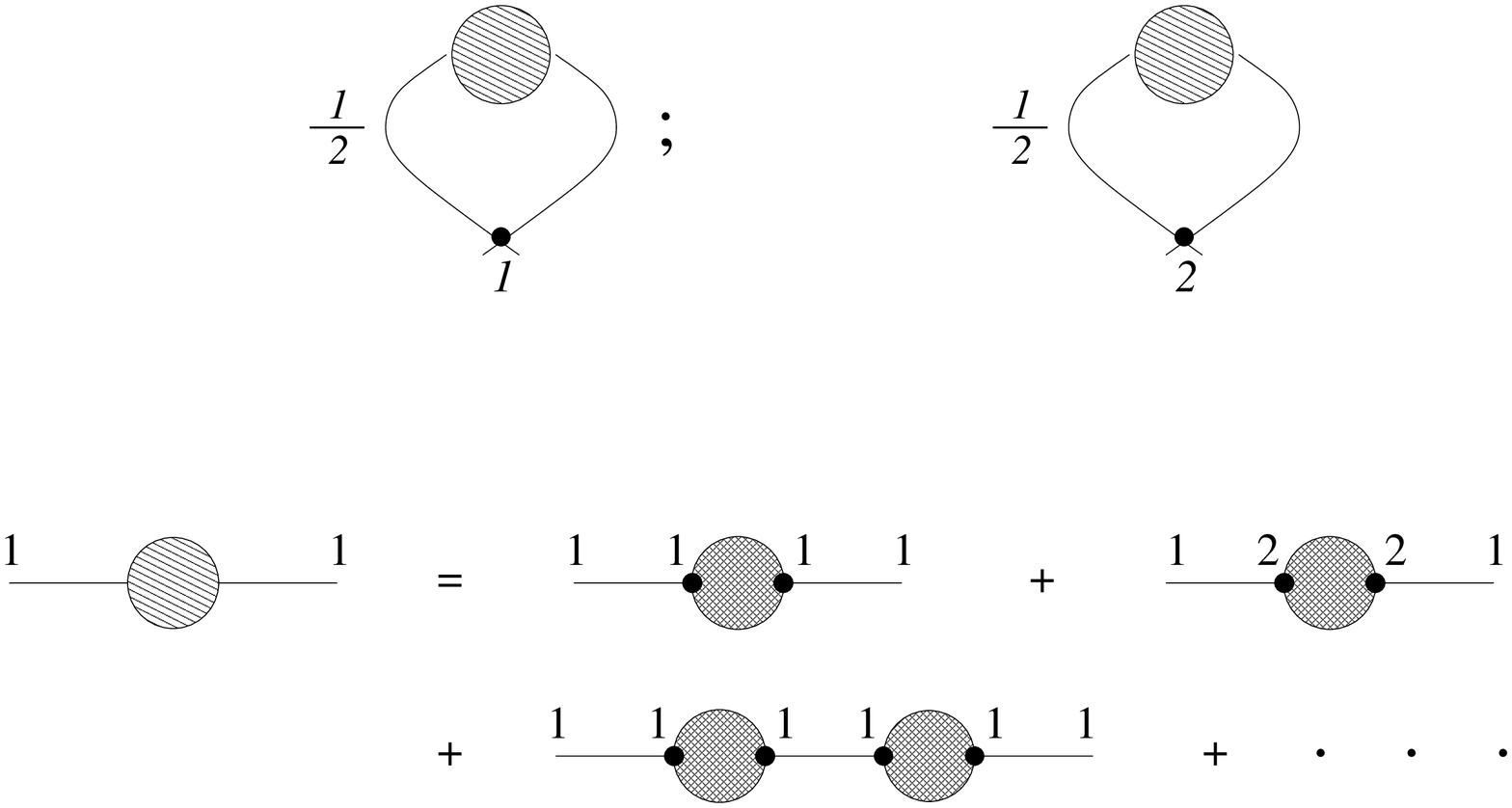}}
\label{fig6}
\end{center}                            
\setlength{\unitlength}{1mm}
\begin{picture}(10,7)
\put(20,66){${-i\vect{\Sigma}}_{11}=$}
\put(76,66){${-i\vect{\Sigma}}_{22}=$}
\put(0,47.5){where}
\put(153,67.5){(3.47)}
\end{picture}
\end{figure}  

\addtocounter{equation}{1}


\noindent and similarly for $\D_{22}$. In (4.49) we have omitted diagrams
which are of order ${\cali{O}}(1/N)$ or less. Note that the fact
that no `setting sun' diagrams are present implies that the off--diagonal
elements of ${\vect{\Sigma}}$ are zero.  Inspection of Eq.(3.47) reveals
that

\begin{equation}
{\vect{\Sigma}}_{11}= \frac{\lambda_{0}}{2}\; \int
\frac{d^{D}q}{(2\pi)^{D}} \; {\D}_{11}(q;T) \; \; \; \mbox{and} \; \; \;
{\vect{\Sigma}}_{22}= -\frac{\lambda_{0}}{2}\; \int
\frac{d^{D}q}{(2\pi)^{D}} \; {\D}_{22}(q;T)\, .
\tseleq{m11}
\end{equation} 

\vspace{3mm}

\noindent It directly follows from Eq.(\tseref{m11}) that both
${\vect{\Sigma}}_{11}$ and ${\vect{\Sigma}}_{22}$ are external--momentum
independent and real\footnote{Reality of ${\vect{\Sigma}}_{11}$ can be
perhaps most easily seen from the largest--time equation. The LTE
states that ${\vect{\Sigma}}_{11} + {\vect{\Sigma}}_{22} +
{\vect{\Sigma}}_{12} + {\vect{\Sigma}}_{21} = 0$.  Because no `setting
sun' graphs are present, ${\vect{\Sigma}}_{12} + {\vect{\Sigma}}_{21} =
0$, on the other hand ${\vect{\Sigma}}_{11} + {\vect{\Sigma}}_{22} =
2i\mbox{Im}{\vect{\Sigma}}_{11}$ (see (\tseref{m8})).}. If we define
$\Sigma_{T}(m_{r}^{2}(T)) = \lambda_{0}\;  {\cali{M}}_{T}(m_{r}^{2}(T))$,
then Eq.(\tseref{m10}) through Eq.(\tseref{m11}) implies that

\begin{equation}
m_{r}^{2}(T) = m^{2}_{0} +  \lambda_{0}\;
{\cali{M}}_{T}(m_{r}^{2}(T))\, .
\tseleq{m12}
\end{equation}  

\vspace{2mm}

\noindent A resumed version of $\D_{11}$ is easily obtainable from
(\tseref{m9}) \cite{LW,LB} and consequently
(\tseref{m11}) yields

\begin{eqnarray}
{\cali{M}}_{T}(m_{r}^{2}(T)) &=& \frac{1}{2}\; \int
\frac{d^{D}q}{(2\pi)^{D}}\; \left\{ \frac{i}{q^{2} - m^{2}_{r}(T) +
i\epsilon} \; \; +\; (4\pi)\; \delta^{+}(q^{2}-m^{2}_{r}(T))\;
\frac{1}{e^{q_{0}\beta}-1} \right\}\nonumber \\
&=&- \int \frac{d^{D}q}{(2\pi)^{D}}\;
\frac{\varepsilon
(q_{0})}{e^{q_{0}\beta}-1}\; \mbox{Im}\frac{1}{q^{2}-m_{r}^{2}(T)
+i\epsilon}\, . \tselea{m14}
\end{eqnarray}

\vspace{3mm}

\noindent Let us remark that (\tseref{m14}) is manifestly independent of
any particular real--time formalism version. This is because the various
real--time formalisms \cite{LW, LB} differ only in the off--diagonal
elements of $\D$.

In passing it may be mentioned that because
${\vect{\Sigma}}_{11}(m^{2}_{r})$ is momentum independent, the wave
function renormalisation $Z_{\Phi} = 1$. (The K{\"a}llen--Lehmann
representation requires the renormalised propagator to have a pole of
residue $i$ at $p^{2}=m^{2}_{r}$. The former in turn implies that
$Z_{\Phi}=(1-{\vect{\Sigma}}_{11}'(p^{2})|_{p^{2}=m^{2}_{r}})^{-1} =1$.)
Trivial consequence of the foregoing fact is that  
$\Gamma^{(2)}_{r}=\Gamma^{(2)}$ and $\Gamma^{(4)}_{r}=\Gamma^{(4)}$.

\vspace{3mm}

\subsection{Coupling constant renormalisation \label{PE32}}

Let us choose the coupling constant to  be defined at $T=0$. This will
have the advantage that the high temperature expansion of the pressure
(see Section \ref{HTE})  will  become more transparent.   In addition,
such   a  choice allows  us    to stay  on  a   safe   ground  as  the
renormalisation of the   coupling constant  at finite temperature   is
rather delicate \cite{EM}.

By assumption the fields $\Phi_{a}$ have non--vanishing masses,
so we can safely choose the renormalisation prescription for $\lambda_{r}$
at $s = 0$ ($s$ is the standard Mandelstam variable). For example, one may
require that for the scattering $aa \rightarrow bb$

\begin{equation}
\Gamma^{(4)\; aabb}(s=0)= -\lambda_{r}/N, \; \; \; \; \; (b \not= a)\,
.
\tseleq{m13}
\end{equation}

\noindent  The formula  (\tseref{m13}) clearly   agrees with the  tree
level value  $\Gamma_{tree}^{(4)\;aabb}(s=0)=  -\lambda_{0}/N$. Let us
also mention   that Ward's  identities corresponding  to  the internal
$O(N)$  symmetry enforce      $\Gamma^{(4)\;  aaaa}$  to  obey     the
constraint\footnote{Actually,    Ward's  identities read  \cite{PC,PR}
$\int        d^{D}x          \;       \frac{\delta\Gamma[\phi]}{\delta
\phi_{a}(x)}\;\phi_{b}(x)       =           \int      d^{D}x        \;
\frac{\delta\Gamma[\phi]}{\delta    \phi_{b}(x)}\;\phi_{a}(x)$   (here
$\phi_{a} =  \frac{\delta   W}{\delta J_{a}}$; $W$  is  the generating
functional  of   connected  Green's functions).  Performing successive
variations with respect to $\phi_{a}(v), \phi_{a}(z), \phi_{a}(y)$ and
$\phi_{b}(w)$, taking the  Fourier transform, and setting the physical
condition $\phi_{c} = 0$, we get directly (\tseref{m134}).} 

\begin{eqnarray}
\Gamma^{(4)\; aaaa}(p_{1}; p_{2}; p_{3}; p_{4}) &=& \Gamma^{(4)\;
bbaa}(p_{1}; p_{2}; p_{3}; p_{4}) + \Gamma^{(4)\; baba}(p_{1}; p_{2};
p_{3}; p_{4})\nonumber \\
 &+& \Gamma^{(4)\; baab}(p_{1}; p_{2}; p_{3};
p_{4})\, ,
\tselea{m134}   
\end{eqnarray}

\noindent for any  $b \not=  a$.  The structure  of $\Gamma^{(4)}$  is
encoded in the Dyson--Schwinger equation which reads \cite{IZ,PC}:

\vspace{6mm}

\begin{figure}[h]
\begin{center}
\leavevmode
\hbox{%
\epsfxsize=14cm
\epsffile{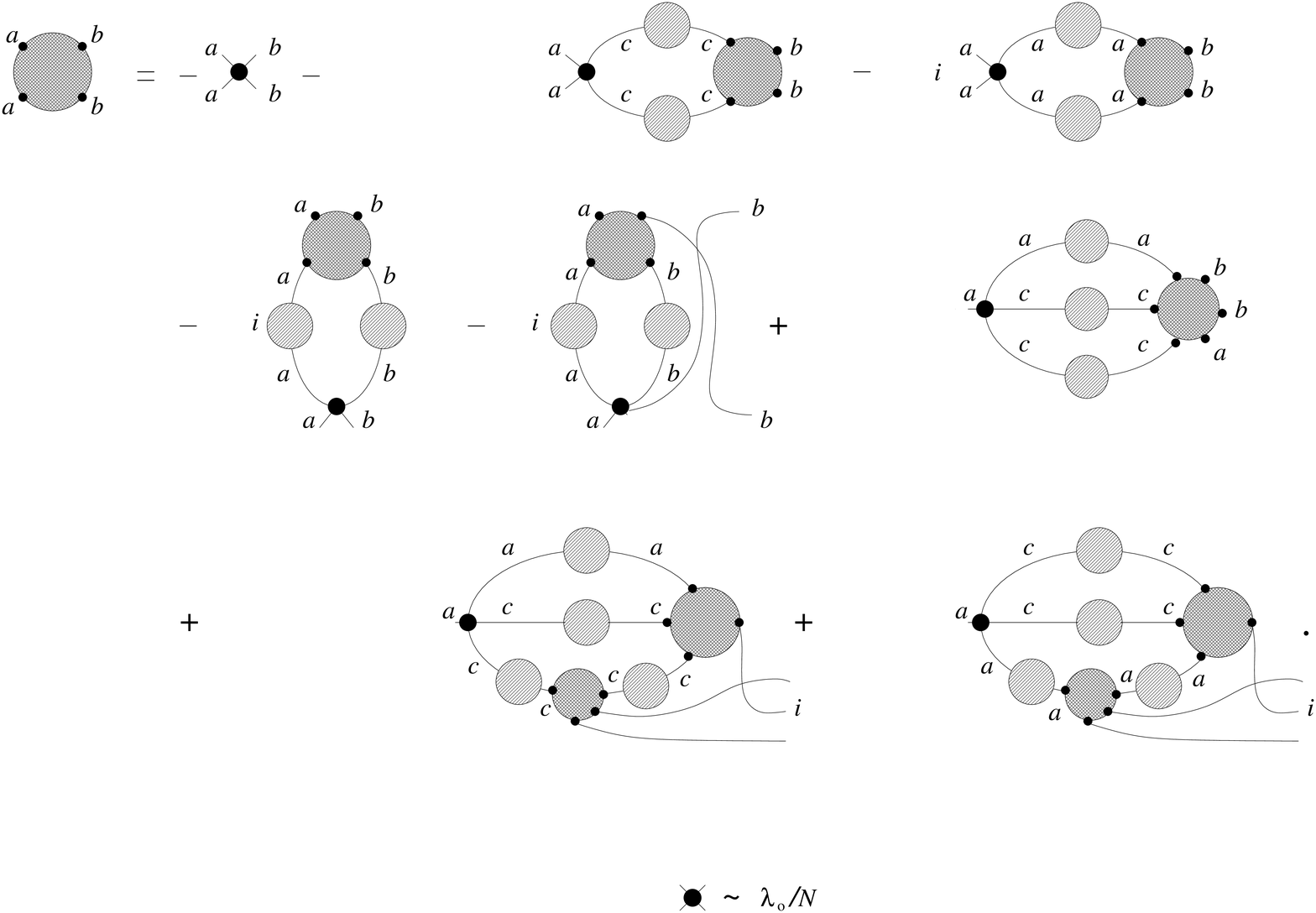}}
\end{center}
\setlength{\unitlength}{1mm}
\begin{picture}(20,7)
\put(51.5,101.5){$\frac{i}{2}\sum_{c=1}^{N}$}
\put(97.5,74.5){$\frac{i}{2}\sum_{c=1}^{N}$}   
\put(34.5,42.5){$\sum_{i=1}^{3}\sum_{c=1}^{N}$}
\put(99.5,42){$\frac{1}{2}\sum_{c;i}$}
\put(152,13.5){(3.53)}
\end{picture}
\end{figure}

\addtocounter{equation}{1}

\noindent The sum $\sum_{i=1}^{3}$ schematically represents a summation
over various scattering channels. Similarly as before, we can argue that
the last three graphs contribute at most $N^{-2}$, whilst the second
(`fish') graph may contribute up to order $N^{-1}$. So in the large--$N$
limit the last three diagrams may be neglected, provided we keep in the
$4$--point vertex function only graphs proportional to $N^{-1}$.  However,
the former can be only fulfilled if we retain such a `fish' graph where
summation over internal index on the loop is allowed. Remaining `fish'
graphs (describing $t$ and $u$ scattering channel interactions)  are
suppressed by the factor $N^{-1}$ as the internal index on the loop is
fixed. In this way we are left with the relation
\begin{eqnarray}
&&\Gamma^{(4)\; aabb}(s=0)\nonumber \\
&&\mbox{\hspace{6 mm}} = -\frac{\lambda_{0}}{N} -
\left. \frac{i\lambda_{0}}{2N} \sum_{c \not= b}\int
\frac{d^{D}q}{(2\pi)^{D}}\;
\Gamma^{(4)\; bbcc}(s) \;
\frac{i}{\left( q^{2} - m_{r}^{2} + i\epsilon \right)}\; \frac{i}{\left(
(q -Q)^{2} - m_{r}^{2} + i\epsilon \right)}\right|_{s=0}
\nonumber \\
&&\mbox{\hspace{6 mm}} =- \frac{\lambda_{0}}{N} - \frac{\lambda_{0}
\lambda_{r}(N-1)}{2N^{2}} 
\int^{1}_{0}dx \int
\frac{d^{D}q}{(2\pi)^{D}}\;\left. \frac{i}{\left( q^{2} -
m_{r}^{2} + x(1-x)s + i\epsilon \right)^{2}} \right|_{s=0}\, ,\nonumber\\ 
\tseleq{m145}
\end{eqnarray}

\noindent with $Q=p_{1}+p_{2}$ and $s=Q^{2}$, $p_{1}, p_{2}$ are
the external momenta. To leading order in $1/N$ we may
equivalently write

\begin{equation}
\lambda_{r} = \lambda_{0} + \lambda_{0}\lambda_{r}\;
{\cali{M}}'(m^{2}_{r})\, ,
\tseleq{m146}
\end{equation}  

\noindent the prime means differentiation with respect to $m_{r}^{2}$;
${\cali{M}}(m^{2}_{r})$ is defined by (\tseref{b45}). Evaluating
explicitly ${\cali{M}}'(m^{2}_{r})$, we get from (\tseref{m146})

\begin{equation}
\lambda_{0} =
\frac{\lambda_{r}}{1-\lambda_{r}\Gamma(2-\frac{D}{2})\;(m_{r})^{D-4}/2\;(4
\pi)^{\frac{D}{2}}}\, .
\tseleq{m1465}
\end{equation}
\vspace{1mm}

\noindent Assuming that both $\lambda_{0} \geq 0$ and $\lambda_{r} \geq
0$, we can infer from (\tseref{m1465}) that

\begin{equation}
0 \geq \lambda_{r} \geq
\frac{2(4\pi)^{\frac{D}{2}}(m_{r})^{4-D}}{\Gamma(2-\frac{D}{2})}\, ,
\end{equation}    

\noindent so for $D=4$ we inevitably get that $\lambda_{r}= 0$. The latter
indicates that the theory is trivial \cite{ID1, PR, BM}, or, in other
words, the $O(N)\; \Phi^{4}$ theory is a renormalised free theory in the
large--$N$ limit. This conclusion is also consistent with the observation
that the theory does not posses any non--trivial UV fixed point in the
large--$N$ limit \cite{PR,BM}.

On the other hand, if we were assuming that $\lambda_{0} < 0$,  
we would get a non--trivial renormalised field theory in $D=4$ (actually,
from (\tseref{m1465}) we see that $\lambda_{0} \rightarrow 0_{-}$ ,
provided that $\lambda_{r}$ is fixed and positive and $D \rightarrow
4_{-}$).  However, as it was pointed out in refs.\cite{ID1,PR,BM,Abb}, such a theory is intrinsically unstable as the ground--state energy
is unbounded from below. This is reflected, for instance, in the
existence of tachyons in the theory \cite{ID1,BM,Abb}, therefore the
case with negative $\lambda_{0}$ is clearly inconsistent.

The straightforward remedy for this situation was suggested by
Bardeen and Moshe \cite{BM}. They showed that the only meaningful
(stable) $O(N)\; \Phi^{4}$ theory in the large--$N$ limit is that with
$\lambda_{r}, \lambda_{0} \geq 0$. This is provided that we view it as an
effective field theory at momenta scale small compared to a fixed UV
cut--off $\Lambda$. The cut--off itself is further determined by
(\tseref{m146}) because in that case (assuming $m_{r} \ll \Lambda$)

\begin{equation}
\lambda_{0} = \frac{\lambda_{r}}{1- \frac{\lambda_{r}}{32
\pi^{2}}\mbox{ln}(\frac{\Lambda^{2}}{m^{2}_{r}})}\, ,
\end{equation}
\vspace{1mm}

\noindent which implies that for $\lambda_{r}, \lambda_{0} \geq 0$ we have
$\Lambda^{2} < m^{2}_{r}\;\mbox{exp}(\frac{32 \pi^{2}}{\lambda_{r}})$. The
case $\Lambda^{2} = m^{2}_{r}\;\mbox{exp}(\frac{32 \pi^{2}}{\lambda_{r}})$
corresponds to the Landau ghost \cite{AC2} (tachyon pole \cite{ID1,BM}). For reasonably small $\lambda_{r}$, $\Lambda$ is truly
huge\footnote{For example, if $\lambda_{r}=1$ and $m_{r} \approx 100$MeV,
we get $\Lambda < 10^{141}$MeV or equivalently $\Lambda < 10^{131}$K (this
is far beyond the Planck temperature - $10^{32}$K).} and so it does not
represent any significant restriction. The following discussion will be
confined to such an effective theory.

\vspace{3mm}

\subsection{The pressure \label{PE33}}

The partition function $Z$ has a well known path--integral
representation at finite temperature, namely

\begin{eqnarray}
&&Z[T] = \mbox{exp}({\textswab{M}}[T]) = \int{\cali{D}}\phi\;
\mbox{exp}(iS[\phi;T])\nonumber \\ &&S[\phi;T] = \int_{C}d^{D}x\;
{\cali{L}}(x)\, . \tselea{b1}
\end{eqnarray}
\vspace{1mm}

\begin{figure}[h]
\vspace{4mm}
\epsfxsize=11cm
\centerline{\epsffile{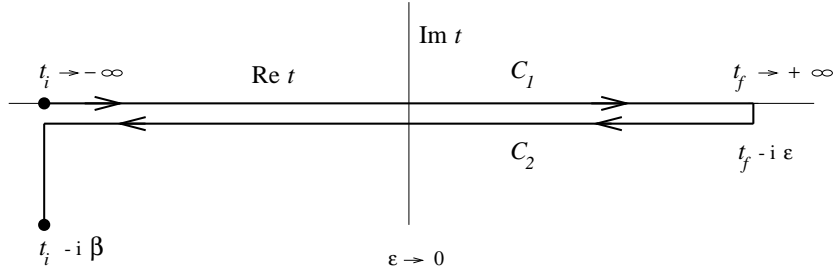}}
\caption{The Keldysh--Schwinger time path.}
\label{fig18}
\vspace{4mm}
\end{figure}

\noindent   Here ${\textswab{M}}= -\beta  \Omega$   is the so called
Massieu       function         \cite{LW,RBa}   and      $\int_{C}d^{D}x  =
\int_{C}dx_{0}\int_{V}d^{D-1}{\vect{x}}$   with  the   subscript   $C$
suggesting    that the time  runs along   some contour  in the complex
plane.  In the real--time   formalism, which we  adopt throughout, the
most  natural version    is   the so called  Keldysh--Schwinger    one
\cite{LW,LB}, which is represented by contour in FIG.\ref{fig18} 

\noindent Let us mention that the fields within the path--integral
(\tseref{b1}) are further restricted by the periodic boundary condition (KMS
condition) \cite{LW,LB,TA} which in our case reads:

\begin{displaymath}
\phi_{a}(t_{i}-i\beta, {\vect{x}})= \phi_{a}(t_{i}, {\vect{x}})\, .
\end{displaymath}

\noindent As explained in Section \ref{PE2}, we can use for a pressure
calculation    the   canonical energy--momentum  tensor   $\Theta^{\mu
\nu}_{c}$.  Employing for  $\Theta^{\mu \nu}_{c}(x)$ its explicit form
(\tseref{7}) together with (\tseref{11}), one may write

\begin{equation}
\langle \Theta^{\mu \nu}_{c} \rangle = \frac{N}{2}\; \int
\frac{d^{D}q}{(2\pi)^{D}}(2q^{\mu}q^{\nu}-g^{\mu
\nu}(q^{2}-m^{2}_{0}))\; \D_{11}(q;T)
\; \; + \frac{\lambda_{0}}{8N} g^{\mu \nu} \left\langle \left(
\sum_{a=1}^{N}\phi_{a}^{2}(0)\right)^{2} \right\rangle \, ,
\tseleq{b2} 
\end{equation}
\vspace{1mm}

\noindent where $\D_{11}$ is the Dyson--resumed thermal propagator
\cite{LW,LB}, i.e.

\begin{equation} 
\D_{11}(q;T) = \frac{i}{q^{2}-m^{2}_{r}(T)+i\epsilon} \; \; + (2\pi) \;
\delta (q^{2}-m^{2}_{r}(T))\; \frac{1}{e^{|q_{0}|\beta}-1}\, .
\tseleq{b3} 
\end{equation}
\vspace{1mm}

\noindent Note that we have exploited in (\tseref{b2}) the fact that
the expectation value of $\Theta^{\mu \nu}_{c}(x)$ is $x$
independent. On the other hand, in (\tseref{b3}) we have used the fact
that $m^{2}_{r}$ is $q$ independent. In order to calculate the
expectation value of the quartic term in Eq.(\tseref{b2}), let us
observe (c.f. (\tseref{b1})) that the derivative of ${\textswab{M}}$ with
respect to the bare coupling $\lambda_{0}$ (taken at fixed $m_{0}$)
gives

\begin{equation}
\frac{\partial{\textswab{M}}[T]}{\partial \lambda_{0}}= - \frac{i}{8\; N}\; \int_{C}
d^{D} x \left\langle \left(\sum_{a=1}^{N}\Phi_{a}^{2}(0) \right)^{2}
\right\rangle \nonumber \nonumber \, ,
\end{equation}

\noindent which implies that

\begin{equation} 
\left\langle \left(\sum_{a=1}^{N}\Phi_{a}^{2}(0)
\right)^{2} \right\rangle
= - \frac{N8}{\beta V}\;\frac{\partial{\textswab{M}}[T]}{\partial
\lambda_{0}}\, .
\tselea{bc444}  
\end{equation}
\vspace{1mm}

\noindent The key point now is that we can calculate ${\textswab{M}}[T]$ in a
non--perturbative form. (The latter is based on the fact that we know the
Dyson--resumed propagator $\D_{11}(q;T)$ (see (\tseref{b3}).) Indeed,
taking derivative of ${\textswab{M}}$ with respect to $m_{0}^{2}$
(keeping $\lambda_{0}$ fixed) we obtain

\begin{eqnarray}
\frac{\partial {\textswab{M}}[T]}{\partial m_{0}^{2}}&=& -\frac{iN}{2}\int_{C}d^{D}x
\left\langle \phi^{2}(0) \right\rangle = -\frac{\beta V N}{2} \int
\frac{d^{D}q}{(2\pi)^{D}}\; \D_{11}(q;T)\nonumber \\
&=& - \beta V N\; {\cali{M}}_{T}(m_{r}^{2}(T))\, ,
\tselea{b555}
\end{eqnarray}
\noindent thus

\begin{equation}
{\textswab{M}}[T;\lambda_{0}; m_{0}^{2}] = \beta V N \; \int_{m_{0}^{2}}^{\infty}
d{\hat{m}}_{0}^{2}\;{\cali{M}}_{T}({\hat{m}}_{r}^{2}(T))\; \; +
{\textswab{M}}[T;\lambda_{0}; \infty]\, .
\tseleq{b6}
\end{equation}
\vspace{1mm}

\begin{figure}[h]
\begin{center}  
\leavevmode
\hbox{%
\epsfxsize=11cm
\epsffile{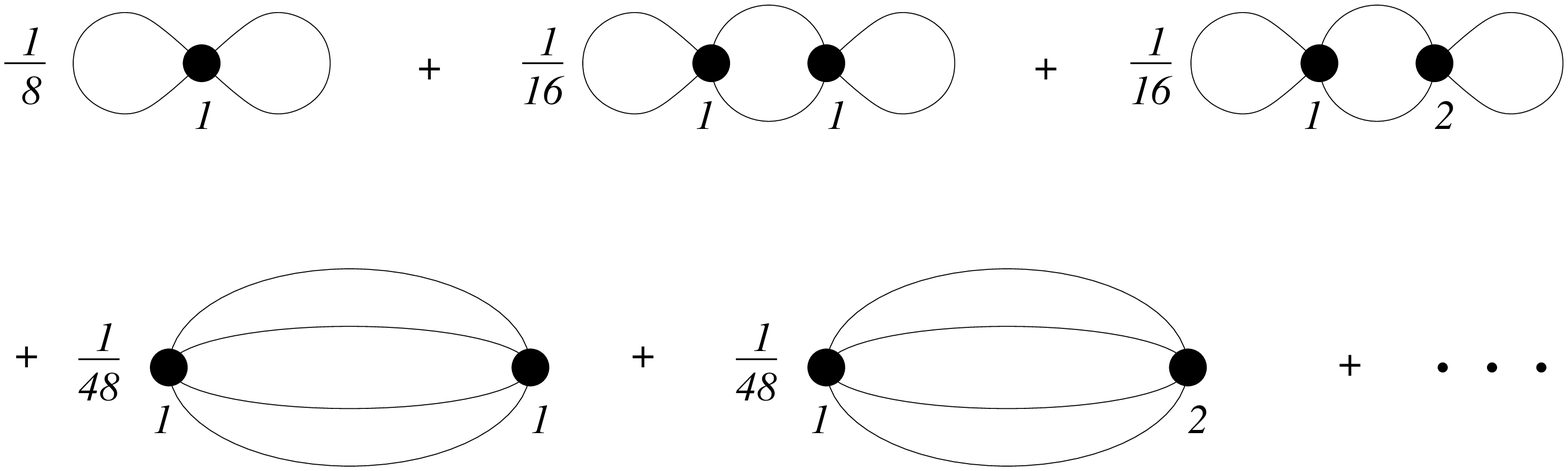}}
\caption{First few bubble diagrams in the ${\textswab{M}}$ expansion.}
\label{N3}
\end{center}
\begin{picture}(10,10)
\end{picture}
\end{figure}

\noindent Let us  note that ${\textswab{M}}[T;\lambda_{0}; \infty]$ is
actually   zero\footnote{To be  precise,  we  should  also  include in
FIG.\ref{N3} an (infinite)  circle  diagram corresponding to  the free
pressure \cite{LW,EM}. However  the later is $\lambda_{0}$ independent
(although $m_{0}$  dependent) and  so    it  is irrelevant  for    the
successive  discussion  (c.f.      Eq.(\tseref{bc444})).}      because
${\textswab{M}}[T;\lambda_{0};m^{2}_{0}]$ has the     standard    loop
expansion \cite{LW,PC} depicted in FIG.\ref{N3}.  

\vspace{6mm}

\noindent It  is worth mentioning  that in the  previous expansion one
must  always have  at least one  type--1 vertex  \cite{LW}. The RHS of
FIG.\ref{N3} clearly tends to  zero for $m_{0} \rightarrow  \infty$ as
all the (free) thermal propagators  from which the individual diagrams
are constructed  tend to zero in this  limit. The former result can be
also  deduced from the CJT   effective action formalism \cite{CJT}  or
from  a  heuristic  argumentation  based on   a thermodynamic pressure
\cite{ID1}.  Note  that in  the large--$N$ limit the  fourth and fifth
diagrams in FIG.\ref{N3} must be omitted.

The expectation value (\tseref{bc444}) can be now explicitly
written as

\begin{equation}
\left\langle \left( \sum_{a=1}^{N}\Phi_{a}^{2}(0) \right)^{2}
\right\rangle =
8N^{2}\; \int_{m_{0}^{2}}^{\infty} d{\hat{m}}_{0}^{2}\; \int
\frac{d^{D}q}{(2\pi)^{D}}\; \frac{\varepsilon(q_{0})}{e^{q_{0}\beta}-1}\;
\mbox{Im} \left(\frac{\frac{\partial
\Sigma_{T}({\hat{m}}_{r}^{2}(T))}{\partial
\lambda_{0}}}{(q^{2} - {\hat{m}}^{2}_{r} +i\epsilon)^{2}}\right)\, .
\tseleq{b65}
\end{equation}
\vspace{1mm}

\noindent In fact, the differentiation of the proper self--energy in
(\tseref{b65}) can be carried out easily. Using (\tseref{m12}), we get

\begin{displaymath}
\frac{\partial \Sigma_{T}}{\partial \lambda_{0}} =
\frac{\Sigma_{T}}{\lambda_{0}}
+ \lambda_{0}{\cali{M}}_{T}' \; \frac{\partial \Sigma_{T}}{\partial
\lambda_{0}} \;\; \Rightarrow \;\; \frac{\partial \Sigma_{T}}{\partial
\lambda_{0}} =
\frac{\Sigma_{T}}{\lambda_{0}(1-\lambda_{0}{\cali{M}}'_{T})}\, .
\end{displaymath}
\vspace{1mm}

\noindent From Eq.(\tseref{m12}) it directly follows that

\begin{displaymath}
\frac{dm_{r}^{2}(T)}{dm_{0}^{2}} = \frac{1}{(1-
\lambda_{0}\;{\cali{M}}'_{T})}\, ,
\end{displaymath}

\noindent which, together with the definition of ${\cali{M}}_{T}$, gives

\begin{eqnarray}
\left\langle \left( \sum_{a=1}^{N} \Phi_{a}^{2}(0) \right)^{2}
\right\rangle &=& 8N^{2}\; \int_{m_{r}^{2}(T)}^{\infty}
d{\hat{m}}_{r}^{2}\; \int \frac{d^{D}q}{(2 \pi)^{D}} \;
\frac{\varepsilon(q_{0})}{e^{q_{0}\beta} -1} \;
\mbox{Im}\frac{{\cali{M}}_{T}({\hat{m}}_{r}^{2})}{(q^{2}-
{\hat{m}}_{r}^{2} +i\epsilon)^{2}}\nonumber \\
&=& -\; 8N^{2} \; \int_{m_{r}^{2}(T)}^{\infty} d{\hat{m}}_{r}^{2}\;
{\cali{M}}_{T}({\hat{m}}_{r}^{2})
\; \frac{\partial {\cali{M}}_{T}({\hat{m}}_{r}^{2})}{\partial
{\hat{m}}_{r}^{2}}\nonumber \\
&=& 4N^{2}\; {\cali{M}}_{T}^{2}(m_{r}^{2}(T))\, ,
\tselea{b8}
\end{eqnarray}
\vspace{1mm}

\noindent where we have exploited in the last line the fact that
${\cali{M}}_{T}^{2}(m_{r}^{2} \rightarrow \infty)=0$. Let us mention that
the crucial point in the previous manipulations was that $m_{r}$ is both
real and momentum independent. Collecting our results together, we can
write for the hydrostatic pressure per particle (cf. Eq.(\tseref{ppp9}))

\begin{eqnarray}
{\cali{P}}(T)-{\cali{P}}(0) &=& -\; \frac{1}{(D-1)N}\; \left( \langle
\Theta^{i}_{c \;i} \rangle - \langle 0| \Theta^{i}_{c\; i} | 0 \rangle
\right)\nonumber \\
&=& +\; \frac{1}{2}\; \int \frac{d^{D}q}{(2 \pi)^{D-1}}\;
\left( \frac{2 {\vect{q}}^{2}}{(D-1)}  \right)\;
\frac{\varepsilon(q_{0})}{e^{q_{0}\beta}-1}\;  
\delta(q^{2}-m_{r}^{2}(T))\nonumber \\
&& -\; \frac{1}{2}\; \int \frac{d^{D}q}{(2
\pi)^{D-1}}\;
\left( \frac{2 {\vect{q}}^{2}}{(D-1)} \right)\;
\delta^{+}(q^{2}-m_{r}^{2}(0))\nonumber \\
&&+ \frac{1}{2\lambda_{0}}\left( \Sigma^{2}_{T}(m_{r}^{2}(T)) -
\Sigma^{2}(m_{r}^{2}(0)) \right)\, .
\tselea{b9}
\end{eqnarray}
\vspace{1mm}

\noindent Applying the Green theorem to the last two integrals and
eliminating the surface terms (for details see Appendix \ref{A2}) we find

\begin{eqnarray} {\cali{P}}(T) - {\cali{P}}(0) &=& \frac{1}{2}\; \int  
\frac{d^{D}q}{(2 \pi)^{D-1}}\; \frac{\varepsilon(q_{0})}{e^{q_{0}\beta}-1}\;
\theta(q^{2}-m^{2}_{r}(T))\nonumber \\ &&- \frac{1}{2} \; \int
\frac{d^{D}q}{(2 \pi)^{D-1}}\; \theta(q_{0}) \;
\theta(q^{2}-m^{2}_{r}(0))\nonumber \\
&&+\frac{1}{2\lambda_{0}}\left( \Sigma^{2}_{T}(m_{r}^{2}(T))-
\Sigma^{2}(m_{r}^{2}(0))\right) \nonumber \\ &&\nonumber \\ =&&
{\cali{N}}_{T}(m_{r}^{2}(T)) - {\cali{N}}(m_{r}^{2}(0)) +
\frac{1}{2\lambda_{0}} \left( \Sigma^{2}_{T}(m_{r}^{2}(T))-
\Sigma^{2}(m_{r}^{2}(0))\right)\, ,
\tseleq{b10}
\end{eqnarray}

\noindent where we have introduced new functions
${\cali{N}}_{T}(m^{2}_{r}(T)$ and ${\cali{N}}(m^{2}_{r})$;

\begin{eqnarray}
{\cali{N}}_{T}(m^{2}_{r}(T)) &=& \frac{1}{2}\; \int
\frac{d^{D}q}{(2 \pi)^{D-1}}\; \frac{\varepsilon(q_{0})}{e^{q_{0}\beta}-1}\;
\theta(q^{2}-m^{2}_{r}(T)) \nonumber \\
{\cali{N}}(m^{2}_{r}) &=& \lim_{T
\rightarrow
\; 0}{\cali{N}}_{T}(m^{2}_{r}(T))\, . \tselea{b101}
\end{eqnarray}

\noindent Eq.(\tseref{b10}) can be rephrased into a form which exhibits an
explicit independence of bar quantities. Using the trivial identity:

\begin{eqnarray}
&&\frac{1}{2\lambda_{0}}\; \left( \Sigma^{2}_{T}(m_{r}^{2}(T))-
\Sigma^{2}(m_{r}(0))\right)\nonumber \\
&& \mbox{\hspace{1.5cm}}= \frac{1}{2\lambda_{0}}\; \left(
\Sigma_{T}(m_{r}^{2}(T))-\Sigma(m^{2}_{r}(0))\right)\left(
\Sigma_{T}(m_{r}^{2}(T))+\Sigma(m^{2}_{r}(0))\right) \nonumber \\
&& \mbox{\hspace{1.5cm}}= \frac{\delta m^{2}(T)}{2}\;\left(
{\cali{M}}_{T}(m_{r}^{2}(T))+{\cali{M}}(m^{2}_{r}(0))\right)\, .
\tselea{b1111}
\end{eqnarray}

\noindent we get

\begin{equation} {\cali{P}}(T) - {\cali{P}}(0) =   
{\cali{N}}_{T}(m_{r}^{2}(T)) -
{\cali{N}}(m_{r}^{2}(0))+ \frac{\delta m^{2}(T)}{2}\;\left(
{\cali{M}}_{T}(m_{r}^{2}(T))+{\cali{M}}(m^{2}_{r}(0))\right)\, ,
\tseleq{b11}
\end{equation}

\noindent where $\delta m^{2}(T)= m_{r}^{2}(T)-m_{r}^{2}(0)$. The result  
(\tseref{b11}) has been previously obtained by authors \cite{ID1} in the
purely thermodynamic pressure framework.

\vspace{5mm}

\section{Hydrostatic pressure in $D=4$ (high--temperature expansion) \label{HTE}}

In order to obtain the high--temperature expansion of the pressure in
$D=4$, it is presumably the easiest to go back to equation (\tseref{b9})
and employ identity (\tseref{b1111}). Let us split this task into two
parts.  We firstly evaluate the integrals with potentially UV divergent
parts using the dimensional regularisation. The remaining integrals, with
the Bose--Einstein distribution insertion, are safe of UV singularities and
can be computed by means of the Mellin transform technique.

Inspecting (\tseref{b9})  and (\tseref{b1111}), we observe that
the only UV divergent contributions come from the integrals: 

\begin{eqnarray}
&+& \frac{1}{(D-1)} \int \frac{d^{D}q}{(2\pi)^{D-1}} \; {\vect{q}}^{2}\;
\left( \delta^{+}(q^{2}-m^{2}_{r}(T)) - \delta^{+}(q^{2}-m^{2}_{r}(0))
\right)\nonumber \\
&+& \frac{\delta m^{2}(T)}{4} \int \frac{d^{D}q}{(2\pi)^{D}} \; \left(    
\frac{i}{q^{2} - m^{2}_{r}(T) + i\epsilon } +
\frac{i}{q^{2}-m^{2}_{r}(0)+i\epsilon}\right)\, ,
\tselea{4.1}
\end{eqnarray}
\vspace{1mm}

\noindent which, if integrated over, give

\begin{eqnarray}
\mbox{(\tseref{4.1})} = &+&
\frac{\Gamma(\frac{-D}{2})\Gamma(\frac{D}{2}+\frac{1}{2})}{(D-1)
\Gamma(\frac{D-1}{2}) (4 \pi)^{\frac{D}{2}}} \left(
(m^{2}_{r}(T))^{\frac{D}{2}} -
(m^{2}_{r}(0))^{\frac{D}{2}}\right)\nonumber \\
&+&\frac{\delta m^{2}(T)\; \Gamma(1-\frac{D}{2})}{4
(4\pi)^{\frac{D}{2}}} \left( (m^{2}_{r}(T))^{\frac{D}{2}-1} +
(m^{2}_{r}(0))^{\frac{D}{2}-1}\right)\, .
\tselea{4.2}
\end{eqnarray}
\vspace{1mm}

\noindent Taking the limit $D=4-2\varepsilon \rightarrow 4$ and using
expansions

\begin{eqnarray}
\Gamma(-n + \varepsilon) &=& \frac{(-1)^{n}}{n!} \left(\frac{1}{
\varepsilon} + \sum_{k=1}^{n} \frac{1}{k} - \gamma +
{\cali{O}}(\varepsilon) \right)\nonumber \\
a^{x+\varepsilon} &=&  a^{x} \left( 1 + \varepsilon \;\mbox{ln}a +
{\cali{O}}(\varepsilon^{2}) \right)\, ,
\tselea{4.2P}
\end{eqnarray}

\noindent ($\gamma$ is the Euler--Mascheroni constant) we are finally left
with

\begin{equation}
\left.\mbox{(\tseref{4.1})}\right|_{D \rightarrow 4} =
- \frac{m^{2}_{r}(0) m^{2}_{r}(T)}{64 \;\pi^{2}} \; \mbox{ln}\left(
\frac{m^{2}_{r}(T)}{m^{2}_{r}(0)}\right) + \delta m^{2}(T)\;(m_{r}^{2}(T)
+ m^{2}_{r}(0))\; \frac{1}{128\; \pi^{2}}\, . \tseleq{4.3}
\end{equation}
\vspace{1mm}

\noindent The fact that we get finite result should not be
surprising as entire analysis of Section \ref{PE2} was made to show that
${\cali{P}}(T)-{\cali{P}}(0)$ defined via $\Theta^{\mu \nu}_{c}$ is
finite in $D=4$.

We may now concentrate on the remaining terms in (\tseref{b9}),
the latter read (we might, and we shall, from now on work in $D=4$)

\begin{equation}
\frac{1}{3} \int \frac{d^{4}q}{(2 \pi)^{3}} \; {\vect{q}}^{2}\;
\frac{1}{e^{|q_{0}|\beta}-1}\; \delta(q^{2}-m_{r}^{2}(T))
+ \frac{\delta m^{2}(T)}{4} \int \frac{d^{4}q}{(2\pi)^{3}} \;
\frac{1}{e^{|q_{0}|\beta}-1}  \delta(q^{2}-m_{r}^{2}(T))\, .
\tseleq{4.4}
\end{equation}
\vspace{1mm}

\noindent Our following strategy is based on the observation that the
previous integrals have generic form:

\begin{eqnarray}
I_{2\nu}(m_{r}) &=& \int \frac{d^{4}q}{(2\pi)^{3}}\; {\vect{q}}^{2\nu}\;
\frac{1}{e^{|q_{0}|\beta}-1}\;
\delta(q^{2}-m^{2}_{r}),\nonumber\\
&=& \frac{m_{r}^{2+2\nu}}{2\; \pi^{2}}\; \int_{1}^{\infty} dx
\;(x^{2}-1)^{\frac{1+2\nu}{2}}\; \frac{1}{e^{xy}-1}\, ,
\tseleq{c14}
\end{eqnarray}

\noindent with $\nu =0,1$ and $y = m_{r}\beta$.  Unfortunately, the
integral (\tseref{c14}) can not be evaluated exactly, however, its small
$y$ (i.e. high--temperature) behaviour can be successfully analysed by means of
the Mellin transform technique \cite{LW, EM}. Before going further, let us
briefly outline the basic steps needed for such a small $y$ expansion.

The Mellin transform $\hat{f}(s)$ is done by the prescription
\cite{LW,EM,B,Br,WH,BB}:

\begin{equation}
\hat{f}(s)= \int_{0}^{\infty} dx\; x^{s-1}\; f(x)\, ,
\tseleq{c12}
\end{equation}  

\noindent with $s$ being a complex number. One can easily check that the
inverse Mellin transform reads

\begin{equation}
f(x)= \frac{1}{i(2\pi)} \int_{-i\infty + a}^{i\infty +a} ds\; x^{-s} \;
\hat{f}(s)\, , \tseleq{c13}
\end{equation}
\vspace{1mm}

\noindent where the real constant `$a$' is chosen in such a way that
$\hat{f}(s)$ is convergent in the neighbourhood of a straight line
($-i\infty +a,\; i\infty +a$). So particularly if $f(x) =
\frac{1}{e^{xy}-1}$ one can find (\cite{B}; formula I.3.19) that

\begin{equation}
\hat{f}(s) = \Gamma(s)\zeta(s) y^{-s} \mbox{\hspace{1cm}} (\mbox{Re}s
> 1)\, ,
\tseleq{4.6}
\end{equation}  

\noindent where $\zeta$ is the Riemann zeta function ($\zeta(s)=
\sum_{n=1}^{\infty}n^{-s}$). Now we insert the Mellin transform of $f(x) =
\frac{1}{e^{xy}-1}$ to (\tseref{c14}) and interchange integrals (this is
legitimate only if the integrals are convergent before the
interchange). As a result we have

\vspace{1mm}
\begin{equation}
\int_{0}^{\infty} dx \; g(x)\ \frac{1}{e^{xy}-1} = \int_{-i\infty
+a}^{i\infty + a} \frac{ds}{i(2\pi)} \; \Gamma(s)\zeta(s) y^{-s}
\hat{g}(1-s)\, , \tseleq{4.7}
\end{equation}
\vspace{1mm}

\noindent with $g(x) = \theta(x-1)\;(x^{2}-1)^{\frac{1+2\nu}{2}}$. Using
the tabulated result (\cite{BB}; formula 6.2.32) we find

\begin{equation}
\hat{g}(1-s) = \frac{1}{2} B(-\nu -1 +
\mbox{$\frac{1}{2}s$};
\mbox{$\frac{3}{2}$} + \nu) \mbox{\hspace{1.5cm}} (\mbox{Re}s
>2+2\nu)\, ,
\tseleq{4.8}
\end{equation}

\noindent with $B(\;;\;)$ being the beta function. Because the integrand
on the RHS of (\tseref{4.7}) is analytic for $\mbox{Re}s > 2 + 2\nu$ and
the LHS is finite, we must choose such $a$ that the integration is
defined. The foregoing is achieved choosing $\mbox{a} > 2+2\nu$. Another
useful expressions for $\hat{g}(1-s)$ are (\cite{BB}; formula I.2.34 or
I.2.37)

\begin{eqnarray*}
\hat{g}(1-s) &=& B(\mbox{$\frac{3}{2}$}+\nu; -2 - 2\nu +s) \;   
{~}_{2}F_{1}[-\mbox{$\frac{1}{2}$} -\nu ; -2 - 2\nu +s;
-\mbox{$\frac{1}{2}$} -\nu +s; -1] \nonumber \\
&=& 2^{\frac{1}{2} +\nu} \; B(\mbox{$\frac{3}{2}$}+\nu; -2 - 2\nu +s)\;
{~}_{2}F_{1}[-\mbox{$\frac{1}{2}$} -\nu ; \mbox{$\frac{3}{2}$} +\nu;
-\mbox{$\frac{1}{2}$} -\nu +s; \mbox{$\frac{1}{2}$}]\, ,
\end{eqnarray*}

\noindent where ${~}_{2}F_{1}$ is the (Gauss) hypergeometric
function \cite{BB}.  Using identity

\vspace{1mm}
\begin{displaymath}
\Gamma(2x) = \frac{2^{2x-1}}{\sqrt{\pi}}\;
\Gamma(x)\Gamma(x+\mbox{$\frac{1}{2}$})\, ,
\end{displaymath}
\vspace{-3mm}
\noindent we can write

\begin{equation}
\mbox{(\tseref{4.7})} = \frac{\Gamma(\mbox{$\frac{3}{2}$} +\nu)}{4
\sqrt{\pi}}\; \int_{-i\infty + a}^{i\infty +a} \frac{ds}{i(2\pi)}
\Gamma(\mbox{$\frac{1}{2}$}s) \zeta(s) \left(\mbox{$\frac{1}{2}$}y
\right)^{-s}\Gamma(-\nu - 1 + \mbox{$\frac{1}{2}$}s)\, .
\tselea{4.91}
\end{equation}
\vspace{1mm}

\noindent The integrand of (\tseref{4.91}) has simple poles in $s=-2n \;
(n=1,2,\ldots)$, $s=1$, $s=-2n +2\nu +2 \;(n= 0,1, \ldots, \nu)$
 and double pole in $s=0$. An important point in the former pole analysis
was the fact that $\zeta(s)$ has simple zeros in $-2m$ ($m>0$) and only
one simple pole in $s=1$. The former together with identity

\begin{displaymath}
\Gamma\left( \frac{x}{2} \right)\; \pi^{- \frac{x}{2}}\; \zeta(x) =
\Gamma \left( \frac{1-x}{2} \right) \; \pi^{ \frac{x-1}{2}} \;
\zeta(1-x)\, ,
\end{displaymath}

\noindent shows that no double pole except for $s=0$ is present in
(\tseref{4.91}). Now, we can close the contour to the left as the value of
the contour integral around the large arc is zero in the limit of infinite
radius (c.f. \cite{B} and \cite{GR};  formula 8.328.1). Using
successively the Cauchy theorem we obtain

\begin{eqnarray}
\lefteqn{\frac{4\sqrt{\pi}\;\mbox{(\tseref{4.7})}}{\Gamma(\mbox{$\frac{3}{2}$}+\nu)}}\nonumber \\
&=& \sum_{n=0}^{\nu} y^{2n - 2\nu -2} \;
\frac{\pi^{-2n +2\nu +2}(-n+\nu)!\; (-1)^{n} |B_{-2n + 2\nu +2}|}{n! \;
(-2n +2\nu +2)!\; 2^{4n -4\nu -4}}\nonumber \\
&+& \sum_{n=1}^{\infty} y^{2n} \; \frac{\pi^{-2n} \;(2n)!
\;\zeta(1+2n)\; (-1)^{n + \nu +1}}{n!\; (n + 1 + \nu)!
\;2^{4n-1}}\nonumber \\
&+& y^{-1} \; \frac{\pi\; (-1)^{\nu + 1} \; (\nu + 1)! \;2^{2\nu +3}}{(2\nu
+ 2)!} + \frac{2\;
(-1)^{\nu + 1}}{(\nu + 1)!}\left\{ \mbox{ln} \left( \frac{y}{4\pi}\right) +
\gamma - \mbox{$\frac{1}{2}$} \sum_{k=1}^{\nu + 1}\frac{1}{k}
\right\}\, ,
\tselea{4.101}
\end{eqnarray}
\vspace{1mm}

\noindent where $B_{\alpha}$'s are the Bernoulli numbers. Let us mention
that for $\zeta(2n + 1)$ only numerical values are available.

Inserting (\tseref{4.101}) back to (\tseref{4.4}), we get 
for ${\cali{P}}(T) - {\cali{P}}(0)$

\begin{eqnarray}
&&{\cali{P}}(T) - {\cali{P}}(0) = (\mbox{\tseref{4.3}}) +
\frac{1}{3}\; I_{2}(m_{r}(T)) + \frac{\delta
m^{2}(T)}{4} \;I_{0}(m_{r}(T))\nonumber \\
&&\mbox{}\nonumber\\
&& \mbox{\hspace{5mm}}=\frac{T^{4}\; \pi^{2}}{90} -
\frac{T^{2}}{24} \left( m^{2}_{r}(T) - \frac{\delta m^{2}(T)}{2}
\right) + \frac{T\; m_{r}(T)}{4\; \pi} \left(
\frac{m^{2}_{r}(T)}{3} - \frac{\delta m^{2}(T)}{4} \right)\nonumber \\
&&\mbox{}\nonumber\\
&& \mbox{\hspace{5mm}} +\; \frac{m_{r}^{2}(T)\;m^{2}_{r}(0)}{32\; \pi^{2}}\;
\left(\mbox{ln}\left( \frac{m_{r}(0)}{T\; 4\pi}  \right) + \gamma -
\frac{1}{2} \right) - \frac{m^{4}_{r}(0)}{128 \; \pi^{2}} \nonumber \\
&&\mbox{}\nonumber\\
&& \mbox{\hspace{5mm}} - \;
\sum_{n=1}^{\infty} \left( m^{2}_{r}(T) - \mbox{$\frac{(n+2)}{2}$}\; \delta
m^{2}(T)\right) \; \frac{m_{r}^{2n+2}(T) \; \pi^{-2n-2}\;
(2n)!\; \zeta (1+2n) \; (-1)^{1+n}}{T^{2n}\;n!\;(n+2)!\;
2^{4n+4}}\, .\nonumber\\
\tselea{4.111}
\end{eqnarray}
\vspace{1mm}

\noindent Note that (\tseref{4.3}) cancelled against the same term in
$\frac{1}{3}\; I_{2}(m_{r}(T)) + \frac{\delta m^{2}(T)}{4}
\;I_{0}(m_{r}(T))$. One can see that (\tseref{4.111}) rapidly converges for
large $T$, so that only first four terms dominate at sufficiently high
temperature. The aforementioned terms come from the poles nearby the
straight line $(-i\infty + a, \; i\infty +a)$ (the more dominant
contribution the closer pole). It is a typical feature of the Mellin
transform technique that integrals of type

\begin{displaymath}
\int_{0}^{\infty}dx \; g(x) \; \frac{1}{e^{xy}-1}\, ,
\end{displaymath}   

\noindent can be expressed as an expansion which rapidly converges for
small $y$ (high--temperature expansion) or large $y$ (low--temperature
expansion)\footnote{By the same token we get the low--temperature
expansion if the integral contour must be closed to the right.}.

For a sufficiently large $T$ we can use the high--temperature
expansion of $\delta m^{2}(T)$ found in Appendix \ref{A3}. Inserting
(\tseref{bb52})  to (\tseref{4.111}) we obtain

\begin{eqnarray}
&&{\cali{P}}(T) - {\cali{P}}(0)
= \frac{T^{4}\; \pi^{2}}{90} -
\frac{T^{2}\;m^{2}_{r}(T)}{24} + \frac{T^{3}\;m_{r}(T)}{12 \pi}\nonumber
\\
&& \nonumber \\
&&\mbox{\hspace{2mm}}+ \frac{\lambda_{r}}{8}\left( \frac{T^{4}}{144} -
\frac{T^{3}\;m_{r}(T)}{24 \pi} + \frac{T^{2}\;m^{2}_{r}(T)}{16
\pi^{2}}\right)
+ {\cali{O}}\left( m^{4}_{r}(T)\; \mbox{ln}\left( \frac{m_{r}(T)}{T4 \pi
}\right) \right)\, .
\tselea{pj1}
\end{eqnarray}
\vspace{1mm}

\noindent Up to a sign, the  result (\tseref{pj1}) coincides with that
found by Amelino--Camelia    and  Pi  \cite{ACP}  for  the   effective
potential\footnote{Let  us  remind\cite{HS,CJT,ACP}  that   from   the
definition of  $V_{eff}$ the thermodynamic  pressure is $-V_{eff}$. In
order to obtain (\tseref{pj1}) from  $V_{eff}$ in \cite{ACP}, one must
subtract the zero temperature value  of $V_{eff}$ and restrict oneself
to   vanishing field  expectation   value    and positive bare    mass
squared.}. Actually,  they used instead of  the $N \rightarrow \infty$
limit the  Hartree--Fock approximation which is  supposed to  give the
same $V_{eff}$ as the leading $1/N$ approximation \cite{AC2}.

Note that the discussion of the mass renormalisation in Section
3.1 can be directly extended to the case when $m_{r}(0) = 0$ (this does
not apply to our discussion of $\lambda_{r}$!). Latter can be also seen
from the fact that (\tseref{4.111}) is continuous in $m_{r}(0)=0$ (however
not analytic). The foregoing implies that the original massless scalar
particles acquire the thermal mass $m_{r}^{2}(T)= \delta m^{2}(T) $ . From
(\tseref{4.111})  one then may immediately deduce the pressure for massless
fields $\Phi_{a}$ in terms of $\delta m(T)$. The latter reads

\begin{eqnarray}
{\cali{P}}(T) - {\cali{P}}(0) &=& \frac{T^{4} \; \pi^{2}}{90}  -
\frac{T^{2}\; (\delta m(T))^{2}}{48} + \frac{T\; (\delta m(T))^{3}}{48
\; \pi} \nonumber \\
&& \mbox{} \nonumber \\
&+& \; \sum_{n=1}^{\infty} \frac{(\delta m(T))^{2n+4}
\;\pi^{-2n-2}\; (2n)!\; \zeta(1+2n)\; (-1)^{n+1}}{T^{2n} \;(n-1)!\;
(n+2)!\; 2^{4n+5}}\, . \tselea{4.121}
\end{eqnarray}
\vspace{1mm}

\noindent This result is identical to that found by Drummond {\em et al.} in
\cite{ID1}.

A noteworthy observation is that when the energy of a thermal
motion is much higher then the mass of particles in the rest, then the 
massive theory approaches the massless one. This is justified in the   
first (high--temperature dominant) term of (\tseref{4.111}) and
(\tseref{4.121}).  This term is nothing but a half of the black body   
radiation pressure for photons \cite{GM,Cub} (photons have two
degrees of freedom connected with two transverse polarisations). One
could also obtain the temperature dominant contributions directly from
the Stefan--Boltzmann law \cite{LW,GM,Cub} for the density energy
(i.e. $\langle \Theta^{00} \rangle$). The formal argument leading to
this statement is based on the noticing that at high energy
(temperature) the theory at hand is (classically) approximately conformally
invariant, which in turn implies that the energy--momentum tensor is
traceless \cite{CCR}.  Taking into account the definition of the
hydrostatic pressure (\tseref{EMT24}), we can with a little effort
recover the leading high--temperature contributions for the massive 
case.

\vspace{5mm}

\section{Conclusions \label{C2}}

In the present chapter we have clarified the status of the hydrostatic
pressure in (equilibrium) thermal QFT. The former is explained in terms of
the thermal expectation value of the `weighted' space--like trace of the
energy--momentum tensor $\Theta^{\mu \nu}$. In classical field theory there
is a clear microscopic picture of the hydrostatic pressure which is
further enhanced by a mathematical connection (through the virial theorem)
with the thermodynamic pressure. In addition, it is the hydrostatic    
pressure which can be naturally extended to a non--equilibrium medium.  
Quantum theoretic treatment of the hydrostatic pressure is however pretty
delicate. In order to get a sensible, finite answer we must give up the
idea of total hydrostatic pressure. Instead, thermal interaction pressure
or/and interaction pressure must be used (see (\tseref{ppp9}) and   
(\tseref{ppp10})). We have established this result for a special case when
the theory in question is the scalar $\Phi^{4}$ theory with $O(N)$   
internal symmetry; but it can be easily extended to more complex
situations. Moreover, due to a lucky interplay between the conservation of
$\Theta^{\mu \nu}$ and the space--time translational invariance of an
equilibrium (and $T=0$) expectation value we can use the simple canonical
(i.e. unrenormalised) energy--momentum tensor. In the course of our 
treatment in Section \ref{PE2} we heavily relied on the counterterm
renormalisation, which seems to be the most natural when one discusses
renormalisation of composite Green's functions. To be specific, we have
resorted to the minimal subtraction scheme which has proved useful in
several technical points.

We have   applied the prescriptions  obtained  for the QFT hydrostatic
pressure to $\Phi^{4}$ theory in the--large $N$  limit. The former has
the undeniable advantage of being exactly  soluble. This is because of
the fact  that  the large--$N$  limit  eliminates `nasty'  classes  of
diagrams in the Dyson--Schwinger  expansion.  The surviving  class  of
diagrams  (superdaisy  diagrams) can  be  exactly resumed, because the
(thermal)  proper  self--energy  ${\vect{\Sigma}}$,  as  well  as  the
renormalised   coupling   constant    $\lambda_{r}$    are    momentum
independent. We have also stressed that the $O(N)\;\Phi^{4}$ theory in
the large--$N$ limit is consistent only if  we view it as an effective
field theory. Fortunately, the upper bound on the UV cut--off is truly
huge,  and it does not represent  any significant restriction. For the
model at  hand the resumed form of  the pressure with $m_{r}(0)=0$ was
firstly  derived (in  the  purely  thermodynamic pressure  context) by
Drummond {\em  et  al.}  in  \cite{ID1}.  We  have  checked, using the
prescription  (\tseref{ppp9}) for   the thermal interaction  pressure,
that their  results are in agreement with  ours. The former is  a nice
vindication of  the validity of  the  virial theorem  even in  the QFT
context.

The expression for the pressure obtained  was in a suitable form which
allowed  us to take  advantage of the   Mellin transform technique. We
were  then able to write  down the high--temperature  expansion for the
pressure in $D=4$  (both for massive  and massless fields) in terms of
renormalised  masses $m_{r}(T)$  and $m_{r}(0)$.  We  have  explicitly
checked  that all UV   divergences present in  the  individual thermal
diagrams `miraculously' cancel in accordance with  our analysis of the
composite operators in Section \ref{PE2}.

\chapter{Pressure in out--of--equilibrium media \label{PN}}

\noindent In  recent   years  significant progress  has been    made  in
understanding  the  behaviour   of    QFT  systems away   from   thermal
equilibrium. Motivation for the study  of such  systems comes both  from
the early  universe as well as  from  the quark--gluon plasma (deconfined
phase  quarks and gluons). Non--equilibrium   effects are expected to  be
relevant in the relativistic heavy--ion collisions planed at RHIC and LHC
in the near future \cite{TDL,SNW,HH,Hag}.

One  of    the   significant physical variables,   in     the context of
non--equilibrium  QFT, is  pressure.   Pressure, as  an easily measurable
parameter\footnote{In  this    connection  we may    mention  the piezo
resistive  silicon pressure sensors   used, for instance,  in superfluid
helium\cite{TH1,TH2} or  neutron (X--ray) diffraction technique used in
solid state physics\cite{SH,JSO}.}, is  expected to play  an important
role in a detection  of phase transitions.   This is usually ascribed to
the fact that the   pressure  should  exhibit  a discontinuity   in  its
derivative(s) when  the      local  phase  transition   occurs.      The
aforementioned has found its vindication in solid state physics and
fluid mechanics, and may  play a crucial role,  for instance, in various
baryogenesis scenarios.

It is well known that for systems in thermal equilibrium, the pressure
may  be calculated     via the  partition  function   \cite{LW,LB,ID}.
However, this procedure cannot be extended to off--equilibrium systems
where   is no such thing  as  the grand--canonical potential.  In this
chapter we consider  an alternative definition  of pressure,  based on
the energy--momentum tensor.  This, so called, hydrostatic pressure is
defined as the space--like  trace of the energy--momentum tensor  (see
Chapter  \ref{PE}), and   in  equilibrium, it  is identified  with the
thermodynamical pressure via  the virial theorem \cite{LW}.  There are
several  problems with  the  validity of   this  identification on the
quantum  level, indeed gauge theories suffer  from a conformal (trace)
anomaly and  require special  care\cite{LW}.   However, we  will avoid
such difficulties by focusing on a scalar theory  which is free of the
mentioned  complications   \cite{LW,PJ1}.   The  major   advantage  of
defining pressure through the  energy--momentum tensor stems  from the
fact that one  may  effortlessly extend  the hydrostatic  pressure  to
non--equilibrium systems (for discussion see Chapter \ref{PE}). 

The aim of  this chapter is to provide  a systematic prescription  for
the calculation of the hydrostatic pressure in non--equilibrium media.
This requires three  concepts; the Jaynes--Gibbs  principle of maximal
entropy, the  Dyson--Schwinger equations and the hydrostatic pressure.
In order to  keep our discussion as simple  as possible we  illustrate
the   whole procedure  on  a toy   model  system,  namely  the $O(N)\;
\Phi^{4}$ theory.  The latter  has advantage of being exactly solvable
in the large--$N$ limit, provided that we  deal with a translationally
invariant  medium.  As  a   result the  hydrostatic   pressure may  be
expressed in a closed form. 

In order  to provide meaningful results  also for readers not entirely
familiar with the  off--equilibrium Dyson--Schwinger equations and  the
Jaynes--Gibbs principle of maximal   entropy, we briefly  summarise in
Section \ref{PN1} the basic  essentials.  (Detailed discussion of the
equilibrium  case may be found in  Appendices \ref{A1}--\ref{A12} or in
Refs.\cite{CH,Jayn2,J-T}.)    As   a   byproduct   we   construct  the
generalised  Kubo--Martin--Schwinger  (KMS)    conditions.     Section
\ref{PN2} is devoted to the  study of the (canonical--) energy  momentum
tensor in the  $O(N)\; \Phi^{4}$ theory.  If  both  the density matrix
and the  full   Hamiltonian are  invariant under   $O(N)$ symmetry one
obtains Ward's identities in a  similar manner as  in equilibrium.  We
show how these  drastically simplify the  expression for the pressure.
In Section \ref{PN3}  we concentrate  our analysis  on  the large--$N$
limit.   In this setting  we derive  a very  simple expression for the
pressure   - pressure depends  only    on two--point Green  functions.
Section \ref{PN4} then  forms the vital part  of this paper.  Owing to
the fact that the infinite hierarchy of the Dyson--Schwinger equations
is  closed (basically by     chance) we obtain  simple  equations  for
two--point  Green   functions -  Kadanoff--Baym  equations.  These are
solved exactly for three   illustrative density matrices  $\rho$.   We
choose deliberately translationally invariant $\rho$'s.  The reason is
twofold.   Firstly,  for a  non--translationally  invariant medium one
must use the improved energy momentum  tensor instead of the canonical
one \cite{PJ1}.  This is rather involved and it will be subject of our
future work.  Secondly, the   Kadanoff--Baym equations turn out  to be
hyperbolic equations  whose fundamental solution  is well known.  As a
result  we   may  evaluate,  for the  density  matrices   at hand, the
hydrostatic pressure explicitly.  The chapter ends with a discussion.

\vspace{5mm}

\section{Basic formalism \label{PN1}}

\noindent  The key object of  our interest is the energy--momentum tensor
$\Theta_{\mu \nu}(x)$. A  typical contribution to $\Theta_{\mu  \nu}(x)$
can be written as

\begin{displaymath}
{\cali{D}}_{\mu_{1}}\Phi(x)       \;  {\cali{D}}_{\mu_{2}}\Phi(x) \ldots
{\cali{D}}_{\mu_{n}}\Phi(x)\, .  
\end{displaymath}

\noindent Here $\Phi$     is a field in   the    Heisenberg picture  and
${\cali{D}}_{\mu_{i}}$ stands for a corresponding differential operator.
Since   ${\cali{D}}_{\mu_{i}}\Phi(x)$ and  ${\cali{D}}_{\mu_{j}}\Phi(x)$
generally do not commute for $i\not= j$, one must prescribe the ordering
in $\Theta_{\mu \nu}$. Our strategy is based on the observation that one
can conveniently define such ordering via the non--local operator

\vspace{-1mm}
\begin{eqnarray}
&&{\lim_   {x_i\to   x}{\cali{T}}^{*}\{   {\cali{D}}_{\mu_{1}}\Phi  (x_{1})\cdots
{\cali{D}}_{\mu_{n}}\Phi     (x_{n})      \}           =}\nonumber    \\
&&\mbox{\hspace{1.3cm}} \lim_ {x_i\to x} {\cali{D}}(i\partial_{\{\mu\}})
{\cali{T}}\{\Phi (x_{1})\cdots \Phi (x_{n})\}\, , 
\tseleq{TP11}
\end{eqnarray}

\noindent where  ${\cali{D}}(i\partial_{\{\mu\}})$  is  just a  useful
short--hand                       notation                         for
${\cali{D}}_{\mu_{1}}{\cali{D}}_{\mu_{2}}\ldots {\cali{D}}_{\mu_{n}}$,
and ${\cali{T}}^{*}$ is defined in Section \ref{PE2}. It is clear that
both ${\cali{T}}^{*}$ and  ${\cali{T}}$ coincide if  all the arguments
$x_{i}$ are different,  so $({\cali{T}}^{*}-{\cali{T}})(\ldots)$ is an
operator with a support   at the contact points. The  ${\cali{T}}^{*}$
ordering is in  general  a very  useful tool whenever  one deals  with
composite operators.  In the sequel we shall repeatedly use this fact. 

\vspace{3mm}

\subsection{Off--equilibrium Dyson--Schwinger equations \label{PN11}}

\noindent  Let   us   now briefly   outline   the   derivation  of   the
Dyson--Schwinger equations  for  systems  away  from  equilibrium.    For
simplicity we illustrate this on a single scalar field $\Phi$.

We start with the action  $S[\Phi]$ where $\Phi$  is linearly coupled to
an external  source $J(x)$.  Working with the  fields  in the Heisenberg
picture, the operator equation of motion can be written as
\begin{equation}
\frac{\delta S}{\delta \Phi(x)}[\Phi = \Phi^{J}] + J(x) = 0\, , \tseleq{4.0}
\end{equation}

\noindent where  the  index $J$ emphasises    that $\Phi$ is  implicitly
$J$--dependent.   This dependence  will be  made  explicit  via a unitary
transformation connecting $\Phi^{J}$  (governed  by  $H - J\Phi$)   with
$\Phi$ (governed by $H$). If $J(x)$ is switched  on at time $t=t_{i}$ we
have   

\vspace{-1mm}
\begin{eqnarray}
\Phi^{J}(x) &=& B^{-1}(t;t_{i})\Phi(x)B(t;t_{i})\nonumber \\
&=& B(t_{i};t_{f})B(t_{f};t)\; \Phi(x) \; B(t;t_{i})\nonumber \\
&=& T_{C} \left( \Phi(x)\; \mbox{exp}\;(i\int_{C}d^{4}y \; J(y)\Phi(y))
\right)\, .
\tselea{4.A}
\end{eqnarray}

\noindent Here we have used the fact that

\begin{equation}
B(t_{1};t_{2}) = T\left( \mbox{exp}(i\int_{t_{2}}^{t_{1}}d^{4}y
\;J(y)\Phi(y) )\right)\,\,\,\;\;\;\; t_{1} > t_{2}\, .
\end{equation}

\noindent The close--time path $C$ runs along the real axis from $t_i$
to  $t_{f}$ ($t_{f}$ is  arbitrary, $t_{f}>t_{i}$)  and  then back  to
$t_i$. With this setting we can rewrite (\tseref{4.0}) as

\begin{equation}
{\cali{T}}^{*}_{C}\left(   \left\{ \frac{\delta  S[\Phi_{\pm}]}{\delta\Phi}   +
J_{\pm}\right\}\, \exp(i \int d^{4}y
\; (J_{+}(y)\Phi_{+}(y) - J_{-}(y)\Phi_{-}(y))) \right) = 0\, ,
\end{equation}

\noindent where,  as it is usual \cite{LW,KCC}, we  have labelled  the field
(source) with the time argument on the upper branch of $C$ as $\Phi_{+}$
($J_{+}$) and that with the time argument on the  lower branch of $C$ as
$\Phi_{-}$ ($J_{-}$).    Introducing   the metric  $(\sigma_{3})_{\alpha
\beta}$ ($\sigma_{3}$ is the   usual Pauli matrix  and $\alpha,  \beta =
\{+;-\}$)     we  can    write   $J_{+}\Phi_{+}     -  J_{-}\Phi_{-}   =
J_{\alpha}\;(\sigma_{3})^{\alpha         \beta}\;    \Phi_{\beta}      =
J^{\alpha}\;(\sigma_{3})_{\alpha \beta}\; \Phi^{\beta}$. For the  raised
and lowered indices we simply read:  $\Phi_{+} = \Phi^{+}$ and $\Phi_{-}
= -\Phi^{-}$ (similarly for $J_{\alpha}$). Taking $\mbox{Tr}(\rho \ldots
)$ with  $\rho  = \rho[\Phi, \partial  \Phi,  \ldots]$ being the density
matrix, we obtain

\begin{eqnarray}
\frac{1}{Z[J]}\frac{\delta S}{\delta \Phi(x)} \left[\Phi^{\alpha}(x) = -
i\frac{\delta}{\delta J_{\alpha}(x)}\right]Z[J]   = -J^{\alpha}(x)  \, ,
\tselea{4.15}
\end{eqnarray}
\vspace{1mm}

\noindent with   $Z[J]=    {\rm Tr}\left\{   \rho   \;  {\cali{T}}_{C}
\,\mbox{exp}\;\left(    i\int_{C}d^{4}y \; J(y)\Phi(y)  \right)\right\}$
being  the generating functional of  Green's functions (which in the
non--equilibrium context we shall denote az $\G$).  Employing the
commutation   relation: $-i\frac{\delta  }{\delta J_{\alpha}}Z =  Z\,(
\phi^{\alpha} - i \frac{\delta}{\delta   J_{\alpha}})$, we may  recast
(\ref{4.15}) into more convenient form, namely

\begin{equation}
-J^{\alpha}(x)   =   \frac{\delta         S}{\delta     \Phi(x)}\bigg[
\phi^{\alpha}(x)  +      i    \int     d^{4}y\;
\G_{\beta}^{\alpha}(x,y)\;          (\sigma_{3})^{\beta           \gamma}
\;\frac{\delta}{\delta\phi^{\gamma} (y)} \bigg]\,\,\1 \, , \tselea{4.16} 
\end{equation}
\vspace{1mm}

\noindent where the  symbol $\1$ indicates  the unit. As usual, the mean
field, $\phi_{\alpha}$, is defined as the expectation value of the field
operator:    $\phi_{\alpha}(x)\equiv\langle   \Phi_{\alpha}(x) \rangle$.
Defining $Z[J] = \mbox{exp}(iW[J])$,  two--point Green functions are then
defined     as           $\G_{\alpha\beta}(x,y)=      -\delta^{2}W/\delta
J^{\alpha}(x)\delta          J^{\beta}(y)$.     Setting    $J$        in
(\tseref{4.15})--(\tseref{4.16}) to $0$ (i.e. physical condition) we obtain
a first out of  infinite hierarchy of  equations for  Green's functions.
Successive   $J$ variations  of  (\tseref{4.15})--(\tseref{4.16})  generate
higher order equations in the hierarchy.  The  system of these equations
is usually referred to as the Dyson--Schwinger equations.

For the future  reference is convenient to have  the expression  for the
effective action $\Gamma [\phi_{c}]$.  This is connected with $W[J]$ via
the Legendre transform:

\begin{equation}
\Gamma[\phi_{c}] =  W[J]  -  \int_{C}  d^{4}y  \; J(y)   \phi_{c}(y)\, .
\tselea{letrans}
\end{equation}

\noindent Following  the  previous reasonings,  one can  easily persuade
oneself that the expectation value of $\Theta^{\mu \nu}$ reads
\begin{equation}
\langle     \Theta^{\mu  \nu}(x)   \rangle   =  \langle  \Theta^{\mu
\nu}[\Phi(x)] \rangle = \Theta^{\mu \nu}[\phi_{  +}(x)  + i
\int           d^{4}y\;         \G_{+    \beta}(x,y)\;(\sigma_{3})^{\beta
\gamma}\;\frac{\delta}{\delta  \phi^{\gamma}   (y)}]\,\,   \1    \,    .
\tselea{exvaten}
\end{equation}

\noindent  We have automatically used  the sub--index `$+$' as the fields
involved in $\Theta^{\mu \nu}$ have, by definition, the time argument on
the upper branch of $C$.

\vspace{3mm}

\subsection{The Jaynes--Gibbs principle of maximal entropy \label{PN12}}

In this section we would like to review  the Jaynes--Gibbs principle of
maximal        entropy               (also          maximum    calibre
principle)\cite{Jayn,Jayn2,Tol,Gibbs}, which  we shall  employ in  the
following. The formalism is  a generalisation from an ordinary Gibbs's
principle of maximal  entropy to the  systems out of equilibrium.  The
objective of the principle is to construct the `most probable' density
matrix  which       fulfils    the      constraints    imposed     by
experimental/theoretical data. 

The standard rules of statistical physics allows us to define   
the expectation value via the density matrix $\rho$ as
\begin{equation}
\langle \ldots \rangle = \mbox{Tr}(\rho \ldots)\, ,
\tseleq{W15}
\end{equation}

\noindent with the trace running over a complete set of {\em physical}
states describing the ensemble in question at some initial time $t_{i}$.
The specification (or reasonable approximation) of the density matrix is
thus crucial for determining the macroscopic properties of a given system.

The usual approaches \cite{CHKMP,M1,M2,EJY} trying to determine $\rho$
start with the Schr{\"o}dinger picture. The merit of this procedure is
that one transforms the whole time dependence of the expectation value
into the density  matrix itself. Thus the   time evolution of  such an
expectation value is  equivalent to the time  evolution of $\rho$, and
we     need   not   to     solve   separate   equations     for   each
observable\footnote{This advantage does  not  seem to be  relevant for
Green's functions with different time arguments.}. 

Suppose that the system is prepared in such a way that at time
$t$ the probability of the system being in the state $|\psi_{n}; t\rangle$
is $p_{n}(t)$.  The density matrix has then the standard form   

\begin{equation}
\rho_{s}(t)= \sum_{n} p_{n}(t)\; | \psi_{n};t \rangle \langle \psi_{n};
t|\, .
\tseleq{W155}
\end{equation}

\vspace{2mm}

\noindent (by $\sum_{n}$ we mean; sum over discrete spectrum and integrate
over continuous one). We should stress that the ensemble
$\{|\psi_{n};t\rangle \}$ in (\tseref{W155}) not necessary consist of   
mutually orthogonal states, although the density matrix can always be
formally diagonalised by selecting its eigenbasis (polar basis). Applying
the Schr{\"o}dinger equation to (\tseref{W155}), the evolution of
$\rho_{s}$ reads

\begin{equation}
\frac{d \rho_{s}(t)}{dt} = i[\rho_{s}(t),H] + \frac{\partial
\rho_{s}(t)}{\partial t}\, ,
\tseleq{W156}
\end{equation}

\noindent where

\begin{equation}
\frac{\partial \rho_{s}(t)}{\partial t} = \sum_{n} \frac{d
p_{n}(t)}{dt} \; |\psi_{n};t\rangle \langle \psi_{n};t|\, .
\end{equation}

\vspace{2mm}

\noindent For a closed Hamiltonian system $p_{n}(t)$ cannot be changed,
and so (\tseref{W156}) reduces to the celebrated von Neumann--Liouville  
equation

\begin{equation}
\frac{d \rho_{s}(t)}{dt} = i[\rho_{s}(t),H]\, .
\tseleq{W157}
\end{equation}

\vspace{2mm}  

\noindent Let us mention that the vanishing of the time derivative of
$p_{n}(t)$ implies that the von Neumann--Gibbs entropy $S_{G} =
-\mbox{Tr}(\rho_{s}\mbox{ln}\rho_{s})$ is time independent. Indeed, in the
latter case the evolution of $\rho_{s}$ can be formally written (c.f.
(\tseref{W156})) as

\begin{equation}
\rho_{s}(t) = U(t;t')\rho_{s}(t')U(t;t')^{-1}\, ,
\end{equation}  

\noindent where the evolution operator is determined from the
Schr{\"o}dinger equation 

\begin{displaymath}
i\frac{\partial}{\partial t} U(t;t') = H(t)U(t;t'), \;\;\;\;\; U(t',t') =
1\, ,
\end{displaymath}

\noindent and so using the property of a trace operation we have,

\begin{equation}
S_{G}(t') = -\mbox{Tr}(\rho_{s}(t')\mbox{ln}\rho_{s}(t')) = -
\mbox{Tr}(\rho_{s}(t)\mbox{ln}\rho_{s}(t)) = S_{G}(t)\, .
\end{equation}

\noindent The time dependence of $p_{n}(t)$ reflects the fact that the
system in question interacts with an exterior -- a heat bath (whose
structure and dynamics are usually unknown). In order to determine
$p_{n}(t)$ one would need to resort to some physical model \cite{EJY}
describing the dynamics of the environmental system -- task by no means
easy. In order to avoid these difficulties it is customary to apply a
stochastic description of a system--bath interaction \cite{DB, CWG}, or
equivalently use an irreversible ``master equation" \cite{CWG} for
$\rho$.

To keep our ideas simple, we shall from the very outset assume
that $p_{n}$ is constant.  This means that either no external environment
interacts with our system or that the interaction is reflected only in the
Hamiltonian $H$ (e.g. effective external fields or time dependences) but
the probability of the population of any state stays unchanged -
adiabatic interaction.

Solving the von   Neumann--Liouville equation  would be a   formidable
task. One usually resorts either to  variational method \cite{EJY, FJ}
with several trial $\rho$'s, or one may recast (\tseref{W157}) into an
infinite     hierarchy    of  integro--differential   equations    for
two-particle,       etc.   distribution   functions   \cite{RB}  (i.e.
Bogoliubov--Born--Green--Kirkwood--Yvon or BBGKY hierarchy).  In   the
latter approach one hopes that one can  device an effective truncation
of  the hierarchy allowing to close  the system of equations, and then
solve it  (usually  perturbatively).  However, in  both aforementioned
approaches  we must specify the initial  value data  in order to solve
uniquely the evolution equation(s).   This is quite delicate task, and
one  usually uses  the most  simple  (and  not always  physically well
motivated)   choices   of   $\rho$ (e.g.     Gibbs  (grand--)canonical
distribution \cite{EJY}, Gaussian distribution \cite{CH,M1,M2}).

Rather than following the previous path, we shall use the
Heisenberg picture instead. This seems to be more suitable for our
purpose. In our particular case the polar form of $\rho_{H}$ reads
\vspace{-1mm}
\begin{equation}
\rho_{H}= \sum_{n}p_{n} |\psi_{n}\rangle \langle \psi_{n}|\, .
\end{equation}

\noindent In the following we shall denote $\rho_{H}$ simply as $\rho$. In
order to determine $\rho$ explicitly we shall resort to the Jaynes--Gibbs  
principle of maximal entropy \cite{Jayn,Jayn2,Tol,Gibbs}. The latter
allows one to construct $\rho$ in a way which naturally extends the     
original Gibbs prescription \cite{Tol,Gibbs} which refers only to
equilibrium. The basic idea is `borrowed' from the information theory. Let
us assume that we have criterion of how to characterise the informative
content of $\rho$. The most ``probable'' $\rho$ is then selected out of
those $\rho$ which are consistent with `whatever' we know about the system
and which have the last informative content (Laplace's principle of
insufficient reasons). Consistency with anything known about $\rho$ must
be kept; while to chose more informative $\rho$ is to presume an extra
information which we do not control, and to make unjustified implicit
assumption concerning the information we do not know about.

It remains to characterise the {\em{information content}} (measure)
${\textswab{H}}[\rho]$ of $\rho$. This was done by C.E.Shannon \cite{SH1},
L.Brillouin \cite{LBr}  and L.Szilard \cite{LSz}  with the   
result

\begin{displaymath}
{\textswab{H}}[\rho] = \mbox{Tr}(\rho \; {\mbox{log$_{2}$}}\,\rho)\; 
\tseleq{sh} 
\end{displaymath}

\noindent      The density   matrix  is   then    chosen to   minimise
${\textswab{H}}[\rho]$. It is quite surprising that up to a (negative)
multiplicative  constant    ${\textswab{H}}[\rho]$    coincides   with
$S_{G}[\rho]$,    and   thus the  principle    of insufficient reasons
basically turns out to  be the maximal entropy principle\footnote{  It
Appendix    \ref{A4}  we show that    the  Shannon  entropy  (which is
proportional to  $S_{G}$)  equals (in  base  2  of  logarithm) to  the
expected number  of binary (yes$/$no)  questions whose  answer take us
from our  current state of  knowledge to  the one  certainty.  So, the
bigger $S_{G}$, the  more   questions  one has  and  consequently  one
possess a higher ignorance about  a system.}.  Note that no assumption
about the nature  of $\rho$ was made; namely  there was  no assumption
whether  $\rho$  describes equilibrium  or non--equilibrium situation.
(We were interested only in the information content of $\rho$, and not
in the underlying  dynamics of a  system).   To put more  flesh on the
bones,   let   us rephrase   the former   into  more physical language
\cite{Jayn,Jayn2}.  What we actually need to do is to maximise $S_{G}$
subject  to  the constraints imposed  by our  knowledge of expectation
values  of  certain operators  $P_{1}[\Phi,  \partial   \Phi], \ldots,
P_{n}[\Phi, \partial \Phi]$.  This   yields  a density  matrix  $\rho$
which incorporates the   fact that all the  quantum   states which are
permitted by  the  constraints have equal  probabilities \cite{Jayn2}.
In   contrast to equilibrium,  all $P_{k}[\ldots]$'s  may be operators
which are not the constants of the motion  (both the position and time
dependences are allowed). So namely if one knows that 

\begin{equation}
\langle P_{k}[ \Phi, \partial \Phi] \rangle = g_{k}(x_{1}, x_{2},
\ldots)\, ,
\tseleq{ivd}
\end{equation}

\noindent the entropy maximalisation leads to

\begin{equation}
\rho = \frac{1}{{\cali{Z}}[\lambda_{1}, \ldots, \lambda_{n}]}
\;\mbox{exp}\left(-\sum_{i=1}^{n} \int \prod_{j}d^{4}x_{j} \;
\lambda_{i}(x_{1},\ldots)P_{i}[ \Phi,
\partial \Phi]\right)\, ,
\tseleq{dm1}
\end{equation}
\vspace{1mm}

\noindent with the `partition function'

\begin{equation}
{\cali{Z}}[\lambda_{1}, \ldots , \lambda_{n}] =
\mbox{Tr}\left(\mbox{exp}(-\sum_{i=1}^{n} \int \prod_{j}d^{4}x_{j} \;   
\lambda_{i}(x_{1},\ldots)P_{i}[ \Phi,
\partial \Phi]) \right)\, .
\tseleq{dm2}
\end{equation}
\vspace{1mm}

\noindent  It is  possible to  show  that  the stationary solution  of
$S_{G}$ is  unique  \cite{BM1} and maximal.   The latter  goes  on the
account of the fact that $S_{G}$ is a concave functional (see Appendix
\ref{A4}).   Both   in  (\tseref{dm1})  and (\tseref{dm2})   the  time
integration is not either present at all (so $f_{k}$ is specified only
in the initial time $t_{i}$), or is  taken over the gathering interval
$(-\tau , t_{i})$  (i.e.   as  $\int_{\tau}^{t_{i}}dt \ldots$).    If,
instead, one has certain partial knowledge about the expectation value
of $P_{k}[\Phi, \partial \Phi]$ at  some discrete times prior $t_{i}$,
the corresponding   integration    must be  replaced  by   a  discrete
summation. Note  that if we have no  prior knowledge about the system,
then  $\rho =  1/{\cali{W}} $, where   ${\cali{W}} $ is  the number of
accessible quantum states.  Equivalently we may  say that the no prior
restrictions  mean  our total  ignorance  about  the system  and  as a
consequence  we   must  affiliate  with each  quantum   state an equal
probability. 

The Lagrange multipliers $\lambda_{k}$ might be eliminated if
one solves $n$ simultaneous equations

\begin{equation}
g_{k}(x_{1}, \ldots) = - \frac{\delta \; \mbox{ln}{\cali{Z}}}{\delta
\lambda_{k}(x_{1},
\ldots )}\, .
\tseleq{jg3}
\end{equation}  
\vspace{1mm}

\noindent Using the definition of the von Neumann--Gibbs entropy together
with (\tseref{dm1}) we get
\begin{eqnarray}
S_{G}[g_{1},\ldots,g_{n}]|_{max} &=&
-\mbox{Tr}(\rho\; \mbox{ln}\rho)|_{max} \nonumber \\
&=& \mbox{ln}{\cali{Z}}[\lambda_{1}, \ldots , \lambda_{n}] + \sum_{i=1}^{n} \int
\prod_{j}d^{4}x_{j} \;
\lambda_{i}(x_{1},\ldots)g_{i}(x_{1},\ldots)\,.
\tselea{jg4}
\end{eqnarray}

\noindent So one may view $S_{G\;max}$ as the Legendre transformation of
$\mbox{ln}{\cali{Z}}$. In the equilibrium case the former is the standard relation
between entropy and the grand--canonical potential. Having (\tseref{jg4}),
the explicit solution of (\tseref{jg3}) may be formally written as

\begin{equation}
\lambda_{k}(x_{1},\ldots ) = \frac{\delta \; S_{G}[g_{1},
\ldots , g_{n}]|{{}_{max}}}{\delta g_{k}(x_{1}, \ldots)}\,.
\tselea{ent1}
\end{equation}

\noindent Now, in order to reflect  the density matrix (\tseref{dm1}) in
the  Dyson--Schwinger equations, we   need to construct the corresponding
boundary conditions.  This   may be done quite straightforwardly.  Using
the cyclicity of $\mbox{Tr}(\ldots)$ together with the relation

\begin{displaymath}
e^{A}Be^{-A}       =       \sum_{n=0}^{\infty}\frac{1}{n!}C_{n},\;\;\;\;
 C_{0}=B,C_{n}= [A,C_{n-1}]\, ,
\end{displaymath}

\noindent we can write the  generalised KMS conditions for the $n$--point
Green function as

\begin{equation}
\langle   \Phi(x_{1})\ldots \Phi(x_{n}) \rangle  = \langle \Phi(x_{2})
\ldots        \Phi(x_{n})\Phi(x_{1})\rangle  +     \sum_{k=1}^{\infty}\frac{1}{    k!}\langle
\Phi(x_{2})\ldots \Phi(x_{n})C_{k}(x_{1}) \rangle \,, \tselea{CMS11}
\end{equation}
\vspace{1mm}

\noindent  where  $A=  {\mbox{ln}}(\rho)$, $B   = \Phi(x_{1})$, $x_{10}=
t_{i}$.   So   for the two--point  Green   functions  we have\footnote{In
special cases when  $\rho = |0\rangle \langle  0|$ or  $\rho = e^{-\beta
H}/{\cali{Z}}(\beta)$ the boundary conditions are the well known Feynman
and KMS boundary conditions, respectively.}

\begin{displaymath}
\G_{+-}(x,y)                    =          \G_{-+}(x,y)                  +
\sum_{n=1}^{\infty}\frac{1}{n!}\mbox{Tr}(\rho \;\Phi(x)C_{n}(y))\, .
\end{displaymath}  

\noindent   In  this chapter we   aim   to demonstrate that conditions
(\tseref{CMS11}) together with the  causality condition are sufficient
to determine Green's functions uniquely. 

\vspace{5mm}

\section{The $O(N)\; \Phi^{4}$ theory \label{PN2}}

\noindent  The  $O(N)\;  \Phi^{4}$ theory   is   described by   the bare
Lagrangian

\begin{displaymath}
{\cali{L}}=         \frac{1}{2}\sum_{a=1}^{N}\left(            (\partial
\Phi^{a})^{2}-m_{0}^{2}(\Phi^{a})^{2}              \right)             -
\frac{\lambda_{0}}{8N}\left(  \sum_{a=1}^{N} (\Phi^{a})^{2}
\right)^{2}\, .
\end{displaymath}

\noindent   The   corresponding   canonical   energy--momentum  tensor is
$\Theta^{\mu \nu}_{c}  =\sum_{a}\partial^{\mu}\Phi^{a}\partial^{\nu}\Phi^{a}
- g^{\mu  \nu}{\cali{L}}$, and from  Eqs.(\ref{TP11}) and (\ref{exvaten})
its expectation value is
\begin{eqnarray}
&&\Theta^{\mu\nu}_{c}(\Phi_{+}(x))=\langle \Theta^{\mu \nu}_{c}(x)
\rangle \nonumber \\
&& \mbox{\hspace{1cm}}=
i\sum_{c}\partial_{x}^{\mu}\partial_{y}^{\nu}\;\G^{cc}_{++}(x,y)|_{x=y}
-\frac{g^{\mu\nu}i}{2}\sum_{c}\bigg\{\partial_{x}^{\alpha}\partial_{\alpha
y}\;\G_{++}^{cc}(x,y)|_{x=y}            
- m_{0}^{2}\G_{++}^{cc}(x,x)\bigg\}\nonumber \\
&& \mbox{\hspace{1cm}} +    \frac{g^{\mu   \nu}\lambda_{0}}{8N}
\sum_{c,d}        \bigg   \{\bigg((\phi^{c}_{+}(x))^{2} +  
i\G^{cc}_{++}(x,x)\bigg)\bigg(
(\phi^{d}_{+}(x))^{2}       +      i\G^{dd}_{++}(x,x)\bigg)   \nonumber\\
&& \mbox{\hspace{1cm}} + 2    \bigg(    \phi^{c}_{+}(x)\phi^{d}_{+}(x)   +
i\G^{cd}_{++}(x,x)\bigg)i\G^{cd}_{++}(x,x)\bigg\} \nonumber \\
&& \mbox{\hspace{1cm}}+  
\mbox{terms  with $\Gamma^{3}$ and $\Gamma^{4}$}
\tseleq{valten} \, .
\end{eqnarray}

\noindent  Before  proceeding further  with  our development, we want to
show how  one  can  significantly simplify  Eq.(\tseref{valten}) provided
that both the   density matrix and the  Hamiltonian  are invariant under
rotations in the $N$--dimensional vector  space of fields. This situation
would occur if the system was  initially prepared in such  a way that no
field $\Phi^{a}$  was  favoured over another.  The  fact  that $\rho$ is
invariant under $O(N)$ transformations means that
\begin{equation}
U(\epsilon)\rho[\Phi,      \partial     \Phi,  \ldots]U^{-1}(\epsilon) =
\rho[\Phi, \partial \Phi, \ldots]\;, \tselea{e33}
\end{equation}

\noindent where  the  fields $\Phi^{a}$  transform under $N$--dimensional
rotations: $U(\epsilon)\Phi^{a}U^{-1}(\epsilon)   =  R^{-1}(\epsilon)^{a
b}\Phi^{b}     =[\mbox{exp}(\epsilon_{i}T_{i})]^{a b}\Phi^{b}$,    where
$R(\epsilon)$ is the   rotation  matrix in the  $N$--dimensional   vector
space,  and the generators $T_{i}$  are real and antisymmetric $N \times
N$ matrices.   It is obvious  that the  previous relation for $\Phi^{c}$
can be satisfied  for all  times only  if  the full  Hamiltonian,  which
governs the evolution  of   $\Phi^{a}$, is  also  invariant against  the
$N$--dimensional rotations.

Let us now  consider the generating  functional $Z[J]$ corresponding  to
the $O(N)$  symmetric density matrix.  Employing  the cyclic property of
$\mbox{Tr}(\ldots)$  together  with the    infinitesimal transformation,
$\delta     R(\epsilon)= 1 +    \epsilon_{i}T_{i}$,  we  obtain that the
variation   of $Z$   must  vanish.   The  latter  implies  the following
(unrenormalised) Ward's identities:

\begin{eqnarray}
\int_{C} d^{4}y \;J^{a}(y)    \frac{\delta   W[J]}{\delta  J^{b}(y)}   =
\int_{C}  d^{4}y\;   J^{b}(y) \frac{\delta  W[J]}{\delta  J^{a}(y)}\,  .
\tselea{e7}
\end{eqnarray}
\vspace{1mm}

\noindent Taking successive variations with   respect to source $J$,  we
obtain  the following results   (see also \cite{J-T}): $n$--point Green
functions with $n$ odd vanish, while for $n$ even ($n=2k, k=1,2,\ldots$)
one has

\vspace{-2mm}
\begin{equation}
\G^{a_1a_2\dots        a_{2k}}_{\alpha_1\cdots\alpha_{(2k)}}(x_1,\ldots
,x_{2k}) = \sum_{{\rm  all\,\, pairings}}\,
\prod_{i<j}                                              \delta_{a_ia_j}
\G^{(2k)}_{\alpha_1\cdots\alpha_{2k}}(x_1,\ldots      ,x_{2k})   \,     ,
\tselea{apc8}
\end{equation}
\vspace{1mm}

\noindent where $\G^{(2k)}$ is   a universal $2k$--point  Green  function.
Similar      results     can       be      obtained  for
$\Gamma^{a_{1}a_{2}\ldots            a_{2k}}_{\alpha_{1}\alpha_{2}\ldots
\alpha_{2k}}(\ldots)$.

Finally  note    that  these results  enable     the expression  for the
expectation   value   of the energy   momentum   tensor to be simplified
significantly to

\begin{eqnarray}
\langle    \Theta^{\mu\nu}_{c}(x)\rangle      &=&         iN\partial^{\mu}_x
\partial^{\nu}_y\,\,  \G_{++}(x,y)\big\vert_{x=y}  - {N+2
\over  8}\lambda_0 g^{\mu\nu}(\G_{++}(x,x))^2  \nonumber\\ 
&-&   i{N\over
2}g^{\mu\nu}\bigg( \partial^{\mu}_x \partial^{\nu}_y\,\, \G_{++}(x,y)\big
\vert_{x=y} -   m_0^2  \,\,\G_{++}(x,x) \bigg)\nonumber \\
   &+&   {\rm
terms\,\,\, with}\,\,\,\, \Gamma^{(4)}\, .  \tselea{me}
\end{eqnarray}  

\noindent In the rest  of this chapter  we shall confine ourselves only to
situations where both $\rho$ and $H$ are $O(N)$ invariant.

\vspace{5mm}

\section{The large--$N$ limit \label{PN3}}

\noindent  Let us now examine  behaviour  of  (\tseref{me}) to the  order
${\cali{O}}(1/N)$.  For this purpose it is important  to know how either
$G^{(n)}$ or $\Gamma^{(n)}$ behave  in the $N \rightarrow \infty$ limit.
At  $T=0$ or in  equilibrium     the  Feynman diagrams are     available
\cite{ID,PJ}  and the corresponding combinatorics   can be worked  out
quite simply.  On the other hand, the situation in off--equilibrium cases
is more difficult as we do not have at our disposal Wick's theorem.  One
may  devise    various  diagrammatic   approaches, e.g.    kernel method
\cite{CH},    cumulant  expansion  \cite{GM},    correlation diagrams
\cite{PH}, etc.  Instead of relying on any graphical representation
(as we done in Section \ref{PE3}) we
show   that for  both  equilibrium   and   off--equilibrium systems,  the
situation may be approached from far more  general standpoint using only
Ward's identities and properties of $\Gamma$ and $W$.

In order to find the leading behaviour at large $N$ it is presumably the
easiest to consider    the Legendre transform  (\tseref{letrans}).    The
explicit $N$ dependence  may be obtained by  setting $\phi^{c} =  \phi$,
which implies $J^{c}=J$  for all the group indices. Eq.(\tseref{letrans})
then indicates that both $\Gamma[\phi]$ and $W[J]$ are of order $N$.  If
we expand $\Gamma[\phi^{a}]$ in terms  of $\phi^{c}$ taking into account
Ward's identities we get
\begin{eqnarray}
\Gamma[\phi]  = \Gamma[0] &+& \frac{1}{2}  N \int_{C} d^{4}x\; d^{4}y \;
\Gamma^{(2)}(x,y) \; \phi(x)\phi(y) \nonumber  \\ &+& \frac{3}{4!} N^{2}
\int_{C} d^{4}x\;  d^{4}y\;  d^{4}z\;  d^{4}q  \;  \Gamma^{(4)}(x,y,z,q)
\nonumber\\ &\times  & \phi(x)  \phi(y)  \phi(z) \phi(q) + \, \cdots  \, .
\tselea{W2}
\end{eqnarray}

\noindent Since the LHS of (\tseref{W2})  is of order $N$, $\Gamma^{(2)}$
must be  of order $N^{0}$,  $\Gamma^{(4)}$  of order  $N^{-1}$, and,  in
general, $\Gamma^{(2n)}$ must be of order $N^{1-n}$.  This suggests that
in the expression  for  the energy--momentum tensor (\tseref{me}),   terms
containing  $\Gamma^{(4)}$ can be  ignored.  The  above  argument can be
repeated in a similar way for $W$.

Hence,  collecting our results together,   the expectation value of  the
energy--momentum tensor to leading order in $N$ may be written as
\begin{eqnarray}
\langle            \Theta^{\mu\nu}_{c}(x)\rangle              
&=&    iN\partial^{\mu}_x    \partial^{\nu}_y\,\,
\G_{++}(x,y)\big\vert_{x=y}       -  {N      \over            8}\lambda_0
g^{\mu\nu}(\G_{++}(x,x))^2 \nonumber    \\     &-&
i{N\over   2}g^{\mu\nu}\bigg(     \partial^{\mu}_x  \partial^{\nu}_y\,\,
\G_{++}(x,y)\big \vert_{x=y}- m_0^2 \,\,\G_{++}(x,x) \bigg)\, .
\tseleq{me2}
\end{eqnarray}

\noindent This  result is surprisingly  simple: the expectation value of
the  energy--momentum   tensor, and  thus  the hydrostatic   pressure, is
expressed purely in terms  of  two--point Green's functions.  The  latter
can be  calculated through the Dyson--Schwinger equations (\tseref{4.16}).
Furthermore, these equations have a very  simple form provided that both
the large--$N$ limit and Ward's identities are applied.   If we perform a
variation of (\tseref{4.16}) with respect to $J_{\beta}(y)$ we obtain, to
order ${\cali{O}}(1/N)$,   the following  Dyson--Schwinger  equations for
two--point Green functions:
\begin{eqnarray}
&&\left(  \Box   +  m_{0}^{2} +  \frac{i  \lambda_{0}}{2}   \; \G_{\alpha
\alpha}(x,   x)   \right)    \G_{\alpha  \beta}(x, y)    \nonumber\\   &&
\mbox{\vspace{2cm}}  =  -\delta(y-x)(\sigma_{3})_{\alpha \beta}   \,   .
\tseleq{em5}
\end{eqnarray}

\noindent These dynamical equations  for  two--point Green functions  are
better known as the Kadanoff--Baym equations \cite{KB}.

Let  us mention  one more   point.  The  generalised KMS conditions  for
$\G_{\pm \mp}$ are significantly simple in  the large--$N$ limit.  This is
because in sum  (\tseref{CMS11}) only terms of  order ${\cali{O}}(N^{0})$
contribute.   This  implies  that only quadratic  operators $P_{i}[\Phi,
\partial \Phi]$ in  the density matrix  are relevant.  As a  result, the
Jaynes--Gibbs  principle naturally provides  a vindication of the popular
Gaussian Ansatz \cite{EJY,M1,M2}.

\vspace{5mm}

\section{Out--of--equilibrium pressure \label{PN4}}

\noindent The objective  of  the present  section  is to show how  the
outlined mathematical machinery works  in the case of the  hydrostatic
pressure.  In  order  to gain   some insight  we  start with    rather
pedagogical,  but   by  no means  trivial   examples;  translationally
invariant,  non-equilibrium density  matrices.  We  consider  the more
difficult  case  of  translationally  non--invariant  density matrices
in our future work. 

\subsection{Equilibrium \label{PN41}}

\noindent As an important special case we can choose the constraints

\begin{equation}
\langle P_{k}[\Phi,    \partial \Phi]  \rangle    |_{t_{i}} =   g_{k}  =
\mbox{constant}\, , \tselea{eq1}
\end{equation}

\noindent where $t_{i}$ is arbitrary. Eq.(\tseref{eq1}) then implies that
$P_{k}$'s are integrals of motion.   Since in the finite volume  systems
the spatial translational invariance is destroyed,  the only integral of
motion  (apart  from conserved charges)  is  the  Hamiltonian.  Thus the
system  is   in  thermal equilibrium   and the   laws  of thermodynamics
\cite{Cub,GM} prescribe that $g  = \int^{T}_{0} dT' C_{V}(T')$  ($C_{V}$ is
the heat capacity at  constant volume $V$  and $T$ is temperature).  Eq.
(\tseref{ent1}) determines the   Lagrange  multiplier; $\lambda =  1/T  =
\beta$.  The density matrix maximising  the  $S_{G}$ is then the density
matrix  of   the  canonical  ensemble: $\rho  =  \frac{\mbox{exp}(-\beta
H)}{{\cali{Z}[\beta]}}$.  Due to the translational invariance of $\rho$,
the Kadanoff--Baym equations read

\begin{equation}
\left(    \Box_{x}  +    m_{r}^{2}(T)\right)   \G_{\alpha   \beta}(x-y) =
-\delta(x-y)(\sigma_{3})_{\alpha \beta} \, , \tselea{em7}
\end{equation}

\noindent where the temperature--dependent renormalised mass is (see, for
example    \cite{ID,PJ});  $m_{r}^{2}(T)  =    m_{0}^{2}   +   \frac{i
\lambda_{0}}{2} \;    \G_{\alpha  \alpha}(0)$.   The    corresponding KMS
boundary condition is

\begin{equation}
\G_{+-}({\vect{x}},t_{i}; {\vect{y}},0) = \G_{-+}({\vect{x}},t_{i}-i\beta;
{\vect{y}},0)\, .  \tselea{em8}
\end{equation}

\noindent Because $m_{r}(T)$ is a  spatial constant, a Fourier transform
solves equations  (\tseref{em7}).  The solutions of (\tseref{em7}) subject
to   condition (\tseref{em8}) are   then the  resumed  propagators in the
Keldysh--Schwinger formalism
\begin{eqnarray}
i\G_{\pm \pm}(k) &=& \frac{\pm  i}{k^{2} - m^{2}_{r} \pm  i\varepsilon} +
2\pi  f_{B}(|k_{0}|)\delta(k^{2}-m^{2}_{r})\nonumber \\ i\G_{\pm \mp}(k)  &=&
2\pi  \left\{  \theta(\mp k_{0})   + f_{B}(|k_{0}|)   \right\}\delta(k^{2} -
m^{2}_{r})\,  , \tselea{sol1}
\end{eqnarray}

\noindent   with  $f_{B}(x)  =   (\mbox{exp}(\beta   x)-1)^{-1}$  being  the
Bose--Einstein distribution.

Now,  the   total hydrostatic pressure  in  $D$  dimensions is
classically defined as \cite{LW,GMW,PJ}
\begin{displaymath}
p(x,T) = -\frac{1}{(D-1)}\langle \Theta^{i}_{\; i}(x) \rangle \, .
\end{displaymath}

\noindent Because $\Theta^{\mu \nu}_{c}$  is a composite operator, a special
renormalisation is  required \cite{LW,PJ1,Collins,Brown}.   As  we have
shown    in   \cite{PJ}, for   translationally  invariant  systems the
renormalised pressure coincides with the, so called, thermal interaction
pressure ${\cali{P}}$ (see Eq.(\tseref{ppp9})).  The latter reads

\vspace{-1mm}
\begin{equation}
{\cali{P}}(T) = p(x,T) -  p(x,0) = -\frac{1}{(D-1)}
\left\{ \langle \Theta_{c\; i}^{i} \rangle  - \langle 0| \Theta_{c\;i}^{i}
| 0 \rangle \right\}\, .  \tselea{ppp101}
\end{equation}

\noindent Let us remind that  the energy--momentum tensor $\Theta^{\mu
\nu}$ need not to be the canonical one  (however, the canonical one is
usually the simplest one), c.f.  Section \ref{PE2}.  As a second  point
we should   mention  that prescription  (\tseref{ppp101})  retains  its
validity for non--equilibrium media   as  well.  This is because   the
short--distance behaviour of  $G_{+ +}(x,y)$, which is responsible for
the singular    behaviour of   $\Theta^{\mu \nu}$,  comes    from  the
particular solution of   the Kadanoff--Baym  equation  (\tseref{em7}).
The latter can be chosen to be  completely independent of the boundary
conditions  (actually it is    useful to  chose  the  Feynman   causal
solution).   On the    other  hand,   the  homogeneous   solution   of
(\tseref{em7}), which  is  regular at $|x-y| \rightarrow  0$, reflects
all   the  boundary conditions.    One   may   see  then that  the  UV
singularities which trouble $\Theta^{\mu  \nu}$ may be treated in  the
same way as at the $T=0$.  Incidentally, the former is an extension of
the well known fact that in order  to renormalise a finite temperature
QFT, it suffices to renormalise it at $T=0$.

Inserting the  solution  (\tseref{sol1}) into  the expression for  the
energy--momentum tensor   (\tseref{me2})  we   arrive at  the  thermal
interaction pressure per particle (compare Section \ref{HTE}, see also
\cite{PJ1,ACP}) 
\begin{eqnarray}
{\cali{P}}(T) &=& \frac{T^{4}\;  \pi^{2}}{90} - \frac{T^{2}\;m^{2}_{r}(T)}{24}  +
\frac{T\;m^{3}_{r}(T)}{12  \pi} +
\frac{\lambda_{r}}{8}\left(            \frac{T^{4}}{144}               -
\frac{T^{3}\;m_{r}(T)}{24   \pi}   +       \frac{T^{2}\;m^{2}_{r}(T)}{16
\pi^{2}}\right)\nonumber     \\   &&\nonumber \\   &+&  {\cali{O}}\left(
m^{4}_{r}(T)\;    \mbox{ln}\left(  \frac{m_{r}(T)}{T4   \pi     }\right)
\right)\, , \tseleq{b91}
\end{eqnarray}

\noindent where the renormalised  coupling constant  $\lambda_{r}$ comes
from  the  $T=0$  renormalisation prescription:   $\Gamma^{(4)}_{aa  \to
bb}(s=0)=-\lambda_{r} /N$ ($s$  is  the usual  Mandelstam variable).   A
direct   calculation  of   $\Gamma^{(4)}$ in     the  large--$N$   limit
were performed in Section \ref{PE2} with the result:

\begin{equation}
\lambda_{r}   =    \lambda_0  -{1\over 2}    \lambda_0\lambda_{r}{1\over
(4\pi)^{D/2}} \Gamma\left[2-D/2\right] \left(m_{r}^{2}\right)^{D/2-2} \,
.  \tseleq{re6}
\end{equation}

\noindent To regularise the theory we have used the usual MS scheme with
the dimensional regularisation ($D  \not=   4$).  The high   temperature
expansion of the pressure  (\tseref{b91}) to all   orders can be  found in
Section \ref{HTE}  where  the calculations, however,  were  approached  from a
different standpoint.

\vspace{3mm}

\subsection{Off--equilibrium I \label{PN42}}

\noindent  The next  question   to   be  addressed  is  how the    above
calculations are modified in the  non--equilibrium case.  To see that let
us choose the following constraint

\begin{equation}
g({\bf  k})=\langle   {\cali{H}}({\bf  k})\rangle|_{t_{i}}    =  \langle
{\tilde{\cali{H}}}({\bf {k}}) \rangle|_{t_{i}}\, .  \tselea{ppp91}
\end{equation}

\noindent   Here  $t_{i}$ is  arbitrary,  and  function  $g({\bf k})$ is
specified by  theory/experiment.  The ${\tilde{\cali{H}}}({\bf {k}})$ is
a     quadratic  operator      fulfilling   the    condition    $\langle
{\cali{H}}\rangle|_{N       \rightarrow      \infty}       =     \langle
{\tilde{\cali{H}}}\rangle$.  As    $g({\bf{k}})$      is  finite,   both
${\cali{H}}$  and ${\tilde{\cali{H}}}$  must  be renormalised  (i.e.  we
must subtract the zero temperature part). Obviously

\begin{displaymath}
{\tilde{\cali{H}}}({\bf{k}})                                           =
\omega_{k}a^{\dagger}({\bf{k}})a({\bf{k}})\, ,
\end{displaymath}

\noindent with $\omega_{k} = \sqrt{{\bf{k}}^{2}+ {\cali{M}}^{2}}$ and

\begin{equation}
{\cali{M}}^2  =m^2_0+{i\over   2}\lambda_0    \int {d^{d}k^{\prime}\over
(2\pi)^{d}} \G_{++}(k^{\prime}) \, .  \tseleq{re3}
\end{equation}

\noindent In   the large--$N$ limit   the  corresponding  density matrix
(\tseref{dm1}) reads

\begin{eqnarray}
\rho=   \frac{1}{{\cali{Z}}(\beta)}\exp\left(-\int   {d^{3}{\bf{k}}\over
(2\pi)^3 2\omega_k}\beta ({\bf k}){\tilde{\cal H}}({\bf k}) \right) \, ,
\tseleq{inicon1}
\end{eqnarray}

\noindent   with   $\beta({\vect{k}})/2(2\pi)^{3}\omega_{k}$   being the
Lagrange multiplier to  be determined.  Using  Eq.(\tseref{jg3}), we find
that
\begin{eqnarray}
g({\bf      k})=  {V\over       (2\pi)^3}    \,      {\omega_k     \over
e^{\beta({\bf{k}})\omega_k}-1} \, , \tseleq{inicon2}
\end{eqnarray}

\noindent where  $V$  denotes   the volume of   the   system.   Clearly,
expression (\tseref{inicon2}) can be interpreted as the density of energy
per mode.   The  fact that $\beta({\bf{k}})$  is  not constant indicates
that different modes have different `temperatures'.

The  Kadanoff--Baym equations  coincide  in   this  case with  those   in
(\tseref{em7}) provided  $m_{r}(T) \rightarrow {\cali{M}}$.  The boundary
condition  can  be    worked    out simply   using   the    prescription
(\tseref{CMS11}). This gives
\begin{equation}
\G_{+-}(k) = e^{-\beta({\bf{k}})k_{0}}\G_{-+}(k)\, .  \tseleq{CMS2}
\end{equation}

\noindent The fundamental solution of the Kadanoff--Baym equation is

\begin{eqnarray}
&&\G_{\alpha  \beta}(k)= {(\sigma_{3})_{\alpha  \beta} \over  k^2 - {\cal
M}^2 +   i\epsilon_{\alpha \beta}} -  2\pi  i \delta ( k^2  -{\cal M}^2)
f_{\alpha     \beta}(k)\,  ,     \nonumber    \\    &&   \nonumber    \\
&&\mbox{\hspace{3cm}}   \epsilon_{\alpha      \beta}       =    \epsilon
(\sigma_{3})_{\alpha \beta}\, .  \tselea{re2}
\end{eqnarray}

\noindent Let    us  mention one more   point.   The  boundary condition
(\tseref{CMS2}) is  not  by  itself  sufficient to  determine  $f_{\alpha
\beta}$'s. (This fact is often overlooked  even in the equilibrium QFT.)
It is actually necessary to substitute this condition with an additional
condition, namely the condition of  causality.  The causality condition,
i.e.  vanishing of the commutator $[\Phi(x),  \Phi(y)]$ for $(x-y)^{2} <
0$, importantly restricts  the  class of possible $f_{\alpha  \beta}$'s.
To    see  this, let us    look at  the  Pauli--Jordan  function $\langle
\,[\Phi(x), \Phi(y)]\rangle$.  The latter is the homogeneous solution of
the Kadanoff--Baym equation with the initial--time value data: $\langle \,
[\Phi(x), \Phi(y)]\rangle|_{x_{0}=y_{0}}  =0$ (i.e. causality condition)
and $\partial_{0}\langle   \, [\Phi(x), \Phi(y)]\rangle|_{x_{0}=y_{0}} =
-\delta^{3}({\bf{x}}-{\bf{y}})$.  Thus    the Pauli--Jordan function   is
uniquely determined and its Fourier transform reads

\begin{equation}
\mbox{F.T.}(\langle   [\Phi(x),\Phi(y)]     \rangle      )(k)  =     -i2\pi
\delta(k^{2}-{\cali{M}}^{2})\; \varepsilon(k_{0})\, .  \tseleq{ft1}
\end{equation}

\noindent Relation (\tseref{ft1}) immediately implies that

\begin{eqnarray}
&&f_{+-}(k) = \theta(-k_{0}) + {\tilde{f}}(k) \nonumber \\ &&f_{-+}(k) =
\theta(k_{0}) + {\tilde{f}}(k)\, , \tselea{f1}
\end{eqnarray}

\noindent with ${\tilde{f}}$  being,   so far  arbitrary, and  for  both
$f_{+-}$ and $f_{-+}$ identical,   function. The ${\tilde{f}}$ is   then
fixed via the generalised  KMS condition (\tseref{CMS2}).  Similarly, the
causality   condition  specifies   that    $\G_{++}(k)  -  \G_{--}(k)    =
\mbox{PP}\{\, 1/(k^{2} - {\cali{M}}^{2})\}$ (the symbol `PP' denotes the
principal part).  By inspection of  the definition of $\G_{\alpha \beta}$
one may easily realise that
\begin{eqnarray}
&\G_{++} +  \G_{--}   =   \G_{+-} + \G_{-+}&\nonumber   \\  &\G_{+-}(k)  =
-(\G_{+-}(-k))^{*}& \nonumber \\ &\G_{++}(k) = -(\G_{--}(-k))^{*}&\,.
\end{eqnarray}

\noindent  From   these  relations  follows  that  $f_{++}   =  f_{--} =
{\tilde{f}}$   and ${\tilde{f}}(k)    =  ({\tilde{f}}(-k))^{*}$.     The
${\tilde{f}}$ is the same as in (\tseref{f1}).

Since the divergences in $\G_{\alpha \beta}$  come from the first term in
(\tseref{re2}) (i.e.  from   the particular solution),  we can  shift the
corresponding (zero  temperature)  poles at  $D=4$ to  the bare mass. In
this case we can write
\begin{equation}
{\cal M}^2 \equiv m^{2}_{r}+\delta m^2 \, , \tseleq{re33}
\end{equation}

\noindent where  $m_{r}$ is  the  renormalised mass   in the vacuum  and
$\delta   m$   is  the mass    shift due   to  an  interaction  with the
non--equilibrium medium.  Inserting the  `$++$' components of (\ref{re2})
into (\ref{re3}), we obtain
\begin{eqnarray}
{\cal M}^2=m^2_0+  {1\over   2}\lambda_0  \left[{{\cal M}^{{D-4}}  \over
(4\pi)^{D\over   2}} \Gamma\left[  {1-D/    2}   \right] \, +    N({\cal
M}^2)\right] \, , \tseleq{re4}
\end{eqnarray}

\noindent where

\begin{equation}
N({\cal M}^2)= \int {d^D k\over (2\pi)^D} 2\pi \delta ( k^2 -{\cal M}^2 )
f_{++}(k) \, . \tseleq{re5}
\end{equation}
\vspace{1mm}

\noindent   In   an  equilibrium   system    $f_{++}$   would be   the
Bose--Einstein distribution $f_{B}$.  Note that because (\tseref{re5})
corresponds to  a homogeneous solution  of the Kadanoff--Baym equation
at $|x-y|=0$ it is  automatically finite.  Thus for a  translationally
invariant medium (both equilibrium and non--equilibrium) $f_{++}$ must
act as a regulator in the UV region.

Let    us  consider  the expression     for   the expectation  value  of
$\Theta^{\mu\nu}$ given in (\ref{me2}). It is a matter of a few lines to
show that

\begin{equation}
\langle  \Theta^{\mu\nu}\rangle_{ r} = iN    \int    {d^D k\over
(2\pi)^D}k^{\mu}k^{\nu}\left[\G_{++}(k)-\G(k)\right] -   i
{N\over      4}g^{\mu\nu}  \delta  m^2    \int    {d^D k\over   (2\pi)^D}
\left[\G_{++}(k)+\G(k) \right] \, , \tselea{precal1}
\end{equation}

\noindent  with  $\G$  being the  $T=0$   causal  Green function.   Using
(\tseref{re6}),  (\tseref{re3}) and  (\tseref{re2})  one may directly check
that  Eq.(\tseref{precal1}) does not   contain UV singularities when  the
limit  $D\to 4$ is  taken.  This verifies  our introductory remark, that
(\tseref{ppp101})   is finite even in   non--equilibrium  context. From the
generalised   KMS condition  (\tseref{CMS2})   and  from (\tseref{re2}) we
obtain that $f_{++}$ is
\begin{eqnarray}
f_{++}(k)={1\over e^{\beta({\bf{k}})\omega_k}-1} \, .  \tseleq{precal2}  
\end{eqnarray}

\noindent  So  far the  results  obtained were  completely general and
valid for any translationally invariant system.  Let us now consider a
system  in  which    $g({\bf    k})= {V    \omega_k\over     (2\pi)^3}
\exp\left(-{\omega_k/\sigma}\right)$.  As we shall see, this condition
corresponds to  systems where the  lowest  frequency modes depart from
strict    equilibrium, whilst the  high    energy  ones obey  standard
Bose--Einstein   statistics.   This  behaviour   is  typical  of  many
off--equilibrium  systems,  eg.  ionised  atmosphere \cite{VPe}, laser
stimulated  plasma \cite{SIc},  hot   fusion \cite{mihai},  etc.   The
$\sigma$ is  usually referred to as the  ionisation half--width of the
energetical  spectrum.     From  (\ref{inicon2})    we  can  determine
$\beta({\bf k})$ as a function of the physical parameter $\sigma$: 

\vspace{-1mm}
\begin{eqnarray}
\beta({\bf                                                          k})=
{1\over\sigma}+{1\over\omega_k}\sum_{n=1}^{\infty}{(-1)^{n+1}\over    n}
e^{-n\omega_k/\sigma} \,.  \tseleq{precal2a1}
\end{eqnarray}
\vspace{0.05mm}

\begin{figure}[h]
\vspace{4mm} 
\epsfxsize=9cm 
\centerline{\epsffile{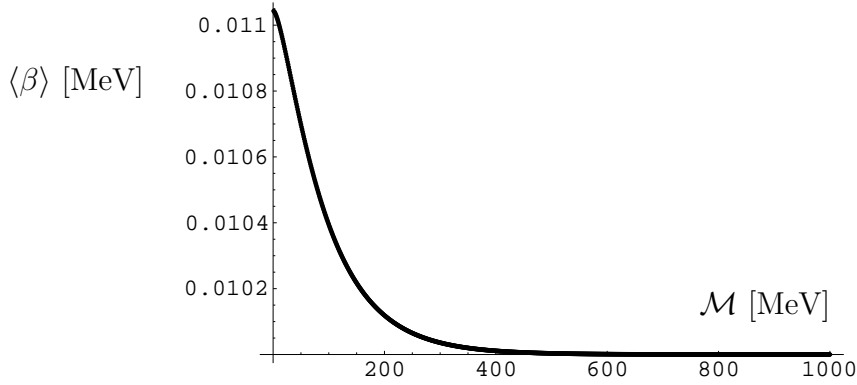}}
\caption{  A plot  of  $\langle  \beta\rangle$ vs.  ${\cali{M}}$   at
$\sigma = 100$ \mbox{MeV}.}
\label{fig2}
\begin{picture}(20,7)
\put(38,150){$\langle  \beta \rangle $ [MeV] } 
\put(300,65){${\cali{M}}$ [MeV]}
\end{picture}
\end{figure}

\noindent   To   proceed,   some  remarks    on  the   interpretation of
$\beta({\bf{k}})$   are   necessary.   Firstly,   Eq.(\tseref{precal2a1})
implies   that  for a sufficiently    large $\omega_{k}$ ($\omega_{k}\gg
\sigma$) the  function $\beta({\bf{k}})$ is  approximately constant, and
equals $1/\sigma$.    Thus  at high  energies  the distribution $f_{++}$
approaches  the  Bose--Einstein distribution with the  global temperature
$T\approx \sigma$.  In  other words, only the  soft modes were sensitive
to   processes which    created   the non--equilibrium  situation.    The
interpretation  of  $\sigma $  as  an  equilibrium temperature, however,
fails whenever $\sigma  \approx \omega_{k}$.  Instead  of $\sigma $  one
may alternatively work with the  expectation value of $\beta({\bf{k}})$,
i.e.

\begin{eqnarray}
\langle\beta\rangle   &=&  {\int    d^3{\bf    k}\,   \beta({\bf  k})e^{
-{\omega_{k}/\sigma}}\over     \int d^3{\bf k} \,e^{-{\omega_k/\sigma}}}
\nonumber\\    &=&   {1\over    \sigma}    +  {    {\sum}_{n=1}^{\infty}
{(-1)^{n+1}\over    n(n+1)}K_1({(n+1){\cali{M}}/\sigma})\over {\cali{M}}
K_2({{\cali{M}}/\sigma})} \, , \tseleq{precal2a}
\end{eqnarray}
\vspace{0.01mm}

\noindent where $K_{n}$ is the Bessel  function of imaginary argument of
order $n$.   Because  the system  is  for  the significant part   of the
energetical  spectrum in thermal  equilibrium, $1/\langle \beta \rangle$
approximates the    corresponding    equilibrium  temperature   to  high
precision.  An interesting  feature of (\tseref{precal2a})  is that it is
quite insensitive to the  actual  value of ${\cali{M}}$.  Dependence  of
$\langle  \beta \rangle$ on  ${\cali{M}}$ for a  sample  value $\sigma =
100$MeV is depicted in FIG.\ref{fig2}.  An important observation is  that the asymptotic behaviour of
$\langle     \beta      \rangle$    at  $\sigma\to\infty$    goes   like
$\langle\beta\rangle\approx    a/\sigma$,        where      $a=1+{1\over
2}{\sum}_{n=1}^{\infty}  {(-1)^{n+1}\over   n(n+1)^2}=   \mbox{ln}2    +
\pi^{2}/24 \approx 1.1$.

\vspace{15mm}

\noindent Using (\tseref{ppp101}) and (\tseref{precal1}) we  get for the pressure per
particle

\begin{equation}
{\cali{P}}(\sigma)  =  {1\over    2\pi^2}\left[   {\cali{M}}^2 \sigma^2  K_2({{\cali{M}}/
\sigma})        +            {\delta       m^2\over   4}{\cali{M}}\sigma
K_1({{\cali{M}}/\sigma})\right]- {\cali{P}}_0 \, , \tselea{precal2b}
\end{equation}
\noindent where

\begin{equation}
{\cali{P}}_0  ={1\over 384\pi^2}  (\delta   m^2)m_{r}^2\left(2+{\delta
m^2\over       m_{r}^2}\right)    +      {1\over
64\pi^2}m_{r}^4\left(1+{\delta           m^2\over        m_{r}^2}\right)
\ln\left(1+{\delta m^2\over m_{r}^2}\right) \, .  \tselea{precal2c}
\end{equation}

\noindent  Note that ${\cali{P}}_{0}$  comes from the UV divergent parts
of   (\tseref{precal1}).    Whilst the  separate    contributions  are UV
divergent, they  cancel  between themselves  leaving  behind the  finite
${\cali{P}}_{0}$.  As we already  emphasised, the divergences  come from
the  particular solutions of  the Kadanoff--Baym  equations.  Because the
former  do  not directly reflect the  boundary  conditions, the  form of
${\cali{P}}_{0}$ must  be identical  for  any translationally  invariant
medium.  The non--trivial  information about the non--equilibrium pressure
is then encoded in terms in the brace $[\ldots]$ in (\tseref{precal2b}).

Let us now perform the ``high--temperature'' expansion of the
pressure (\ref{precal2b}). As a ``temperature'' parameter we may chose
$\sigma$. In this case we have

\begin{eqnarray}
P(\sigma) &=&  {\sigma^4 \over \pi^2}  -  { \sigma^2 \over
2\pi^2}\left({\cali{M}}^{2} - {\delta m^{2}\over 4} \right)\nonumber \\
&-& \mbox{ln}\left( {\cali{M}}\over 2\sigma \right) \sum_{k=0}^{\infty}
C_{k}(2{\cali{M}}^{2} - \delta m^{2}\,(k+2))\nonumber \\
&-& \sum_{k=0}^{\infty} C_{k} \left\{ \delta m^{2}\left(\psi(k+1)+ {1\over
2(k+1)}\right)(k+2) \right.\nonumber \\
&-& \left.{\cali{M}}^2 \left(2 \psi(k+1) + {(2k +3)\over(k+1)(k+2)}\right)
\right\} - {\cali{P}}_0
\, ,\nonumber \\
\tseleq{precal2d}
\end{eqnarray}
\noindent  where
\begin{displaymath}
C_{k} =
\frac{{\cali{M}}^{2k+2}}{\pi^{2}2^{2k+4}k!(k+2)!\,\sigma^{2k}}\, .
\end{displaymath}

\noindent   The  $\psi(\ldots)$ is     Euler's psi  function.   For  a
sufficiently large $\sigma$  the leading  behaviour of $\delta  m^{2}$
may  be     easily evaluated.  To do  this,     let us  first assemble
(\tseref{re6})  and (\tseref{re33})--(\tseref{re5})  together.    This
gives us the (renormalised) gap equation 

\begin{eqnarray}
\delta      m^{2}    &=&         \frac{1}{2          \lambda_{0}}\left\{
\frac{\Gamma[1-\frac{D}{2}]}{(4\pi)^{\frac{D}{2}}}\left({\cali{M}}^{D-4}
- m_{r}^{D-4}\right) + N({\cali{M}}^{2})\right\}\nonumber \\
&=& \frac{ \lambda_{r}}{2}\left\{   {\tilde{\Sigma}}(m_{r}^{2},   \delta
m^{2})                 +                        \frac{1}{2\pi^{2}}\sigma
{\cali{M}}K_{1}[{\cali{M}}/\sigma]\right\}\, , \tseleq{precal2e}
\end{eqnarray}

\noindent with

\begin{displaymath}
\frac{1}{2}   {\tilde{\Sigma}}(\ldots) =   \frac{1}{32 \pi^{2}}  \left\{
(m_{r}^{2}+      \delta      m^{2})\mbox{ln}\left(    1+    \frac{\delta
m^{2}}{m_{r}^{2}}\right) - \delta m^{2}_{r} \right\}\, .
\end{displaymath}

\noindent Setting $x= \delta m^{2}/ m^{2}_{r}$ and $s  = \sigma / m_{r}$
we obtain the following transcendental equation for $x$

\begin{equation}
\lambda_{r}^{-1}  =  \frac{1}{32  \pi^{2}}  \left\{ \frac{1}{x} \left[
(1+x) \mbox{ln}(1+x) - x + 8  s \sqrt{1+x}\;K_{1}[\sqrt{1+x}/s]\;  
\right]\right\}\, .
\tselea{trans1}
\end{equation}
\vspace{1mm}

\begin{figure}[h]
\vspace{4mm} \epsfxsize=11cm \centerline{\epsffile{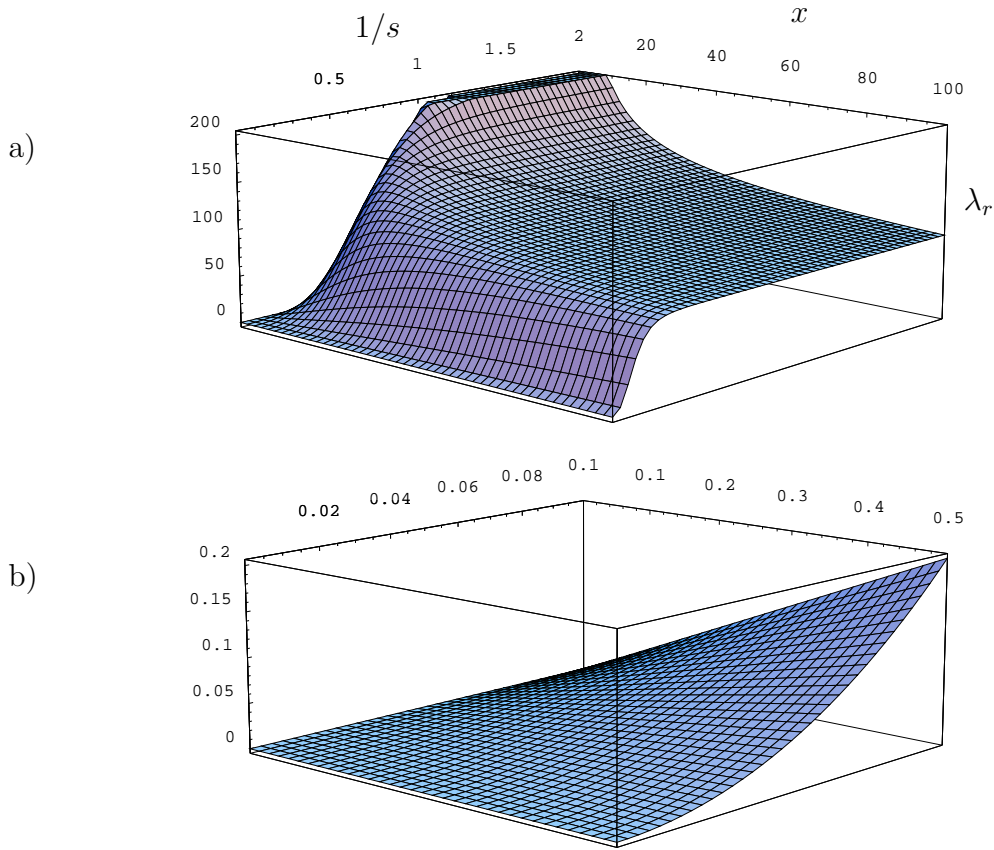}}
\caption{ A plot of the Eq.(\tseref{trans1}): a) the general shape, b)
a small $x$ behaviour.}
\label{fig3}
\setlength{\unitlength}{1mm}
\begin{picture}(20,7)
\put(6,112){a)}
\put(133,105){$\lambda_{r}$}
\put(110,130){$x$}
\put(52,128){$1/s$}
\put(6,55){b)}
\end{picture}
\end{figure}

\noindent   The   graphical representation of   Eq.(\tseref{trans1})   is
depicted in  FIG.\ref{fig3}.  

\vspace{8mm}

\vspace{3mm}

\noindent From this one may  read off that for large
$x$  ($x > 50$)  there exists  a critical value  of  $\lambda_{r}$ above
which the gap  equation has no solution.   (The plateau is actually bent
downward with a very gentle  slope.)  FIG.\ref{fig3}b clearly shows that
if $\lambda_{r} \ll 1/s   < 1$ then $x  \ll  1$.  Using   the asymptotic
behaviour of $K_{1}[z]$ for $z \rightarrow 0$ ($K_{1} \sim (z)^{-1}$) we
arrive at more  precise estimate of $\lambda_{r}$  for which  $x \ll 1$;
namely $\lambda_{r}\approx  1/s^{2}   = m_{r}^{2}/  \sigma^{2}$.    This
estimate is very helpful for the asymptotic expansion of $\delta m^{2}$.
For a sufficiently high $\sigma$ we may write

\begin{eqnarray}
\delta m^{2} &=& \lambda_{r}\left\{ \frac{\frac{(\delta
m^{2})^{2}}{2m^{2}_{r}} - \frac{(\delta m^{2})^{3}}{6
m^{4}_{r}}
+ \frac{(\delta m^{2})^{4}}{12m^{6}_{r}} + \ldots
}{32 \pi^{2}} +  \frac{1}{4\pi^{2}}(\sigma {\cali{M}}
\;K_{1}[{\cali{M}}/\sigma])|_{\sigma \rightarrow \infty} \right\}\nonumber \\
&&\nonumber \\
&=& \frac{\lambda_{r} \sigma^{2}}{4 \pi^{2}} +
{\cali{O}}({\cali{M}}
\mbox{ln}({\cali{M}}/\sigma))\, .
\tseleq{bb51} \end{eqnarray}

\noindent  Inserting this back into  (\tseref{precal2b})  we get

\begin{equation}
{\cali{P}}(\sigma)    =     \frac{\sigma^{4}}{\pi^{2}}    -  \frac{\sigma^{2}\,
{\cali{M}}^{2}}{2\pi^{2}}          +         \frac{\lambda_{r}}{8}\left(
\frac{\sigma^{2}{\mathcal{M}}^{2}}{64  \pi^{4}}  - \frac{\sigma^{4} 3}{4
\pi^{4}} \right) + {\mathcal{O}}\left(
{\mathcal{M}}^{2}{\mbox{ln}}({\mathcal{M}}/\sigma);      \lambda_{r}^{2}
\right)\, .  \tselea{bb6}
\end{equation}

\begin{figure}[h]
\vspace{4mm} \epsfxsize=9cm \centerline{\epsffile{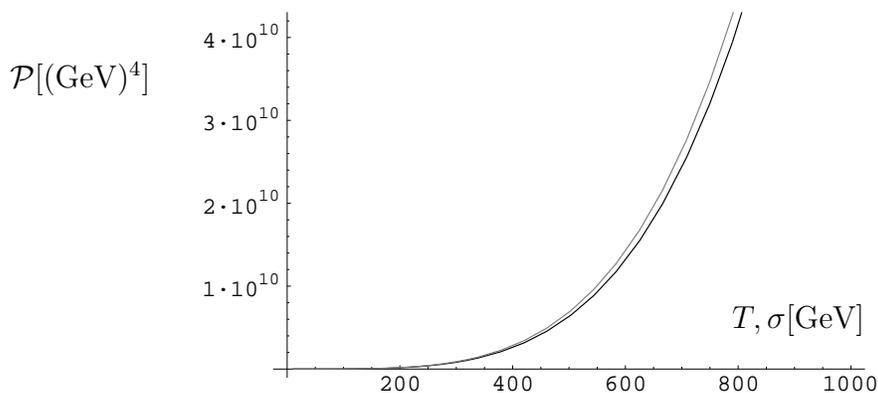}}
\vspace{3mm}
\caption{A plot  of pressure  as a function   of $T$,  $\sigma$  for
$m_{r}=  100\mbox{MeV}$.   The   gray line  corresponds   to equilibrium
pressure, the black line corresponds to pressure (\tseref{bb6}).}
\label{fig511}
\setlength{\unitlength}{1mm}
\begin{picture}(20,7)
\put(107,34){$T,\sigma \mbox{[GeV]}$}
\put(11,66){${\cali{P}}[(\mbox{GeV})^{4}]$}
\end{picture}
\end{figure}

\noindent    It is   interesting to   note   that $1/\pi^{2}   \approx
\pi^{2}/97$ which is almost the Stefan--Boltzmann constant.  This once
more   vindicates    our    interpretation of      $\sigma   $   as  a
``temperature''. A plot of the pressure as  a function of $\sigma $ is
depicted in  FIG.\ref{fig511}.   

\vspace{4mm}

\begin{figure}[h]
\vspace{4mm} \epsfxsize=9cm \centerline{\epsffile{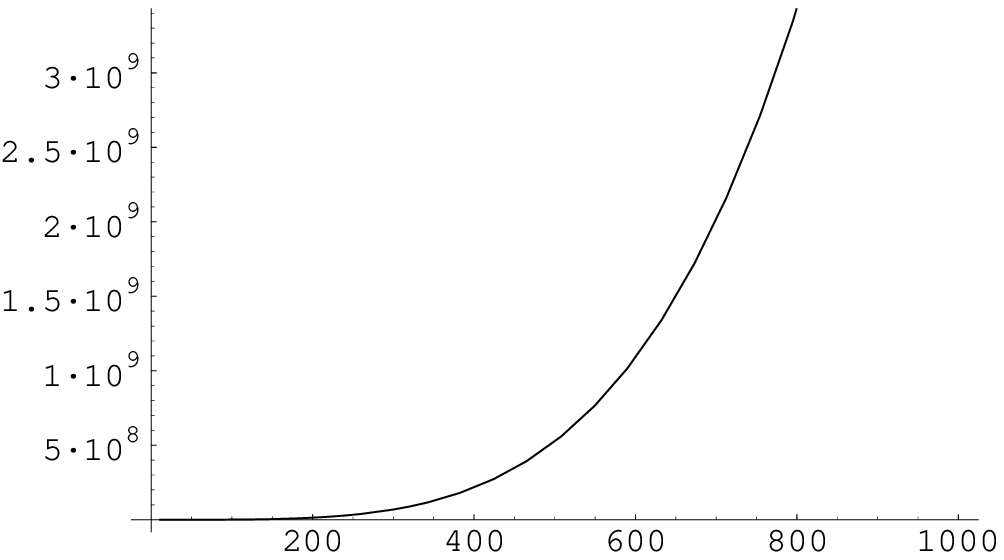}}
\vspace{3mm}
\caption{A  plot   showing   the  difference of   equilibrium    and
non--equilibrium pressures for $m_{r}=100\mbox{MeV}$.}
\label{fig55}
\setlength{\unitlength}{1mm}
\begin{picture}(20,7)
\put(107,32){$T,\sigma $[GeV]}
\put(0,61){$\Delta{\cali{P}}[(\mbox{GeV})^{4}]$}
\end{picture}
\end{figure}

\noindent It is due to   the low frequence  modes
contribution to  the  pressure that  $P(\sigma) < P(T)  $.   This is a
direct   result of our choice  of  $g({\bf{k}})$, namely that $\sigma$
cannot be interpreted   as  temperature for  the  low frequence  modes
(i.e. $\omega_{k} <  \sigma$).    The smaller  $\sigma$ is  the   less
important  contribution from non--equilibrium  soft  modes and so  the
smaller difference between both pressures. 

The  result  (\tseref{bb6}) can be  alternatively  rewritten in  terms of
 $\langle  \beta  \rangle$  .     Using,  for instance,   the   Pad\'{e}
 approximation \cite{bak} for $\langle \beta \rangle$, we arrive at

\begin{eqnarray}
{\cali{P}}(\langle \beta  \rangle )  &=& 0.0681122\,{\langle \beta  \rangle^{-4}}
-0.0415368\,{\langle \beta \rangle^{-2}}{\cali{M}}^{2}\nonumber   \\ &+&
\lambda_{r}\,\left(      -0.000647\,{\langle    \beta   \rangle^{-4}}  +
0.0000164\,{\langle          \beta      \rangle^{-2}}\,{\mathcal{M}}^{2}
\right)\nonumber             \\         &+&          {\mathcal{O}}\left(
{\mathcal{M}}^{2}{\mbox{ln}}({\mathcal{M}}\langle     \beta    \rangle);
\lambda_{r}^{2} \right)\, .
\end{eqnarray}

\noindent The coefficient $0.0681122 \approx \pi^{2}/145$ is $1.6$ times
smaller than  the required value  for the Stefan--Boltzmann  constant, so
the parameter $\langle \beta  \rangle$ is a slightly worse approximation
of the equilibrium temperature than  $\sigma$.  In practice, however, it
is a matter of taste and/or a particular context  which of the above two
descriptions is more useful.

\vspace{3mm}

\subsection{Off--equilibrium II \label{PN43}}

\noindent As was noted above, it is the  specific form of the constraint
$g({\bf{k}})$ which prescribes  the behaviour of $\beta({\bf{k}})$.  Let
us  now   turn our  attention to  systems  which depart  `slightly' from
equilibrium, i.e. when $g({\bf{k}})$ in (\tseref{ppp91}) deviates a little
from  the equilibrium  density of energy  per  mode.   In this  case the
constraint (\tseref{ppp91}) reads

\begin{equation}
g({\bf{k}}) = \frac{V}{(2 \pi)^{3}}\, 
\frac{\omega_{k}}{e^{\beta_{0}\omega_{k}}-1} + \delta g({\bf{k}}); \;\;
\delta g({\bf{k}}) \ll g({\bf{k}})\, ,
\tselea{cond1}
\end{equation}

\noindent with $\beta_{0}=1/T_{0}$ being an inverse of the equilibrium
temerature. As a special example of (\tseref{cond1}) we choose

\begin{equation}
g({\bf{k}})      =                 \frac{V}{(2               \pi)^{3}}\,
\frac{\omega_{k}}{e^{\beta_{0}\omega_{k}}\alpha^{-1}({\bf{k}})-1};\;\;\; 
\alpha({\bf{k}}, \beta_{0}) \approx 1\, .
\tselea{cond2}
\end{equation}

\noindent The inverse   mode ``temperature'' $\beta({\bf{k}})$ is   then
$\beta_{0}-          \mbox{ln}(\alpha({\bf{k}}))/\omega_{k}$.         So
$\mbox{ln}(\alpha)$ measures (in units of $\omega_{k}$) the deviation of
the mode temperature from the equilibrium  one.  The particular value of
$\alpha({\bf{k}}, \beta_{0})$  depends on the  actual  way in  which the
system is  prepared.  To  avoid unnecessary technical  complications, we
select  $\alpha ({\bf{k}},\beta_{0})$  to  be a  momentum--space constant
(generalisation is, however,  straightforward).  This choice  represents
the change in the  mode temperature which  is now inversely proportional
to $\omega_{k}$; the deviation is  bigger for soft  modes and is rapidly
suppressed for higher    modes.  Obviously, $T_{0}$ becomes the   global
temperature if  $\omega_{k} \gg \mbox{ln}(\alpha)$.  The generalised KMS
conditions      (\tseref{CMS2})   together  with  solutions (\tseref{re2})
determine $f_{++}$ as
\begin{eqnarray}
f_{++}(k)    ={\alpha  \over   {e^{\beta_0\omega_k}-\alpha}}    \,     .
\tseleq{precal3}
\end{eqnarray}

\noindent Eq.(\tseref{precal3})  is   a reminiscent of   the, so  called,
J\"uttner   distribution\footnote{It   should  be   mentioned  that this
similarity is rather superficial.  The J\"{u}tner distribution describes
systems which are in thermal but not chemical equilibrium. (As we do not
have a chemical potential, chemical equilibrium is ill defined concept.)
The fugacity then parametrises  the deviation from chemical equilibrium.
} with   fugacity   $\alpha$  \cite{bairetal,strickland}.   Now,  using
(\tseref{ppp101}) and (\tseref{precal1})   we  get  for  the pressure   per
particle

\begin{equation}
{\cali{P}}(T_{0})    +   {\cal  P}_0  =    {1\over   2\pi^2}\left[  {{\cali{M}}^2\over
\beta_0^2}\sum_{n=1}^{\infty}       {\alpha^{n}\over     n^2}      K_2(n
{\cali{M}}\beta_0) + {{\cali{M}}\delta m^2
\over    4\beta_0}\sum_{n=1}^{\infty}      {\alpha^{n}\over    n}  K_1(n
{\cali{M}}\beta_0)\right] \, , \tselea{precal4}
\end{equation}

\noindent where $\delta m^2$ satisfies the gap equation

\begin{equation}
\delta     m^2  ={\lambda_r\over 2}\left(  {\tilde{\Sigma}}(m^{2}_{r},
\delta  m^{2}) +   {1\over    2\pi^2}\int_0^{\infty}{k^2dk\over\omega_k}
{\alpha \over  {e^{\beta_0\omega_k}-\alpha}}\right) \, .
\tselea{precal5}
\end{equation}

\noindent If we  set,  as  before,  $x=\delta m^{2}/m_{r}^{2}$ and   $s=
1/\beta_{0}m_{r}$ we get the transcendental equation for $x$

\begin{equation}
\lambda^{-1}_{r} =  \frac{1}{32\pi^{2} x}\left\{ (1+x)\mbox{ln}(1+x) -x
+ 8(1+x)\; \int^{\infty}_{1}\,   dz \,
\sqrt{z^{2}-1}  \, \frac{\alpha}{e^{z\,\sqrt{x+1}/s}- \alpha}
\right\}\, .
\tselea{precal54}
\end{equation}

\begin{figure}[h]
\vspace{4mm} \epsfxsize=11cm \centerline{\epsffile{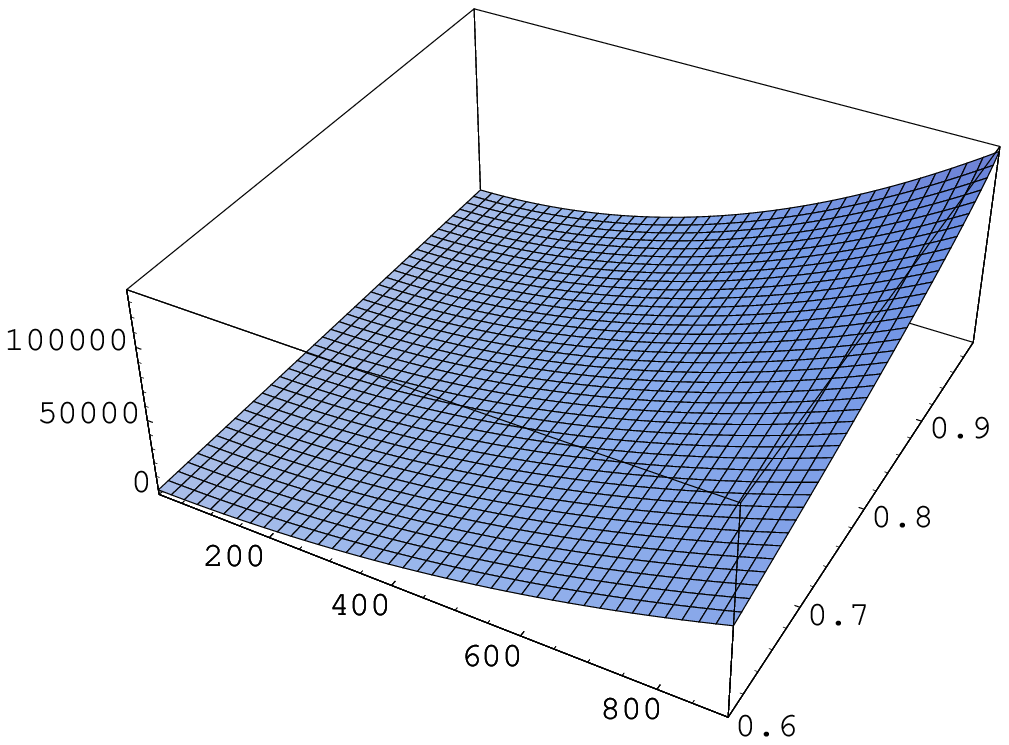}}
\vspace{2mm}
\caption{Behaviour of the pressure  (\tseref{precal4}) as a  function
of $\alpha$ and $T_{0}$ at $m_r=100 \mbox{MeV}$.}
\label{fig76}
\setlength{\unitlength}{1mm}
\begin{picture}(20,7)
\put(80,88){\small{${\cali{P}} [(\mbox{GeV})^4]$}}    
\put(130,33){$\alpha$}
\put(55,24.5){\small{$T_0 [\mbox{GeV}]$}}
\end{picture}
\end{figure}

\noindent  The corresponding numerical analysis of (\tseref{precal54})
reveals  that   for $x \ll  1$ ,   $1/s \ll  1$.   So   at $x  \ll 1$,
$\lambda_{r}  \approx  \delta   m^{2}/T_{0}^{2}$.   This  estimate  is
important for the asymptotic expansion of $\delta m^{2}$.  However, in
order to carry out the asymptotic expansion of (\tseref{precal5}) (and
consequently  (\tseref{precal4})) we need to   cope first with the sum
$\sum_{n=1}^{\infty}
\alpha^{n}K_{k}(n{\cali{M}}\beta_{0})/n^{k}\,\,\,(k=1,2)$ (also called
Braden's function).  Expansion    of $K_{k}(\ldots)$ yields  a  double
series which is  very slowly convergent, and so  it does not allow one
to easily grasp the leading  behaviour in $T_{0}$.  In this case it is
useful to  resume  (\tseref{precal4})--(\tseref{precal5})  by virtue of
the  Mellin   summation  technique \cite{LW}.     (It   is well  known
\cite{LW,PJ,WH} that at equilibrium this  resummation leads to a rapid
convergence for  high temperatures.) As  a result,  for a sufficiently
large $T_{0}$ we get 

\begin{eqnarray}
\delta  m^2   &=&  \frac{\lambda_{r}\,T^2_{0}}{24}  -{\lambda_rT_0 {\cal
M}\over     4\pi}   \left[{1\over 2}(1-r^2)^{1/2}  \right.   \nonumber\\
&-&\left.                 r\left(1-\ln\left({{\cal               M}\over
2T_0}\right)\right)-(1-r^2)^{1/2}\arcsin    (r)\right] \nonumber\\   &+&
{\cal O}(\ln T_0) \, , \tseleq{precal6}
\end{eqnarray}

\noindent where we have  set $r= \mbox{ln}(\alpha)T_{0}/{\cali{M}}$. The
corresponding expansion of the pressure (\tseref{precal4}) reads
\begin{eqnarray}
{\cali{P}}(T_{0})  &= & {\pi^2  T_0^4\over 90} +{{\cal M}\zeta  (3) \over \pi^2} T_0^3 r
-{T_0^2 {\cal  M}^2 \over 24}\left(  1-2 r^{2}\right)  - {{\cal M}^3 T_0
\over  4\pi^2}   \nonumber\\  &\times &  \left[-  {1  \over  3} \left(1-
r^{2}\right)^{\frac{3}{2}}+          {r\over           2}\left(1-{2\over
3}r^2\right)\left(1   -    2\ln\left({{\cal            M}\over         2
T_0}\right)\right)\right.   \nonumber\\   &-   &   {2\over    9}\left(1-
r^{2}\right)\bigg(-r^3+3 - \left. 3 \left(1-  r^{2}\right)^{\frac{1}{2}}
\bigg)\arcsin    (r)\right]   \nonumber\\ &+&  {(\delta   m^2)^2   \over
2\lambda_r}+ {\cal O}(\ln T_0) \, .  \tseleq{precal6a}
\end{eqnarray}

\noindent  where $\zeta(3)\approx 1.202$. Note that  for $\alpha = 1$ we
regain the equilibrium expansion (\tseref{b91}). The corresponding plot of
${\cali{P}}$  as  a     function  of  $T_{0}$   and   $\alpha$ is    depicted  in
FIG.\ref{fig76}.

\noindent In    passing    it may be    mentioned    that the  expansion
 (\ref{precal6a}) is mathematically justifiable only for $\alpha \approx
 e^{\sqrt{\lambda_{r}/24}\, r} \approx 1$.

\vspace{5mm}

\section{Conclusions \label{PN5}}

\noindent
In   order  to  get a   workable   recipe for  non--equilibrium  pressure
Jaynes--Gibbs   principle  of  maximal   entropy,   the   Dyson--Schwinger
equations,  and   the hydrostatic pressure.   The  basic  steps   are as
follows.

To find quantitative results for pressure one needs to know the explicit
form of  the Green's functions involved. These  may be find  if we solve
the Dyson--Schwinger  equations.  The  corresponding solutions are unique
provided  we specify the  density   matrix $\rho$ (the construction   of
$\rho$ is one of  the cornerstones of our approach,  and we tackled this
problem   using the Jaynes--Gibbs  principle  of  maximal entropy).  With
$\rho$  at  our disposal  we   showed how to  formulate  the generalised
Kubo--Martin--Schwinger (KMS) boundary conditions.

To show how the   outlined method works  we  have illustrated  the whole
procedure on an exactly  solvable model, namely $O(N)\; \Phi^{4}$ theory
in the large--$N$ limit.   This model is  sufficiently simple yet complex
enough  to  serve as  an illustration  of   basic characteristics of the
presented method in contrast to other ones in use.  In order to find the
constraint  conditions we  have  considered two  gedanken experiments in
which  the system in question  was prepared in such  a way that only low
frequence modes  departed from the  strict  equilibrium behaviour.  Such
processes can   be found, for example, in   ionised atmosphere, in laser
stimulated   plasma  or in   hot  fusion.    In both alluded
gedanken experiments we were  able to work  out the hydrostatic pressure
exactly.    Furthermore,  after identifying  a ``temperature'' parameter
(virtually temperature of  high modes) we  carried out the corresponding
high--temperature expansions.

As it  was  discussed, one  of  the main advantages  of the Jaynes--Gibbs
construction is that one  starts  with the (physical) constraints  (i.e.
parameters    which     are    really   controlled     and measured   in
experiments). These constraints   directly determine the  density matrix
and  hence   the  generalised KMS   conditions for   the Dyson--Schwinger
equations.  This contrasts with the  usual approaches where the  density
matrix is treated as the primary object.  In these cases it is necessary
to solve the von Neumann--Liouville equation.  This is usually circumvent
using  either  variational methods  \cite{EJY,FJ}  with several  trial
$\rho$'s or reformulating the problem in  terms of the quantum transport
equations  for Wigner's functions  \cite{Bal}.   It is, however,  well
known that  the inclusion of constraints into  transport equations  is a
very delicate  and rather complicated task (the  same  is basically true
about the variational methods) \cite{Bal,Nem}.

\chapter{Summary and outlooks}

In this dissertation we have explored various aspects of equilibrium
and non--equilibrium quantum field theory.

In  Chapter  \ref{dcoset}  we dealt  with  certain  aspects of infrared
effects in finite--temperature QFT. It is well known \cite{LB,LW} that
these effects are far more subtle at finite temperature than at $T=0$.
The  latter    is    ascribed   to the    fact that     there   is  no
finite--temperature Bloch--Nordsieck  mechanism \cite{LB2,IZ}   at our
disposal.  Intuitively, however, one might  expect that if thermal QFT
is   a  well  posed   theory,   no  infrared    divergences should  be
present.  This point was extensively   studied on particles that decay
and  scatter inside   a plasma (or   heat bath).   As  a byproduct  we
proposed an easy mathematical formalism  allowing one to calculate the
plasma particle number    spectrum formed when a   particle(s)  decays
(scatter) within  the plasma.   This  formalism which  is based on the
largest--time equation and the Dyson--Schwinger equations, is embodied
in a one--line modification to the corresponding thermal cut diagrams.
Such diagrams  arise  naturally when the  imaginary   part of  a Green
function   is   computed, and the    whole  calculations are immensely
simplified  by the  observation that   the thermal  cut diagrams   are
virtually the  same as  the Kobes--Semenoff  diagrams in  the  Keldysh
formalism. 

The vital part  of Chapter \ref{dcoset} was dedicated  to the study of
the modifications which must be made in various quantum field contexts
(scalar fields, spinor fields and  gauge fields in temporal gauge). It
was found that the modified propagators  have an easy interpretation
in  terms of emission,   absorption   and fluctuation of the    plasma
particles. To  demonstrate how the  method works, we used various heat
baths  (photon, and electron--photon)   in which two  uncharged scalar
particles scatter into a pair of charged particles.  We calculated the
leading  $\omega$  behaviour for the resulting   changes in the plasma
particles number spectra. We also found that the energy density $\omega
dN/d\omega$ is finite as $\omega$ tends to zero.

In Chapter \ref{large-N1} we have   turned our attention to  pressure.
Pressure is undoubtedly one of the important parameters characterising
quantum  media at finite  temperature. An  extension  of the  pressure
calculations    to   off--equilibrium   systems should   enhance   our
predicative ability  in  such areas as  (realistic) phase transitions,
early universe cosmology  or hot fusion dynamics.   Unfortunately, the
standard approaches based on the  partition function do not allow such
an extension  because there is  no such  thing as the grand--canonical
potential  away  from equilibrium.     We considered    an alternative
approach based on the, so  called, hydrostatic pressure. As a  warming
up exercise we started with  thermal equilibrium.  The whole procedure
was  then illustrated on a toy  model ($\lambda  \Phi^{4}$ theory with
$O(N)$ internal symmetry in the large--$N$ limit). The result obtained
matches    that  found   in    \cite{ID1}     for  the   thermodynamic
pressure. (Whilst these two pressures  agree in equilibrium , there is
$no$ thermodynamic pressure out of equilibrium.)

We took advantage to probe various mathematical techniques
(composite operators renormalisation, Dyson--Schwinger diagrammatic
equations, finite temperature coupling renormalisation, Mellin transform)
which are indispensable in calculations of the hydrostatic pressure (most
of them also being applicable to non--equilibrium as well).  One of the 
key results obtained was the prescription for the renormalised
pressure (Eqs.(\tseref{ppp3}), (\tseref{ppp91}) and (\tseref{ppp101})). 
This was achieved by means of the Zimmerman forest formula.

The model in question has the undeniable merit of being exactly
solvable.  In particular, one could find the pressure by means of a
non--perturbative treatment. This is because of the fact that the
large--$N$ limit eliminates `nasty' classes of diagrams in the
Dyson--Schwinger expansions.  The surviving class of diagrams (superdaisy
diagrams) could be exactly resumed because the (thermal)  proper
self--energy ${\bf{\Sigma}}$ as well as the renormalised coupling constant
$\lambda_{r}$ were momentum independent.  The resulting expression for the
pressure obtained was then in a suitable form which allowed us to take
advantage of the Mellin transform technique, and we were able to evaluate 
the pressure in $D=4$ (both for massive and massless fields)  to all   
orders in the high--temperature expansion.

As we have already mentioned, there are various motivations to be
interested in the non--equilibrium pressure. One of them is based on the
belief that the pressure (as an easily measurable parameter) should   
exhibit a discontinuity in its derivative(s) when a local phase transition
occurs.  This has been observed in solid state physics and fluid dynamics
and it may play a central role in, for example, detection of the
quark--gluon plasma or in various baryogenesis scenarios.

In  Chapter \ref{PN} we  approached  the  problem  of  pressure in   a
non--equilibrium  quantum  plasma by   extending   the notion  of  the
hydrostatic pressure  to   systems    out of equilibrium.      On  the
equilibrium   level   the     hydrostatic  pressure;   i.e.    $p    =
-\frac{1}{3}\langle   \Theta^{i}_{\;\;i}  \rangle$ (here  $\Theta^{\mu
\nu}$  is the energy--momentum tensor of  a system in question), has a
well defined  microscopic picture,  which can  be carried  over  to an
off--equilibrium as we have showed in  Chapter \ref{large-N1}.  Unlike
in equilibrium, the non--equilibrium expectation value is sensitive to
the  particular choice  of  the  energy--momentum tensor  (see Chapter
\ref{PE}), though  fortunately in translationally invariant media this
sensitivity  is not relevant.    Our strategy for the  calculation  of
pressure  was simple.  To find  quantitative results  for pressure one
needs   to know the explicit form   of the Green's functions involved.
These may be   found if we  solve  the  corresponding Dyson--Schwinger
equations.  The   solution is unique provided  we  specify the density
matrix $\rho$, and hence the  generalised KMS conditions.  In order to
construct the density matrix we invoke  the Jaynes--Gibbs principle of
maximal   entropy.   The   Jaynes--Gibbs  principle is   basically the
Bayesian inference  about   the most probable density   matrix $\rho$.
This is  based on the maximalisation of   the Shannon (or information)
entropy, subject to the prior knowledge which one has about the system
(usually specified at some initial time $t_{i}$).

To show how  the outlined method  works we have illustrated the  whole
method on our favoured   $O(N)\;\Phi^{4}$   theory in the  large--$N$
limit. In   order  to specify the  initial--time  constraints  we have
considered two gedanken experiments   in which the system in  question
was prepared in such a way that only low frequency modes departed from
the strict equilibrium behaviour.   Such  processes can be found,  for
example, in  ionised atmosphere, in laser stimulated  plasma or in hot
fusion heated up by ultra--sound waves. Furthermore, after identifying
a ``temperature parameter''  (virtually the  temperature of  the  high
modes)   we carried     out  the    corresponding    high--temperature
expansions.  The leading high--temperature  coefficients coincided  to a
good approximation   with  the Stefan--Boltzmann  constant  which   is
precisely what one would  expect: at high ``temperature'' particles do
no feel   non--equilibrium   ``background''  and  consequently   their
distribution approaches the equilibrium  one.

The next logical step  in our investigation  would be to calculate the
pressure  in a     non--equilibrium,   non--translationally  invariant
medium. In  contrast to the previous  situation, one is confronted now
with   two conceptual  difficulties.    As we  have   shown in Section
\ref{PN42}, in order  to  get the renormalised   pressure one  has  to
consistently    use   the   renormalised    energy--momentum    tensor
($\Theta^{\mu \nu}$ is a composite operator).  Unfortunately there are
infinitely  many renormalised energy--momentum tensors (generated from
each other via Pauli's  transformation), and these give rise to
different pressures.  The choice of the ``correct'' $\Theta^{\mu \nu}$
is then crucial  for  the next investigation.    (As we  discussed  in
Sections \ref{PE2} and \ref{PN4},  in translationally  invariant media
pressure does not depend on  a particular form of $\Theta^{\mu \nu}$.)
Our future strategy   is  based on  the observation  that  because the
pressure  is  an  observable,  one  should  obviously   work  with  an
energy--momentum  tensor  which is  an   observable  as  well (or  the
corresponding  Hermitian operator). We propose   to use the Belinfante
energy--momentum tensor.  The fact  that the physical significance  of
the Belinfante tensor   (both in  classical  and  quantum systems)  is
greater than that  of other energy--momentum  tensors was  pointed out
successively by Belinfante \cite{BF},   Rosenfeld \cite{R} and  Jackiw
\cite{RJ}.    In  scalar QFT   the   Belinfante energy-momentum tensor
coincides  with the,   so  called, improved   energy--momentum  tensor
\cite{RJ}. The generic   form of the improved  energy--momentum tensor
for  the  $O(N)\; \Phi^{4}$ theory  was  derived  in Section \ref{PE2}
Eq.(\tseref{ppp6}). Because   the corresponding $\beta$   function  is
exactly solvable  in the--large $N$ limit  \cite{AS}  we  expect that
constant  $c(\lambda_{r};D)$   is    exactly   solvable  as well,   and
consequently the improved  energy--momentum tensor  can be written  in
closed form. 

The second problem  to be solved  is connected with  the fact that the
corresponding Dyson--Schwinger equations are far more involved than in
the  translationally invariant case.   Even,  for the $O(N)\;\Phi^{4}$
theory  in the large--$N$  limit the  Kadanoff--Baym are  not any more
hyperbolic  equations.    This  fact,  among others,   complicates the
renormalisation   program    (one cannot    use   the  momentum  space
renormalisation).  For   this    purpose  we intend to resort   to 
differential renormalisation  \cite{Lat}.     Solving the renormalised
Kadanoff--Baym equations  (at least numerically)  is the next  step in
our study.

\appendix
\chapter{Finite--temperature Dyson--Schwinger equations \label{A}}
\section{Functional formalism \label{A1}}

Eq.(\tseref{S-D2}) gives  us   an  alternative definition   of  Wick's
theorem in terms of the ``functional derivation" $\frac{\delta}{\delta
\psi(x)}$.  We  refer to   Eq.(\tseref{S-D2}) as the  Dyson--Schwinger
equation because  the classical  $T=0$ Dyson--Schwinger equations  are
implied by  it  (see Appendix  \ref{A12}). Let  us  first  show  that
(\tseref{S-D2})  is       consistent        with  Wick's       theorem
(\tseref{wick})--(\tseref{wick2}). To be  specific, let us consider an
ensemble of      non--interacting      particles  in   thermodynamical
equilibrium. In order to keep the work  transparent, we shall suppress
all   the   internal indices. There  is    no difficulty whatsoever in
reintroducing the necessary details.   Let  us first realize that  for
any (well behaved) functional  the following Taylor's expansion  holds
\cite{PR} 

\begin{equation} X[\psi] = \sum_{n} \int dx_{1} \ldots \int dx_{n}
\alpha^{n}(x_{1}\ldots x_{n}) \psi(x_{1})\ldots \psi(x_{n})\, , \tseleq{Hin2}
\vspace{1mm} \end{equation}

\noindent The same is true if $\psi$ is an operator instead. In the latter
case the $\alpha^{n}(\ldots)$ are not generally symmetric in the $x$'s
{\footnote{If $X = X[\psi, \partial\psi]$, the $\alpha^{n}$ may also
contain derivations working on the various fields.}}. When Fermi fields
are involved, we might, for the sake of compactness, include in the
argument of $\psi$ the space-time coordinate, the Dirac index, and a
discrete index which distinguishes $\psi_{\alpha}$ from
${\overline{\psi}}_{\alpha}$. In the latter case $\int dx \rightarrow \sum
\int dx$, where summation runs over the discrete indices. With this
convention, the expansion (\tseref{Hin2}) holds even for the Fermi fields.
An extension of (\tseref{Hin2}) to the case where different fields are
present is natural. Particularly important is the case when $\psi$ is a   
field in the interaction picture, using Wick's theorem and decomposition
(\tseref{Hin2}) one can then write
\begin{eqnarray}
\lefteqn{\langle G[\psi] \psi(x) F[\psi] \rangle =}\nonumber\\
&=&\sum_{m,n} \left(\int dx \right)^{n}
\left( \int dy \right)^{m} \alpha^{n}(x_{1} \ldots x_{n}) \beta^{m}(y_{1}
\ldots  y_{m})
\left\langle \left(\prod_{k}^{n}\psi(x_{k})\right) \psi(x)
\prod_{k'}^{m}\psi(y_{k'}) \right\rangle\nonumber\\
&=& \sum_{n} \left(\int dx \right)^{n} \alpha^{n}(x_{1}
\ldots x_{n})  \sum_{l}^{n} (\pm 1)^{n-l} \langle
\psi(x_{l}) \psi(x) \rangle \left\langle \prod_{k \not= l}^{n}
\psi(x_{k})
F[\psi] \right\rangle  \nonumber\\
&+&  \sum_{m} \left(\int dy
\right)^{m} \beta^{m}(y_{1} \ldots y_{m}) \sum_{l}^{m} (\pm 1)^{l-1} \langle
\psi(x) \psi(x_{l})\rangle \left\langle
G[\psi] \prod_{k' \not= l}^{m} \psi(y_{k'} )
\right\rangle\, .
\tselea{D-S3}
\end{eqnarray}

\noindent with $\left( \int dx \right)^{n}=\int dx_{1} \ldots \int dx_{n}$.
The `$-$' stands for fermions and `$+$' for bosons. On the other hand, using
the formal prescriptions (\tseref{var1}) and
(\tseref{var2}) for  $\frac{{\stackrel{\rightarrow}{\delta}}}{\delta
\psi(x)}$ one can read

\begin{eqnarray}
\lefteqn{\int dz \langle \psi(x) \psi(z) \rangle \left\langle G[\psi]
\frac{{\stackrel{\rightarrow}{\delta}} F[\psi]}{\delta \psi (z)}
\right\rangle =}\nonumber\\
&=&\mbox{\hspace{5mm}} \sum_{m} \left( \int dy \right)^{m} \beta^{m}(y_{1}
\ldots y_{m})
\int dz \langle \psi(x) \psi(z) \rangle \sum_{l}^{m} (\pm 1)^{l-1}
\delta(z-y_{l}) \left\langle G[\psi] \prod_{k' \not= l}^{m} \psi(y_{k'})
\right\rangle\nonumber\\
&=&\mbox{\hspace{5mm}} \sum_{m} \left( \int dy \right)^{m} \beta^{m}(y_{1}
\ldots y_{m}) \sum_{l}^{m} (\pm 1)^{l-1}\langle \psi(x) \psi(y_{l})
\rangle \left\langle G[\psi]
\prod_{k' \not= l}^{m} \psi(y_{k'}) \right\rangle \, .
\end{eqnarray}

\noindent Similar expression holds for $\int dz \langle \psi(x) \psi(z)
\rangle \left\langle \frac{ G[\psi]
{\stackrel{\leftarrow}{\delta}}}{\delta \psi (z)} F[\psi]
\right\rangle$. Putting latter two together we get precisely
(\tseref{D-S3}). This confirms the validity of (\tseref{S-D2}). It is easy
to persuade oneself that exactly the same sort of arguments leads to

\begin{eqnarray}
&&\langle \psi(x) F[\psi] \rangle = \int dz \langle \psi(x) \psi(z)
\rangle \left\langle \frac{{\stackrel{\rightarrow}{\delta}}
F[\psi]}{\delta \psi(z)} \right\rangle\\
\tselea{S-D3}   
&&\langle {\cali{T}}(\psi(x) F[\psi]) \rangle =\int dz \langle
{\cali{T}}(\psi(x) \psi(z)) \rangle \left\langle {\cali{T}}
\left( \frac{{\stackrel{\rightarrow}{\delta}} F[\psi]}{\delta \psi(z)}
\right) \right\rangle\\ \tselea{S-D4}
&&\langle G[\psi] {\cali{T}}( \psi(x) F[\psi]) \rangle = \int dz
\langle {\cali{T}}(\psi(x) \psi(z)) \rangle \left\langle
G[\psi] {\cali{T}} \left( \frac{{\stackrel{\rightarrow}{\delta}}
F[\psi]}{\delta \psi(z)} \right) \right\rangle +  \nonumber\\
&& \mbox{\hspace{3.7cm}} + \int dz \langle \psi(z) \psi(x)
\rangle
\left\langle \frac{ G[\psi] {\stackrel{\leftarrow}{\delta}}}{\delta
\psi(z)} {\cali{T}}(F[\psi]) \right\rangle\, ,\\
&&\mbox{etc.}\nonumber
\tselea{S-D6}
\end{eqnarray}

\noindent with ${\cali{T}}$ being either the chronological or
anti--chronological time ordering symbol. At this stage it is important to 
realize that from the definition of
$\frac{{\stackrel{\rightarrow}{\delta}}}{\delta \psi(x)}$ directly follows
that $[\frac{{\stackrel{\rightarrow}{\delta}}}{\delta \psi(x)};
\frac{{\stackrel{\rightarrow}{\delta}}}{\delta \psi(y)}]_{\mp}=0$ ( `$-$'  
holds for bosons and `$+$' for fermions). Indeed,

\vspace{-2mm}   
\begin{eqnarray}
\frac{{\stackrel{\rightarrow}{\delta^{2}}}F[\psi]}{\delta \psi(x) \delta
\psi(y)} &=& \sum_{n=2} \sum_{i < j} \left(\int dx \right)^{n-2}
(\alpha^{n}(x_{1}
\ldots \stackrel{x_{i}}{\stackrel{\downarrow}{x}} \ldots
\stackrel{x_{j}}{\stackrel{\downarrow}{y}} \ldots x_{n})  \pm \nonumber\\
&\pm &~\alpha^{n}(x_{1} \ldots
\stackrel{x_{i}}{\stackrel{\downarrow}{y}} \ldots
\stackrel{x_{j}}{\stackrel{\downarrow}{x}} \ldots
x_{n}))(\pm1)^{i+j} \prod_{m \not=
i,j}^{n} \psi(x_{m}) = \mp
\frac{{\stackrel{\rightarrow}{\delta^{2}}}F[\psi]}{\delta
\psi(y) \delta \psi(x)}\, .\nonumber\\
\tselea{commut5}
\end{eqnarray}

\noindent Similarly $[\frac{{\stackrel{\leftarrow}{\delta}}}{\delta
\psi(x)};  \frac{{\stackrel{\leftarrow}{\delta}}}{\delta
\psi(y)}]_{\mp}=0$. Analogously we might prove

\begin{equation}
\frac{F[\psi]{\stackrel{\leftarrow}{\delta^{2}}}}{\delta \psi(x)\delta     
\psi(y)}=
\frac{{\stackrel{\rightarrow}{\delta^{2}}}F[\psi]}{\delta \psi(x)\delta
\psi(y)}\, , \end{equation}

\noindent and

\begin{eqnarray} 
\frac{{\stackrel{\rightarrow}{\delta^{2}}}(F[\psi]G[\psi])}{\delta
\psi(x)\delta
\psi(y)} &=& \frac{F[\psi]{\stackrel{\leftarrow}{\delta^{2}}}}{\delta
\psi(x)\delta \psi(y)}G[\psi] + (-1)^{p}
\frac{F[\psi]{\stackrel{\leftarrow}{\delta}}}{\delta \psi(x)}
\frac{{\stackrel{\rightarrow}{\delta}}G[\psi]}{\delta \psi(y)}\nonumber\\
&+& \frac{F[\psi]{\stackrel{\leftarrow}{\delta}}}{\delta \psi(y)}
\frac{{\stackrel{\rightarrow}{\delta}}G[\psi]}{\delta \psi(x)}
+F[\psi]\frac{{\stackrel{\rightarrow}{\delta^{2}}}G[\psi]}{\delta
\psi(x)\delta \psi(y)}\, .
\end{eqnarray}

\noindent The ``$p$'' is $0$ for bosons and $1$ for fermions. With 
(\tseref{S-D2}) and (A.4)--(A.6) one can easily construct more
complicated expectation values. For example, using (\tseref{S-D2}) and
(A.4) we get
\begin{eqnarray}
\lefteqn{\langle \psi(x) \psi(y) F[\psi] \rangle =}\nonumber\\
&=& \int
\frac{dz_{1}dz_{2}}{2} (\langle \psi(x) \psi (z_{1}) \rangle
\langle \psi(y) \psi (z_{2})  \rangle + (-1)^{p}\langle \psi(x)
\psi(z_{2}) \rangle \langle \psi(y) \psi(z_{1})
\rangle)\nonumber\\
&& \times \; \left\langle \frac{{\stackrel{\rightarrow}{\delta^{2}}}
F[\psi]}{\delta
\psi(z_{1}) \delta \psi(z_{2})} \right\rangle + \langle  \psi(x) \psi(y) \rangle \langle F[\psi]\rangle \,
.
\tselea{S-D7}
\end{eqnarray}

\noindent Similarly, using (\tseref{S-D2}) and (anti--)commutativity of the
arrowed $\frac{\delta}{\delta \psi(x)}$, we get
\begin{eqnarray}
\lefteqn{\langle G[\psi]  \psi(x) \psi(y) F[\psi] \rangle
=}\nonumber\\
&=& \int \frac{dz_{1} dz_{2}}{2} (\langle \psi(x)
\psi(z_{1})\rangle \langle \psi(y) \psi(z_{2})\rangle +
(-1)^{p}\langle \psi(x)
\psi(z_{2}) \rangle \langle \psi(y) \psi(z_{1})
\rangle)\nonumber\\
&&\times ~ \left\langle G[\psi]
\frac{{\stackrel{\rightarrow}{\delta^{2}}}F[\psi]}{\delta
\psi(z_{1}) \delta \psi(z_{2})} \right\rangle \nonumber\\
&+& \int \frac{dz_{1} dz_{2}}{2} (\langle \psi(z_{1})
\psi(x)\rangle \langle \psi(z_{2}) \psi(y)\rangle +
(-1)^{p}\langle \psi(z_{2})
\psi(x) \rangle \langle \psi(z_{1}) \psi(y)
\rangle)\nonumber\\
&& \times ~\left\langle
\frac{G[\psi]~{\stackrel{\leftarrow}{\delta^{2}}}}{\delta
\psi(z_{1}) \delta \psi(z_{2})} F[\psi]\right\rangle\nonumber\\
&+& \int dz_{1} dz_{2} (\langle \psi(z_{1})
\psi(x)\rangle \langle \psi(y) \psi(z_{2})\rangle +
(-1)^{p} \langle \psi(x) \psi(z_{2})
\rangle\langle \psi(z_{1})
\psi(y) \rangle)\nonumber\\
&& \times ~ \left\langle
\frac{G[\psi]{\stackrel{\leftarrow}{\delta}}}{\delta
\psi(z_{1})} \frac{
{\stackrel{\rightarrow}{\delta}}F[\psi]}{\delta \psi(z_{2})}
\right\rangle + \langle \psi(x) \psi(y) \rangle \langle G[\psi] F[\psi]
\rangle \, .
\tselea{S-D8}
\end{eqnarray}
\vspace{0mm}

\noindent We could proceed further having still higher powers of fields
and variations. However, there is a quite interesting generalisation in
case when we have (anti--)time ordered operators. Let us have $F[\psi]=
{\cali{T}}(F[\psi])$, in this case

\begin{eqnarray}
\langle F[\psi] \rangle &=& \sum_{n} \left(\int dx\right)^{n}
\alpha^{n}(\ldots) \langle {\cali{T}} (\prod_{i=1}^{n} \psi(x_{i})
\rangle \nonumber\\
&=& \sum_{n=1} \left( \int dx \right)^{n}  \frac{\alpha^{n} (\ldots)}{n}
\sum_{i,j} \varepsilon_{P}\langle {\cali{T}}(\psi(x_{i}) \psi(x_{j}))
\rangle
\langle {\cali{T}}(\prod_{m \not= i,j}^{n} \psi(x_{m})) \rangle
+ \alpha^{0}(\ldots)\nonumber\\   
&=& \int dz_{1}dz_{2} \langle {\cali{T}}(\psi(z_{1})
\psi(z_{2}))\rangle \left\langle
\frac{{\stackrel{\rightarrow}{\delta^{2}}}
{\overline{F}}[\psi]}{\delta \psi(z_{2}) \delta \psi(z_{1})}
\right\rangle + \langle F[0]\rangle \, ,
\tselea{S-D9}
\end{eqnarray}
\vspace{1mm}

\noindent where ${\overline{F}}[\psi]$ differs from $F[\psi]$ in the
replacement $\alpha^{n}(\ldots) \rightarrow \frac{\alpha^{n}(\ldots)}{n}$
($n$ starts from $1$ !). In comparison with (A.4)--(A.11), the
$\alpha^{0}(\ldots)$ (i.e. the pure $T=0$ contribution) does matter here.
Note that $\alpha^{0}(\ldots)$ generally involves non--heat--bath fields 
with corresponding space-time integrations.  Similar extension is true if
$F[\psi] = {\cali{T}}_{C}(F[\psi])$, where ${\cali{T}}_{C}$ is the time
path ordering symbol. In that case

\begin{eqnarray}
\langle F[\psi] \rangle
&=& \sum_{n} \left( \int_{C} dx \right)^{n}
\alpha^{n}(\ldots) \langle {\cali{T}}_{C} (\prod_{p=1}^{n} \psi(x_{p})) 
\rangle \nonumber\\
&=& \int_{C} dz_{1}dz_{2} \langle {\cali{T}}_{C}(\psi(z_{1})
\psi(z_{2}))\rangle \left\langle
\frac{{\stackrel{\rightarrow}{\delta^{2}}}
{\overline{F}}[\psi]}{\delta \psi(z_{2}) \delta \psi(z_{1})}
\right\rangle \nonumber\\
&+& \langle F[0] \rangle \, ,
\tselea{S-D10}
\end{eqnarray}

\vspace{2mm}

\noindent with\footnote{A contour  $\delta$-function $\delta_{C}(x-y)$
is  defined as $\int_{C}  dz \delta_{C}(z-z^{'}) f(z) = f(z^{'})$, see
{\cite{Mills, NS2}}.} $\int_{C} dx = \int_{C} dt \int_{V} d{\vect{x}}$
and  $\frac{\delta \psi(x)}{\delta \psi(y)}= \delta_{C}(x-y)$ . Wick's
theorem for the  ${\cali{T}}_{C}$--oriented product  of fields has  an
obvious form 

\begin{equation} \langle {\cali{T}}_{C} ( \psi (x_{1}) \ldots \psi
(x_{2n}) )
{\rangle} = \sum_{\stackrel{j}{j \not= i}} \varepsilon_{P} \langle
{\cali{T}}_{C} ( \psi (x_{i})\psi (x_{j}) ) {\rangle} ~\langle
{\cali{T}}_{C} ( \prod_{k \not= i;j}\psi(x_{k})) {\rangle}\, .
\tseleq{wick6} \end{equation}

\noindent This can be directly derived from Wick's theorem (\tseref{wick2}),
realizing that

\begin{equation}
{\cali{T}}_{C}(\psi(x_{1}) \ldots \psi(x_{m})) = \sum_{P} \varepsilon_{P}
\theta_{C}(t_{P_{1}}, \ldots, t_{P_{m}}) \psi(x_{P_{1}}) \ldots
\psi(x_{P_{m}})\, ,
\end{equation}

\noindent where $P$ refers to the permutation of the indices and
$\theta_{C}(t_{1},\ldots , t_{m})$  being a contour step function
\cite{EM} defined as

\begin{equation}
\theta_{C}(t_{1}, \ldots, t_{m})= \left\{ \begin{array}{ll}
                                    1&\mbox{($t_{1}, \ldots, t_{m}$
\noindent are ${\cali{T}}_{C}$-oriented along $C$)} \\
                                    0&\mbox{(otherwise)}
\end{array}
\right.
\end{equation}

\vspace{2mm}

\noindent So for example (\tseref{S-D4}) may be written as

\begin{equation}
\langle {\cali{T}}_{C}(\psi(x)F[\psi])\rangle = \int_{C}dz\,
\langle {\cali{T}}_{C}(\psi(x)\psi(z))\rangle \left\langle
{\cali{T}}_{C}\left(
\frac{\stackrel{\rightarrow}{\delta}F[\psi]}{\delta
\psi(z)}\right)\right\rangle \,.
\tseleq{S-D234}
\end{equation}
\vspace{0.5mm}

\noindent Particularly important is the Keldysh--Schwinger path \cite{LB,
KL, EM} (see FIG.\ref{fig18}). In the latter case
\vspace{1mm}
\begin{eqnarray}
\langle F[\psi] \rangle &=& \int_{C_{1}} dz_{1} dz_{2}
\langle{\cali{T}}(\psi(z_{1}) \psi(z_{2})) \rangle \left\langle
\frac{{\stackrel{\rightarrow}{\delta^{2}}} {\overline{F}}[\psi]}{\delta
\psi(z_{2}) \delta \psi(z_{1})} \right\rangle \nonumber\\
&+& \int_{C_{2}} dz_{1} dz_{2}
\langle{\overline{\cali{T}}}(\psi(z_{1}) \psi(z_{2})) \rangle
\left\langle
\frac{{\stackrel{\rightarrow}{\delta^{2}}} {\overline{F}}[\psi]}{\delta
\psi(z_{2}) \delta \psi(z_{1})} \right\rangle \nonumber\\
&+& 2 \int_{C_{2}} dz_{1} \int_{C_{1}} dz_{2}
\langle  \psi(z_{1}) \psi(z_{2}) \rangle
\left\langle
\frac{{\stackrel{\rightarrow}{\delta^{2}}} {\overline{F}}[\psi]}{\delta
\psi(z_{2}) \delta \psi(z_{1})} \right\rangle \nonumber\\
&+& \langle F[0] \rangle \, .
\tselea{S-D11} 
\end{eqnarray}

\noindent Application to the product $G[\psi]F[\psi]$ with $F[\psi] =
{\cali{T}}_{C_{1}}(F[\psi])$ and $G[\psi] = {\cali{T}}_{C_{2}}(G[\psi])$ is
straightforward and reads

\begin{eqnarray} \langle G[\psi] F[\psi] \rangle &=&\int
dz_{1}dz_{2} \langle {\overline{\cali{T}}}(\psi(z_{1})
\psi(z_{2}))\rangle \left\langle ~
{\overline{\frac{G[\psi]{\stackrel{\leftarrow}{\delta^{2}}}}{\delta
\psi(z_{2})
\delta \psi(z_{1})}F[\psi]}}~ \right\rangle\nonumber\\ &+&\int 
dz_{1}dz_{2} \langle {\cali{T}}(\psi(z_{1}) \psi(z_{2}))\rangle
\left\langle ~
{\overline{G[\psi]\frac{{\stackrel{\rightarrow}{\delta^{2}}}F[\psi]}{\delta
\psi(z_{2}) \delta \psi(z_{1})}}} ~\right\rangle \nonumber\\ &+& 2
\int dz_{1}dz_{2} \langle \psi(z_{1}) \psi(z_{2}) \rangle
\left\langle ~
{\overline{\frac{G[\psi]{\stackrel{\leftarrow}{\delta}}}{\delta
\psi(z_{1})}
\frac{{\stackrel{\rightarrow}{\delta}} F[\psi]}{\delta \psi(z_{2})}}} ~
\right\rangle\nonumber\\
&+& \langle G[0] F[0] \rangle \, , \tselea{S-D13}
\end{eqnarray}

\noindent where the overlining indicates that we work with
$\frac{\alpha^{n}(\ldots) \beta^{m}(\ldots)}{n+m}$ instead of
$\alpha^{n}(\ldots) \beta^{m}(\ldots)$, we have also abbreviated
$\int_{C_{1}} dz$ to $\int dz \int_{C_{1}} dz$. We should also emphasise   
that $\frac{\delta \psi(x)}{\delta \psi(y)}$ used in (\tseref{S-D13}) is
$\delta (x-y)$ rather than $\delta_{C}(x-y)$.

In Eq.(\tseref{average2}) it has been used the inverted version of
(\tseref{S-D13}), namely

\begin{eqnarray} \langle (G[\psi] F[\psi])^{'} \rangle &=&\int 
\frac{dz_{1}dz_{2}}{2} \langle {\overline{\cali{T}}}(\psi(z_{1})
\psi(z_{2}))\rangle\left\langle
{\frac{G[\psi]{\stackrel{\leftarrow}{\delta^{2}}}}{\delta
\psi(z_{2}) \delta \psi(z_{1})}F[\psi]} \right\rangle\nonumber\\
&+&\int \frac{dz_{1}dz_{2}}{2} \langle {\cali{T}}(\psi(z_{1})
\psi(z_{2}))\rangle \left\langle
G[\psi]\frac{{\stackrel{\rightarrow}{\delta^{2}}}F[\psi]}{\delta
\psi(z_{2}) \delta
\psi(z_{1})}\right\rangle\nonumber\\ &+& \int dz_{1}dz_{2}
\langle \psi(z_{1}) \psi(z_{2} \rangle \left\langle {\frac{
G[\psi]{\stackrel{\leftarrow}{\delta}}}{\delta \psi(z_{1})}
\frac{{\stackrel{\rightarrow}{\delta}} F[\psi]}{\delta
\psi(z_{2})}} \right\rangle \, , \tselea{S-D12}
\end{eqnarray}
\vspace{0mm}

\begin{figure}[h]
\vspace{3mm}
\epsfxsize=15cm
\centerline{\epsffile{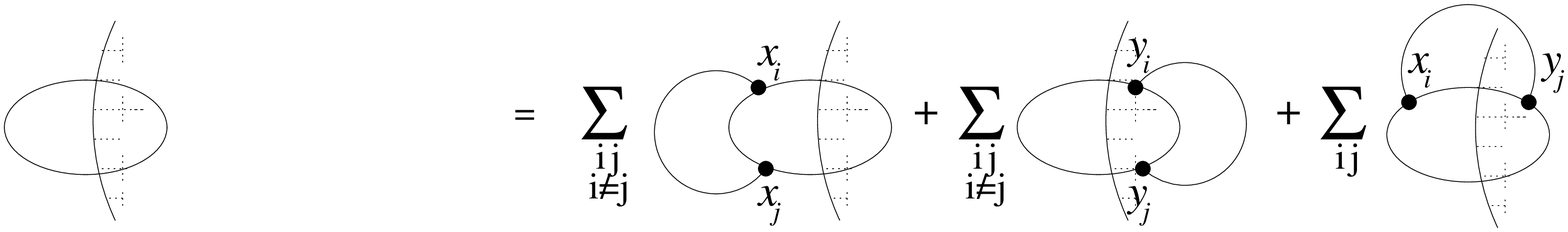}}
\caption{{ Diagrammatic equivalent of Eq.(\tseref{S-D12}). The cut
 separates areas constructed out of $F[\psi]$ and
$G[\psi]$.}} \label{fig19}
\setlength{\unitlength}{1mm}
\begin{picture}(10,7)
\put(22,31.5){ $\times$ \footnotesize{(number of lines)} }
\end{picture}
\end{figure}

\noindent Here $(G[\psi]F[\psi])^{'}$ has the coefficients
$\alpha^{n}(\ldots)\beta^{m}(\ldots)\frac{(n+m)}{2}$ instead of
$\alpha^{n}(\ldots)\beta^{m}(\ldots)$. Note, that the
$\alpha^{0}(\ldots)\beta^{0}(\ldots)$ does not contributes and thus we do
not have any pure $T=0$ contributions. Eq.(\tseref{S-D12}) has a natural
interpretation. Whilst the LHS tells us, that from each thermal diagram
(constructed out of $\langle G[\psi] F[\psi] \rangle$) with
$\frac{n+m}{2}$ internal heat--bath particle lines we must take $n+m$
identical copies, the RHS says, that this is virtually because we sum over
all possible distributions of one heat-bath particle line inside of the
given diagram. The pictorial expression of (\tseref{S-D12}) is depicted in
FIG.\ref{fig19} .

\vspace{3mm}

\section{Graphical formalism \label{A11}}
\begin{center}
{\bf (This section is partially based on refs.\cite{PC,JZ}.)}
\end{center}
In Chapter \ref{PE} we heavily used a graphical representation of the
Dyson--Schwinger   equations.  In this section   we  provide a  short
derivation of  such a representation and in  the section  to  follow we
perform a comparison with the functional Dyson--Schwinger equations of
Section \ref{dcoset} and Appendix \ref{A1}.

The     Dyson--Schwinger          equations      were       originally
constructed\cite{JSc1,FJDa}  with   the   motivation that  they  could
provide some information about  the complete Green's functions outside
the scope of perturbative theory.  The basic philosophy is to directly
use  the equations of  motion in  order  to construct the hierarchy of
(integral) equations  for  full Green's functions.  The  usefulness of
these equations   is  usually   confined   to the  cases   where  some
approximation (truncation) is  available  in order to bring  them into
manageable   form.    In  this appendix    we   aim   to   derive  the
finite--temperature    Dyson--Schwinger  equations   using  the   more
intuitive path--integral formalism.  The alternative derivation in the
non--equilibrium  context (based  purely  on operatorial  approach) is
discussed in Section \ref{PN11}.

Let us start with the ($T \not= 0$) generating functional of Green's functions
$Z[J]$ (the partition functions with an external source $J$)
\vspace{1mm}
\begin{eqnarray}
&&Z[T] = \int{\cali{D}}\phi\;
\mbox{exp}\{i(S[\phi;T] + \int_{C}d^{4}{x}\,J(x)\phi(x))\}\nonumber \\ &&S[\phi;T] = \int_{C}d^{4}x\;
{\cali{L}}(x)\,.
\tselea{A1111}
\end{eqnarray}

\noindent          Here             $\int_{C}d^{4}x                  =
\int_{C}dx_{0}\int_{V}d^{3}{\vect{x}}$     with  the  subscript    $C$
indicating  that the time runs along  some contour  in the complex
plane.  In the  real--time formalism, which  we adopt  throughout, the
most  natural   version  is   the  so called    Keldysh--Schwinger one
\cite{LW,LB},   which     is      represented    by the  contour    in
FIG.\ref{fig18}. Within the  path--integral the $c$--number fields are
further restricted by the periodic  boundary condition (KMS condition)
\cite{LW,LB,TA} which for bosonic fields reads: 
\begin{equation}
\phi(t_{i}-i\beta, {\vect{x}})= \phi(t_{i}, {\vect{x}})\, .
\tselea{A111}
\end{equation}

\noindent The zero temperature  generating functional may be recovered
from (\tseref{A1111}) if we   integrate  over the close--time  path  (no
vertical parts) and omit the  KMS condition (\tseref{A111}).  Now, the
LHS  of (\tseref{A1111}) is independent  of $\phi$, thus particularly it
is invariant under infinitesimal point transformation 

\begin{equation}
\phi(x) \rightarrow \phi(x) + \varepsilon f(x)\, ,
\tseleq{A13}
\end{equation}

\noindent where $f$ is an arbitrary ($\phi$ independent) function
which fulfils the periodic boundary conditions 
\begin{equation}
f(t_{i}-i\beta, {\vect{x}})= f(t_{i}, {\vect{x}})\, .
\tseleq{A131}
\end{equation}

\noindent  From  (\tseref{A13})   is  obvious  that  the  corresponding
functional   Jacobian     is    $1$,   i.e.,    ${\cali{D}}\phi     =
{\cali{D}}\phi'$. This immediately implies that

\begin{eqnarray}
Z[J] &=& \int {\cali{D}}\phi' \, \mbox{exp}\left\{ i(S[\phi' - \varepsilon f]
+    \int_{C}d^{4}x\, J(x)\phi'(x)   - \varepsilon\int_{C}d^{4}x\,    J(x)f(x))\right\}
\nonumber \\
&=&  -i\varepsilon   \int{\cali{D}}\phi'      \left\{   \int_{C}d^{4}x
\left(\frac{\delta S}{\delta \phi'}(x) + J(x)\right)f(x)\right\}
\mbox{exp}(iS[\phi'] + i\int_{C}J\phi') + {\cali{O}}(\varepsilon^{2})
\nonumber\\
&+& Z[J]\, , \nonumber \\
\Rightarrow 0 &=& \int {\cali{D}}\phi \left\{ \int_{C}d^{4}x \left( \frac{\delta
S}{\delta \phi}(x) +  J(x)\right) f(x)\right\}\, \mbox{exp}(iS[\phi] +
i \int_{C}J\phi)\, . 
\tselea{A14}
\end{eqnarray}

\noindent As (\tseref{A14})    is  true for  any  $f(\ldots)$  fulfilling  the
condition (\tseref{A131}) one may write

\begin{eqnarray}
0  &=&   \int_{C} d^{4}x   \left\langle  \frac{\delta  S[\psi_{H}]}{\delta
\phi}(x) + J(x)\right\rangle f(x)\, ,\nonumber \\
0 &=& \left\langle \frac{\delta  S[\psi_{H}]}{\delta
\phi}(x)      +  J(x)\right\rangle    =  \left(   \frac{\delta
S}{\delta  \phi}\left[  \frac{\delta}{i\delta   J(x)}\right]    +
J(x)\right)Z[J]\, . 
\tselea{A15}
\end{eqnarray}

\noindent Employing the commutation relation: $-i \frac{\delta}{\delta
J}\, Z = Z (\phi - i \frac{\delta}{\delta J})$, we may recast
(\tseref{A15}) into more appropriate form, namely

\begin{equation}
-J(x) = \frac{\delta S}{\delta \phi}\left[ \phi(x) - i
\frac{\delta}{\delta J(x)}\right] \1 = \frac{\delta S}{\delta \phi}
\left[ \phi(x) + i \int_{C}d^{4}z\, \D^{c}(x,z)\frac{\delta}{\delta
\phi(z)}\right]\1\, .
\tseleq{A144}
\end{equation}

\noindent So, for instance, for $\lambda\, \Phi^{4}$ theory we have

\begin{eqnarray}
-J_{\alpha}(x) &=& -(\Box + m_{0}^{2})\phi_{\alpha}(x) - \frac{\lambda_{0}}{3!}
\left\{ (\phi_{\alpha}(x))^{3} + i3 \phi_{\alpha}(x)\,\D_{\alpha
\alpha}^{c}(x,x)\right.\nonumber \\
 &&- \left.\int d^{4}y\, d^{4}w\, d^{4}z\; \D^{c}_{\alpha
\beta}(x,y)\,\D^{c}_{\alpha \gamma}(x,w)\,\Gamma^{(3)\; \beta \gamma
\delta}(y,w,z)\,\D^{c}_{\delta \alpha}(z,x)\right\}\nonumber \\
&=& -(\Box + m_{0}^{2})\phi_{\alpha}(x) - \frac{\lambda_{0}}{3!}\left\{
(\phi_{\alpha}(x))^{3} + i3 \phi_{\alpha}(x)\,\D^{c}_{\alpha
\alpha}(x,x) -
\D^{c\;(3)}_{\alpha \alpha \alpha}(x,x,x)\right\}\nonumber \\
&=& - ( \Box + m_{0}^{2} )\phi_{\alpha}(x) +
\frac{\lambda_{0}}{3!}\,\D_{\alpha \alpha \alpha}^{(3)}(x,x,x)\, ,
\tselea{A19}
\end{eqnarray}
\vspace{0mm}

\noindent(no sum over $\alpha$!)\footnote{In (\tseref{A19}) we have
implicitly used the identities;

\begin{eqnarray*}
&& \frac{1}{Z}\frac{\delta^{n}Z}{\delta J(x_{1}) \ldots \delta
J(x_{n})} = (-1)^{n+1}i\,\D^{(n)}(x_{1},\ldots ,x_{n})\nonumber \\
&&\frac{\delta^{n}W}{\delta J(x_{1}) \ldots \delta J(x_{n})} =
(-1)^{n+1}\,\D^{c\;(n)}(x_{1}, \ldots , x_{n})\\
&& \frac{\delta \phi(x)}{\delta J(y)} = - \D^{c}(x,y)\, ; \;\;
\frac{\delta^{n}\Gamma}{\delta \phi(x_{1}) \ldots \phi(x_{n})} =
\Gamma^{(n)}(x_{1}\ldots x_{n})\\
&& \frac{\delta \,\D^{c}_{\alpha \beta}(x,y)}{\delta \phi_{\gamma}(z)} = -
\int dy^{4}_{1}\, dy^{4}_{2}\, \D^{c}_{\alpha
\delta}(x,y_{1})\Gamma^{(3)\;\delta \kappa \gamma}(y_{1}, y_{2}, z)\,
\D^{c}_{\kappa \beta}(y_{2},y)\, .
\end{eqnarray*} 
} or, using identity; $(\Box_{x} + m_{0}^{2})\,\D_{F}^{\alpha
\beta}(x,y) =  -(\sigma_{3})^{\alpha \beta}\delta(x-y)$, we may invert
the differential operator and write equivalently 

\begin{equation}
-\int d^{4}y \, J_{\beta}(y)\,\D_{F}^{\beta \alpha}(y,x) =
\phi^{\alpha}(x) +  \frac{\lambda_{0}}{3!} \int d^{4}y \,
\D_{F}^{\alpha \beta}(x,y)\,\D^{(3)}_{\beta \beta \beta}(y,y,y).
\tseleq{A161}
\end{equation}
\vspace{0mm}

\noindent Eq.(\tseref{A161}) has the following graphical representation

\vspace{4mm}
\begin{figure}[h]
\vspace{3mm}
\epsfxsize=14cm
\centerline{\epsffile{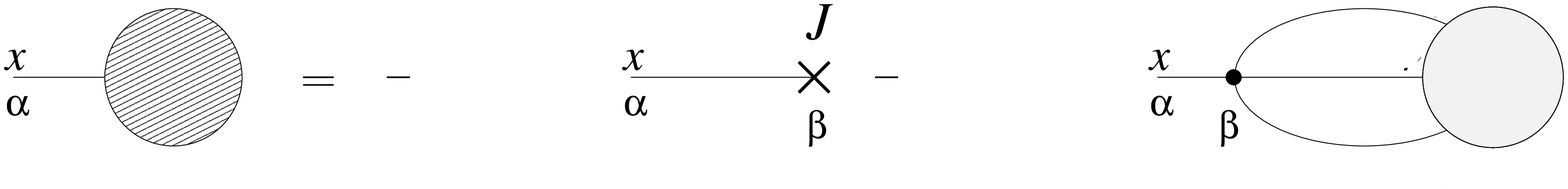}}
\label{figD-S2}
\setlength{\unitlength}{1mm}
\begin{picture}(10,10)
\put(50,18){$\sum_{\beta = 1,2}$}
\put(95,18){$\frac{1}{6}\sum_{\beta = 1,2}$}
\end{picture} 
\end{figure}
\noindent   where the  hatched  blob  refers   to the (full)  1--point
amputated  Green's function,  the  dotted blob  refers  to  the (full)
3--point amputated Green's function, the cross denotes the source $J$,
while the heavy dot without coordinate indicates  that the vertex must
be  integrated  over  all   possible positions.     The  corresponding
graphical representation in terms of  the connected Green's  functions
reads (c.f. Eq.(\tseref{A19}))

\begin{figure}[h]
\vspace{3mm}
\epsfxsize=14cm
\centerline{\epsffile{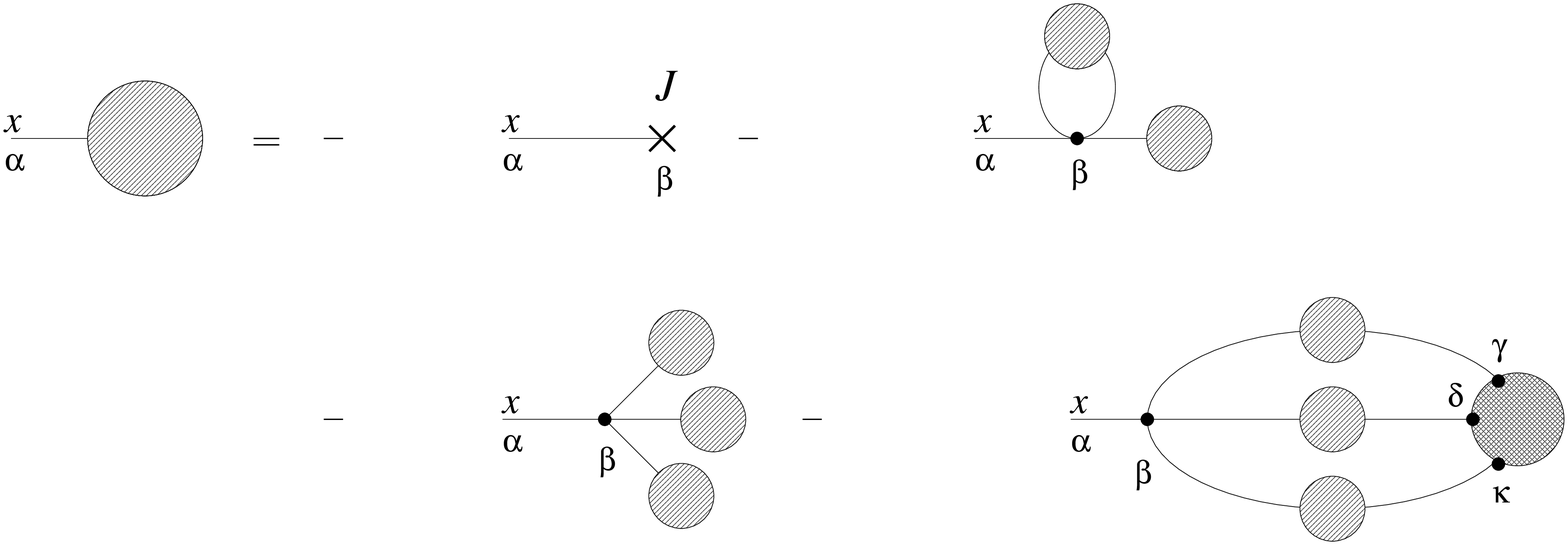}}
\label{figD=S3}
\setlength{\unitlength}{1mm}
\begin{picture}(10,10)
\put(44,45){$\sum_{\beta}$}
\put(83,45){$\frac{i}{2}\sum_{\beta}$}
\put(44,20.5){$\frac{1}{6}\sum_{\beta}$}
\put(86,20.5){$\frac{1}{6}\sum_{\beta, \gamma, \delta, \kappa}$}
\end{picture} 
\vspace{3mm}
\end{figure} 
\noindent Here the  double--hatched blob describe  the 3--point vertex
function and  the hatched  blobs refer,  as before, to  the  connected
(truncated) Green's functions. 

To see how to obtain the corresponding Dyson--Schwinger equations for the
2--point   (connected) Green's    functions,    let  us perform     in
Eq.(\tseref{A144}) (or, for simplicity's sake, in Eq.(\tseref{A19})) 
variation w.r.t.  $J_{\beta}(y)$. This directly gives us 

\vspace{3mm}
\begin{displaymath}
(\Box_{x} + m_{0}^{2})\, \D_{\alpha \beta}^{c}(x,y) -
\frac{\lambda_{0}}{3!}\,\D^{(4)}_{\alpha \alpha \alpha \beta }(x,x,x,y)
+ \frac{\lambda_{0} i}{3!}\,\phi_{\beta}(y)\,\D^{(3)}_{\alpha \alpha \alpha}(x,x,x)= -(\sigma_{3})_{\alpha \beta}\delta(x-y)\, ,
\end{displaymath}
\vspace{0mm}

\noindent or, inverting the differential operator, we may equivalently
write

\begin{eqnarray}
\D_{\alpha \beta}^{c}(x,y) = \D_{F\; \alpha \beta}(x,y) &-& \frac{\lambda_{0}}{3!}\, \int
d^{4}z\, \D_{F\; \alpha}^{\;\, \gamma}(x,z)\, \D^{(4)}_{\gamma \gamma
\gamma \beta}(z,z,z,y)\nonumber \\
&+& \frac{\lambda_{0}i}{3!}\, \int d^{4}z \, \D_{F\; \alpha}^{\;\, \gamma}(x,z)\, \D^{(3)}_{\gamma \gamma
\gamma}(z,z,z) \,\phi_{\beta}(y)
\, .
\tselea{A346}
\end{eqnarray}
\vspace{0.05mm}

\noindent Eq.(\tseref{A346}) is graphically represented as follows

\vspace{8mm}
\begin{figure}[h]
\vspace{3mm}
\epsfxsize=14cm
\centerline{\epsffile{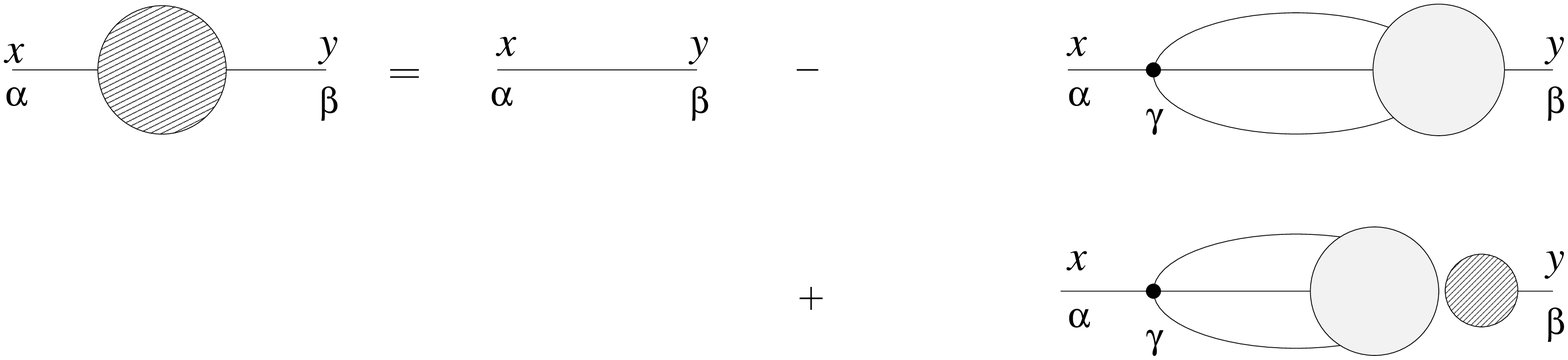}}
\label{D-S4}
\setlength{\unitlength}{1mm}
\begin{picture}(10,10)
\put(87,36){$\frac{1}{6}\sum_{\gamma =1,2}$}
\put(87,15){$\frac{i}{6}\sum_{\gamma =1,2}$}
\end{picture} 
\end{figure}

\noindent This may be reformulated purely in terms of the connected Green's
functions and the vertex functions, indeed, performing variation
w.r.t. $J_{\beta}(y)$ in first two lines in Eq.(\tseref{A19}), and
inverting the differential operator as before we obtain

\begin{figure}[h]
\vspace{3mm}
\epsfxsize=14cm
\centerline{\epsffile{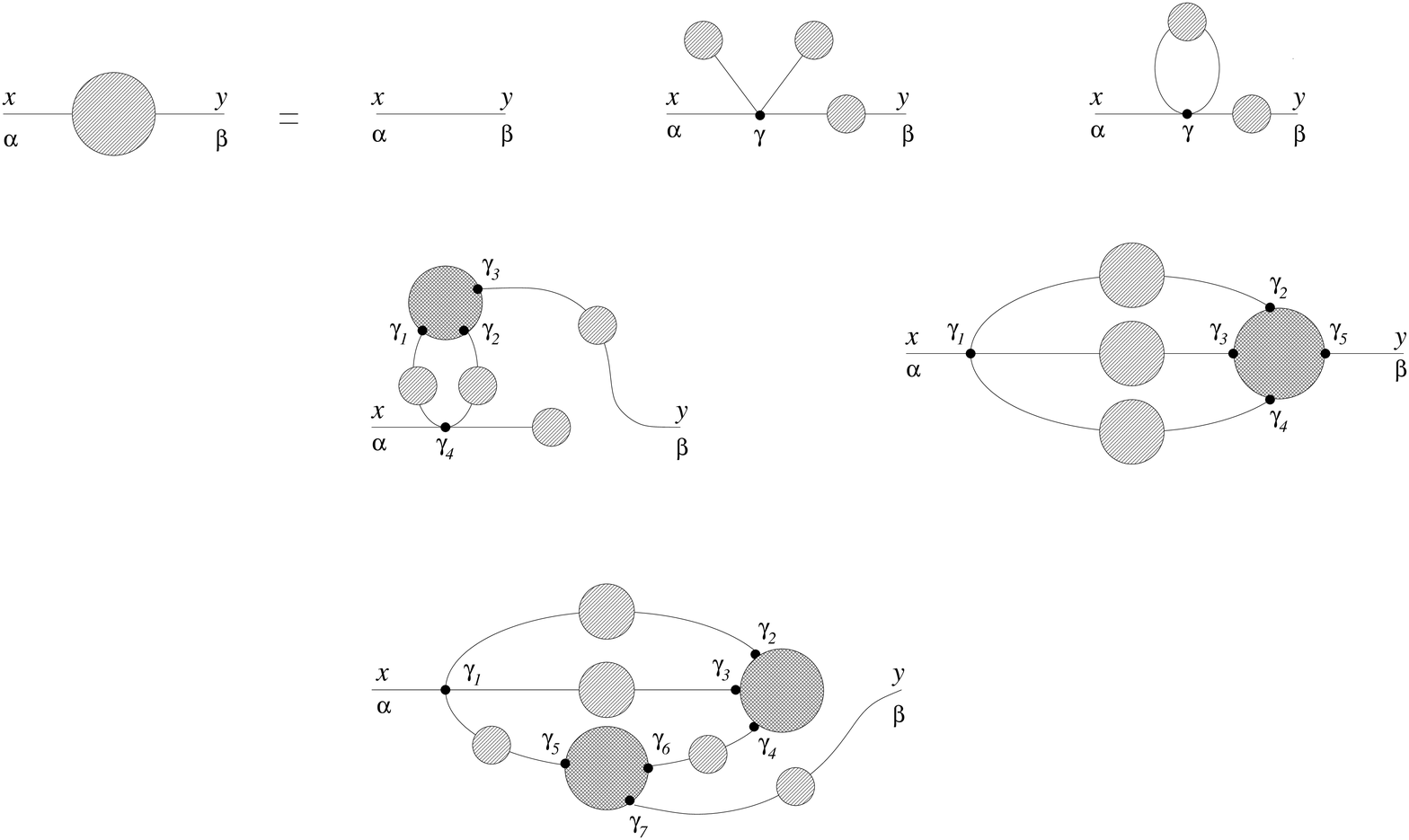}}
\label{D-S5}
\setlength{\unitlength}{1mm}
\begin{picture}(10,10)
\put(105,81.5){$-\;\frac{1}{2}\sum_{\gamma}$}
\put(63,81.5){$-\;\frac{1}{2}\sum_{\gamma}$}
\put(32,57){$-\; \frac{i}{2}\sum_{\gamma_{i}}$}
\put(85,57){$+\; \frac{1}{6}\sum_{\gamma_{i}}$}
\put(32,24){$+\;\frac{1}{2}\sum_{\gamma_{i}}$}
\end{picture} 
\end{figure}

\noindent  Note  that  if  the  symmetry   is  unbroken (i.e.  when
$m_{0}^{2}>0$) then after setting  $J=0$ the second, fourth  and sixth
term on the RHS will vanish. 

It is important to mention that similarly as for the connected Green's
functions we may write  the Dyson--Schwinger equations for the  vertex
functions.   Those  may be  obtained directly  from Eq.(\tseref{A144})
provided that  one  performs the  corresponding number   of variations
w.r.t. $\phi$. So, for instance, for the $\lambda \Phi^{4}$ theory the
1--point   vertex function   $\Gamma^{(1)}_{\alpha}(x)  = \frac{\delta
\Gamma}{\delta \phi^{\alpha}(x)}  =   -J_{\alpha}(x)$  is  graphically
represented (c.f. Eq.(\tseref{A19})) as 

\vspace{8mm}

\begin{figure}[h]
\vspace{3mm}
\epsfxsize=14cm
\centerline{\epsffile{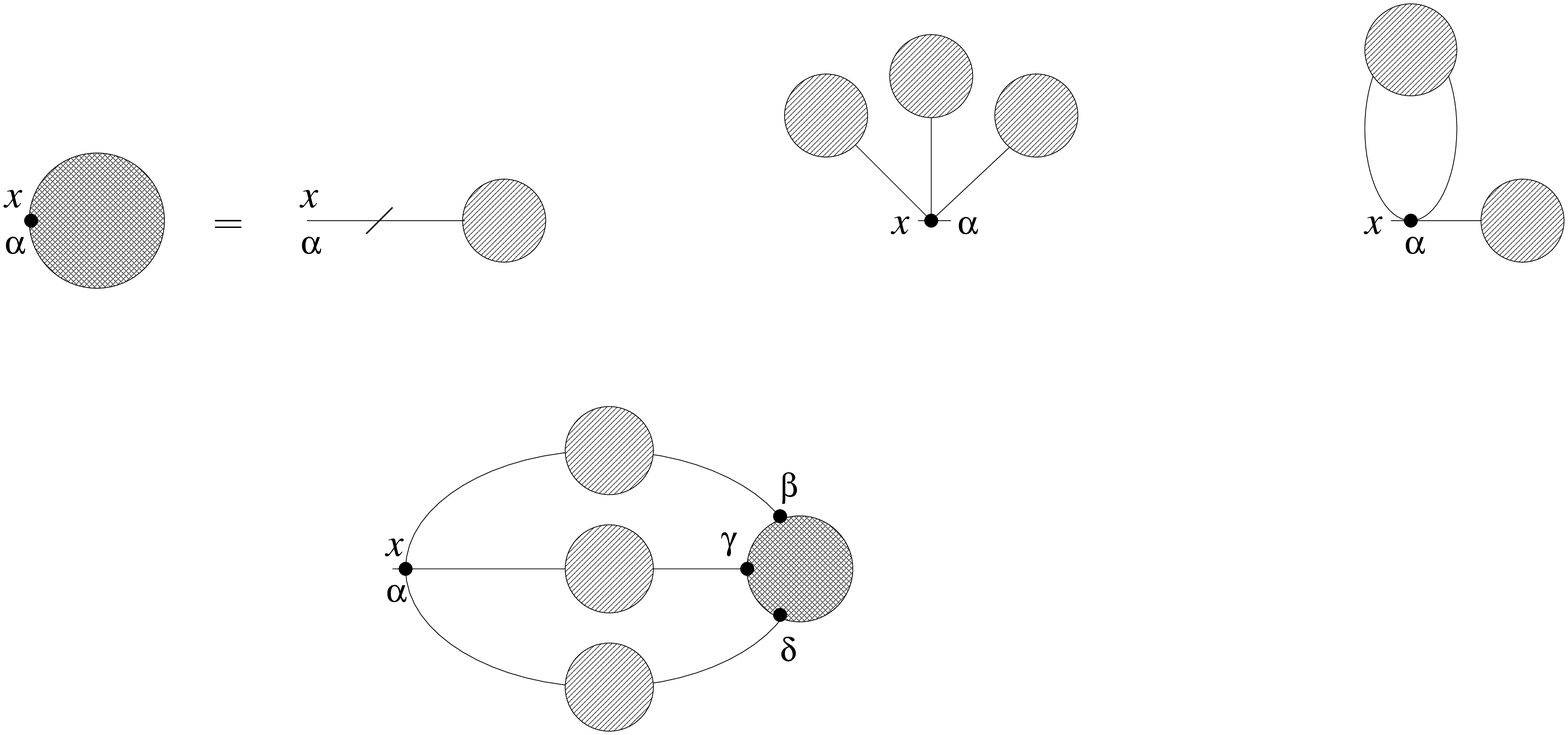}}
\label{D-S6}
\setlength{\unitlength}{1mm}
\begin{picture}(10,10)
\put(115,55){$+\;\;\; \;\;\; \frac{1}{2}$}
\put(70,55){$-\;\;\;\;\;\;\frac{1}{6}$}
\put(25,25){$+\;\;\frac{1}{6}\sum_{\beta \gamma \delta}$}
\end{picture} 
\end{figure}

\noindent   Here the  slash  stands  for   the  inverse  free  thermal
propagator.  Particularly important examples  are the 2-- and 4--point
vertex  functions   $\Gamma^{(2)}$ and   $\Gamma^{(4)}$, respectively,
which  are indispensable   in  the  renormalisation prescription  (see
Chapter \ref{PE}). In the case of the $O(N)\; \lambda \Phi^{4}$ in the
large--$N$  limit the  corresponding   graphical representations  were
explicitly  calculated    in    Chapter    \ref{PE}   (see    Sections
\ref{PE31}--\ref{PE32}).

\section{Comparison \label{A12}}

In order to find  a link between the previous  two approaches,  let us
start with the full $n$--point Green's function  at $T=0$.  It is well
known that in this case, the full Green's function may be expressed by
means of the Gell--Mann and Low formula\cite{IZ,JS,JZ,LB2} 

\begin{equation}
\langle 0|{\cali{T}}(\psi_{H}(x_{1}) \ldots   
\psi_{H}(x_{n}))|0 \rangle =
\frac{\left\langle 0 \left| {\cali{T}}\left( \psi(x_{1}) \ldots \psi(x_{n})
e^{i\int
d^{4}x {\cali{L}}_{I}(x)}
\right)\right| 0 \right\rangle}{\left\langle 0
\left| {\cali{T}}\left( e^{i\int d^{4}x {\cali{L}}_{I}(x)} \right)
\right| 0 \right\rangle}\, , 
\end{equation}
\vspace{0mm}

\noindent which may be equivalently rewritten in the Schwinger form
\cite{JSc} as

\begin{equation}
\langle 0|{\cali{T}}(\psi_{H}(x_{1}) \ldots   
\psi_{H}(x_{n}))|0 \rangle = 
\left\langle 0 \left|{\cali{T}}_{C}\left( \psi(x_{1}) \ldots \psi(x_{n})
e^{i\int_{C}
d^{4}x {\cali{L}}_{I}(x)}
\right)\right| 0 \right\rangle \, ,
\end{equation}
\vspace{0mm}

\noindent where  $\psi_{H}$  is a  field  operator in  the  Heisenberg
picture, while $\psi$ and ${\cali{L}}_{I}$, are operators in
the    interaction picture,       the   sub--index  $C$    denotes   the
Schwinger--Keldysh time contour. The Schwinger (but not the Gell--Mann
and Low!) formula can be directly extended to  finite temperature as we
have  mentioned  in   Chapters \ref{dcoset}   and \ref{PN}, the   only
difference is that $\langle 0  |\ldots | 0 \rangle \rightarrow  \langle
\ldots \rangle$. The  following results will be equally  valid
for  both   $T=0$  and   $T\not=0$  provided   we  take  into   account
corresponding expectation    values. In order   to  be concise,  we shall
restrict ourselves to bosonic fields. The necessary extension to fermions
and gauge fields is straightforward.

For one field $\psi_{H}$ in the presence of an external resource $J(x)$
we have

\begin{equation}
\phi(x) = \langle \psi_{H}(x) \rangle = \left\langle {\cali{T}}_{C}
\left( \psi(x) e^{i\int_{C} d^{4}x ({\cali{L}}_{I}(x)+ J(x)\psi(x))}\right)
\right\rangle.
\end{equation}

\vspace{3mm}

\noindent Setting in (\tseref{S-D234}) $F[\psi]= e^{i\int_{C}
d^{4}x({\cali{L}}_{I}(x)+J(x)\psi(x))}$, we get

\begin{equation}
\langle      \psi_{H}(x)  \rangle      +  \int_{C}      d^{4}z
\,\D_{F}(x,z)\left\langle  {\cali{T}}_{C} \left(\frac{\delta \int_{C} d^{4}y
{\cali{L}}_{I}(y)}{\delta \psi(z)}  \right)_{H}\right\rangle =
-\int_{C} d^{4}z \,\D_{F}(x,z)J(z)\, , 
\end{equation}
\vspace{1mm}

\noindent where $i\,\D_{F}(x,z) = \langle
{\cali{T}}_{C}(\psi(x)\psi(z))\rangle$. For instance, for the $\lambda \Phi^{4}$ theory (${\cali{L}}_{I} =
- \frac{i\lambda_{0}}{4!} \Phi^{4}$) we have

\begin{equation}
\langle \Phi_{H}(x)\rangle - \frac{\lambda_{0}}{3!}\int_{C} d^{4}z
\,\D_{F}(x,z) \langle\Phi^{3}_{H}(z)\rangle = -\int_{C}
d^{4}z \,\D_{F}(x,z)J(z)\, ,
\tselea{D-S8}
\end{equation}
\vspace{0mm}

\noindent or in the thermal--index notation; $\alpha, \beta, \gamma,
\delta = \{ 1, 2 \}$,

\begin{equation}
\phi^{\alpha}(x) + \frac{\lambda_{0}}{3!} \int d^{4}z \,\D_{F}^{\alpha
\delta}(x,z)\, \D^{(3)}_{\delta \delta \delta}(z,z,z) = - \int d^{4}z \,\D_{F}^{\alpha
\delta}(x,z) J_{\delta}(z)\, .
\tselea{D-S81}
\end{equation}
\vspace{0.5mm}

\noindent In (\tseref{D-S81}) we can recognise the Dyson--Schwinger
equation (\tseref{A161}) for the $\lambda \Phi^{4}$ theory from the
previous section.

To  get  further equations in   an infinite tower  of couplet integral
equations we may  perform  successive  variations  in (\tseref{D-S8})
with respect to $J_{\beta}(y)$, or we may
alternatively          set   in (\tseref{S-D234}) $F[\psi] =
\psi(y)\,e^{i\int_{C}d^{4}x \, ({\cali{L}}_{I}(x) + J(x)\psi(x))}$.   In
both cases we obtain for $\lambda \Phi^{4}$ theory the following identity
\begin{eqnarray}
&&\D^{\alpha \beta}(x,y) = \D_{F}^{\alpha \beta}(x,y) -
\frac{i\lambda_{0}}{3!}\int d^{4}z \, \D_{F}^{\alpha
\delta}(x,z)\;(\sigma_{3})_{\delta \gamma}\; \langle
{\cali{T}}_{C}(\Phi^{3}_{H}(z)\Phi_{H}(y))\rangle^{\gamma
\beta}\nonumber \\
&&\mbox{\hspace{3.9cm}} + i \int d^{4}z \, \D^{\alpha
\delta}_{F}(x,z)\,J_{\delta}(z)\, \phi^{\beta}(y)\nonumber \\
\Leftrightarrow &&\D^{c \; \alpha \beta}(x,y) = \D_{F}^{\alpha \beta}(x,y) - \frac{\lambda_{0}}{3!} \int d^{4}z \,
\D^{\alpha \delta}_{F}(x,z)\, \D^{(4)\; \beta}_{\delta \delta
\delta}(z,z,z,y)\nonumber \\
&& \mbox{\hspace{4.1cm}} + \frac{i\lambda_{0}}{3!}\, \int d^{4}z\,
\D^{\alpha \delta}_{F}(x,z)\, \D^{(3)}_{\delta \delta \delta}(z,z,z)\,\phi^{\beta}(y)   \, .
\tselea{D113}
\end{eqnarray}
\vspace{0.05mm}

\noindent  Note  that (\tseref{D113})   precisely  coincides with  the
Dyson--Schwinger equation (\tseref{A346}) derived in the previous section. 

Let     us  mention    one  more  point     in  the    connection with
Eq.(\tseref{D113}).          If,        instead        of      setting
$F[\psi]=\psi(y)e^{i\int_{C}d^{4}x         \,      ({\cali{L}}_{I}(x)+
J(x)\psi(x))}$, as we  have  done in the previous  case,  we would set
$G[\psi]=1$,    $F[\psi]=  e^{i\int_{C}      d^{4}x({\cali{L}}_{I}(x)+
J(x)\psi(x))}$ in (\tseref{S-D7})   or (\tseref{S-D8}),  we  would get
somewhat different identity, namely 

\begin{eqnarray}
\D^{\alpha \beta}(x,y) &=& \D_{F}^{\alpha \beta}(x,y) -
\int d^{4}z_{1}d^{4}z_{2}\; \D_{F}^{\alpha
\gamma}(x,z_{1})(\sigma_{3})_{\gamma \delta}\nonumber \\
& \times & \left\langle   
{\cali{T}}_{C}\left(
\frac{\delta^{2} (i\int_{C} d^{4}x ({\cali{L}}_{I}(x)+ J(x)\psi(x)))_{H}}
{\delta \psi(z_{1}) \delta
\psi(z_{2})} \right) \right\rangle^{\delta \iota}
(\sigma_{3})_{\iota \kappa}\; \D_{F}^{\kappa \beta}(z_{2},y)\, ,
\tselea{D1}
\end{eqnarray}
\vspace{0mm}

\noindent which in the case of $\lambda \Phi^{4}$ theory Eq.(\tseref{D1})
may be explicitly written as

\begin{eqnarray}
&&\D^{\alpha \beta}(x,y)=\D_{F}^{\alpha \beta}(x,y) -
\frac{i\lambda_{0}}{2}\int d^{4}z_{1}\; \D_{F}^{\alpha
\gamma}(x,z_{1})(\sigma_{3})_{\gamma \delta}\left\langle  \Phi_{H}^{2}(z_{1})
\right\rangle^{\delta \iota}
(\sigma_{3})_{\iota \kappa}\; \D_{F}^{\kappa \beta}(z_{1},y)\nonumber
\\
&&\mbox{\hspace{7mm}}- \frac{\lambda^{2}_{0}}{(3!)^{2}}\int d^{4}z_{1}d^{4}z_{2}\; \D_{F}^{\alpha
\gamma}(x,z_{1})(\sigma_{3})_{\gamma \delta} \;\langle {\cali{T}}_{C}(\Phi^{3}_{H}(z_{1})\Phi^{3}_{H}(z_{2}))\rangle^{\delta \iota}\;
(\sigma_{3})_{\iota \kappa}\; \D_{F}^{\kappa \beta}(z_{2},y)\nonumber
\\
&&\mbox{\hspace{7mm}} - \int d^{4}z_{1}d^{4}z_{2}\, \D^{\alpha
\gamma}_{F}(x,z_{1})\, J_{\gamma}(z_{1})\, \D_{F}^{\delta
\beta}(z_{2},y)\,J_{\delta}(z_{2})\, .
\tselea{D112}
\end{eqnarray}
\vspace{0mm}

\noindent It  is,  however, not complicated  to persuade  oneself that
(\tseref{D112})  is  equivalent   to  (\tseref{D113}).   Indeed, using
(\tseref{D-S81}) we   get  the desired   equality (\tseref{D113}).  In
(\tseref{D1}) we can  recognise the Dyson  equation with $\left\langle
{\cali{T}}_{C}\left(            \frac{\delta^{2}(i\int          d^{4}x
({\cali{L}}_{I}(x)+ J(x)\psi(x)))_{H}}{\delta \psi(z_{1}) \delta \psi(z_{2})}
\right)\right\rangle$         being   the        self--energy\cite{FW}
$i{\tilde{\bf{\Sigma}}}(z_{1},    z_{2})$.   Introducing  a     proper
self--energy ${\bf{\Sigma}}$, which  is a self--energy  that cannot be
separated  into two  pieces by cutting  a single   line (i.e.  in  the
matrix      form       ${\tilde{\bf{\Sigma}}}    =  {\bf{\Sigma}}    +
{\bf{\Sigma}}\D_{F} {\bf{\Sigma}} + \ldots $) the series (\tseref{D1})
can be summed formally to yield 

\begin{displaymath}
\D^{\alpha \beta}(x,y) = \D_{F}^{\alpha \beta}(x,y) + \int
d^{4}z_{1}d^{4}z_{2}\, \D_{F}^{\alpha \delta}(x,z_{1})
(-i{\bf{\Sigma}}_{\delta \gamma}(z_{1},z_{2})) \D^{\gamma \beta}(z_{2},y)\, ,
\end{displaymath}

\noindent  which is the  usual  form of  the (thermal) Dyson  equation
(c.f.  equation (\tseref{b5})).  So the second in the hierarchy of the
Dyson--Schwinger  equations is nothing  but the Dyson equation for the
full two--point Green's function.  Analogous procedure can be repeated
for the Dyson--Schwinger  equations involving the  higher--point Green
functions. As a result we can see then that the usual Dyson--Schwinger
equations emerge naturally  as a   special sub--class of    identities
derived in Appendix \ref{A1}. Another  sub--class of identities which
may be derived from the functional  formalism of Appendix \ref{A1} are
thermal Ward's identities.   For  example, for the  $O(N)\,  \Phi^{4}$
theory  we   immediately  get  from  (\tseref{S-D7})  that $\D^{\alpha
\beta}_{a\,a}(x,y) = \D^{\alpha  \beta}_{b \, b}(x,y)$ and $\D^{\alpha
\beta}_{a \,b}(x,y)= 0$ if $a \not= b$.  We, however, do not intend to
dwell on this point more.

We   thus find that   our functional  formalism  derived in  Appendix
\ref{A1} provides a unifying framework embracing such diverse concepts
as   Wick's  theorem,  the   Dyson--Schwinger   equations  and  Ward's
identities.

\chapter{Surface term in Eq.(4.69) \label{A2}}

In this appendix we give some details of the derivation of
Eq.(\tseref{b10}). We particularly show that the surface integrals arisen
during the transition from (\tseref{b9}) to (\tseref{b10}) mutually cancel
among themselves. As usual, the integrals will be evaluated for integer
values of $D$ and corresponding results then analytically continued to a
desired (generally complex) $D$. 

The key quantity in question is

\vspace{2mm}

\begin{eqnarray}
&+& \frac{1}{2}\; \int \frac{d^{D}q}{(2\pi)^{D-1}} \left(
\frac{2{\vect{q}}^{2}}{(D-1)}
\right) \; \frac{\varepsilon(q_{0})}{e^{\beta
q_{0}}-1}\;\delta(q^{2}-m^{2}_{r}(T))\nonumber \\
&-& \frac{1}{2}\; \int \frac{d^{D}q}{(2\pi)^{D-1}} \left(
\frac{2{\vect{q}}^{2}}{(D-1)} \right)\; \delta^{+}(q^{2}-m^{2}_{r}(0)).
\tselea{a1}
\end{eqnarray}

\vspace{2mm}

\begin{eqnarray}
\mbox{(\tseref{a1})}&=& {\cali{N}}_{T}(m^{2}_{r}(T)) -
{\cali{N}}(m^{2}_{r}(0))\nonumber\\
 &+& \lim_{R \rightarrow \infty} \;
\frac{1}{2(D-1)} \int \frac{dq_{0}}{(2\pi)^{D-1}}\; \int_{\partial
S^{D-2}_{R}}
d{\vect{s}}\;{\vect{q}}\; \theta(q^{2}-m^{2}_{r}(T))\; \theta(q_{0})\;
\left( \frac{2}{e^{\beta q_{0}}-1}+1 \right)\nonumber \\
&-&\lim_{R \rightarrow \infty} \; \frac{1}{2(D-1)} \int
\frac{dq_{0}}{(2\pi)^{D-1}}\; \int_{\partial S^{D-2}_{R}}
d{\vect{s}}\;{\vect{q}}\; \theta(q^{2}-m^{2}_{r}(0))\; \theta(q_{0}).
\tseleq{a2}
\end{eqnarray}

\vspace{4mm}

\noindent As usual, ${\vect{a}}{\vect{b}} =
\sum_{i=1}^{D-1}{\vect{a}}_{i}{\vect{b}}_{i}$ and $S^{D-2}_{R}$ is a
$(D-2)$-sphere with the radius $R$. The expressions for ${\cali{N}}_{T}$ and
${\cali{N}}$ are done by (\tseref{b101}).

With the relation (\tseref{a2}) we can show that the surface
terms cancel in the large $R$ limit. Let us first observe that

\begin{eqnarray}
&&\lim_{R \rightarrow \infty} \; \int
\frac{dq_{0}}{(2\pi)^{D-1}}\; \int_{\partial S^{D-2}_{R}}
d{\vect{s}}\;{\vect{q}}\; \theta(q^{2}-m^{2}_{r}(T))\;
\frac{2 \theta(q_{0})}{e^{\beta q_{0}}-1}\nonumber \\
&=& \lim_{R \rightarrow \infty}\;
\frac{2\pi^{\frac{D-1}{2}}R^{D-1}}{\Gamma \left(\frac{D-1}{2}\right)} \;
\int \frac{dq_{0}}{(2\pi)^{D-1}} \; \theta(q_{0}^{2}-R^{2}-m^{2}_{r}(T))\;
\frac{2 \theta(q_{0})}{e^{\beta q_{0}}-1}\nonumber \\  
&=& \lim_{R \rightarrow \infty}
\frac{\pi^{\frac{1-D}{2}}R^{D-1}}{2^{D-2}\Gamma
\left( \frac{D-1}{2} \right)}\int_{\sqrt{R^{2}+m_{r}^{2}(T)}}^{\infty}
dq_{0}\; \frac{2}{e^{\beta q_{0}}-1} \; = \;0.
\tseleq{a3}
\end{eqnarray}

\vspace{3mm}

\noindent In 2-nd line we have exploited Gauss's theorem and in the last line
we have used L'H{\^{o}}pital's rule as the expression is in the
indeterminate form $0/0$. The remaining surface terms in (\tseref{a2})
read

\begin{eqnarray}
&& \lim_{R \rightarrow \infty} \int
\frac{dq_{0}}{(2\pi)^{D-1}} \; \int_{\partial S^{D-2}_{R}}
d{\vect{s}}\;{\vect{q}}\; \left\{ \theta(q^{2}-m^{2}_{r}(T)) -
\theta(q^{2}-m^{2}_{r}(0)) \right\}\; \theta(q_{0}) \nonumber\\
&=& \lim_{R \rightarrow \infty}\;
\frac{\pi^{\frac{1-D}{2}}R^{D-1}}{2^{D-2}\Gamma \left( \frac{D-1}{2}
\right)}\; \left\{ \int_{\sqrt{R^{2}+m_{r}^{2}(T)}}^{\infty} -
\int_{\sqrt{R^{2}+m_{r}^{2}(0)}}^{\infty} \right\}dq_{0}\; = \; 0.
\tseleq{a4}
\end{eqnarray}

\vspace{2mm}

\noindent The last identity follows either by applying L'H{\^{o}}spital's
rule or by a simple transformation of variables which renders both
integrals inside of $\{ \ldots \}$ equal. Expressions on the last lines in
(\tseref{a3}) and (\tseref{a4}) can be clearly (single--valuedly) continued
to the region $\mbox{Re}D > 1$ as they are analytic there. We
thus end up with the statement that

\vspace{2mm}

\begin{displaymath}
\mbox{(\tseref{a1})}= {\cali{N}}_{T}(m^{2}_{r}(T))- {\cali{N}}(m^{2}_{r}(0)).
\end{displaymath}

\chapter{High--temperature expansion of the gap equation \label{A3}}

In this appendix we shall derive the high--temperature expansion of the
mass shift $\delta m^{2}(T)$ in the case when fields $\Phi_{a}$ are
massive (i.e. $m^{2}_{r}(0) \not= 0$).

Consider Eqs.(\tseref{c1}) and (\tseref{m12}). If we combine
them together, we get easily the following transcendental equation for   
$\delta m^{2}(T)$

\begin{equation}
\delta m^{2}(T) = \lambda_{0} \left\{ {\cali{M}}(m^{2}_{r}(T)) -
{\cali{M}}(m^{2}_{r}(0)) + \frac{1}{2} I_{0}(m^{2}_{r}(0) + \delta
m^{2}(T))\right\}.
\tseleq{bb1}
\end{equation}

\vspace{2mm}

\noindent Here ${\cali{M}}$ and $I_{0}$ are done by (\tseref{b45}) and
(\tseref{c14}), respectively.

Now, both $\lambda_{0}$ and ${\cali{M}}$ are divergent in $D=4$.
If we reexpress
$\lambda_{0}$ in terms of $\lambda_{r}$, divergences must cancel, as
$\delta m^{2}(T)$ is finite in $D=4$. The latter can be easily seen if we
Taylor expand ${\cali{M}}$,
i.e.

\begin{equation}
{\cali{M}}(m^{2}_{r}(T)) = {\cali{M}}(m^{2}_{r}(0)) + \delta m^{2}(T) \;
{\cali{M}}'(m^{2}_{r}(0)) + \hat{{\cali{M}}}(m^{2}_{r}(0); \delta
m^{2}(T)).
\tseleq{bb2}
\end{equation}

\vspace{2mm}

\noindent Obviously, $\hat{{\cali{M}}}$ is finite in $D=4$ as ${\cali{M}}$
is quadratically divergent. Inserting (\tseref{bb2}) to (\tseref{bb1})
and employing Eq.(\tseref{m146}) we get

\begin{equation}
\delta m^{2}(T) = \lambda_{r} \left\{ \hat{{\cali{M}}}(m^{2}_{r}(0);
\delta m^{2}(T)) + \frac{1}{2} I_{0}(m^{2}_{r}(0) + \delta
m^{2}(T))\right\}.
\tseleq{bb3}
\end{equation}

\noindent This is sometimes referred to as the renormalised gap equation.
In order to determine $\hat{{\cali{M}}}$ we must go back to
(\tseref{bb2}). From the former we read off that

\begin{eqnarray}
&&\hat{{\cali{M}}}(m^{2}_{r}(T); \delta m^{2}(T))\nonumber \\
&&\nonumber \\
&& \mbox{\hspace{1.2cm}}=
{\cali{M}}(m^{2}_{r}(T))
- {\cali{M}}(m^{2}_{r}(0)) - \delta m^{2}(T)\;
{\cali{M}}'(m^{2}_{r}(0))\nonumber \\
&& \mbox{\hspace{1.2cm}}= \frac{\Gamma(1-
\frac{D}{2})}{2(4\pi)^{\frac{D}{2}}} \left\{
(m^{2}_{r}(T))^{\frac{D}{2}-1} - (m^{2}_{r}(0))^{\mbox{$\frac{D}{2}$}-1} -
\delta
m^{2}(T)(\frac{D}{2}-1)\; (m^{2}_{r}(0))^{\frac{D}{2}-2} \right\}
\nonumber \\
&& \mbox{\hspace{1.2cm}}
 \stackrel{D \rightarrow 4}{=}  \frac{1}{32
\pi^{2}} \left\{
m^{2}_{r}(T) \; \mbox{ln}\left( \frac{m^{2}_{r}(T)}{m^{2}_{r}(0)}\right) -
\delta m^{2}(T) \right\}.
\tseleq{b4}    
\end{eqnarray}
\noindent So
\begin{equation}
\delta m^{2}(T) = \lambda_{r} \left\{ \frac{(m^{2}(0) + \delta
m^{2}(T))\;\mbox{ln}\left( 1 + \frac{\delta m^{2}(T)}{m^{2}_{r}(0)}\right)
- \delta m^{2}(T)}{32 \pi^{2}} + \frac{1}{2}I_{0} \right\}.
\tseleq{bb4}
\end{equation}  

\noindent Eq.(\tseref{bb4}) was firstly obtained and numerically solved in
\cite{ID1}. It was shown that the solution is double valued. The former
behaviour was also observed by Abbott {\em{et al.}} \cite{Abb} at
$T=0$, and by Bardeen and Moshe \cite{BM} at both $T=0$ and $T\not=0$.
The relevant solution is only that which fulfils the condition $\delta
m^{2}(T) \rightarrow 0$ when $T \rightarrow 0$. For such a solution it can
be shown (c.f.  \cite{ID1}, FIG.3) that $\frac{\delta
m^{2}(T)}{m^{2}_{r}(0)}\ll 1$ for a sufficiently high $T$. So the
high--temperature expansion of (\tseref{bb4}) reads

\begin{eqnarray}
\delta m^{2}(T) &=& \lambda_{r}\left\{ \frac{\frac{(\delta
m^{2}(T))^{2}}{2m^{2}_{r}(0)} - \frac{(\delta m^{2}(T))^{3}}{6
m^{4}_{r}(0)}   
+ \frac{(\delta m^{2}(T))^{4}}{12m^{6}_{r}(0)} + \ldots
}{32 \pi^{2}}  + \frac{1}{2}I_{0} \right\}\nonumber \\
&&\nonumber \\ 
&\doteq& \frac{\lambda_{r}}{2} I_{0} = \frac{\lambda_{r}T^{2}}{24} -
\frac{\lambda_{r}m_{r}(T)}{8 \pi}\;T + {\cali{O}}\left( m_{r}^{2}(T)
\;\mbox{ln}\left(\frac{m_{r}(T)}{T4 \pi} \right) \right).
\tselea{bb52} 
\end{eqnarray}

\chapter{Derivation of the Shannon (information) entropy \label{A4}}
\begin{center}
{\bf{(This section is based on refs.\cite{Jayn3,RBa,SH1,MRa}.)}} 
\end{center}
In Chapter \ref{PN} we intensively used the concept of the Shannon (or
information)    entropy which  we     identified  (up  to a   negative
multiplicative  constant) with an  ``information content''  associated
with  a (macro--) system  in question.  The  objective  of the present
appendix is to formulate mathematically  more clearly the rather vague
notion   of ``informative  content''   and  to find   a link   between
statistical physics and information  theory. As a  next step we derive
the basic properties of the Shannon entropy. 

Let us  first introduce the concept of  information in the  context of
probability calculus and of  information theory.  We shall  consider a
set of events (or ensemble of all possible messages) $\{ x_{1}, \ldots
,   x_{N}\}$  with   respective  probabilities  $\{p_{1},   \ldots   ,
p_{N}\}$.    So, for example,   if  only  single  letter messages  are
transmitted 

\begin{displaymath}
\{ x_{1}, \ldots , x_{N} \} = \{ A,B, \ldots, Z \}\, ,
\end{displaymath}

\noindent  then  the  corresponding   set of   $p$'s   characterises a
particular  language\footnote{For   instance, $p(A)$  in  English   is
0.0703, 0.0645 in French is and 0.0693 in Czech, the least frequent
letter $Z$  has  $p(Z)=0.0005$  in English,   $0.0006$ in  French  and
$0.0008$ in Czech. In information theory is usual to call a set of
$p$'s  a  {\em{langue}}  even if  a message itself is  not composed of actual
letters}. Because events $\{ x_{i} \}$ are all possible different messages
which  might be send from a  sender  to a recipient, the corresponding
probabilities must sum up to one: $\sum_{i}^{N}p_{i} = 1$. 

One says that ${\Im}_{m}$ is the amount  of information conveyed by the
message   $x_{m}$ if ${\Im}_{m}$  is  a  non--negative and  continuous
function  $\Im$   of  $p_{m}$  defined  on  the    range  $0<p_{m}\leq
1$.  Moreover,  the more likely  a  message  is, the less  information
is conveyed by the knowledge of its actual occurrence (the
more stereotypical a message is   (i.e. the larger is the  probability
for  being received), the less   information it imparts). This implies
that $\Im_{m} = \Im(p_{m})$ must be decreasing  function of $p_{m}$ on
the interval $[0,1]$, and particularly ${\Im}(p_{m}=1)=0$. 

In order to get more qualitative results let us consider first a simple
system with only two possible   messages $x_{1}$ and $x_{2}$ and   the
corresponding language  $\{p_{1},  p_{2}\}$. Since  the  probability of
receiving the amount   of information ${\Im}_{1}$ is  $p_{1}$  and
that of ${\Im}_{2}$ is $p_{2}$, the expected averaged amount of information
received is 

\begin{displaymath}
I(p_{1},p_{2}) = p_{1}{\Im}(p_{1}) + p_{2}{\Im}(p_{2})\, ,
\end{displaymath}

\noindent or, more generally, for  $N$ mutually different messages  $\{
x_{1}, \ldots , x_{N}\}$ with the language $\{ p_{i}\}$
the   average  amount   of  information   received  is\footnote{We may
also say that the averaged amount of received information is
the averaged gain of information associated with the transmitted message
or  equivalently,  the ignorance  of  a receiver,  that is removed by
recipe of the message.} 

\begin{equation}
I(p_{1},   \ldots ,   p_{N})    = p_{1}{\Im}(p_{1})   +  \ldots  +
p_{N}{\Im}(p_{n})\, . 
\tseleq{A41}
\end{equation}

\noindent Let us note  that (\tseref{A41}) implies that $I(\ldots)$ is
symmetric and continuous in all its  arguments. As we shall see, these
conditions  will strongly    restrict a  possible  class of   feasible
$I$'s. A further, sever restriction on the  possible form of $I(p_{1},
\ldots  ,  p_{N})$ is  imposed by the,  so  called, additivity law for
information\cite{SH}. To understand the basic  features of the latter,
let us start with $N=3$. 

In order to determine  the degree of uncertainty  about the system (or
equivalently, the expected amount  of  information received), we  may,
e.g., first evaluate the degree of uncertainty connected with the fact
that  we  do    not  know whether  $x_{1}$    or  ${\overline{x}}_{1}$
(complement  of $x_{1}$) occured  and   add the degree  of uncertainty
whether the message in ${\overline{x}}_{1}$ is $x_{2}$ or $x_{3}$. The
amount of uncertainty due to  (or information  conveyed by) the  first
determination is obviously 

\begin{displaymath}
I(p_{1},{\overline{p}}_{1})           =      p_{1}{\Im}(p_{1})   +
{\overline{p}}_{1}\Im({\overline{p}}_{1})\, . 
\end{displaymath}

\noindent The amount of uncertainty due to the second determination is

\begin{displaymath}
I(x_{1},x_{2}|{\overline{x}}_{1})    :=  I(p(x_{2}|{\overline{x}}_{1}),
p(x_{3}|{\overline{x}}_{1}))                                         =
p(x_{2}|{\overline{x}}_{1})\Im(p(x_{2}|{\overline{x}}_{1})) +
p(x_{3}|{\overline{x}}_{1})\Im(p(x_{3}|{\overline{x}}_{1})) \, .
\end{displaymath}

\noindent Using the  Bayes's formula for the conditional probabilities
($p(XY) = p(X)p(Y|X)$) and the fact that if $X\cap Y = X$ then
$p(XY)=p(X)$ ($X$ and $Y$ denote sets of events) we may write
\begin{equation}
I(x_{1},x_{2}|{\overline{x}}_{1})                                    =
\frac{p(x_{2})}{p({\overline{x}}_{1})}\Im\left(\frac{p(x_{2})}{p({\overline{x}}_{1})}\right)
+
\frac{p(x_{3})}{p({\overline{x}}_{1})}\Im\left(\frac{p(x_{3})}{p({\overline{x}}_{1})}\right)\,.
\tseleq{A42}
\end{equation}
\vspace{0.5mm}

\noindent However (\tseref{A42}) is the amount of information conveyed
only when the  ${\overline{x}}_{1}$ event occurs,  so the total amount
of information conveyed on average must be 

\begin{eqnarray}
I(p_{1},p_{2},p_{3})       &=&       I(p_{1},{\overline{p}}_{1})       +
{\overline{p}}_{1}I(x_{2},x_{3}|{\overline{x}}_{1})\nonumber \\
&=& I(p_{1},{\overline{p}}_{1})       +
{\overline{p}}_{1}\, I \left( \frac{p_{2}}{{\overline{p}}_{1}}, \frac{p_{3}}{{\overline{p}}_{1}}\right)\, . 
\tselea{A43}
\end{eqnarray}
\vspace{0.5mm}

\noindent   We shall now show   that the additivity law (\tseref{A43})
together with symmetry  and continuity of $I(\ldots)$  suffice to
determine  $I(\ldots)$ uniquely  up to a  multiplicative constant. The
proof will be done in three steps. 

\vspace{1cm}
\rule[0.05in]{12cm}{0.1mm}~~~~{\bf{\large{Step 1}}}
\vspace{2mm}

Let  us assume that we   have events  $\{x_{1},  \ldots , x_{N-1},
y_{1} , \ldots , y_{M}  \}$ and the  corresponding alphabet $\{ p_{1},
\ldots , p_{M}, q_{1},   \ldots, q_{M}\}$. (The   way  we  perform
a splitting into $x$'s and $y$'s is actually irrelevant.) We shall show
now that the following important identity holds

\begin{equation}
I(p_{1},  \ldots, p_{N-1}, q_{1}, \ldots,  q_{M})  = I(p_{1}, \ldots ,
p_{N})    +   p_{N}\,I\left(     \frac{q_{1}}{p_{N}}, \ldots           ,
\frac{q_{M}}{p_{N}}\right)\, ,
\tseleq{A44}
\end{equation}
\vspace{1mm}

\noindent where $p_{N} =  \sum_{i=1}^{M}q_{i}$. We may think of events
$\{ x_{1}, \ldots , x_{N-1}\}$ as a one composite event $x$. For $M=1$
Eq.(\tseref{A44}) is then trivially  fulfilled. If $M=2$, Eq.(\tseref{A44})
coincides with  Eq.(\tseref{A43}) and so is  true as  well. Let us now
take an induction step and let assume that  (\tseref{A44}) is true for
a   general $M>2$  and  let  us prove  that  this must   hold also for
$M+1$. Actually, due to the induction hypothesis we may directly write
for $M+1$ 

\begin{equation}
I(p_{1}, \ldots, p_{N-1}, q_{1}, \ldots, q_{M+1}) = I (p_{1}, \ldots ,
p_{N-1}, q_{1},  p'_{N})   +   p'_{N}\, I\left(  \frac{q_{2}}{p'_{N}},
\ldots , \frac{q_{M+1}}{p'_{N}}\right)\, ,
\tseleq{A45}
\end{equation}
\vspace{1mm}

\noindent with $p'_{N} = \sum_{i=2}^{M+1}q_{i}$. Using now relation
for $M=2$ we may write the first term on the RHS of (\tseref{A45}) as 
\begin{equation}
I (p_{1}, \ldots , p_{N-1}, q_{1},  p'_{N})= I(p_{1}, \ldots , p_{N}) 
+ p_{N}I\left( \frac{q_{1}}{p_{N}} ,
\frac{p'_{N}}{p_{N}}\right)\, ,
\tseleq{A47}
\end{equation}

\noindent with $p_{N}= q_{1} + p'_{N} = \sum_{i=1}^{M+1}q_{i}$. Using the
symmetry of $I(\ldots)$ and the induction hypothesis we may write
\begin{eqnarray}
I\left( \frac{q_{1}}{p_{N}}, \ldots, \frac{q_{M+1}}{p_{N}}\right) &=& I\left(
\frac{q_{1}}{p_{N}},\frac{p'_{N}}{p_{N}}\right) +
\frac{p'_{N}}{p_{N}}\,I\left(\frac{q_{2}}{p'_{N}}, \ldots ,
\frac{q_{M+1}}{p'_{N}}\right)\nonumber \\
\Leftrightarrow \;\;p'_{N}\, I\left( \frac{q_{2}}{p'_{N}}, \ldots ,
\frac{q_{M+1}}{p'_{N}}\right) &=& p_{N}\, I\left( \frac{q_{1}}{p_{N}},
\ldots , \frac{q_{N+1}}{p_{N}}\right) - p_{N}\,I \left(
\frac{q_{1}}{p_{N}}, \frac{p'_{N}}{p_{N}} \right)\, .
\tselea{A46}
\end{eqnarray}
\vspace{1mm}

\noindent Plugging both (\tseref{A47}) into (\tseref{A46}) into
(\tseref{A45}) we receive the desired relation for $M+1$, and so this
proves that Eq.(\tseref{A44}) holds.

\vspace{5mm}
\rule[0.05in]{12cm}{0.1mm}~~~~{\bf{\large{Step 2}}}
\vspace{2mm}

In this step we extend our reasoning to an arbitrary number,
say $n$, of groups of the messages, i.e., we want to evaluate the
amount of information 

\begin{displaymath}
I(p_{1,1},\ldots , p_{1,N_{1}}, p_{2,1}, \ldots, p_{2,N_{2}}, \ldots ,
p_{n,1}, \ldots , p_{n,N_{n}})\, .
\end{displaymath}

\noindent Using the result (\tseref{A44}) we may directly write 
the former as

\begin{displaymath}
I(p_{1,1}, \ldots , p_{1,N_{1}}, \ldots , p_{n-1,1}, \ldots , p_{n-1,
N_{n-1}}, p_{n} ) + p_{n}\, I \left( \frac{p_{n,1}}{p_{n}}, \ldots ,
\frac{p_{n,N_{n}}}{p_{n}} \right)\, ,
\end{displaymath}

\noindent where $p_{n}= \sum_{i=1}^{N_{n}}p_{n,i}$. If we now shift
$p_{N}$ to the very left in $I(\ldots)$ and iterate further we get
\begin{eqnarray*}
I(p_{n}, p_{1,1}, \ldots , p_{1,N_{1}}, \ldots , p_{n-1,1}, \ldots ,
p_{n-1,N_{n-1}}) &=& I(p_{n}, p_{1,1}, \ldots ,
p_{n-2,N_{n-2}},p_{n-1})\nonumber \\
&+& p_{n-1}\, I \, \left( \frac{p_{n-1,1}}{p_{n-1}}, \ldots ,
\frac{p_{n-1,N_{n-1}}}{p_{n-1}} \right)\, ,
\end{eqnarray*}

\noindent with $p_{n-1} = \sum_{i=1}^{N_{n-1}}p_{n-1,i}$. Shifting
$p_{n-1}$ to the very left and repeating iteration we get

\begin{eqnarray*}
I(p_{n}, p_{n-1}, p_{1,1}, \ldots , p_{1,N_{1}}, \ldots ,
p_{n-2,N_{n-2}}) &=& I(p_{n}, p_{n-1}, p_{1,1}, \ldots ,
p_{n-3,N_{n-3}}p_{n-2})\nonumber \\
&+& p_{n-2}\, I\left( \frac{p_{n-2,1}}{p_{n-2}}, \ldots ,
\frac{p_{n-2,N_{n-2}}}{p_{n-2}}\right)\, ,
\end{eqnarray*}

\noindent with $p_{n-2} = \sum _{i=1}^{N_{n-2}}p_{n-2,i}$. Shifting
$p_{n-2}$ to the very left and repeating iteration, etc. up to $p_{1}$
we finally obtain

\begin{equation}
I(p_{1,1}, \ldots , p_{n,N_{n}}) = I(p_{1}, p_{2}, \ldots , p_{n}) +
\sum _{i=1}^{n} p_{i} \, I \left( \frac{p_{i,1}}{p_{i}}, \ldots ,
\frac{p_{i,N_{i}}}{p_{i}}\right)\, .
\tseleq{A49}
\end{equation}

\noindent Eq.(\tseref{A49}) is the desired result.

\vspace{5mm}
\rule[0.05in]{12cm}{0.1mm}~~~~{\bf{\large{Step 3}}}
\vspace{2mm}

In this last step we  shall actually solve Eq.(\tseref{A49}). Before
we start let us define one useful function.   If all the messages were
equiprobable,  i.e., all $p_{i}=1/n$ with $n$  being the number of all
possible messages, then we define 

\begin{displaymath}
\sigma(n) := I\left( \frac{1}{n}, \ldots , \frac{1}{n} \right); \;\,
n\geq 2; \; \; I(1) = 0\, .
\end{displaymath}

\noindent The first noticeable fact about $\sigma(x)$ is that it
fulfils a very simple functional relation, namely

\begin{eqnarray}
\sigma(mn) &=& I\left( \frac{1}{mn}, \ldots , \frac{1}{mn} \right) =
I\left( \frac{1}{n}, \ldots , \frac{1}{n} \right) + \sum_{i=1}^{n}\frac{1}{n}\,
I\left( \frac{1}{m}, \ldots , \frac{1}{m} \right)\nonumber \\  
&=& \sigma(n) + \sigma(m)\, ,
\tselea{A491}
\end{eqnarray}

\noindent where in the first line we have used Eq.(\tseref{A49}). The
functional identity (\tseref{A491}) has the well known
solution\footnote{A simple way how to solve (\tseref{A491}) is to
assume that $\sigma(x)$ is a continuous function. If we were able to
find the solution of the relation $\sigma(xy) =
\sigma(x) + \sigma(y)$ for continuous arguments we could at the end
restrict our attention only to discrete ones. Assuming this continuity
we may perform derivation w.r.t. $x$ and get

\begin{displaymath}
\frac{d\sigma(yx)}{dx} = y\,\sigma'(yx) = \sigma'(x)\,.
\end{displaymath}

\noindent Setting $yx=z$ we obtain the differential equation 

\begin{displaymath}
z\,\sigma'(z) = x\,\sigma'(x) = k\, ,
\end{displaymath}

\noindent where $k$ is constant.  So the solution is obvious: $\sigma(x)=
k\,{\mbox{ln}}(x)$.};  $\sigma(x) = k\,\mbox{ln}(x)$.  The  constant $k$ is
the only ambiguity  which the solution  possesses. We shall specify $k$
latter on. 

Let us now assume that probabilities $\{p_{1}, \ldots , p_{n}\}$
on the RHS of (\tseref{A49}) are rational numbers, i.e., we may write
$p_{1} = N_{1}/N,\, p_{2}=N_{2}/N, \ldots ,\, p_{n} = N_{n}/N$ with $N=
\sum_{i=1}^{n}N_{i}$. In this case

\begin{displaymath}
I(p_{1}, p_{2}, \ldots, p_{n}) = I \left( \frac{N_{1}}{N}, \ldots ,
\frac{N_{n}}{N} \right)\, .
\end{displaymath}

\noindent Inserting this back into Eq.(\tseref{A49}) we may rewrite
(\tseref{A49}) as

\begin{eqnarray}
&&I\left( \frac{1}{N}, \ldots , \frac{1}{N}\right) = \sigma(N) = I\left(
\frac{N_{1}}{N}, \ldots , \frac{N_{n}}{N} \right) + \sum_{i=1}^{n}
\frac{N_{i}}{N}\, I\left(\frac{1}{N_{i}}, \ldots ,
\frac{1}{N_{i}}\right)\nonumber \\
&&\mbox{\hspace{2.8cm}}= I\left(\frac{N_{1}}{N}, \ldots , \frac{N_{n}}{N} \right) +
\sum_{i=1}^{n}\frac{N_{i}}{N}\, \sigma(N_{i})\nonumber \\
\Leftrightarrow &&  I\left(
\frac{N_{1}}{N}, \ldots , \frac{N_{n}}{N} \right) = \sigma(N) -
\sum_{i=1}^{n}\frac{N_{i}}{N}\sigma(N_{i}) = - k \sum_{i=1}^{n}
\frac{N_{i}}{N}\, \mbox{ln}\left( \frac{N_{i}}{N} \right)\, .
\tselea{A492}
\end{eqnarray}

\noindent As $I(\ldots)$ is by assumption continuous function we may
analytically continue result (\tseref{A492}) to irrational
probabilities, and so generally

\begin{equation}
I(p_{1}, \ldots , p_{n}) = -k \sum_{i=1}^{n} p_{i}\,\mbox{ln}(p_{i})\, .
\tseleq{A493}
\end{equation}
  
\noindent Note that  the structure of Eq.(\tseref{A493}) is  precisely
that    as  in  (\tseref{A41}).   Comparing both    (\tseref{A41}) and
(\tseref{A493})  we may identify  the amount of information $\Im_{m}$ of
the message $x_{m}$  as $\Im(p_{m}) = -k\,\mbox{ln}(p_{m})$.   Because
$\Im(\ldots)$ is a non--negative, decreasing function of its argument,
the constant $k$ must be positive\footnote{Note: If $k
=\mbox{log}_{2}e$ the amount of information (\tseref{A493}) is called
both the Shannon and information entropy. For a different choice of $k$
only notion of the Shannon entropy is used.}. 

\vspace{5mm}
\rule[0.05in]{13.5cm}{0.1mm}~~~~{\bf{\large{}}}
\vspace{2mm}

The  constant  $k$  clearly  fixes   the  units of   information.   In
information theory $k$ is  chosen to be 1   with the logarithm  in the
base 2, so $I(\ldots) = -\sum_{i}p_{i}\,{\mbox{log}}_{2}(p_{i})$.  The
reason for choice this is  quite pragmatic.  The messages are  usually
written    in the binary  characters    ($\{0,1\}$, yes/no).  The most
elementary  message is composed of   two equiprobable events (unbiased
choice between two possible  messages about which  a receiver does not
posses any further information).  In this case the information entropy
is  $I(1/2,1/2)=\sigma(1/2)=1$.   Thus  the  most elementary   message
carries the unit amount of information.  This unit is called ``bit'' 
\footnote{If the base   of  the logarithm  is  $e$  ($k=1$), then  the
information   is measured   in  ``nats''.}.   If   one transmits  only
$n$--letter messages (i.e.  $111\ldots11, \; 111\ldots 10, \, \ldots ,
\;   000\ldots  00$) then    the  amount  of information  is   clearly
$\sigma(2^{n})=n$.  Consequently such  messages   convey
information with $n$ bits. 

As it was  mentioned, in information theory  Eq.(\tseref{A493}) refers
to the situation {\em{after}} the reception of a message.  Accordingly
we interpret (\tseref{A493}) as the   average gain in the  information
associated   with  the transmitted  message   (the greater the initial
uncertainty, the greater the amount of information conveyed.  If there
is  no initial uncertainty  or doubt to be  resolved, the alphabet $\{
p_{i} \}$ shrinks to a  single case $\{ p_{j}  =1, p_{k\not= j} =0 \}$
and hence the Shannon entropy (gained information) is zero).
Let  us note   that  if   events  $\{x_{i}\}$  are   equiprobable  the
information entropy  equals to the expected  number of binary (yes/no)
questions whose  answer take us from  our current knowledge to the one
of  certainty\footnote{For instance,    for $I(1/2,1/2)$  the   binary
question may sound: is it  $1$ which is transmitted ?   In the case of
the  $n$--letter messages we may  ask  whether the transmitted message
has   on the  first position 1   (yes/no), on  the  second position  1
(yes/no), $\ldots$ , at the $n$-th position 1 (yes/no).}. 

In physical systems, however, no message is send, so to speak. We may,
nonetheless,   in  accordance   with   the   previous  derivation  view
(\tseref{A493}) as the mean information about the  system which is not
transmitted yet   (i.e. the recipient is waiting   for it).  From this
standpoint $I(\ldots)$ appears as entropy of the  language, that is as
uncertainty (or ignorance) about the ensemble of all possible messages
which may be received.  The  only  difference is that the  uncertainty
(ignorance) cannot never be completely  removed,  as in   the  case  when the
message is received, but it may be removed partially,
namely when one performs a measurement on the system.  We shall return
to this point in a while. 

Now, question  stands what $k$ we should  choose in real (macroscopic)
physical systems. It is clear that in statistical physics, the systems
are   too   complex  and the    amount   of all possible transmittable
information is so vast that $k$ must be chosen very small in order for
one  to get judiciously large numerical  values of the Shannon entropy
in typical processes.  Because $k$ should be  a constant valid for all
systems, its numerical  value depends only on  the  choice of physical
units and hence  may be determined  via arbitrary, but suitably chosen
system.  For  example, we may   define as a   unit entropy the entropy
corresponding to one mole  of  spins of  free valence electrons  in  a
piece of iron.  Assuming that all  the $2^{N_{A}}$ spin configurations
are equiprobable (Avogadro's  number $N_{A} = 0.6024 \times 10^{24}\,$
mol$^{-1}$), then this  yields $I(\ldots) = \sigma(2^{N_{A}}) = kN_{A}
= 1\,\mbox{mol}^{-1}$.  From  this reasoning  we   would get $k   \sim
10^{-24}$.    In  order to  obtain  a   connection  with  the  usual von
Neumann--Gibbs entropy  of  statistical physics we   may note that  the
Shannon entropy coincides with the von  Neumann--Gibbs one provided we
fix $k/\mbox{ln}(2)=k_{B}= 1.3804 \times 10^{-23}\mbox{JK$^{-1}$}$.

In  connection  with statistical  physics  we  may define a  notion of
{\em{information content}} ${\textswab{H}}$  inherent  to a system  of
interest. Let us assume that the system had originally the information
entropy $I_{0}(\ldots)$ (i.e., entropy of the ensemble of all possible
transmittable     results or     all      possible  results    of    a
measurement). Entropy $I_{0}$ is  often called a--priory entropy. Then
measurement       is made but  because     of   experimental errors or
impossibility to measure all phenomena  there is a whole new  ensemble
of values, each of  which could give  rise  to the one observed.   The
information entropy, say $I_{1}(\ldots)$  may be defined also for this
a--posteriori ensemble.   The  latter expresses how   much uncertainty
still  left  unresolved after   measurement.  Let  us now  define  the
informative content ${\textswab{H}}$   as  an  amount by  which    the
uncertainty about a system has been reduced, i.e. 

\begin{displaymath}
{\textswab{H}} = I_{0} - I_{1}\, , 
\end{displaymath}

\noindent  or equivalently   $I_{1} =  I_{0} -{\textswab{H}}$.   After
discarding the (constant) additive entropy $I_{0}$ the latter leads to
the statement that the  informative content is equivalent to  negative
entropy (``negatropy'').  That is, as our information about a physical
system increases,  its  entropy  must  decrease\footnote{Note that  in
information  theory   $I_{1}(\ldots)=0$.} .  This   result  is due  to Szilard
\cite{LSz} and Brillouin \cite{LBr}. 

The passage to quantum mechanics is simple. If the macro--state of a
system is represented by the density matrix
\begin{equation}
\rho_{H}= \sum_{n}p_{n} |\psi_{n}\rangle \langle \psi_{n}|\, ,
\tseleq{A489}
\end{equation}

\noindent then the information entropy turns out to be (for simplicity
we omit from now on the sub--index $H$)
\begin{equation}
I(\rho) = - \mbox{Tr}(\rho \,\mbox{log}_{2}\,\rho )\, ,
\end{equation}

\noindent and the information content
\begin{equation}
\textswab{H} = \mbox{Tr}(\rho\, \mbox{log}_{2}\,\rho )\, .
\end{equation}

\noindent  In the language of information  theory  the ensemble of all
possible messages is set of all possible results  of a measurement of a
given system of observables.  The corresponding alphabet is $\{ p_{1},
p_{2}, \ldots \}$. 

\vspace{7mm}

\noindent {\bf{Some properties of the Shannon entropy}}

\begin{itemize} 
\item {\bf{Concavity:}} The Shannon entropy  is concave on the set  of
$\rho$'s on a given Hilbert space.
\end{itemize}

\noindent Concavity of  $I(\rho)$ means that  for any pair  $\rho_{1}$
and $\rho_{2}$ and $0< \lambda <1$ we have

\begin{displaymath}
I(\lambda \rho_{1}  + (1-\lambda)\rho_{2})  \geq \lambda I(\rho_{1}) +
(1-\lambda)I(\rho_{2})\, . 
\end{displaymath}

\noindent This    may be proven    very simply with   a help of the
inequality

\begin{displaymath}
\mbox{Tr}(X\mbox{log}_{2}\;Y)   -  \mbox{Tr}(X\mbox{log}_{2}\;X)  \leq
(\mbox{Tr}X - \mbox{Tr}Y)\mbox{log}_{2}e\, . 
\end{displaymath}

\noindent  (use  the  spectral  decomposition  of   $X$   and $Y$  and
the inequality $\mbox{ln}x \leq (1-x)$) The generalised concavity identity
reads
\begin{displaymath}
I(\sum_{i}\lambda_{i} \rho) \geq \sum_{i} \lambda_{i} I(\rho) \, ,
\end{displaymath}

\noindent where $\lambda_{i} >0$ and $\sum_{i} \lambda_{i} =1$. 

\vspace{7mm}

\begin{itemize}
\item{\bf{Maximum:}}    If   the  possible   kets   in   the  spectral
decomposition (\tseref{A489})  span  a  finite  ${\cal{W}}$--dimensional
subspace of the Hilbert space then
\end{itemize}
\begin{displaymath}
I(\rho) \leq {\mbox{log}}_{2}{\cal{W}}\, ,
\end{displaymath}

\noindent with  the equality only  in the  case when all probabilities
in (\tseref{A489}) are equal, i.e. $p_{i} = p = 1/{\cal{W}}$. 

To prove this we may look at the difference

\begin{displaymath}
I(\rho) - \mbox{log}_{2}{\cal{W}} = \mbox{Tr}(\rho \,(\mbox{log}_{2}\,
\rho^{-1}{\cal{W}}))\,. 
\end{displaymath}

\noindent Taking the spectral decomposition of $\rho$ together with the
inequality $\mbox{ln}x \leq (1-x)$ (note the equality is fulfilled only when
$x=1$)  we obtain $I(\rho)  - \mbox{log}_{2}{\cal{W}} \leq   0$. The latter is
equal to $0$ if and only if ${\cal{W}}/p_{i} =1$ for all $p_{i}$.

\vspace{7mm}

\begin{itemize}
\item{\bf{Minimum:}} The   Shannon  entropy  has a minimum    equal to
zero. This happen only when $\rho$ describes a pure state. 
\end{itemize}

Because $-\mbox{log}_{2}(\ldots)$ is   a convex function, one  may use
Jensens's  inequality  of statistical  mathematics  \cite{GM,Cub}:  if
$f(\ldots)$ is a  convex   function then  $\langle f(X)  \rangle  \geq
f(\langle    X   \rangle)$.  Thus   $I(\rho)   = -\langle \mbox{log}_{2}
\rho\rangle \geq - \mbox{log}_{2}\langle \rho \rangle =0$, so 
\begin{displaymath}
I(\rho) \geq 0\, . 
\end{displaymath}

\noindent  $I(\rho)  =  0$ only if   there is  no  uncertainty about  a
message, i.e. when alphabet  $\{ p_{i} \}$  shrinks to a single  case
$\{ p_{j}=1, p_{i\not= j} =0 \}$, i.e. when $\rho$ describes the pure state.

\chapter{Some mathematical formulae \label{MF}}
\section{Integrals in $D$ dimensions} 
\begin{center}
{\bf{(This section is based on
refs.\cite{V,TV,IZ,PR}.)}}\footnote{Note: All the quantities entering 
formulae bellow are dimensionless!!}
\end{center}
\vspace{5mm}
\noindent Throughout  our dissertation we frequently apply dimensional
regularisation; i.e.  we replace the dimension  4 by a lower dimension
$D$ where the corresponding (loop) integrals are convergent. Bellow we
provide a   short list of  integrals which   we  found  useful  during
our calculations (cf.  Sections \ref{PE2}, \ref{PE3}, \ref{HTE} and
\ref{PN4}, Appendices \ref{A2} and \ref{A3}). 

The paradigmatic integral of the dimensional regularisation
is

\begin{displaymath}
\int \frac{d^{D}q_{E}}{(q^{2}_{E} + X )^{n}} =
\pi^{D/2}\frac{\Gamma(n -
\mbox{$\frac{D}{2}$})}{\Gamma(n)} X^{-n + \mbox{$\frac{D}{2}$}}\, .
\;\;\;\;\;\;\;\; n < 2 \,.
\end{displaymath}
\noindent (use $D$--dimensional polar coordinates and the fact
that $\int d\Omega = S^{D-1} =
2\pi^{\frac{D}{2}}/\Gamma (\mbox{$\frac{D}{2}$})$ ) Euclidean
regime is defined via Wick's rotation as; $q^{0} = iq^{0}_{E}, {\vect{q}} =
{\vect{q}}_{E}, d^{D}q = id^{D}q_{E}$. Although the LHS as a
$D$--dimensional integral is senseful only for integer values of $D$,
the RHS has an analytic continuation for all $D \in \Com$ with
$D\not=2n$ (so namely for $D=4-2\varepsilon \; (\varepsilon >0, \varepsilon
\rightarrow 0)$). Performing the change of variables $q_{E}
\rightarrow q_{E} + l_{E}$ we get 

\begin{displaymath}
\int \frac{d^{D}q_{E}}{(q^{2}_{E} + 2l_{E}q_{E} + X)^{n}} = 
\pi^{D/2}\frac{\Gamma(n -
\mbox{$\frac{D}{2}$})}{\Gamma(n)} (X-l_{E}^{2})^{-n + \mbox{$\frac{D}{2}$}}\,.
\end{displaymath}

\noindent Successive derivatives with respect to $l_{E}^{\alpha}$ then yield

\begin{eqnarray*}
\int d^{D}q_{E}\, \frac{q^{\mu}}{(q^{2}+ 2l_{E}q_{E} + X)^{n}} &=&   
-\pi^{D/2}\,l_{E}^{\mu}\,\frac{\Gamma(n -
\mbox{$\frac{D}{2}$})}{\Gamma(n)} (X-l_{E}^{2})^{-n +
\mbox{$\frac{D}{2}$}}\, . \\
\int d^{D}q_{E}\, \frac{q^{\mu}q^{\nu}}{(q^{2}+ 2l_{E}q_{E} +
X)^{n}} &=& \frac{\pi^{D/2}}{\Gamma(n)}(X-l_{E}^{2})^{-n+
\mbox{$\frac{D}{2}$}}\\
&\times& \left\{
\Gamma(n-\mbox{$\frac{D}{2}$})l_{E}^{\mu}l_{E}^{\nu} + \frac{1}{2} \delta^{\mu
\nu}\Gamma(n-1 -\mbox{$\frac{D}{2}$})(X-l_{E}^{2})\right\}\, . 
\end{eqnarray*}

\noindent The analytical continuation to Minkowski regime (i.e. Wick's
rotation of both $q_{E}^{\alpha}$ and $l_{E}^{\alpha}$) together with Eq.(\ref{4.2P}) gives 
\vspace{1mm}
\begin{eqnarray*}
\int \frac{d^{D}q}{(2\pi)^{D}} \frac{i}{q^{2}-m^{2}\pm i\epsilon} &=&
\frac{\Gamma(1-\mbox{$\frac{D}{2}$})}{(4\pi)^{D/2}}(m^{2})^{D/2-1}\\
&=&-\frac{m^{2}}{16 \pi^{2}} \,\left( \Delta
-\mbox{ln}m^{2} +1 + {\cal{O}}(\varepsilon) \right)\, ,
\end{eqnarray*}
\noindent with
\begin{displaymath}
\Delta = \frac{1}{\varepsilon} - \gamma + \mbox{ln}4\pi \, .
\end{displaymath}

\noindent   (for  convenience  we     have   introduced   the  usual factor
$1/(2\pi)^{D}$)  Previous results  together with     the Feynman
parametrisation: $1/ab = \int_{0}^{1}dt\, 1/[at + b(1-t)]^{2}$, yield 
\vspace{1mm}
\begin{eqnarray*}
&&\int\frac{d^{D}q}{(2\pi)^{D}} \frac{i}{(q^{2}-m^{2}+
i\epsilon)((q+p)^{2}-m^{2}+i\epsilon)}\\
&&\mbox{\hspace{60mm}}=\frac{-1}{16 \pi^{2}}
\left( \Delta -\int_{0}^{1} dt\, \mbox{ln}(m^{2}+p^{2}(t^{2} -t))
+ {\cal{O}}(\varepsilon) \right)\\
&&\int \frac{d^{D}q}{(2\pi)^{D}} \frac{i\,q^{\mu}}{(k^{2}-m^{2}+
i\epsilon)((q+p)^{2}-m^{2}+i\epsilon)}\\
&&\mbox{\hspace{60mm}}= \frac{p^{\mu}}{32
\pi^{2}}\,
\left( \Delta - \int_{0}^{1} dt\, \mbox{ln}(m^{2}+p^{2}(t^{2}
-t)) 
+ {\cal{O}}(\varepsilon) \right)\\
&&\int \frac{d^{D}q}{(2\pi)^{D}} \frac{i\,q^{\mu}q^{\nu}}{(k^{2}-m^{2}+
i\epsilon)((q+p)^{2}-m^{2}+i\epsilon)} = \frac{1}{16 \pi^{2}}
\left(g^{\mu \nu} A(p^{2},m) + p^{\mu}p^{\nu}B(p^{2},m) \right)\, ,
\end{eqnarray*}

\noindent with $A(p^{2},m)$ and $B(p^{2},m)$:
\begin{eqnarray*}
A(p^{2},m) &=& \frac{1}{3} \left\{ -m^{2}\left[ \mbox{$\frac{3}{2}$}\Delta -
\mbox{$\frac{1}{2}$}\mbox{ln}m^{2} + \mbox{$\frac{3}{2}$} -
\int_{0}^{1}dt\, \mbox{ln}(m^{2}+ p^{2}(t^{2}-t))\right]\right. \\
&& + \left.
\mbox{$\frac{1}{4}$}p^{2}\left[ \Delta + \mbox{$\frac{2}{3}$}- \int_{0}^{1}dt\, \mbox{ln}(m^{2} +
p^{2}(t^{2}-t))\right] + {\cal{O}}(\varepsilon)  
\right\} \\
&&\\
B(p^{2},m) &=& \frac{4}{3p^{2}}\left\{ m^{2}\left[
\mbox{$\frac{3}{2}$} \Delta - \mbox{$\frac{1}{2}$}\mbox{ln}m^{2} + 3
-\int_{0}^{1}dt\,\mbox{ln}(m^{2} + p^{2}(t^{2}-t))\right] \right. \\
&& - \left. \mbox{$\frac{1}{4}$}\left[ \Delta + 1
-\int_{0}^{1}dt\,\mbox{ln}(m^{2} + p^{2}(t^{2}-t))\right] +
{\cal{O}}(\varepsilon) \right\}\, .
\end{eqnarray*}
\vspace{2mm}  

\noindent Note: the integral $\int_{0}^{1}dt \, (\ldots)$ might be
evaluated explicitly, the result reads
\begin{eqnarray*}
\int_{0}^{1}dt\,\mbox{ln}(m^{2} + p^{2}(t^{2}-t)) &=& \mbox{ln}(m^{2})
-2 + 2\,{\sqrt{{\frac{4 - a}{a}}}}\,
   \arctan \left({\frac{{\sqrt{a}}}{{\sqrt{4 - a}}}}\right)\, ,
\end{eqnarray*}
\noindent with $a = p^{2}/m^{2}$.

\vspace{3mm}
\section{Special functions and important relations} 
\begin{center}
{\bf{(This section is based on refs.\cite{LW,IZ,PR,PJ,FW,B,BB,GR}.)}}
\end{center}
\vspace{5mm}
\noindent The gamma function $\Gamma(x)$ and the Riemann zeta function
$\zeta (x)$ are defined as follows: 
\begin{eqnarray*}
&&\Gamma(x) = \int_{0}^{\infty} dt \,
e^{-t}t^{x-1}\mbox{\hspace{15mm}} \mbox{Re}x > 0 \nonumber \\
&&\zeta(x) = \sum_{n=1}^{\infty} n^{-x}\mbox{\hspace{15mm}} \mbox{Re}x
> 1\, . 
\end{eqnarray*}

\noindent The above definitions converge only in the specified regions
of the complex plane, but they  can be analytically (single--valuedly)
continued. The   following   important  relations  (used  in   Section
\ref{HTE}) are valid in the entire complex  plane (save for the points
$x=-n,   (n= 0,  1, 2,  \ldots)$   where  the simple  pole residue  is
$\frac{(-1)^{n}}{n!}$) 
\vspace{-1mm}
\begin{eqnarray*}
&&\Gamma(x+1)= x \Gamma(x)\\ 
&&\Gamma(x)\Gamma(1-x) = \frac{\pi}{\mbox{sin}(\pi x)} \\
&&\Gamma\left( \frac{1}{2} + x \right)\Gamma \left( \frac{1}{2} - x
\right) = \frac{\pi}{\mbox{cos}(\pi x)} \\
&&\Gamma(2x) = \frac{2^{2x-1}}{\sqrt{\pi}}\Gamma(x)\Gamma\left(
x+\frac{1}{2}\right)\\
&&\Gamma\left( \frac{x}{2} \right)\pi^{-\frac{x}{2}}\zeta(x) =
\Gamma\left(\frac{1-x}{2}\right) \pi^{\frac{x-1}{2}}\zeta(1-x)\,.
\end{eqnarray*}

\noindent For $n$ being integer  

\begin{displaymath}
\Gamma\left( \frac{1}{2} - n \right) = (-1)^{n}
\frac{2^{n}\sqrt{\pi}}{(2n - 1)!!}\, .
\end{displaymath}

\noindent Important numerical values of $\zeta (x)$ used in the text
are~:

\vspace{2mm}
\begin{center}
\renewcommand{\arraystretch}{1.5}
\begin{tabular}{|l||l|l|l|l|l|l|l|} \hline
~$x$~ & ~$0$~ & ~$\mbox{$\frac{3}{2}$}$~& ~$2$~ &
~$\mbox{$\frac{5}{2}$}$~& ~$3$~ & ~$4$~&~$5$~\\ \hline
~$\zeta(x)$~ & ~$\mbox{$\frac{1}{2}$}$~ & ~$2.612$~ & ~$\mbox{$\frac{\pi^{2}}{6}$}$~&  ~$1.341$~ &  ~$1.202$~ &
~$\frac{\pi^{4}}{90}$~&~$1.037$~\\ \hline
\end{tabular}
\end{center}
\begin{displaymath}
\zeta' (0) = \left( \frac{d \zeta (x)}{d x} \right)_{x=0} = -
\mbox{$\frac{1}{2}$} \mbox{ln}(2 \pi)\, .
\end{displaymath}

\noindent (note: only numerical values of $\zeta (2n +1)$ are
available)

\vspace{3mm}

\noindent Important numerical values of $\Gamma (x)$ used in the text
are :

\vspace{2mm}
\begin{center}
\renewcommand{\arraystretch}{1.5}
\begin{tabular}{|l||l|l|l|l|l|} \hline
~$x$~ &~ $\mbox{$\frac{1}{2}$}$~ &~ $1$~ & ~$\mbox{$\frac{5}{4}$}$~ &
~$\mbox{$\frac{3}{2}$}$~ &~ $\mbox{$\frac{7}{4}$}$~ \\ \hline
~$\Gamma(x)$~ &~ $\sqrt \pi$~ &~ $1$~ &~ $0.906$~ & ~$\frac{1}{2} \sqrt \pi$~ &
~$0.919$~\\ \hline
\end{tabular}
\end{center}  
\vspace{2mm}

\noindent Both $\zeta(x)$ and $\Gamma(x)$ appear in expansions of the
following definite integrals  used in the text ($n>0$):
\vspace{-1mm}
\begin{eqnarray*}
&& \int_{0}^{\infty} dt\, \frac{t^{n-1}}{e^{t} \pm 1} = (1-(1\mp
1)2^{-n})\Gamma(n)\zeta(n) \;\;\;\;\;\;\; (\mbox{Einstein's integrals})\\
&& \int_{0}^{\infty} dt\, \frac{t^{n-1}}{\mbox{sinh}t} = 2(1-2^{-n})
\Gamma(n)\zeta(n)\\
&& \int_{0}^{\infty} dt\, \frac{t^{n-1}}{\mbox{cosh}t} = 2 \Gamma(n)
\sum_{k=0}^{\infty} (-1)^{k}(2k + 1)^{n}\\
&& \int_{0}^{\infty} dt\, \frac{x^{n-1}}{\mbox{cosh}^{2}t} =
2^{2-n}(1-2^{2-n})\Gamma(n) \zeta(n-1)\, .
\end{eqnarray*}
\vspace{0.5mm}
\noindent Euler's $\psi$ function (or digamma) was used both in
Section \ref{HTE} and Section \ref{PN4}. $\psi(\ldots)$ is defined as the
logarithmic derivation of $\Gamma$ function:
\begin{eqnarray*}
\psi(x) &=& \frac{1}{\Gamma(x)}\frac{d \Gamma(x)}{dx}\\        
&=& \int_{0}^{\infty} \left( \frac{e^{-t}}{t} -
\frac{e^{-zt}}{1-e^{-t}}\right)dt \, .   
\end{eqnarray*}

\noindent The former directly implies that
\vspace{-0.5mm}
\begin{eqnarray*}
\psi(x+1) = \psi(x) + \frac{1}{x}\\
\psi(\mbox{$\frac{1}{2}$}) = \psi(1) - 2\mbox{ln}2\, ,
\end{eqnarray*}

\noindent or recursively ($n$ is integer)

\begin{displaymath}
\psi(x+n) = \psi(x+1) + \frac{1}{1+x} + \frac{1}{2+x} + \ldots +
\frac{1}{(n-1)+x}\, .
\end{displaymath}

\noindent Defining the Euler--Mascheroni constant $\gamma = -\psi(1)$
(the only numerical value is known: $\gamma = 0.5772156649 \ldots$), we get directly
\vspace{-2mm}
\begin{eqnarray*}
&&\psi(n) = - \gamma + \sum_{k=1}^{n-1}\frac{1}{k}\\
&&\psi(\mbox{$\frac{1}{2}$} +n) = - \gamma -2\mbox{ln}2 + 2\left( 1 +
\frac{1}{3} + \ldots + \frac{1}{2n-1}\right)\, .
\end{eqnarray*}

\noindent In Section \ref{PN4} we use functions $K_{n}$, which are the
Bessel   functions of  imaginary argument  of   order $n$ (we deal only with $n$ being integer). $K_{n}$  is
defined as:
\begin{eqnarray*}
K_{n}(x)                        & =&                      \frac{\sqrt{\pi}
(\mbox{$\frac{1}{2}$}x)^{n}}{\Gamma(n+\mbox{$\frac{1}{2}$})}
\int_{0}^{\infty}dt \, e^{-z \mbox{\scriptsize{cosh}} t} \mbox{sinh}^{2n}t\\
&=& \frac{\sqrt{\pi}
(\mbox{$\frac{1}{2}$}x)^{n}}{\Gamma(n+\mbox{$\frac{1}{2}$})}
\int_{0}^{\infty}dt \, e^{-zt}(t^{2}-1)^{n-\frac{1}{2}}\,. 
\end{eqnarray*}

\noindent The former implies the important relation: $K'_{0}(x) =
K_{1}(x)$. In Section \ref{PN4} we use the following relations:
\vspace{-2mm}
\begin{eqnarray*}
&&\int_{0}^{\infty}dt      \,      \frac{e^{-pt}}{\sqrt{t(t+a)}}       =
e^{\frac{ap}{2}}K_{0}\left( \frac{ap}{2} \right)\\
&&\int_{m}^{\infty}  dt  \,  \frac{t e^{-pt}}{\sqrt{t^{2}-m^{2}}}  = m
K_{1}(mp)\\
&&\int_{0}^{\infty}   dt  \, \frac{(t+a)e^{-pt}}{\sqrt{t^{2}+2at}} = a
e^{ap}K_{1}(ap)\,. 
\end{eqnarray*}

\noindent The limiting form for small arguments $x$ ($n$ fixed) reads:
\vspace{-1mm}
\begin{displaymath}
K_{n}(x)                                                          \sim
\mbox{$\frac{1}{2}$}\Gamma(n)(\mbox{$\frac{1}{2}$}x)^{-n}\, . 
\end{displaymath}

\vspace{7mm}
\noindent{\bf{Some miscellaneous functions used in the text:}}

\begin{itemize}
\item Beta function $B(z;y)$ (see Section \ref{HTE}):
\end{itemize}
\begin{displaymath}
B(x;y) = \int^{1}_{0} dt \, t^{x-1}(1-t)^{y-1} = \int^{\infty}_{0} \,
\frac{t^{x-1}}{(1+t)^{x+y}} = \frac{\Gamma(x)\Gamma(y)}{\Gamma(x+y)}\, .
\end{displaymath}

\begin{itemize}
\item Gauss' hypergeometric functions ${~}_{2}F_{1}[\ldots ]$ (see Section \ref{HTE}):
\vspace{-2mm}
\begin{eqnarray*}
{~}_{2}F_{1}[a,b;c;x] &=&
\frac{\Gamma(c)}{\Gamma(a)\Gamma(b)}\sum_{k=0}^{\infty}\frac{\Gamma(a+k)\Gamma(b+k)}{\Gamma(c+k)}\frac{x^{k}}{k!}\\
&=& \frac{1}{B(b; c-b)} \int_{0}^{1} dt \, t^{b-1}(1-t)^{c-b-1}(1-tx)^{-a}\,.
\end{eqnarray*}
\vspace{1mm}
\noindent (note:${~}_{2}F_{1}$  converges for $|x|<1$ with a branch
point at $x=1$, for $c = -n (n = 0,1,\ldots)$ ${~}_{2}F_{1}$ is undetermined)

The following relations for ${~}_{2}F_{1}$ are used in Section \ref{HTE}:
\vspace{-2mm}
\begin{eqnarray*}
&&{~}_{2}F_{1}[a,b;a-b+1;-1] = 2^{-a}\sqrt\pi \frac{\Gamma(1+a-b)}{\Gamma(1
+ \mbox{$\frac{1}{2}$}a -b)\Gamma(\mbox{$\frac{1}{2}$}
+\mbox{$\frac{1}{2}$}a)} \\
&&{~}_{2}F_{1}[a,b;\mbox{$\frac{1}{2}$}a +\mbox{$\frac{1}{2}$}b
+\mbox{$\frac{1}{2}$};\mbox{$\frac{1}{2}$}]= \sqrt \pi
\frac{\Gamma(\mbox{$\frac{1}{2}$} +\mbox{$\frac{1}{2}$}a
+\mbox{$\frac{1}{2}$}b )}{\Gamma(\mbox{$\frac{1}{2}$}
+\mbox{$\frac{1}{2}$}a )\Gamma(\mbox{$\frac{1}{2}$}
+\mbox{$\frac{1}{2}$}b )}\,.
\end{eqnarray*} 
\end{itemize}
\begin{itemize}
\item Bernoulli numbers $B_{\alpha}$ (see Section \ref{HTE}) are
defined through the series :
\vspace{-1mm}

\begin{displaymath}
\frac{x}{e^{x}-1} = \sum_{\alpha = 0}^{\infty} B_{\alpha}\,
\frac{x^{\alpha}}{\alpha !}\;\;\;\;\;\;\;\; |x|< 2\pi\, .
\end{displaymath}

\noindent Important numerical values of $B_{\alpha}$ used in the text
are :

\vspace{2mm}
\begin{center}
\renewcommand{\arraystretch}{1.5}
\begin{tabular}{|l||l|l|l|l|} \hline
~$\alpha$~&~$0$~&~$1$~&~$2$~&~$4$~\\ \hline 
~$B_{\alpha}$~&~$1$~&~$-\frac{1}{2}$~&~$\frac{1}{6}$~&~$-\frac{1}{30}$~\\
\hline
\end{tabular}  
\end{center}
\vspace{2mm}
\end{itemize}

\bibliographystyle{unsrt}
\bibliography{reference}

\begin{thebibliography}{100}

\bibitem{AL}
{D}.~{A}. {K}irzhnitz and {A}. {L}inde.
\newblock {\em {Phys. Lett.}}, {\bf{B42}}:{~471}, {1972}.

\bibitem{Wein}
{S}. {W}einberg.
\newblock {\em {Phys. Rev.}}, {\bf{D9}}:{~3357}, {1974}.

\bibitem{DJ}
{L}. {D}olan and {R}. {J}ackiw.
\newblock {\em {Phys. Rev.}}, {\bf{D9}}:{~3320}, {1974}.

\bibitem{CP}
{J}.~{C}. {C}ollins and {M}. {P}erry.
\newblock {\em {Phys. Rev. Lett.}}, {\bf{34}}:{~1353}, {1975}.

\bibitem{PH}
{P}.{A}. {H}enning.
\newblock {\em {Nucl. Phys.}}, {\bf{B337}}:{~547}, {1990}.

\bibitem{JBe}
{J}. {B}erges and {K}. {R}ajagopal.
\newblock {\texttt{hep-ph/9804233}}.

\bibitem{MAHa}
{{M}. {A}. {H}alasz, {A}. {D}. {J}ackson, {R}. {E}. {S}hrock, {M}. {A}.
  {S}tephanov and {J}. {J}. {M}. {V}erbaarschot}.
\newblock {\em {Phys. Rev.}}, {\bf{D58}}:{~096007}, {1998}.

\bibitem{JIK}
{J}.~{I}. {K}apusta.
\newblock {\em {Nucl. Phys.}}, {\bf{B148}}:{~461}, {1979}.

\bibitem{EWKo}
{E}.~{W}. {K}olb and {M}.~{S}. {T}urner.
\newblock {\em {The Early Universe}}.
\newblock {Addison--Wesley, London}, {1994}.

\bibitem{TWBKi2}
{T}. {W}.~{B}. {K}ibble.
\newblock {\em {Acta Phys. Polon.}}, {\bf{B13}}:{~723}, {1982}.

\bibitem{RuSh}
{V}.~{A}. {R}ubakov and {M}.~{E}. {S}haposnikov.
\newblock {\em {Phys. Usp.}}, {\bf{39}}:{~461}, {1996}.

\bibitem{EWKo1}
{{E}. {W}. {K}olb, {A}. {L}inde and {A}. {R}iotto}.
\newblock {\em {Phys. Rev. Lett.}}, {\bf{77}}:{~4290}, {1996}.

\bibitem{AS}
{A}.~{D}. {S}akharov.
\newblock {\em {Pisma ZhETF}}, {\bf{5}}:{~32}, {1967}.

\bibitem{ArMt}
{A}. {R}iotto and {M}. {T}rodden.
\newblock {\texttt{hep-ph/9901362}}.

\bibitem{TWBKi}
{T}. {W}.~{B}. {K}ibble.
\newblock {\em {J. Phys.}}, {\bf{A9}}:{~1389}, {1976}.

\bibitem{TK}
{{V}. {M}. {H}. {R}uutu, {V}. {B}. {E}ltsov, {A}. {J}. {G}ill, {T}. {W}. {B}.
  {K}ibble, {M}. {K}rusius, {Y}u. {G}. {M}akhlin, {B}. {P}lacais, {G}. {E}.
  {V}olovik and {W}en {X}u}.
\newblock {\em {Nature}}, {\bf{382}}:{~334}, {1996}.

\bibitem{HL}
{{P}. {C}. {H}endry, {N}. {S}. {L}awson, {R}. {A}. {M}. {L}ee, {P}. {V}. {E}.
  {M}cClintock and {C}. {D}. {H}. {W}illiams}.
\newblock {\em {Nature}}, {\bf{368}}:{~315}, {1994}.

\bibitem{NT}
{{I}. {C}huang, {R}. {D}urrer, {N}. {T}urok and {B}. {Y}urke}.
\newblock {\em {Science}}, {\bf{251}}:{~1336}, {1991}.

\bibitem{WZ}
{W}.~{H}. {Z}urek.
\newblock {\em {Acta Phys. Polon.}}, {\bf{B24}}:{1301}, {1993}.

\bibitem{UH}
{U}. {H}einz.
\newblock {\texttt{hep-ph/9902424}}.

\bibitem{JDo}
{J}.~{P}. {D}ougherty.
\newblock {\em {Phil. Trans. Roy. Soc. London}}, {\bf{346}}:{~259}, {1993}.

\bibitem{WTGr}
{W}.~{T}. {G}randy.
\newblock {\em {Principle of maximal entropy and irreversible processes, 2
  vols.}}
\newblock {D.Reidel, Amsterdam}, {1980}.

\bibitem{Bal}
{R}. {B}alescu.
\newblock {\em {Equilibrium and Nonequilibrium Statistical Mechanics}}.
\newblock {John Wiley \& Sons}, {1975}.

\bibitem{Jayn}
{E}.{T}. {J}aynes.
\newblock {\em {Am. J. Phys.}}, {\bf{33}}:{~391}, {1965}.

\bibitem{Jayn2}
{E}.{T}. {J}aynes.
\newblock {\em {Phys. Rev.}}, {\bf{106}, \bf{108}}:{~620,~171}, {1957}.

\bibitem{Jayn3}
{E}.~{T}. {J}aynes.
\newblock {\em {{P}apers on probability, statistics and statistical
  mechanics}}.
\newblock {D.Reidel, Amsterdam}, {1983}.

\bibitem{BM1}
{B}. {B}uck and {V}.{A}. {M}acaulay.
\newblock {\em {Maximum Entropy in Action}}.
\newblock {Oxford Press, Oxford}, {1991}.

\bibitem{RBa}
{R}. {B}alian.
\newblock {\em {From Microphysics to Macrophysics; Methods and Applications of
  Statistical Physics, Vol.II}}.
\newblock {Springer--Verlag, London}, {1992}.

\bibitem{IPr}
{I}. {P}rigogine.
\newblock {\em {Introduction to Thermodynamics of Irreversible Processes}}.
\newblock {Interscience, New York}, {1969}.

\bibitem{RB}
{R}. {B}alescu.
\newblock {\em {Statistical Dynamics, Matter out of Equilibrium}}.
\newblock {Imperial College Press, London}, {1997}.

\bibitem{Tol}
{R}.{C}. {T}olman.
\newblock {\em {The principles of the Statistical Mechanics}}.
\newblock {Clarendon Press, Oxford}, {1938}.

\bibitem{PEh}
{P}. {E}hrenfest and {T}. {E}hrenfest.
\newblock {\em {The conceptual foundations of the statistical approach to
  mechanics}}.
\newblock {Cornell University Press}, {1959}.

\bibitem{NNBo}
{N}.~{N}. {B}ogoliubov.
\newblock {\em {Problems of a dynamical theory in statistical physics}}.
\newblock {Nort-Holland, London}, {1962}.

\bibitem{LB}
{M}. {L}e{B}ellac.
\newblock {\em {{T}hermal {F}ield {T}heory}}.
\newblock {Cambridge University Press, Cambridge}, {1996}.

\bibitem{AD}
{A}. {D}as.
\newblock {\em {{F}inite {T}emperature {F}ield {T}heory}}.
\newblock {World Scientific, London}, {1997}.

\bibitem{LW}
{N}.~{P}. {L}andsman and {C}h. {G}.~van {W}eert.
\newblock {\em {Physics Report}}, {\bf{145}}:{~141}, {1987}.

\bibitem{Cub}
{R}. {K}ubo.
\newblock {\em {Statistical Mechanics}}.
\newblock {North--Holland Publishing Company, Oxford}, {1965}.

\bibitem{TA}
{T}. {A}ltherr.
\newblock {\em {Int. J. Mod. Phys.}}, {\bf{A8}}:{~5605}, {1993}.

\bibitem{Gibbs}
{J}.{W}. {G}ibbs.
\newblock {\em {Elementary Principles in Statistical Mechanics}}.
\newblock {Dover Publications, New York}, {1960}.

\bibitem{KB}
{L}.{P}. {K}adanoff and {G}. {B}aym.
\newblock {\em {Quantum Statistical Mechanics}}.
\newblock {Benjamin, reading}, {1962}.

\bibitem{FW}
{A}.{L}. {F}etter and {J}.{D}. {W}alecka.
\newblock {\em {Quantum Theory of Many Particle Systems}}.
\newblock {McGraw-Hill, New York}, {1971}.

\bibitem{PR}
{P}. {R}amond.
\newblock {\em {Field Theory, A Modern Primer}}.
\newblock {The Benjamin/Cummings Publishing Company, INC, London}, {1981}.

\bibitem{Zub}
{D}.{N}. {Z}ubarev.
\newblock {\em {Nonequilibrium Statistical Thermodynamics}}.
\newblock {Consultants Bureau, London}, {1974}.

\bibitem{KCC}
{{K}.{C}. {C}hou, {Z}.{B}. {S}u, {B}.{L}. {H}ao and {L}.{Y}u}.
\newblock {\em {Phys. Rep.}}, {\bf{118}}:{~1}, {1985}.

\bibitem{PVLJT}
{P}.~{V}. {L}andshoff.
\newblock {\em {Nucl. Phys.}}, {\bf{B430}}:{~683}, {1994}.

\bibitem{PL2}
{P}.~{V}. {L}andshoff.
\newblock {\em {Phys. Lett.}}, {\bf{B386}}:{~291}, {1996}.

\bibitem{K1}
{L}.~{V}. {K}eldysh.
\newblock {\em {Sov. Phys. -- JETP}}, {\bf{20}}:{~1018}, {1964}.

\bibitem{PN}
{P}. van {N}ieuwenhuizen.
\newblock {Canonical methods in quantized gauge field theories}.
\newblock {Teyler's lectures, Leyden Univerity}, {1992}.

\bibitem{V}
{M}. {V}eltman.
\newblock {\em {Diagrammatica -- The Path to Feynman Diagrams}}.
\newblock {Cambridge University Press, Cambridge}, {1994}.

\bibitem{TV}
{G}. `t~{H}ooft and {M}. {V}eltman.
\newblock {Diagrammar}.
\newblock {\em {CERN Jellow Report}}, {\bf{73-9}}, {1973}.

\bibitem{Mills}
{R}. {M}ills.
\newblock {\em {Propagators for Many--Particle Systems}}.
\newblock {Gordon and Breach, New York}, {1969}.

\bibitem{Evans}
{T}.~{S}. {E}vans and {D}.~{A}. {S}teer.
\newblock {\em {Nucl. Phys.}}, {\bf{B476}}:{~481}, {1996}.

\bibitem{IZ}
{C}. {I}tzikson and {J}.~{B}. {Z}uber.
\newblock {\em {Quantum Field Theory}}.
\newblock {McGraw-Hill, New York}, {1980}.

\bibitem{KS}
{R}.~{L}. {K}obes and {G}.~{W}. {S}emenoff.
\newblock {\em {Nucl. Phys.}}, {\bf{B272}}:{~329}, {1986}.

\bibitem{LR}
{P}.~{V}. {L}andshoff and {A}. {R}ebhan.
\newblock {\em {Nucl. Phys.}}, {\bf{B410}}:{~23}, {1993}.

\bibitem{PVLJ}
{M}. {J}acob and {P}.~{V}. {L}andshoff.
\newblock {\em {Phys Lett.}}, {\bf{B281}}:{~114}, {1992}.

\bibitem{NS}
{A}.~{J}. {N}iemi and {G}.~{W}. {S}emenoff.
\newblock {\em {Nucl. Phys.}}, {\bf{B230}}:{~181}, {1984}.

\bibitem{EP}
{T}.~{S}. {E}vans and {A}.~{C}. {P}earson.
\newblock {\em {Phys. Rev.}}, {\bf{D52}}:{~4652}, {1995}.

\bibitem{GM}
{G}. {M}orandi.
\newblock {\em {Statistical Mechanics, An Intermediate Course}}.
\newblock {World Scientific Publishing Co. Pte. Ltd., London}, {1995}.

\bibitem{LL}
{M}.{C}.{J}. {L}eermakers and {C}h.{G}.~van {W}eert.
\newblock {\em {Nucl. Phys.}}, {\bf{B248}}:{~671}, {1984}.

\bibitem{ID1}
{{I}.{T}. {D}rummond, {R}.{R}. {H}organ, {P}.{V}. {L}andshoff and {A}.
  {R}ebhan}.
\newblock {\em {Nucl. Phys.}}, {\bf{B524}}:{~579}, {1998}.

\bibitem{ID}
{{I}.{T}. {D}rummond, {R}.{R}. {H}organ, {P}.{V}. {L}andshoff and {A}.
  {R}ebhan}.
\newblock {\em {Phys. Lett.}}, {\bf{B398}}:{~326}, {1997}.

\bibitem{DG}
{W}.{A}. van~{L}eeuwen {S}.{R}.~de {G}root and {C}h.{G}.~van {W}eert.
\newblock {\em {Relativistic Kinetic Theory. Principles and Applications}}.
\newblock {North-Holland Publishing Company, Oxford}, {1980}.

\bibitem{Bach}
{G}.{K}. {B}atchelor.
\newblock {\em {An Introduction to Fluid Dynamics}}.
\newblock {Cambridge University Press, London}, {1967}.

\bibitem{Collins}
{J}. {C}ollins.
\newblock {\em {Renormalization; An introduction to renormalization, the
  renormalization group, and the operator--product expansion}}.
\newblock {Cambridge University Press, Cambridge}, {1984}.

\bibitem{Brown}
{L}.{S}. {B}rown.
\newblock {\em {Ann. Phys.}}, {\bf{126}}:{~135}, {1979}.

\bibitem{N}
{Y}. {N}ambu.
\newblock {\em {Prog. Theor. Phys. (Kyoto)}}, {\bf{7}}:{~131}, {1952}.

\bibitem{RJ}
{R}. {J}ackiw.
\newblock In {S.B.~Treiman, R.~Jackiw, B.~Zumino and E.~Witten}, editor, {\em
  {Current Algebra and Anomalies}}. {World Scientific Publishing Co. Pte.Ltd.,
  Singapore}, {1985}.

\bibitem{CCR}
{S}.~{C}oleman {C}.{G}.~{C}allan and {R}. {J}ackiw.
\newblock {\em {Ann. Phys.}}, {\bf{59}}:{~42}, {1970}.

\bibitem{JS}
{G}. {S}terman.
\newblock {\em {An Introduction to Quantum Field Theory}}.
\newblock {Cambridge University Press, Cambridge}, {1993}.

\bibitem{Zimm}
{W}. {Z}immermann.
\newblock In {S.Deser et al.}, editor, {\em {Lectures on Elementary Particles
  and Quantum Field Theory}}. {M.I.T. Press, Cambridge (Mass.)}, {1970}.

\bibitem{PC}
{P}. {C}vitanovic.
\newblock {\em {Field Theory}}.
\newblock {Nordita, Copenhagen}, {1983}.

\bibitem{EM}
{M}. van {E}ijck.
\newblock {\em {Thermal Field Theory and the Finite-Temperature Renormalization
  Group}}.
\newblock PhD thesis, {University of Amsterdam}, {1995}.

\bibitem{PJ}
{P}. {J}izba.
\newblock {\em {Phys. Rev.}}, {\bf{D57}}:{~3634}, {1998}.

\bibitem{HS}
{H}.{J}. {S}chnitzer.
\newblock {\em {Phys. Rev.}}, {\bf{D10}}:{~1800}, {1974}.

\bibitem{CJT}
{{J}.{M}. {C}ornwall, {R}. {J}ackiw and {E}. {T}omboulis}.
\newblock {\em {Phys. Rev.}}, {\bf{D10}}:{~2428}, {1974}.

\bibitem{BM}
{W}.{A}. {B}ardeen and {M}. {M}oshe.
\newblock {\em {Phys.Rev.}}, {\bf{D28}}:{~1372}, {1983}.

\bibitem{Abb}
{{L}.{F}. {A}bbot, {J}.{S}. {K}ang and {H}.{J}. {S}chnitzer}.
\newblock {\em {Phys. Rev.}}, {\bf{D13}}:{~2212}, {1976}.

\bibitem{AC2}
{G}. {A}melino {C}amelia.
\newblock {\texttt{hep-th/9811236}}.

\bibitem{B}
{F}. {O}berhettinger.
\newblock {\em {Tables of Mellin Transforms}}.
\newblock {Springer, Berlin}, {1974}.

\bibitem{Br}
{H}.{W}. {B}raden.
\newblock {\em {Phys. Rev.}}, {\bf{D25}}:{~1028}, {1982}.

\bibitem{WH}
{H}.{E}. {H}aber and {H}.{A}. {W}eldon.
\newblock {\em {J. Math. Phys.}}, {\bf{23}}:{~1852}, {1982}.

\bibitem{BB}
{A}. {E}rd{\'e}lyi.
\newblock {\em {Tables of Integral Transformations I}}.
\newblock {McGraw--Hill, London}, {1954}.

\bibitem{GR}
{I}.{S}. {G}radshteyn and {I}.{M}. {R}yzhik.
\newblock {\em {Tables of Integrals, Series, and Products}}.
\newblock {Academic Press, INC., London}, {1980}.

\bibitem{ACP}
{G}. {A}melino {C}amelia and {S}.-{A}. {P}i.
\newblock {\em {Phys. Rev.}}, {\bf{D47}}:{~2356}, {1993}.

\bibitem{TDL}
{T}.{D}.{L}ee.
\newblock {pp 1--13}.
\newblock In {W.C. Haxton and E.M. Hanley}, editor, {\em {Symmetries and
  fundamental interactions in nuclei}}. {Columbia University Press}, {1997}.

\bibitem{SNW}
{S}.{N}. {W}hite.
\newblock {\em {Nucl. Instrum. Methods}}, {\bf{A409}}:{~618}, {1998}.

\bibitem{HH}
{H}. {H}eiselberg and {A}.{D}. {J}ackson.
\newblock {to be published in the proceedings of 3rd Workshop on Continuous
  Advances in QCD (QCD98), Minneapolis, MN, 16--19 Apr. 1998.}

\bibitem{Hag}
{R}. {H}agedorn.
\newblock {\em {Z. Phys.}}, {\bf{C17}}:{~265}, {1983}.

\bibitem{TH1}
{{T}. {H}aruyama, {N}. {K}imura and {T}. {N}akamoto}.
\newblock {to be published in the proceedings of 17th International Conference
  on Cryogenic Engineering (ICEC 17), Bornemounth, England, 14--17 Jul. 1998.}

\bibitem{TH2}
{{T}. {H}aruyama, {N}. {K}imura and {T}. {N}akamoto}.
\newblock {to be published in the proceedings of theCryogenic Engineering
  Conference and International Cryogenic Materials Conference (CEC / ICMC 97),
  Portland, OR, 27 Jul -- 1 Aug. 1997}.

\bibitem{SH}
{{S}. {H}ull, {D}.{A}. {K}een, {R}. {D}one and {C}.{N}. {U}den}.
\newblock {\texttt{RAL-91-089}}.
\newblock {to be published}.

\bibitem{JSO}
{{J}. {S}taun {O}lsen, {L}. {G}erward and {U}. {B}enedict}.
\newblock {\texttt{DESY SR-84-22}}.
\newblock {to be published}.

\bibitem{PJ1}
{P}. {J}izba.
\newblock {\texttt{hep-th/9801197}}.
\newblock {to be published in Phys. Rev. \bf{D}}.

\bibitem{CH}
{E}. {C}alzetta and {B}.{L}. {H}u.
\newblock {\em {Phys. Rev.}}, {\bf{D37}}:{~2878}, {1988}.

\bibitem{J-T}
{P}. {J}izba and {E}.{S}. {T}ututi.
\newblock {work in progress}.

\bibitem{CHKMP}
{{F}. {C}ooper, {S}. {H}abib, {Y}. {K}luger, {E}. {M}ottola and {J}.{P}.
  {P}az}.
\newblock {\texttt{hep-ph/9405352}}.
\newblock {to be published}.

\bibitem{M1}
{F}. {C}ooper and {E}. {M}ottola.
\newblock {\em {Phys. Rev.}}, {\bf{D36}}:{~3114}, {1987}.

\bibitem{M2}
{{F}. {C}ooper, {S}. {H}abib, {Y}. {K}luger and {E}. {M}ottola}.
\newblock {\em {Phys. Rev.}}, {\bf{D55}}:{~6471}, {1997}.

\bibitem{EJY}
{{O}. {{\'E}}boli, {R}. {J}ackiw and {S}o--{Y}oung {P}i}.
\newblock {\em {Phys. Rev.}}, {\bf{D37}}:{~3557}, {1988}.

\bibitem{DB}
{D}.{C}. {B}rody and {L}.{P}. {H}ughston.
\newblock {to be published}.

\bibitem{CWG}
{C}.{W}. {G}ardier.
\newblock {\em {Handbook of Stochastic Methods for Physics, Chemistry and the
  Natural Sciences}}.
\newblock {Springer--Verlag, New York}, {1985}.

\bibitem{FJ}
{F}. {F}loreanini and {R}. {J}ackiw.
\newblock {\em {Phys. Rev.}}, {\bf{D37}}:{~2206}, {1988}.

\bibitem{SH1}
{C}.{E}. {S}hannon and {W}. {W}eaver.
\newblock {\em {The Mathematical Theory of Communication}}.
\newblock {University of Illinois Press, Urbana}, {1949}.

\bibitem{LBr}
{L}. {B}rillouin.
\newblock {\em {J. Applied Phys.}}, {\bf{22}}:{~334}, {1951}.

\bibitem{LSz}
{L}. {S}zilard.
\newblock {\em {Z. Physik}}, {\bf{53}}:{~840}, {1929}.

\bibitem{GMW}
{G}. {M}arc and {W}.{G}. {M}cMillan.
\newblock In {I.~Prigogine and S.~Rice}, editor, {\em {Advances in Chemical
  Physics}}, volume {Vol. LVIII}. {Wiley, New York}, {1985}.

\bibitem{VPe}
{V}. {P}etviashvili and {O}. {P}okhotelov.
\newblock {\em {Solitary Waves in Plasmas and in the Atmosphere}}.
\newblock {Gordon and Breach Science Publishers, Philadelphia}, {1992}.

\bibitem{SIc}
{S}. {I}chimaru.
\newblock {\em {Basic Principles of Plasma Physics, A Statistical Approach}}.
\newblock {W.A.Benjamin, INC. London}, {1973}.

\bibitem{mihai}
{M}. {G}avrila.
\newblock {\em {Atoms in Intense Laser Fields}}.
\newblock {Academic Press, San Diego Ca.}, {1992}.

\bibitem{bak}
{G}.{A}. {B}aker.
\newblock {\em {Essentials of Pad\'e Approximation}}.
\newblock {Academic Press, London}, {1975}.

\bibitem{bairetal}
{{R}. {B}aier, {M}. {D}irks, {K}. {R}edlich, and {D}. {S}chiff}.
\newblock {\em {Phys. Rev.}}, {\bf{D56}}:{~2548}, {1997}.

\bibitem{strickland}
{M}. {S}trickland.
\newblock {\em {Phys. Lett.}}, {\bf{B331}}:{~245}, {1994}.

\bibitem{Nem}
{{J}. {B}aacke, {K}. {H}eitmann and {C}.{P}. {P}\"{a}tzold}.
\newblock {\texttt{hep-th/9711144}}.
\newblock {to be published}.

\bibitem{LB2}
{{M}. {L}eBellac}.
\newblock {\em {Quantum and Statistical Field Theory}}.
\newblock {Clarendon Press, Oxford}, {1991}.

\bibitem{BF}
{{F}. {J}. {B}elinfante}.
\newblock {\em {Physica}}, {\bf{7}}:{~449}, {1940}.

\bibitem{R}
{{L}. {R}osenfeld}.
\newblock {\em {Mem. Roy. Acad. Belg. Cl. Sci.}}, {\bf{6}}:{~18}, {1940}.

\bibitem{Lat}
{{P}. {E}. {H}aagenson and {J}. {I}. {L}atorre}.
\newblock {\texttt{hep-ph/9203207}}.
\newblock {to be published}.

\bibitem{NS2}
{A}.~{J}. {N}iemi and {G}.~{W}. {S}emenoff.
\newblock {\em {Ann. Phys.}}, {\bf{152}}:{~181}, {1984}.

\bibitem{KL}
{J}.~{I}. {K}apusta and {P}.~{V}. {Landshoff}.
\newblock {\em Nucl. Part. Phys.}, {\bf{15}}:~267, 1989.

\bibitem{JZ}
{{J}. {Z}inn--{J}ustin}.
\newblock {\em {Quantum Field Theory and Critical Phenomena}}.
\newblock {Oxford Science Publisher, Oxford}, {1996}.

\bibitem{JSc1}
{J}. {S}chwinger.
\newblock {\em {Proc. Nat. Acad. Sc.}}, {\bf{37}}:{~452}, {1951}.

\bibitem{FJDa}
{F}.~{J}. {D}yson.
\newblock {\em {Phys. Rev.}}, {\bf{75}}:{~1736}, {1949}.

\bibitem{JSc}
{J}. {S}chwinger.
\newblock {\em {J. Math. Phys}}, {\bf{2}}:{~407}, {1961}.

\bibitem{MRa}
{{M}. {R}asetti}.
\newblock {\em {Modern Methods in Equilibrium Statistical Machanics}}.
\newblock {World Scientific Poblishing Co. Pte. Ltd., Singapore}, {1986}.

\end{thebibliography}

\end{document}